\documentclass[draft,11pt,sflabels]{thesis}

\usepackage[final]{graphicx}    
\usepackage{subfigure}   
\usepackage{threeparttable}
\usepackage{amsmath,amssymb,amstext}  
\usepackage{array}

\begin{document}        


\def\beq#1 #2\eeq{\begin{equation}\label{#1}#2\end{equation}}
\def\beqa#1 #2\eeqa{\begin{eqnarray}\label{#1}#2\end{eqnarray}}
\def\be#1\ee{\begin{equation}#1\end{equation}}

\newcommand\aj{AJ}%
\newcommand\apj{ApJ}%
\newcommand\apjl{ApJ Lett.}
\newcommand\aap{Astronomy \& Astrophysics}
\newcommand\nat{Nature}
\newcommand\mnras{Monthly Notices of the RAS}
\newcommand\physrep{Physics Reports}
\newcommand\ao{Appl.~Opt.}
\newcommand\prd{Phys.~Rev.~D}
\newcommand\prl{Phys.~Rev.~Lett.}%
\newcommand\pasp{Publications of the Astronomical Society of the
                Pacific}%
\newcommand\pasj{PASJ}%
\newcommand\ieetm{IEEE.~Trans.~Microwave~Theory~Tech.}
\newcommand\ieetg{IEEE.~Trans.~Geosci.~Remote~Sensing}

\newcommand{\app}{\ensuremath{\sim}}
\newcommand{\sct}[1]{\S\ref{#1}}
\newcommand{\chap}[1]{Chapter \ref{#1}}
\newcommand{\abs}[1]{\ensuremath{|#1|}}
\newcommand{\ka}{\ensuremath{K_a}}
\newcommand{\Ka}{\ensuremath{K_a}}
\newcommand{\polar}{\textsc{polar}}
\newcommand{\polars}{\textsc{polar's}}
\newcommand{\compass}{\textsc{compass}}
\newcommand{\eqn}[1]{Equation (\ref{#1})}
\newcommand{\eqns}[1]{Equations (\ref{#1})}
\newcommand{\eq}[1]{Equation (\ref{#1})}
\newcommand{\tbl}[1]{Table \ref{#1}}
\newcommand{\fig}[1]{Figure \ref{#1}}
\newcommand{\ie}{{\frenchspacing\it i.e.}}
\newcommand{\eg}{{\frenchspacing\it e.g. }}
\newcommand{\etal}{{\frenchspacing\it et.al.}}
\newcommand{\etc}{{\frenchspacing\it etc.}}
\newcommand{\cphi}{\cos{\phi}}
\newcommand{\sphi}{\sin{\phi}}
\newcommand{\dphi}{\ensuremath{\Delta\phi}}
\newcommand{\Q}{{\bf Q}}
\newcommand{\U}{{\bf U}}
\newcommand{\I}{{\bf I}}
\newcommand{\V}{{\bf V}}
\newcommand{\PP}{{\bf P}}
\newcommand{\M}{{\bf M}}
\newcommand{\C}{{\bf C}}
\newcommand{\G}{{\bf G}}
\newcommand{\D}{{\bf D}}
\newcommand{\bra}[1]{\ensuremath{\langle #1 |}}
\newcommand{\ket}[1]{\ensuremath{| #1 \rangle}}
\newcommand{\eval}[1]{\ensuremath{\langle #1 \rangle}}
\newcommand{\voltage}{\ensuremath{\texttt{\textit{v}}}}
\newcommand{\matrixb}[1]{\begin{bmatrix}#1\end{bmatrix}}
\newcommand{\mmb}[4]
{\left[\begin{array}{cc}
#1 & #2 \\
#3 & #4 \end{array}\right]}

\newcommand{\mmbt}[4]
{\ensuremath{ \left[ \begin{array}{cc}
{#1} & {#2} \\
{#3} & {#4} \end{array}\right]}}

\newcommand{\sqinv}{\ensuremath{\frac{1}{\sqrt{2}}}}
\renewcommand{\deg}{\ensuremath{^\circ}}
\renewcommand{\Im}{\mmb{1}{0}{0}{1}}
\newcommand{\Qm}{\mmb{1}{0}{0}{-1}}
\newcommand{\Um}{\mmb{0}{1}{1}{0}}
\newcommand{\Vm}{\mmb{0}{-i}{i}{0}}
\newcommand{\seq}{\; = \;}
\newcommand{\uK}{\ensuremath{\mu K}}
\newcommand{\fixts}{\renewcommand{\baselinestretch}{0.7}\small\normalsize}

\newcommand{\onephi}{\ensuremath{1\phi}}
\newcommand{\twophi}{\ensuremath{2\phi}}
\newcommand{\onephir}{\ensuremath{1\phi_r}}
\newcommand{\twophir}{\ensuremath{2\phi_r}}
\newcommand{\Zeta}{\ensuremath{\zeta}}

\newcommand{\vect}[1]{{\bf #1}}
\newcommand{\mmatrix}[1]{{\bf #1}}

\newcommand{\y}{\vect{y}}
\newcommand{\n}{\vect{n}}
\newcommand{\x}{\vect{x}}
\newcommand{\xt}{\tilde{\x}}
\newcommand{\cv}{\vect{c}}
\newcommand{\sv}{\vect{s}}
\newcommand{\neta}{\mathbf{\eta}}
\newcommand{\eO}{\ensuremath{\vect{e_0}}}
\newcommand{\yeff}{\vect{y}_e}
\renewcommand{\a}{\vec{a}}
\newcommand{\q}{\vect{q}}
\renewcommand{\u}{\vect{u}}

\newcommand{\A}{\mmatrix{A}}
\newcommand{\N}{\mmatrix{N}}
\newcommand{\W}{\mmatrix{W}}
\newcommand{\Z}{\mmatrix{Z}}
\newcommand{\R}{\mmatrix{R}}
\newcommand{\Lc}{\mmatrix{L}}
\newcommand{\NN}{\mmatrix{\Sigma}}
\newcommand{\PI}{\mmatrix{\Pi}}
\newcommand{\Sc}{\mmatrix{S}} 
\newcommand{\Sig}{{\bf \Sigma}}
\newcommand{\Aeff}{\mmatrix{A}_e}
\newcommand{\mm}{Minimum Variance Mapmaking}

\newcommand{\expec}[1]{\ensuremath{\langle #1 \rangle}}

\newcommand{\rodp}{ROD2}

\newcommand{\ct}[1]{\cos{\phi_{#1}}}
\newcommand{\st}[1]{\sin{\phi_{#1}}}
\newcommand{\ctt}[1]{\cos{2\phi_{#1}}}
\newcommand{\stt}[1]{\sin{2\phi_{#1}}}

\newcommand{\ft}{\mathcal{F}}
\newcommand{\fft}[1]{\ft \{ #1 \}}
\newcommand{\fftinv}[1]{\ft^{-1} \{ \, #1 \, \}}
\newcommand{\Cp}{\ensuremath{{\bf \mathcal{C}}^{-1}}}
\renewcommand{\L}{\ensuremath{{\bf \mathcal{L}}}}

\renewcommand{\l}{\ell}
\newcommand{\Cl}{C_\l}
\newcommand{\lan}{\langle}
\newcommand{\ran}{\rangle}

\newcommand{\ep}{\hat{\epsilon}}
\newcommand{\epp}{\hat{\epsilon}'}

\newcommand{\Ctt}{\ensuremath{\Cl^{TT}}}
\newcommand{\Cte}{\ensuremath{\Cl^{TE}}}
\newcommand{\Cee}{\ensuremath{\Cl^{EE}}}
\newcommand{\Cbb}{\ensuremath{\Cl^{BB}}}
\newcommand{\Ceb}{\ensuremath{\Cl^{EB}}}
\newcommand{\Ctb}{\ensuremath{\Cl^{TB}}}

\newcommand{\sm}[1]{\small #1\normalsize}
\newcommand{\smbf}[1]{\small {\bf #1}\normalsize}
\newcommand{\as}{"}
\newcommand{\hs}{\hspace{2ex}}


\title{A New Upper Limit on the Large Angular Scale Polarization
of the Cosmic Microwave Background Radiation}
\author{Christopher W. O'Dell}
\year{2001}

\adviser{Peter Timbie}
\adviserrank{Professor}

\ttlpage                
\cpypage                

\frontmatter            

\pagestyle{thesis}      



\begin{center}
\Large {\bf Abstract} \normalsize
\end{center}
\addcontentsline{toc}{chapter}{Abstract}
 The Cosmic Microwave Background Radiation (CMB) is an
invaluable probe of the conditions of the early universe. Recent
measurements of its spatial anisotropy have allowed accurate
determinations of several fundamental cosmological parameters, such
as the curvature of the universe, the shape of the spectrum of
primordial density fluctuations, and the contribution of baryons,
dark matter, and dark energy to the overall energy density of the
universe.  In addition to being spatially non-uniform, the CMB is
theorized to be slightly polarized. Measurements of this
polarization, particularly at large angular scales, have the
potential to provide information on primordial gravitational waves,
theories of inflation, and the ionization history of the universe, as
well as help further constrain cosmological parameters.

Polarization has not yet been detected in the CMB. This thesis
describes a recent search for CMB polarization at large angular
scales, conducted in the spring of 2000 at the University of
Wisconsin-Madison.  After a general introduction on both CMB
polarization and general microwave polarimetry, details of the
experiment itself are given, as well as a full description of the
data selection and analysis techniques. Using these techniques, our
data lead to a new upper limit on CMB polarization at large angular
scales of 10 $\mu$K in both E- and B-type polarization at 95\%
confidence. If B-polarization is assumed to be zero, the limit for
E-type polarization is lowered to 8 \uK.  This experiment is the
first of a new breed of highly-sensitive instruments that will one
day map out this interesting property of the Cosmic Microwave
Background radiation.

\begin{center}
\Large
{\bf Acknowledgements}
\normalsize
\end{center}
\addcontentsline{toc}{chapter}{Acknowledgements}
There are so many people who have given me their time and patience
throughout my years in grad school that I would like to thank.
First and foremost, I want to thank my advisor, mentor, and friend
Peter Timbie, who has been a truly wonderful
thesis advisor.  Peter always showed patience with me while I explained the newest
problem in my thesis; he had a knack of being able to get to the root of any problem,
for which I was grateful on several occasions.  He also gave me
great freedom as a graduate student and
allowed me to make my own mistakes, and has helped me to truly grow as a
scientist.

Someone who taught me immeasurable quantities of
radiometric knowledge, Brian Keating, also deserves much credit.
Thanks to him I have finally memorized the radiometer equation.
Brian, if I could have a number of ``Meatloaf Carver Combos'' equal to the logarithm
of what you taught me, I would never go hungry.

I would also like to thank all my science ``elders'', who have helped me
with so many of the knotty problems that have come up over the years:
Dan Mccammon, Don Cox, Josh Gundersen, Bob Benjamin, Ed Wollack, and Grant Wilson.
Also, thanks to Max
Tegmark and Angelica de Oliveira-Costa, who provided excellent theory
mentorship for the data analysis segment of my thesis, and who have been
great friends.  Thanks to
Craig Marquardt, David Fanning, and the rest of the IDL newsgroup who
helped with all my IDL issues.  Special thanks to Gail Bayler and
Gary Wade at the UW Space Sciences and Engineering center who
provided all the wonderful GOES satellite data for use in this
project.

Thanks to all the technicians and machine-shop ``artists'' I have had
the privilege to work with: Walt Wigglesworth, who taught me
everything I know about machining; Lee Potratz and the entire
Instrument shop team, who constructed so many of the critical
elements that POLAR utilized; and finally all the guys in the
Stirling Hall electronics shop, who built me so many power supplies
that I could have probably separated every hydrogen atom on Earth.

A big ``shout-out'' to all the people I had the honor of knowing and working with
over the years in the Wisconsin Observational Cosmology Lab.  Slade
Klawikowski, who was a great partner in the construction and running
of POLAR, and was much better than I at keeping POLAR warm and dry;
Karen Lewis, the best cookie tzarina this side of the Mississippi;
the ``wonder twins'' Jodi and Mark Supanich, Nate Stebor, Dan ``Crowbar'' Swetz,
Zak Staniszewski, Kip Hyatt, Marye ``Molly'' Reed, and Josh ``H to
the izz-O'' Friess.  Not only were you all great to work with, you
all made the lab fun, a place I actually \emph{wanted} to be!

Finally, I want to recognize my friends, who over the years
have truly been my family and without
whom I don't know where I'd be.  It's hard to name you all, but you
know who you are.  Jack and Karyn, you always reminded me the right way
to live, that you have to pull your head out of the details and smell
the roses every once in a while.  Melanie Barker, who has been my
support and good friend so long; thanks for giving me so much over the years.  I think
I still owe you many hundreds of Meow-Meow bucks.  John
Wright, the best roommate and biking partner and Latex mentor a guy
could ever ask for.  Rob Haslinger, who was always willing
to go out for a beer, who always wanted to live it up in Madison and was
a great partner-in-crime on so many occasions.
Bruce and Oleg, Jay and Ted, Lisa J, Jen and Stefan, Christine and Steve and the
rest of the ``Jeffer-Gang''; it was great enjoying Wisconsin with you.
And of course, my entire Dayton family, the ``Platt'', who I miss dearly and who
always provided me with a fantastic home-away-from-home: Jeff,
Trevor, Randy, Christine, Curt, Alicia and Joe and Michael and Marie,
Mike Collins, Chris Corn, and all the rest of the Daytonites.  The
times I went home and visited with you were my times of spiritual
healing.
\begin{center}
\Large {\bf Dedication} \normalsize
\end{center}
\addcontentsline{toc}{chapter}{Dedication}

This thesis is dedicated
to my parents, who put so much of themselves into me that this work is
as much a reflection of their hard work as parents as it is mine as a
student.

To my father, who first introduced me to the beauty and wonder of
physics; when he gave me his well-loved copy of \emph{One, Two,
Three..Infinity}; when he climbed our tv-antenna fifty feet in the
air, in what were surely hurricane-force winds, to help me with my
first science project; and when he suffered all my frustrations with
patience and love.  Dad, you have truly made me what I am today.

To my mother, who has always put her children before herself, I can't
imagine what life would have been like without you.  From sitting in
the kitchen helping me with vocabulary, to all your little lessons on
life, I owe a large debt of love to you.

This thesis is also dedicated to Jeff Hessell, who taught me and gave
me so much during my life, who showed me the beauty and enormity of the
night sky. From using your telescope to see the moons of Jupiter and
the rings of Saturn, to wondering how the universe began, those times
we spent on the roof gazing upward at the heavens have never
really stopped.

\tableofcontents        
\listoftables           
\listoffigures          

\pagebreak              


\mainmatter

\renewcommand\arraystretch{1.0} 









\chapter{Introduction}\label{intro}
\section{The Pillars of Cosmology}
\begin{figure}[tb]
\begin{center}
\includegraphics[height=4in,width=4in]{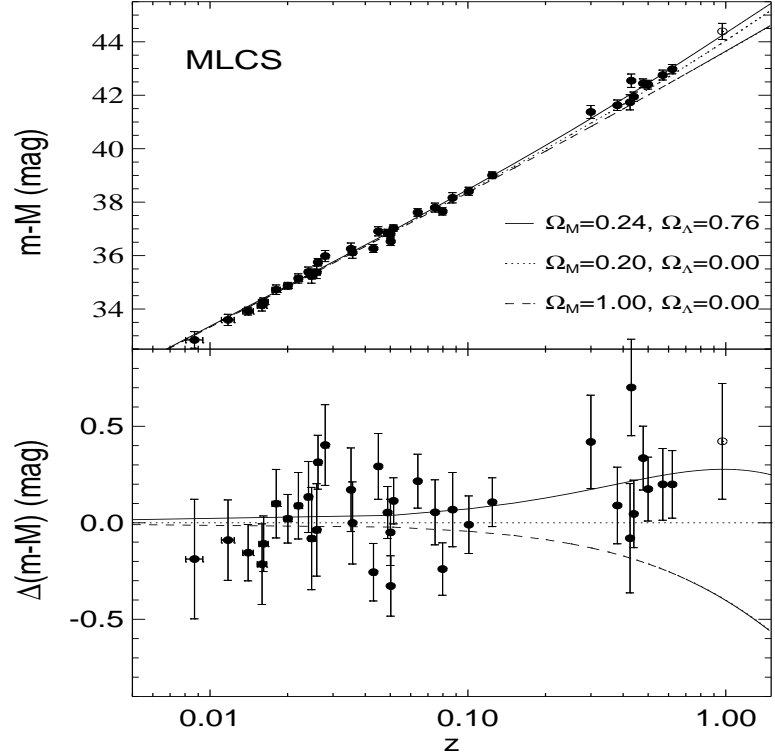}
\caption[Modern Measurements of Hubble's Law] {\label{HubbleLaw}
\fixspacing Modern Measurements of Hubble's Law.  Plotted is the
apparent magnitude minus absolute magnitude (which is proportional to
the logarithm of distance) versus redshift, along with three
cosmological models. In addition to illustrating Hubble's Law, this
work is accurate enough to determine the rate of change of Hubble's
constant, and is consistent with an accelerating universe. Adopted
from \cite{snboth}. }
\end{center}
\end{figure}

The understanding of the origin of the universe has greatly
evolved over the course of human
history.  We initially believed a multitude of religion-based
creation theories.  During the
last century, however, these beliefs were largely superseded by
scientifically-based theories, from
a steady-state picture of a static universe, as was popular
in the early part of the 20th
century, to the current Hot Big Bang model of modern cosmology.
The Big Bang Model rests upon three sturdy ``pillars'' of
observational evidence.  In 1929, Edwin Hubble discovered that
virtually every galaxy he observed was moving away from us, and
the galaxies' recessional velocities were roughly proportional to
their distances from us.  This led to the famous ``Hubble Law'' of
the expansion of the universe, measurements of which have greatly
improved over the last seventy years (see \fig{HubbleLaw}). Today,
the Hubble constant is known with unprecedented accuracy, and
similar observations have led to a measurement of the deceleration
parameter and the conclusion that the expansion appears to be
accelerating \cite{sn1,sn2,snboth}.

\begin{figure}[h!tb]
\begin{center}
\includegraphics[height=5in]{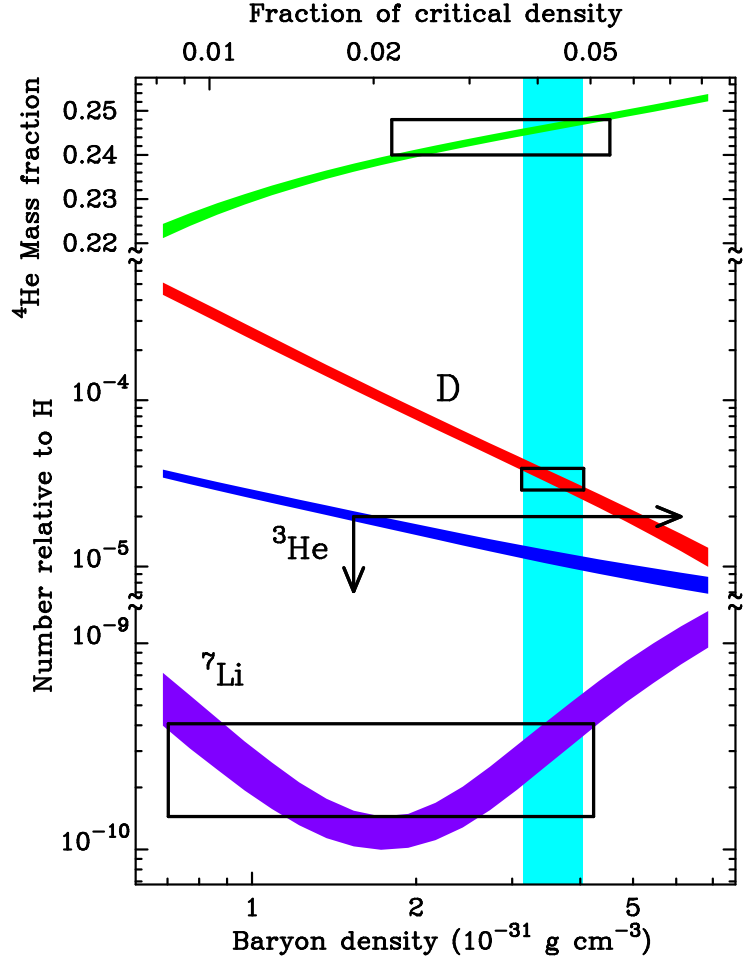}
\caption[Predictions of BBN and Measured Elemental Abundances]
{\label{bbn} \fixspacing
Predictions of BBN for the four most
important light elements versus the baryon density in the universe
(this is an equivalent alternative to the baryon-to-photon ratio, $\eta$,
sometimes used, as well as $\Omega_b h^2$).  The boxes show the
95\% confidence limits of observations of primordial abundances.
There is really only one acceptable value for the baryon density,
and that is $\Omega_b h^2 = 0.020 \pm 0.002$ (95\% confidence) \cite{turner01}.
Figure adopted from \cite{tytler00}.
}
\end{center}
\end{figure}

The discovery of galaxy redshifts
in and of itself led to the idea of a universe in
which space itself is expanding, and necessarily requires
that the universe was much smaller, hotter and denser in the past.
Running from the beginning of the universe forward, there would
therefore be a time when the universe cooled enough to create
nuclei from free protons and neutrons, themselves in a virtual
``soup'' of particles with electrons, photons, and other less
abundant species.  This era of ``Big Bang Nucleosynthesis'' (BBN) is
entirely calculable, and can predict elemental abundances of the
light nuclei.  These predictions closely match current
observations, and have led to a measurement of the primordial
entropy of the universe, expressed in the baryon-to-photon ratio
(see \fig{bbn}).  To date, the match between observations and
theory is remarkable \cite{tytler00,turner01}.

The final observational pillar of the Big Bang is the Cosmic
Microwave Background radiation (CMB), discovered in 1965 by Arno
Penzias and Robert Wilson of Bell Telephone Laboratories \cite{pw65}.
If the Big Bang actually happened, this background of microwaves
necessarily exists.  Its existence was first postulated by George
Gamow and his students in 1948 \cite{gamow48}, and the theory describing
the CMB was later refined
by others, notably Robert Dicke and James Peebles of Princeton
\cite{dprw65}.

\section{Origin and Characteristics of the CMB}
During the first 100,000 years, the universe was
completely ionized, its primary matter components being free protons
and electrons.  When the universe had cooled to a temperature of
about 3000 Kelvin, it was cold enough for electrons and
protons to stick together as hydrogen atoms, without an energetic
photon immediately reionizing them.  This was the beginning of the
era of ``recombination'' \footnote{\fixspacing \emph{Recombination} is a
well-known misnomer. The protons and electrons had never been
``combined'' before, so far as we know.} during which the universe
rapidly went from ionized to neutral, and led to the release of CMB
photons. This can be thought of as happening on a ``surface in
redshift space'', which is commonly called  the \emph{last scattering
surface} (LSS).

\subsection{Spectrum}
\begin{figure}[tb]
\begin{center}
\includegraphics[height=4in]{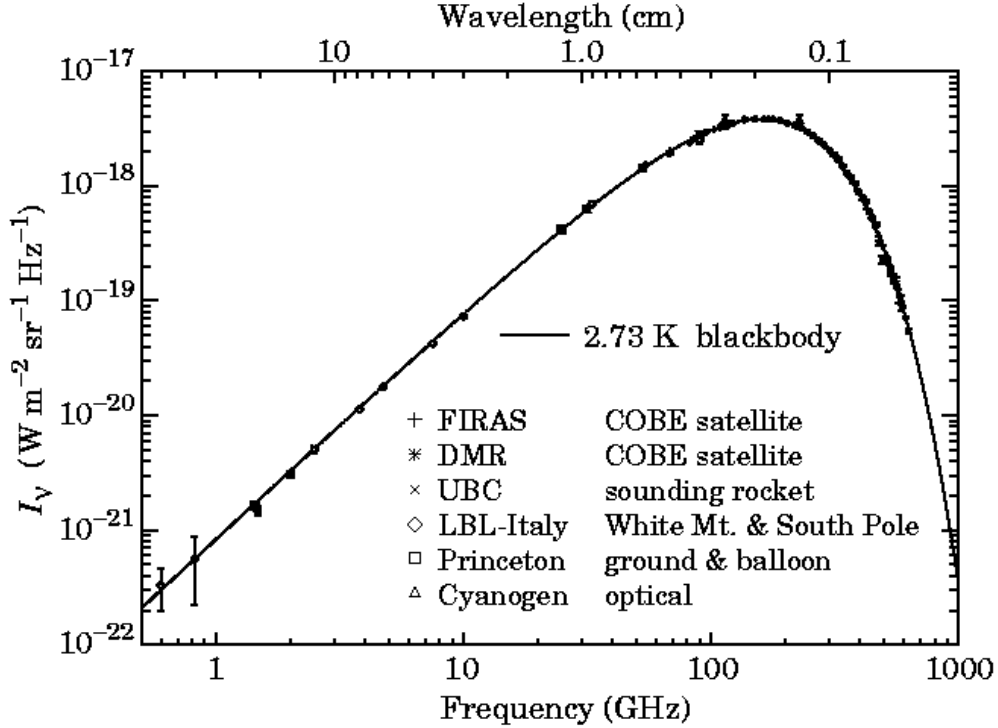}
\caption[Measurements of the CMB Spectrum] {\label{firas}
\fixspacing Measurements of the CMB spectrum.  The most
constraining measurements come from the FIRAS instrument aboard
COBE \cite{mather94}, although the spectrum has since been
measured at a variety of frequencies well away from the blackbody
peak. Adopted from \cite{smoot96}. }
\end{center}
\end{figure}

Because the universe had undergone a period of thermal equilibrium in
its history, the CMB was initially created with a blackbody spectrum.
It can be shown that a blackbody radiation field in an expanding
universe retains its blackbody spectrum, but its characteristic
temperature decreases in proportion to the scale factor of the
universe \cite{peebles93}.  The spectrum of the CMB was measured by
the FIRAS (Far-Infrared Absolute Spectrophotometer) instrument aboard
the COBE (Cosmic Background Explorer) COBE satellite in 1989, during
its very first days in space.  As shown in \fig{firas}, it was
consistent with a perfect blackbody; in fact, the FIRAS measurement
is the most perfect measurement of a blackbody ever performed.  It
corresponds to a temperature of 2.7253 $K$ $\pm$ 0.66 $mK$
\cite{mather99}.

Deviations from a blackbody are expected at low levels due to various
forms of ``energy injections'' into the universe at times during and
since recombination.  Energy released via starlight would heat up
whatever free electrons there were, and these (now hotter) electrons
would transfer some energy to the CMB via Compton scattering.  This
would cause the CMB spectrum to be an admixture of different
blackbody curves, causing a distortion in the spectrum due to a
decrement of photons at lower energies and and increment of photons
at higher energies.  This distortion is characterized by the Compton
$y$-parameter, defined by $y = \int k_B (T_e - T_{cmb})/m_ec^2
d\tau$, where $T_e$ is the kinetic temperature of the electrons and
$\tau$ is the optical depth of free electrons to last scattering. The
best upper limits on $y$ are from COBE, and suggest $|y| < 15 \times
10^{-6}$ \cite{fixsen96}.  A proposed satellite mission may further
illuminate the history of spectrum-distorting mechanisms in the
universe \cite{dimes}.

\subsection{Spatial Anisotropy}
The CMB is not uniform in its intensity on the sky, but rather varies
from place to place.  The primary \emph{anisotropy} in the CMB is due
to the Earth's relative motion with respect to the rest frame of the
CMB; this leads to the so-called ``dipole anisotropy'', which is a
temperature difference of 3.353 $\pm$ 0.024 mK in the CMB sky
(corresponding to an Earth velocity of about 370 km/sec with respect
to this ``cosmic rest frame'') \cite{fixsen96}.

Initial seeds of structure, present in the early universe in the form
of perturbations to the Robertson-Walker metric, imprinted themselves
upon the CMB during last scattering.  They are imprinted on the CMB
primarily via the Sachs-Wolfe mechanism \cite{sw67}, in which CMB
photons originating from an overdense region are forced to climb out
of the local gravitational well; therefore, areas of the sky with
less CMB intensity than average correspond to overdense regions, and
likewise increased CMB intensity corresponds to underdense regions.
There are other effects that couple density perturbations to the CMB;
for an excellent introduction to these processes, see \cite{hu96}.

\begin{figure}
\begin{center}
\subfigure[]{\label{powerspec-a}
\includegraphics[height=3.2in,width=4in]{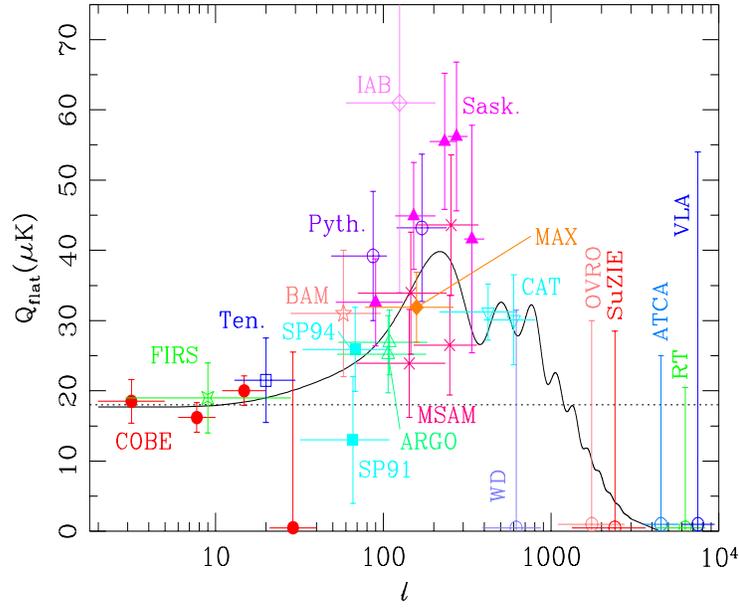}
} \\
\subfigure[]{\label{powerspec-b}
\includegraphics[height=3.2in,width=4in]{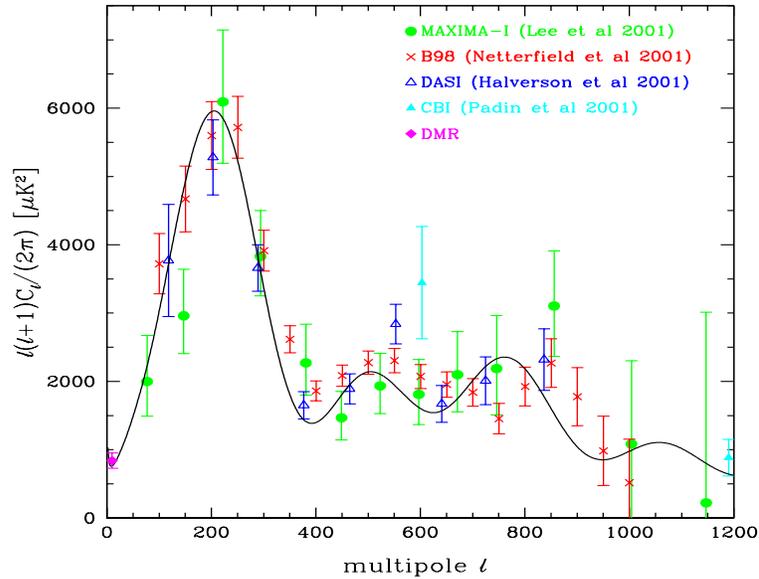}
}
\caption[Anisotropy Power Spectrum Measurements: Old and New]
{\label{powerspec}\fixspacing
Anisotropy Power Spectrum Measurements: Old and New.  Panel (a) shows
the state of power spectrum measurements circa 1997
(adopted from \cite{smoot96}).  The quantity shown, $Q_{rms}$ is the angular
power spectrum normalized to the COBE measurement of the
quadrupole moment.  Clearly, the spectrum is not well-constrained,
and at one time the error bars on most cosmological parameters
were huge or non-existent.  Panel (b) shows the situation as of
2001; with measurements by DASI, Maxima-I, and BOOMERANG, the
spectrum is reasonably well-sampled out to the second doppler peak
at $\l \approx 600$ (adopted from\cite{church01}).
}
\end{center}
\end{figure}

If the fluctuations in the microwave background intensity are a
Gaussian random variable, then all the information in anisotropy is
included in the \emph{two-point autocorrelation function},
$C(\theta)$, of the fluctuations \cite{novikov}:
\beq{T_autocor}
C(\theta) \equiv \lan \frac{\Delta T(\x)}{T_{cmb}} \frac{\Delta
T(\x')}{T_{cmb}} \ran
\eeq
where $\Delta T(\x)/T_{cmb}$ is the fractional deviation in the CMB temperature
in the direction $\x$, and
the average is over all pairs of
directions on the sky $\x$ and $\x'$ such that $\x \cdot \x' =
\cos{\theta}$. This function $C(\theta)$ then contains all the
information contained in the anisotropy, and can be expanded in terms
of Legendre polynomials, where the information is in turn kept in the
$\Cl$ coefficients: \beq{Cleq} C(\theta) \ = \ \frac{1}{4\pi}\sum_\l
\Cl P_\l(\cos{\theta}) \ . \eeq The $\Cl$'s comprise the
\emph{angular power spectrum} of CMB anisotropy. The power spectrum
is truly ``powerful''; it depends sensitively on the fundamental
parameters of cosmology, and by measuring it accurately, over the
years we have been able to rule out many theories of structure
formation in the universe, and constrain the current theories of
cosmology rather tightly. \fig{powerspec} shows the state of
anisotropy observations, and is a testament to technology, dedication
and driving curiosity.  The upper panel shows the state of the
angular power spectrum as it was just four years ago (when I started
working on the CMB), and the lower panel shows the situation today.

The most recent measurements by the BOOMERANG, MAXIMA-1, and DASI
teams have led to an excellent determination of the CMB power
spectrum for multipoles less than about 600
\cite{barth01,dasi01b,maxima01}.  These measurements, along with
recent supernova-1a results  (such as shown in \fig{HubbleLaw}),
observations of large-scale structure, and measurements of the
primordial abundance of light elements have helped not only to
confirm the Hot Big Bang paradigm that now dominates cosmology,
but also to nail down several of the fundamental parameters that
characterize the theory.  Indeed, we now have good measurements of
many fundamental parameters:
the Hubble constant, the age of the universe, the relative content
of baryons, dark matter, and dark energy in the universe,
the overall curvature of the universe,
a strong upper bound on the epoch of reionization, and the
spectral index of primordial scalar fluctuations.
\tbl{compilation} shows a compilation of the most recent
measurements of these quantities.
When I began graduate school six years ago, most of these quantities
were known to at best within a factor of two!

\begin{table}
\begin{center}
\begin{tabular}{||c|c|c|c||} \hline \hline
{\bf Parameter} & {\bf Value} & {\bf Measurement(s)} & {\bf Reference}
\\ \hline
$H_0$ & 68 $\pm$ 3 $km/s Mpc^{-1}$& Multiple Methods Combined & \cite{krauss01} \\
$T_{universe}$ &  12.3 $\pm$ 1.6 $Gyr$& CMB+HST-Key & \cite{maxp_1} \\
$\Omega_{tot}$ &  1.0 $\pm$ 0.06 & CMB+HST-Key & \cite{maxp_1}
\\
$\Omega_b h^2$ & 0.02 $\pm$ 0.002 & Deuterium+BBN & \cite{maxp_1}
\\
$\Omega_m h^2$ & $0.13^{+.04}_{-.02}$ & CMB+BBN+HST-Key &
\cite{maxp_2} \\
$\Omega_\Lambda$ & $0.66^{+.10}_{-.17}$ & CMB+HST-Key &
\cite{maxp_1} \\
$n_s$ & $0.93^{+.12}_{-.10}$ & CMB+HST-Key & \cite{maxp_1} \\
$\tau$ & $ < 0.17 $ & CMB+PSCz & \cite{maxp_1} \\
\hline
$z_{re}$ & $7^{+14}_{-2}$ & CMB+PSCz+GP & \cite{maxp_1} \\
\hline
\hline
\end{tabular}
\caption[Current Constraints On Selected Cosmological Parameters]
{\fixspacing \label{compilation} Current constraints on selected
cosmological parameters. The techniques used to constrain the data
are: CMB anisotropy (see references in text), the IRAS
Point-Source Survey (PSCz) \cite{PSCz}, Big-Bang Nucleosynthesis
calculations \cite{turner01}, deuterium abundance measurements
\cite{tytler00}, the HST-Key project \cite{hstkey}, and the Gunn-Peterson
Measurement \cite{gunn65}.}
\end{center}
\end{table}

\subsection{Polarization}
Though anisotropy studies have been remarkably successful, there
is a third characteristic of the CMB
that we have not yet discussed: its polarization.
In contrast to the spectrum and spatial anisotropy of the CMB,
not only has polarization not been well measured, it has never
been detected at all!  \tbl{pol_limits} shows a history of the
measurements aimed at detecting CMB polarization and the
limits they reached.  The first realistic measurements of the
polarization of the CMB were undertaken by Lubin and Smoot
\cite{ls79} in the late 1970s; since then limits have improved by
about an order of magnitude, and experimentalists are
inching ever closer to a detection.  In the following sections,
I will first review the mathematical description of CMB polarization,
and then discuss what can be learned from it.

\begin{table}[tb]
\begin{center}
\begin{tabular}{||c|c|c|c||} \hline \hline
\smbf{Name} & \smbf{Frequency [GHz]}& \smbf{Angular Res.} & \smbf{Limit [$\frac{T_{pol}}{T_{cmb}}$]} \\
\hline
\sm{Penzias and Wilson 1965 \cite{pw65}} & 4.0 & 30\deg & 0.1 \\
\sm{Caderni \etal\ 1978 \cite{caderni78}} & 100-600 & 40\deg\ & $\sim 1 \times 10^{-3}$ (65\%)\\
\sm{Nanos 1979 \cite{nanos79}} & 9.3 & 15\deg\ & $6 \times 10^{-4}$ (90\%)\\
\sm{Lubin \& Smoot 1979 \cite{ls79}} & 33 & 7\deg\ & $3 \times 10^{-4}$ \\
\sm{Lubin \& Smoot 1981 \cite{ls81}} & 33 & 7\deg\ & $1 \times 10^{-4}$ \\
\sm{Lubin \etal\ 1983 \cite{lms83}} & 33 & 7\deg\ & $7 \times 10^{-5}$ \\
\sm{Partridge \etal\ 1988 \cite{partridge88}} & 5 & 18\as--160\as\ & $6 \times 10^{-5}$ \\
\sm{Fomalont \etal\ 1993 \cite{fom93}} & 8.44 & 10\as--80\as & $2.6 \times 10^{-5}$ \\
\sm{Wollack \etal\ 1993 \cite{wollack93}} & 26--36 & 1.4\deg\ & $9 \times 10^{-6}$ \\
\sm{Netterfield \etal\ 1995 \cite{net95}} & 26--46 & 1.05\deg\ & $7 \times 10^{-6}$ \\
\sm{Partridge \etal\ 1997 \cite{partridge97}} & 8.44 & 30\as & $1 \times 10^{-5}$ \\
\sm{Sironi \etal\ 1998 \cite{sironi98}} & 33 & 30\deg\ & $7 \times 10^{-5}$ \\
\sm{Subrahmanyan \etal\ 2000 \cite{sub00}} & 8.7 & 2' & $3.7 \times 10^{-6}$ \\
\sm{Hedman \etal\ 2000 \cite{hed00}} & 84--100 & 0.24\deg\ & $3.7 \times 10^{-6}$ \\
\sm{This Work \cite{kea01}} & 26--36 & 7\deg\ & $2.9 \times 10^{-6}$ \\
\hline \hline
\end{tabular}
\caption[Results of Searches for CMB Polarization]
{\fixspacing \label{pol_limits}
Results of searches for CMB polarization.
Limits are at 95\% confidence unless otherwise noted.  Data graciously compiled by Josh Gundersen.
}
\end{center}
\end{table}

\section{CMB Polarization - Background and Theory}
\subsection{Review of Polarization Description}
Let us briefly review the mathematical description of
polarization.  An electromagnetic wave can in general be written
in terms of its electric field as
\begin{equation}
\label{Efield}
\vec{E}\ = E_x \hat{x} + E_y \hat{y}
\end{equation}
where
\begin{eqnarray}
E_x & = E_{x_0} e^{i (kz - \omega t + \phi_x)} \notag \\
E_y & = E_{y_0} e^{i (kz - \omega t + \phi_y)} \notag \; \; \text{.}
\end{eqnarray}
It is implicit that one takes the real part of $\vec{E}$ to obtain
the physical field.  We can equally well describe this radiation
by four \emph{scalar} quantities, the \emph{Stokes parameters}
which are defined as follows \cite{rohlfs96}:
\begin{subequations}
\label{stokes}
\begin{align}
I &= \langle E_{x_0}^2 + E_{y_0}^2 \rangle \\
Q &= \langle E_{x_0}^2 - E_{y_0}^2 \rangle \\
U &= 2 \langle E_{x_0} E_{y_0} \cos(\phi_x - \phi_y) \rangle \\
V &= 2 \langle E_{x_0} E_{y_0} \sin(\phi_x - \phi_y) \rangle ,
\end{align}
\end{subequations}
where $\lan \ldots \ran$ denotes a time average.
For quasimonochromatic light, each component of
\eqn{stokes} is understood to be averaged over the entire frequency band.

The Stokes parameters $I$ and $V$ are unchanged under rotations of
the $\hat{x}-\hat{y}$ plane, but $Q$ and $U$ are not.  If we
rotate the $\hat{x}-\hat{y}$ axes through an angle $\theta$, the
Stokes parameters change as
\begin{eqnarray}
    Q' = Q\cos{2\theta} + U\sin{2\theta} \notag \\
    U' = -Q\sin{2\theta} + U\cos{2\theta}
\end{eqnarray}
The angle $\alpha \equiv \frac{1}{2}\arctan{\frac{U}{Q}}$ transforms to
$\alpha-\theta$ under the rotation; hence it defines a constant
direction in space, which is interpreted as the axis of polarization.
Finally, the magnitude of polarization is typically denoted by
\beq{}
P \equiv \sqrt{Q^2 + U^2 + V^2} \ .
\eeq
For a fully polarized signal,
$I = P$.  A partially polarized signal
is said to have a fractional polarization  $\Pi \equiv
\frac{P}{I}$.

\subsection{Why is the CMB polarized?}\label{s:thomson}
\begin{figure}[tb]
\begin{center}
\includegraphics[height=3.5in]{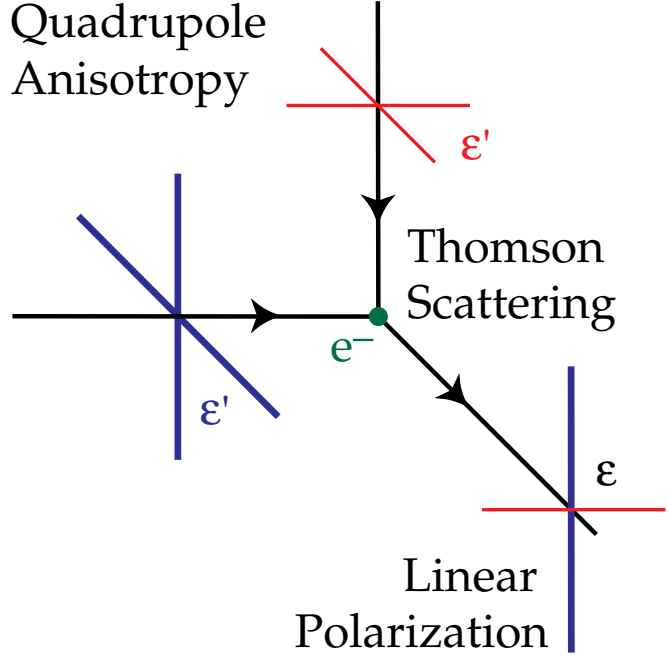}
\caption[Generation of CMB Polarization via Thomson Scattering]
{\label{thomson} \fixspacing
Generation of CMB Polarization via Thomson Scattering.
An electron viewing an anisotropic radiation distribution with a
non-zero quadrupole moment will produce a polarized radiation
pattern.  Figure originally from reference \cite{primer}.
}
\end{center}
\end{figure}

The CMB is partially polarized via Thomson scattering of CMB photons
by free electrons at the last scattering surface \cite{rees68}.
This situation is shown schematically in \fig{thomson}.
An electron on the last scattering surface viewing an anisotropic
distribution will generate a polarized radiation distribution if
there is a non-zero quadrupole moment to the anisotropic
distribution \cite{kosowsky}.  Following Kosowsky,
the cross-section for Thomson scattering when an incident wave
with polarization $\epp$ is scattered into a wave with polarization
$\ep$ is given by
\beq{e:thomson}
\frac{d\sigma}{d\Omega} = \frac{3\sigma_T}{8\pi} | \epp \cdot \ep
|^2 \ ,
\eeq
where $\sigma_T$ is the total Thomson cross section.
Let us now integrate this quantity over the input radiation field.
This field, $I'$, can be expanded into spherical harmonics such
that
\beq{}
I'(\theta,\phi) = \sum_{\l m} a_{\l m} Y_{\l m}(\theta,\phi) \
.
\eeq
It can be shown the that output Stokes parameters by the
re-radiating electron are given by \cite{kosowsky}
\begin{subequations}
\begin{eqnarray}
I & = & \frac{3\sigma_T}{16\pi} \left[\frac{8}{3}\sqrt{\pi}a_{00}
+\frac{4}{3}\sqrt{\frac{\pi}{5}}a_{20}\right] \ , \\
Q & = & \frac{3\sigma_T}{4\pi} \sqrt{\frac{2\pi}{15}} \, \text{Re}
\, a_{22} \ ,\\
U & = & -\frac{3\sigma_T}{4\pi} \sqrt{\frac{2\pi}{15}} \, \text{Im}
\, a_{22} \ ,\\
V & = & 0 \ .
\end{eqnarray}
\end{subequations}
Thus, we see that both $Q$ and $U$ are generated by the radiation
process, but they are entirely due to the quadrupole of the
incident radiation field.  $V$ is not generated through this
process, and though it can be generated through certain types of
galactic foregrounds, we will in general ignore it through the
rest of this thesis.

Because the source of the polarization \emph{is} the anisotropy,
the polarization fraction can be at most \app $1\cdot 10^{-5}$;
theoretical studies show that really it can be at most about 10\%
of this level.  Thus, the polarization signal is truly small, and
represents a significant challenge for experimentalists.

\subsection{Description of CMB Polarization}
Experimentalists seeking CMB polarization are attempting
to measure the scalar fields $Q$ and $U$ at every point on the
sky.  Because $Q$ and $U$ are dependent on the coordinate system chosen,
the universally accepted convention is to use the meridian passing
through both the celestial pole and the observation point as the defining
axis.  This meridian represents the N-S direction for the
observation point.  If the experiment is such that it measures the
temperature of the CMB along a given axis, then the linear Stokes
parameters in this universal coordinate system are given by
\cite{ls81}
\beqa{}
Q & = & T_{NS} \ - \ T_{EW} \notag \\
U & = & T_{NE,SW} \ - \ T_{NW,SE} \ .
\eeqa

However, this is still not the most natural way to express CMB
polarization; a much better way uses the fact that a polarization
field can be decomposed into two components with special
symmetry properties, analogous to the fact that a vector field
can be decomposed into a curl-free component and a gradient-free
component \cite{kosowsky,zs97}.
These components are typically denoted $E$ and $B$,
though some authors also use $C$ and $G$.  They each are
symmetric under rotations, but $E$ is symmetric under parity
(reflection)
as well, whereas $B$ is anti-symmetric under parity \cite{primer}.

\begin{figure}[tb]
\begin{center}
\subfigure[Pure +E Hot Spot]{\label{pluse}
\includegraphics[width=2in]{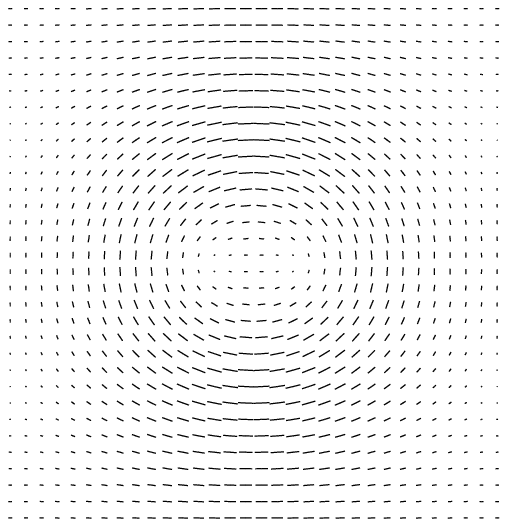}
}
\subfigure[Pure -E Hot Spot]{\label{minuse}
\includegraphics[width=2in]{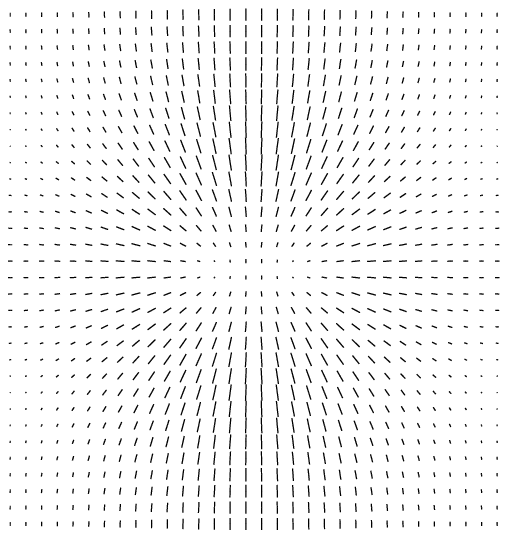}
} \\ \vspace{2mm}
\subfigure[Pure +B Hot Spot]{\label{plusb}
\includegraphics[width=2in]{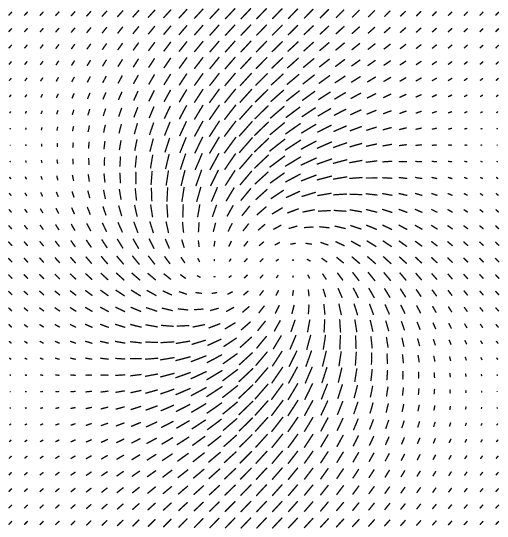}
}
\subfigure[Pure -B Hot Spot]{\label{minusb}
\includegraphics[width=2in]{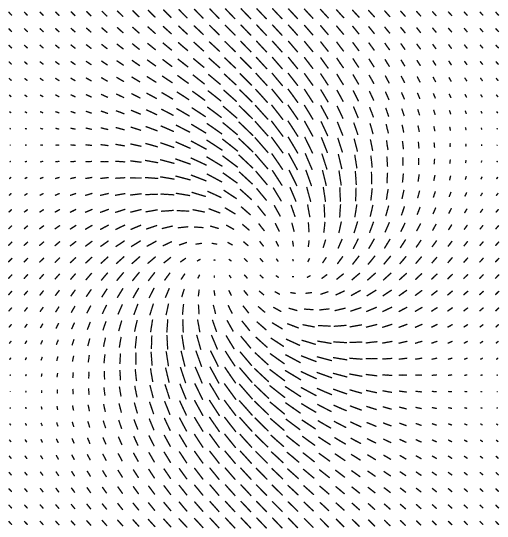}
} \\
\caption[Hot Spots of $E$ and $B$ Polarization]
{\label{hotspots}\fixspacing Hot Spots in E and B.
Notice that reflections about any
axis through the center of the hot spots
leave $E$ unchanged, but take $B\rightarrow -B$.
All patterns are symmetric under rotation.
Courtesy of Ted Bunn \cite{bunn01}.
}
\end{center}
\end{figure}

There are standard formulae to construct $E$ and $B$ maps from $Q$
and $U$ maps; for a good description of this process, see
\cite{matias01}. In general the relationship is \emph{nonlocal}; that
is, the values of $E$ and $B$ at a specific point on the sky are
functions of $Q$ and $U$ everywhere on the sky, although areas close
to the point in question matter more than areas far from the point in
question.  \fig{hotspots} shows typical pure $E$ and $B$ patterns.
The E-modes are symmetric under both rotations and reflections about
an axis (parity), while the B-modes, possessing a handedness, are
symmetric under rotations, but are anti-symmetric under parity.

\subsection{Polarization Power Spectra}
The power spectra for polarization are most naturally expressed in
terms of the $E$ and $B$ modes.  The mathematics needed to write down
these power spectra will be briefly covered in \chap{likes}, where we
will see that the power spectra are written directly in terms of the
Stokes parameters; in the present section we restrict the discussion
to a qualitative description of the power spectra.

There are four power spectra used to characterize the CMB
radiation field: $\Ctt$,$\Cee$, $\Cbb$, and $\Cte$.  Their
construction and description is discussed extensively in
\cite{zs97,kks,max01}.  $\Ctt$ is the usual temperature anisotropy
power spectrum, formed from the temperature two-point correlation
function.  $\Cee$ is related to the autocorrelation of the
$E$-field; because it has the same parity as $\Ctt$, there is a
non-vanishing correlation between temperature anisotropy
and $E$-mode polarization, leading to the
cross-correlation power spectrum $\Cte$.  Finally, there is the
autocorrelation of the $B$-modes, yielding the $\Cbb$ spectrum.
Because the symmetry under parity is opposite for $B$ as compared
to $E$ or $T$, the last two power spectra, $\Ctb$ and $\Ceb$ are
zero for the CMB.  However, these two power spectra may be
non-zero in the presence of foregrounds and hence should still be
calculated for real data if possible.

\section{What do we learn from CMB polarization?}
This section will attempt to motivate observations of CMB
polarization by reviewing what we can learn from it. However, I only
scratch the surface of the mass of literature on the subject. For
more interested readers, the following reviews are suggested:
Kamionkowski and Kosowsky (1999) review how the CMB relates to
particle physics, and discuss the clues to understanding inflation
left in CMB polarization \cite{kk99}; Peterson \etal\ (1999) discuss
the CMB in the post-Planck era, and present a very good initial
review of CMB polarization \cite{cmbpp};  Hu and White (1997) present
an excellent primer on the physics of CMB polarization \cite{primer}.

\begin{figure}
\begin{center}
\includegraphics[height=6in]{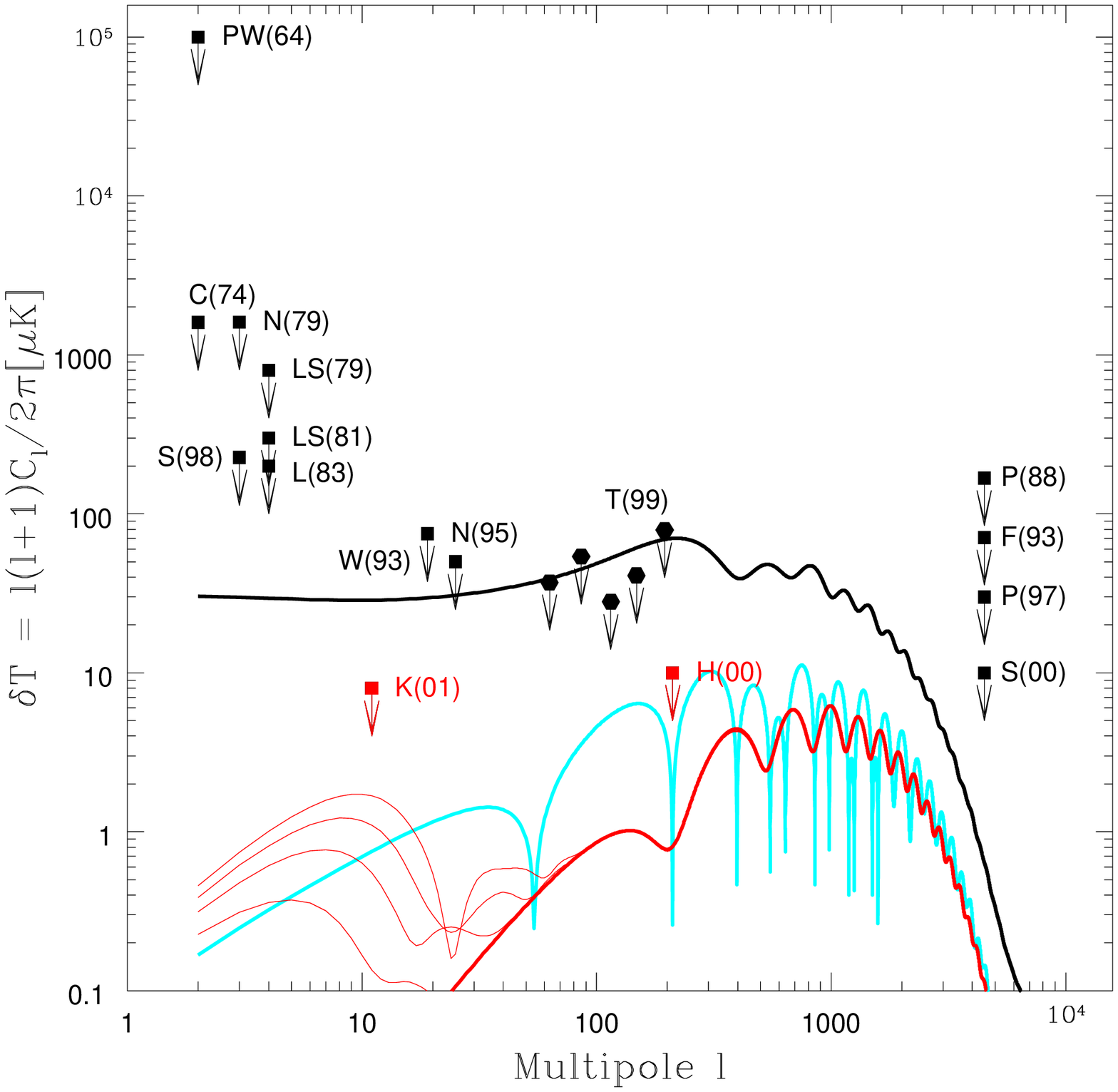}
\caption[The State of CMB Polarization] {\label{polspectra}
\fixspacing The State of CMB Polarization:  approximate experimental
limits on the CMB polarization. The unpolarized power spectrum T is
shown in black, while the E and cross-polarization TE are shown in
red and blue, respectively. The thin red lines (from botton to top)
are for reionization models with $\tau$ values of 0, 0.1, 0.2, 0.3
and 0.4. Notice the generation of the ``reionization peak'' at $\l =
5-10$ in the models with non-zero $\tau$. The fiducial power spectra
T, E and TE are computed using the CMBFAST software designed by
Zaldarriaga and Seljak \cite{cmbfast}, employing cosmological
parameters from the concordance model of Wang, Tegmark \& Zaldarriaga
\cite{maxp_2}. The historical limits represent decades of work upon
which this thesis builds, and are
\cite{pw65,caderni78,nanos79,ls79,ls81,sironi98,lms83,wollack93,net95,torbet99,
partridge88,partridge97,sironi98,sub00,hed00}. Reference ``K(01)'' is
this work, originally published in \cite{kea01}. The compilation of
upper limits was kindly provided by Josh Gundersen. Figure provided
by Angelica de Oliveira-Costa. }
\end{center}
\end{figure}

The CMB power spectra depend sensitively on the particular choice
of cosmological model and the parameters within it.
\fig{polspectra} shows sample power spectra calculated for a
typical ``concordance'' model \cite{wang00}, along with the
history of polarization upper limits obtained to date. The power
spectra were computed with CMBFAST \cite{cmbfast}.  There are several
important pieces of cosmological information that are difficult or
impossible to probe with anisotropy alone, but are revealed via
polarization studies. We will now discuss a few of the most
important of these cosmological questions.

\subsection{The Effect of Reionization}
The universe is currently ionized.  By studying the absorption spectra along
the line of sight to distant quasars and
noticing the lack of HI absorption lines, Gunn and Peterson
concluded that the universe is ionized out to a redshift of at
least $z \sim 5$ \cite{gunn65}.  Very recent results suggest a
Lyman-alpha trough in several extremely high redshift systems, implying
that the \emph{reionization epoch}\footnote{\fixspacing
The reionization epoch is the transition period during which
the universe passed from neutral to ionized.}, ended at a redshift of
6--7 \cite{becker01}.

Reionization is a very important process in understanding
polarization.  After the CMB was released during the era of
recombination, it was slightly polarized via Thomson scattering off
free electrons at the LSS, as discussed in \sct{s:thomson}. However,
the free electrons at last scattering are nothing special; any
subsequent free electrons encountered by the CMB will additionally
polarize the CMB. After recombination, the universe underwent a
cosmic ``dark age'', in which the CMB propagated freely, but there
was no light from stars.  A first generation of stars presumably
created the radiation that ionized the universe, and led to the
Gunn-Peterson observation.  We currently believe this happened
between a redshift of 6--20.

Reionization has the effect of inducing a ``reionization peak'' in
the E-mode power spectrum \cite{matias97,bk98} at $\l \lesssim 20$.
The true figure of merit is the optical depth to reionization,
$\tau$, induced by reionization.  The more free electrons between us
(as observers) and the LSS, the higher is $\tau$, and hence the more
polarized is the CMB.  This effect happens primarily at large angular
scales, because clumps of free electrons that are closer to us take
up a larger angular size on the sky.  In models of reionization, it
is typically assumed that reionization occurred very quickly, and led
to a completely ionized universe; however, it is possible to
incorporate a ``fraction-polarized'' parameter into the models,
typically denoted $x$.  To convert between the reionization optical
depth and the redshift of reionization, $z_{re}$, and $x$ we use
\cite{bk98} \beq{zreion} \tau \ = \ 0.0015(x)\frac{\Omega_B}{0.05}
(\Omega_{tot})^{-1/2}\left(\frac{h}{0.65}\right)(1+z_{re})^{3/2} \ ,
\eeq where $x$ is the fractional ionization ratio, and $\Omega_B$,
$\Omega_{tot}$ and $h$ have their usual definitions.  Recent
constraints from anisotropy measurements place $\tau \lesssim 0.20$
\cite{maxp_2}, although this is a bit deceiving because the effect of
reionization is easily mimicked by changing the tensor
content of primordial metric perturbations, or the baryonic or
cosmological constant contributions to the energy density of the
universe. Observations of the $E$-mode CMB polarization can
\emph{directly} constrain the epoch of reionization, because
\emph{only reionization can generate a low-$\l$ peak in the $E$-mode
polarization power spectrum} (see \fig{polspectra}).  Thus, the CMB
is a powerful probe of the reionization epoch which ended the cosmic
``dark ages''.

\subsection{A Window on Inflation}
\subsubsection{The Origin of Structure}
The CMB temperature anisotropy is strong evidence for the formation
of structure in the universe.  We currently believe that large-scale
structure arose from tiny density fluctuations in spacetime, that
grew via gravitational instability. This growth of structure began
perhaps just a little before recombination, in which the universe
passed from radiation to matter-dominated; at this time,
structures were able to grow.  Before that time, structure formation
had been inhibited by the photon pressure smoothing out any
structures before they could form. As was stated previously, various
physical effects, primarily the Sachs-Wolfe effect \cite{sw67}, led
to the imprinting of this structure on the CMB during recombination.
The level of anisotropy of the CMB today tells us indirectly about
the size of density fluctuations at the time of last scattering.
However, density fluctuations are just one type of perturbation to
the metric of spacetime; along with them, there also could have been
gravitational waves. Gravitational waves represent
tensor perturbations to the metric. These additional metric
fluctuations leave their own imprint on the CMB, but anisotropy
measurements alone cannot constrain their relative magnitudes;  as we
will see, only CMB polarization will help us to determine this
information.

Of course, this discussion has only characterized the types of
spacetime fluctuations that led to the formation of structure; it
begs the question, ``Where did the primordial seeds of cosmic
structure come from?''  Many theories have been proposed, and they
all have come from particle physics, such as: primordial adiabatic
perturbations due to inflation, topological defects (such as cosmic
strings, domain wells, etc), superconducting cosmic strings, axion
fluctuations, and many more (see \cite{kk99} and references therein).
These different theories lead to different spectra of metric
perturbations, which can be studied through CMB observations. The
dominant model is inflation with primordial adiabatic fluctuations.
Current observations of the CMB anisotropy, along with observations
of large-scale structure via galaxy surveys, have mostly ruled out
alternative models to inflation.  These models generally require
larger temperature anisotropy in the CMB than we see, in order to
give rise to the amount of structure observed in the present.
Adiabatic fluctuations due to inflation, on the other hand, are quite
consistent with the current data.

A full description of inflation is well-beyond the scope of this
text; several recent reviews are given in \cite{tenthings,lyth99,
liddle00}. The basic idea of inflation is that quantum fluctuations
grew to cosmological size during a period of exponential expansion
very early in the universe (approximately $10^{-38}$ seconds after
the Big Bang). Theorists ascribe a field, called the \emph{inflaton
field}, as the cause of inflation. Typically, this field is not in
its lowest-energy state at the time of the Big Bang; instead, it
``rolls'' from this initial state down its potential curve into a
lower energy state. This changing of states by the inflaton field
releases massive quantities of energy that drives the rapid expansion
of the universe.

\begin{figure}
\begin{center}
\includegraphics[height=6in]{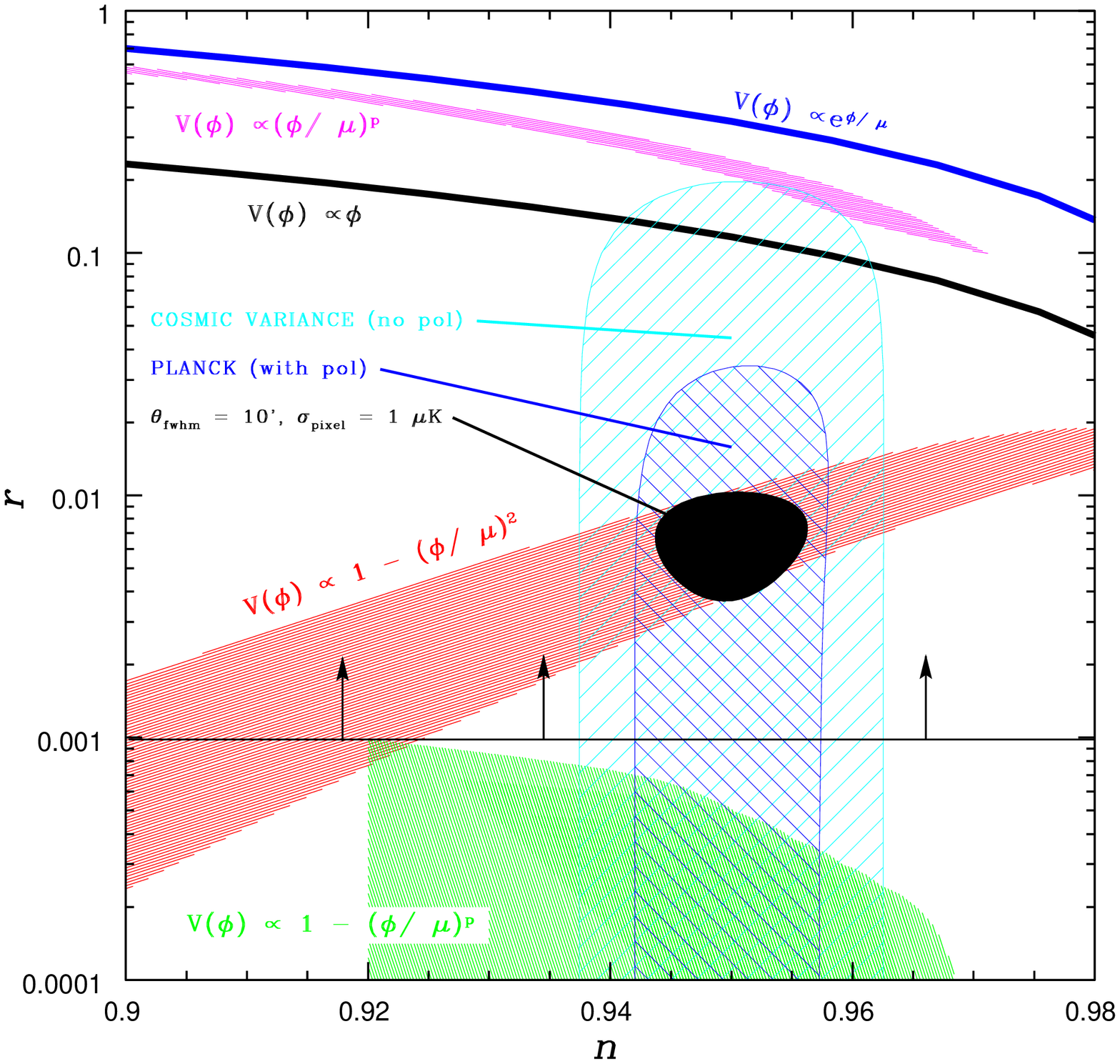}
\caption[Constraints on Inflation from CMB Polarization]
{\label{kinney} \fixspacing
    Constraints on the Inflaton Potential from CMB polarization, in
    the $r$-$n_s$ parameter space, where $n_s$ is the
    spectral index for scalar fluctuations, and $r$ is the
    tensor-to-scalar amplitude ratio, measured at the quadrupole.
    Simulated $2\sigma$ error ellipses that would be
    obtained by
     the Planck Surveyor (without and with polarization), and an
     experiment with three times the sensitivity of Planck.
     This assumes an inflationary model with $r=0.01$
     and $n_s=0.95$ and an optical depth to the surface of
     last scatter of $\tau=0.05$.  \textit{Shaded regions}
     indicate the
     predictions of various inflationary models.  \textit{Solid
     horizontal line} indicates the regions of this
     parameter space that would be accessible
     with a polarization experiment with 30 times
     the sensitivity of Planck\cite{kk98}.
     This figure is illustrative of the gains CMB polarization
     may provide over the next decade.
     Courtesy of William Kinney \cite{kin98}.
}
\end{center}
\end{figure}

\subsubsection{Constraining the Inflaton Potential}
One of the most important aspects of any inflationary theory is the
shape of the inflaton potential. This characteristic shape is related
to the tensor-to-scalar amplitude ratio, $r$, one of the standard
parameters for inflation. \fig{kinney} shows a typical inflation
parameter space, $r$ vs. $n_s$ (the spectral index of scalar
perturbations \footnote{The spectrum of primordial density
fluctuations is typically characterized as a power law, with exponent
$n_s$.}). This figure illustrates the utility of polarization
measurements; without polarization measurements, inflation can barely
be constrained at all, but with polarization, large classes of
inflationary models can be excluded.  This is because gravitational
waves leave a characteristic signature in the CMB polarization.
Gravitational waves are tensors and hence have a ``handedness''
(unlike scalar perturbations), which will lead directly to $B$-mode
CMB polarization (recall, $E$-modes have no handedness).  In
actuality, tensors lead to roughly equal parts $E$ and $B$ modes, but
scalar perturbations generate only $E$-modes. Thus, measurement of
$B$-modes in the CMB not only is a ``smoking gun'' discovery of
primordial gravitational waves, but also it helps us directly
constrain theories of inflation, which set the ratio of
tensor-to-scalar metric fluctuations.

%

\subsubsection{The Energy Scale of Inflation}
However, we can even go a little further with $B$-modes. It turns out
that the energy scale at which inflation occurs imprints itself in
the amplitude of $B$-modes; this amplitude is roughly proportional to
the square-root of the energy scale of inflation. Most scientists
guess that this energy is roughly at the Planck scale, and is due to
some Grand-Unified Theory (GUT) phase transition, which sets the
energy scale at \app\ $10^{19}$ $GeV$. However, some new physics at
lower energies could also have led to inflation, making the amplitude
of $B$-modes correspondingly lower.  Generally, it is believed that a
next-generation satellite designed specifically for CMB polarization
will be able to see the $B$-modes if inflation occurred at the GUT
scale; their absence would indicate new physics at lower energies
\cite{kk99}.

\subsubsection{Adiabatic vs. Isocurvature Fluctuations}
The most common inflationary theory holds that the scalar density
perturbations were \emph{adiabatic}; that is, density perturbations
with an equal fraction in number-density for each particle species in
the Universe.  However, other types of scalar fluctuations are also
possible, namely isocurvature perturbations which lead to differences
in the number densities of different species in the universe.
Isocurvature fluctuations generate a universe where different patches
have different ratios of particle densities, which directly leads to
pressure differences in the universe, although the universe every
maintains a flat geometry (hence the name ``isocurvature''). When two
initially causally disconnected regions come into contact with each
other, their pressure differences will drive the motions of matter,
which in turn seeds large-scale structure. This structure will
imprint itself on the CMB in the same way that adiabatic density
fluctuations will, but in general they lead to slightly different CMB
power spectra. Observations of CMB anisotropy have ruled out
isocurvature fluctuations as the \emph{only} source of density
fluctuations, but it is still possible that there was a mixture of
adiabatic and isocurvature fluctuations.

In general, observations of CMB polarization can help characterize
the amount of isocurvature fluctuations, because these fluctuations
lead to somewhat different polarization patterns than do adiabatic
fluctuations, and these differences are primarily evident at large
angular scales \cite{bucher01}. Current parameter constraints from
CMB anisotropy assume purely adiabatic density fluctuations; however,
it has recently been shown that constraints on cosmological
parameters become \emph{significantly worsened} when a mixture of
adiabatic and isocurvature perturbations is allowed \cite{durrer01}.
Thus, observations of CMB polarization will constrain the type of
initial density fluctuation, and greatly aid in the process of
parameter extraction from anisotropy data.

\section{Summary} In this chapter, we have seen that the CMB
polarization potentially provides a wealth of information.  In
addition to helping further constrain the usual cosmological
parameters, CMB polarization can also constrain reionization, thus
opening up a window on the cosmic ``dark ages''. However, not only
does the CMB open up this window, but by viewing gravitational waves
via polarization $B$-modes, we potentially have a direct view of
events occurring at the beginning of inflation, at $10^{-38}$ seconds
after the Big Bang itself!  This is because gravitational waves are
not affected by ionized matter, and survive the expansion unaffected
until they are imprinted upon the CMB. Hence, CMB polarization is
truly a new window on the universe.

This work describes a ground-based experiment, \polar (Polarization
Observations of Large Angular Regions), which searched
for CMB polarization during the spring of
2000 from Madison, Wisconsin.
The rest of this thesis is organized as follows.  In Chapter
2, I describe the problem of galactic foregrounds and their potential
impact on CMB polarization experiments.  Chapter 3 introduces a new
formalism for analyzing microwave polarimeters, and motivates the
design of the \polar\ radiometer.  In Chapter 4, I present an
overview of the instrument. Chapter 5 discusses the calibration of
the instrument, while Chapter 6 gives an overview of the year 2000
observations and the climate conditions of the telescope location.
Chapter 7 discusses data selection, Chapter 8 presents the mapmaking
procedure that was used to construct maps from the data, and finally,
Chapter 9 covers the analysis used to set limits on CMB polarization.
Finally, let me point the reader's attention to the \polar\
\emph{glossary}
(Appendix A), which defines several terms used in this work.


\chapter{Foreground Radiation}\label{foregrounds}

The potential accomplishments of CMB polarization studies must be
taken with a grain of salt, however, due to the possible obscuration
of CMB polarization by polarized galactic foregrounds.  This is
especially likely at larger angular scales where the CMB signal is
expected to be very small, and several foregrounds are expected to
have a falling power angular spectrum (thus being worse at larger
angular scales).

While there are many dark sections in the sky where CMB anisotropy
dominates galactic emission at the relevant microwave frequencies, it
is simply not known the extent to which polarized foregrounds will
pose a problem to CMB polarization searches. To truly characterize
foregrounds for polarization, we must understand them in terms of
their $E$ and $B$ mode contributions, their behavior in $\l$-space,
and their intensity and polarization dependence on frequency. The
primary foregrounds to be concerned with at microwave frequencies are
dust emission, bremsstrahlung, and synchrotron radiation.

There is also the question of possible polarized emission by the
Earth's atmosphere.  As shown in \chap{observations}, there are
strong features in the atmosphere that lead to significant emission
at microwave frequencies. A polarization fraction of even one part in
one million would lead to an emission of 20 \uK\ which would entirely
swamp the tiny cosmological signal from the CMB. As far as is known,
the mechanism producing the highest polarization level from the
Earth's atmosphere is the Zeeman splitting of oxygen lines by the
Earth's \app\ 0.5 Gauss magnetic field, but Keating (2000) shows that
this leads to less than $10^{-8}$ fractional polarization.  There is
also the possibility of Faraday rotation of the plane of CMB (or
foreground) polarization due to the magnetic field of the Earth, but
this can be shown to be less than 0.01\deg\ at frequencies above 25
GHz \cite{bk98}, where most CMB observations occur. Thus, although
the atmosphere adds noise to our experiment, we can neglect the
concerns of polarized emission and Faraday rotation by the
atmosphere.

\begin{figure}
\begin{center}
\includegraphics[height=4in]{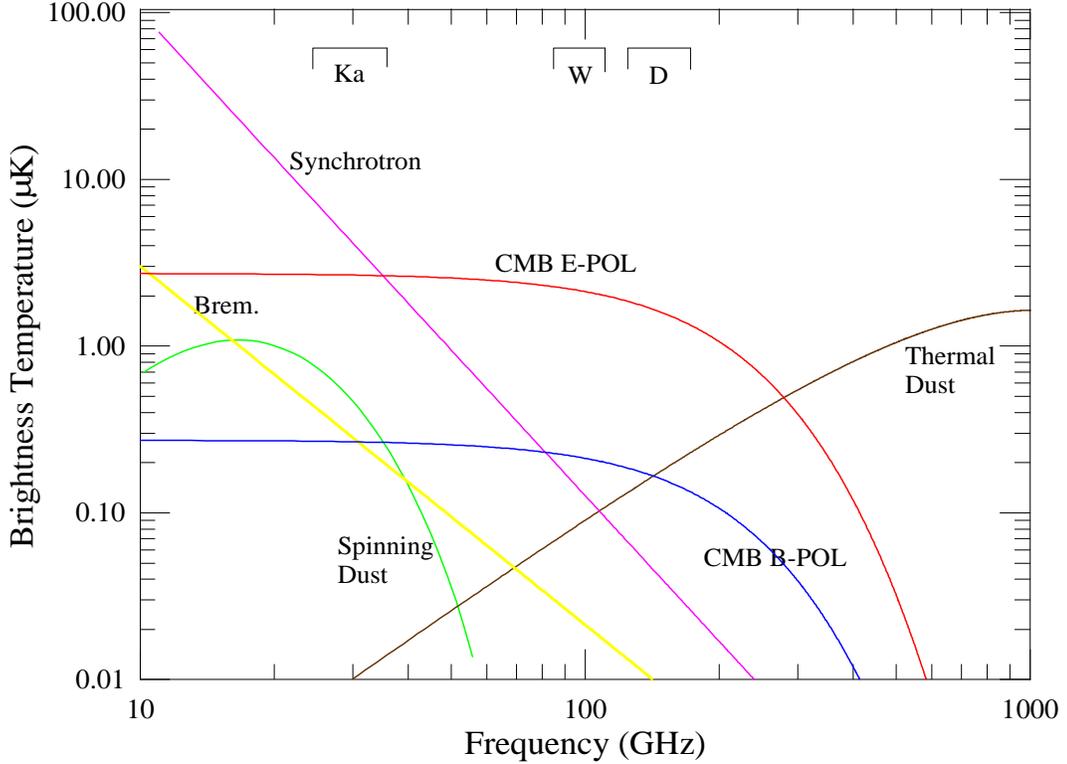}
\caption[Estimated Spectra of Polarized Microwave
Foregrounds]{\label{forspectra} \fixspacing Estimated Spectra of
Polarized Microwave Foregrounds.  The synchrotron spectrum is
normalized to the rms brightness temperature of synchrotron at 19 GHz
(de Oliveira-Costa et al., 1998) and assumes 30\% polarization. The
bremsstrahlung spectrum is normalized to 30 \uK\ at 10 GHz (Davies
and Wilkinson, 1999) and assumes a 10\% polarization. The spinning
dust spectrum proposed by Draine and Lazarian (1998) is shifted by
2/3 to lower frequencies as prescribed by (de Oliveira-Costa, 1999)
and assumes 3\% polarization. The thermal dust spectrum assumes 5\%
polarization (Prunet and Lazarian, 1999), a dust temperature of 18 K,
an emissivity index of 1.8 (Kogut et al., 1996) and uses 3 \uK/MJy/sr
to scale typical degree scale rms values (of 0.5 MJy/sr) at 100
microns to 90 GHz. The CMB E-polarization spectrum is assumed to be
$10^{-6}$ of the CMB brightness spectrum and the CMB B-polarization
is assumed to be 0.1 of the E-polarization spectrum.  Three frequency
bands (\Ka, W and D) are shown above the spectra. }
\end{center}
\end{figure}

\fig{forspectra} is a good introduction to polarized microwave
foregrounds.  It shows that polarized synchrotron and dust are the
most troublesome foregrounds. Dust exhibits three possible sources of
polarized emission, thermal emission due to the vibration modes in
the dust, non-thermal electric dipole emission due to rotating dust
grains, and magnetic dipole-emission due to magnetic grains. At 30
GHz, the dominant foreground is expected to be synchrotron, although
in certain areas spinning dust grains could also pose a threat.
Bremsstrahlung (free-free) radiation is not expected to be polarized
at any significant level \cite{bk98}, and studies show that in most
areas, its intensity is completely dominated by that of synchrotron
(\eg, de Oliveira-Costa (2001) \cite{doc00}).  We will now discuss
each of these foregrounds, as they relate to \polar, in more detail.

\section{Galactic Synchrotron} As is shown in \fig{forspectra},
synchrotron will likely be the dominant polarized foreground at 30
GHz.  For a fairly comprehensive review to the issue of synchrotron
as a foreground, the reader is referred to Cortiglioni and Spoelstra
(1995) \cite{cort95}. Synchrotron emission occurs when a charged,
relativistic particle is travelling through a local magnetic field
\footnote{\fixspacing When the particle is not relativistic, it emits \emph{cyclotron}
radiation, which is in general negligible compared to synchrotron.}.
The charged particle will not have its energy changed by interaction
with the field (magnetic fields do no work!), but it will be caused
to spiral about the magnetic field lines. This radial acceleration
causes the charged particle to radiate, and the resulting radiation
is in general polarized, and has a frequency distribution that
depends on both the charge and speed of the particle, as well as the
magnetic field strength. A collection of particles moving at
different speeds will give rise to a broad spectrum of radiation.

Thus, synchrotron is highest in areas of high magnetic field, as well
as high concentrations of charged particles along the line-of-sight.
It is strongest in the galactic plane, but is present even at high
galactic latitudes \cite{smoot99}. Synchrotron radiation at high
galactic latitudes follows a rough power-law form: \beq{synchspec}
T_{ant} \propto \nu^{\alpha} \ , \eeq where $\alpha$ is called the
\emph{spectral index} of synchrotron. The primary species dominating
synchrotron radiation at higher frequencies (\ie\ greater than 1 GHz)
is cosmic-ray electrons.  The energy distribution of these electrons
largely determines the spectral index \cite{rl79}.  At lower
frequencies, the spectral index of synchrotron is roughly $-2.8 \pm
0.1$ \cite{platania99}. Above 10 GHz there is strong evidence for a
steepening spectral index, to as high as -3.5 or so
\cite{platania99,doc00}; this is due to the rapid drop-off in
relativistic electrons above about 15 GeV \cite{lawson87,bw90,bw91}.
It should also be noted that many surveys indicate a significant
variation (up to 0.3) of the value of the synchrotron spectral index
\cite{bw91,smoot99}.

Synchrotron is elliptically polarized by its very nature.  This
polarization has a maximum of about 75\%, and
this maximum polarization fraction is related to the
spectral index via $\Pi = \frac{3\alpha+3}{3\alpha+1}$ \cite{cort95}.
Brouw and Spoelstra (1976) attempted to map polarized synchrotron
emission at 1411 MHz \cite{bs76}.
The resulting maps are very under-sampled and are
subject to non-negligible Faraday depolarization (which goes as
$\nu^{-2}$), but still indicate 20-30\% polarization
at high galactic latitudes, $|b|>30\deg$ \cite{bs76}.  Some studies have found
regions with up to 72\% polarization; 30-50\% is a common maximum
polarization in various high-latitude features \cite{dw98}.

In terms of power spectra, the spatial variation of galactic
synchrotron is also modelled as a power law, with $\Cl \propto
\l^{-\beta}$. Data sets such as the Haslam 408 MHz survey and the
Parkes southern hemisphere survey indicate  $2.4 < \beta < 3.0$
(\cite{tehd00} and references therein).  For \emph{polarized}
synchrotron, much less is known.  Polarization maps at 2.4 GHz by the
Parkes survey \cite{duncan97} indicate $\beta \app 1.0$, a much less
steep index that would imply polarized synchrotron is not as bad at
large angular scales as small \cite{tehd00}. Recent work by Baccigalupi
\etal\ (2001) analyzing polarization surveys indicates a steeper
index, $\beta = 1.8 \pm 0.3$ for $\l > 100$, and steepening at lower
$\l$-values, towards a $\beta$ of \app\ 3 \cite{bacc01a}.

\begin{figure}[tb]
\begin{center}
\includegraphics[width=6.2in]{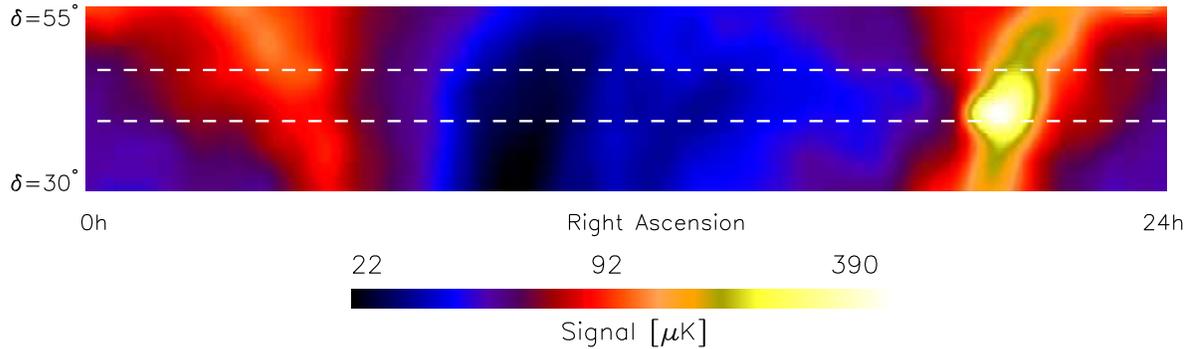}
\caption[The Haslam 408 MHz data, convolved with a 7 \deg\ beam, between
declinations 30\deg\ and 55\deg]
{\label{has_strip} \fixspacing
The Haslam 408 MHz data \cite{haslam82}, convolved with a 7\deg\ beam, between
declinations 30\deg\ and 55\deg.  The strip covered by \polars\ FWHM
is shown between the dashed lines.  The map was obtained by
extrapolating to 31 GHz with a spectral index of -2.8 up to 7.5 GHz,
and -3.0 at higher frequencies, based on \cite{platania99}.  50\%
polarization has been assumed. Intensity scaling is logarithmic.
The bright feature  at \app\ 20\emph{h} is Cygnus A.
Data from \emph{http://skyview.gsfc.nasa.gov}.
}
\end{center}
\end{figure}

Clearly, much work remains to be done in this area, but surely
synchrotron could be a big problem for \polar.  In order to further
understand the potential effects of synchrotron on our experiment, we
extrapolated the Haslam data to higher frequencies using the
knowledge available of the spectral index, and assumed a fairly
high polarization of 50\%; the results are shown in
\fig{has_strip}.  The map has been smoothed with our 7\deg\ beam. The
bright source at RA 20\emph{h} is Cygnus A, the second brightest
radio source in the sky.  There is obviously a large chance we should
see synchrotron.  In order to limit our sensitivity to synchrotron,
we restricted our CMB analysis to right ascensions in the range $7.5h
< RA < 18h$, where polarized synchrotron should be 20-40 \uK\ or less
in our beam.  The reader should note that this is much larger than
the \app\ 0.1-2 \uK\ signal we are attempting to see!

The usual technique to subtract foregrounds is to understand their
spectral behavior, and make multifrequency measurements of the same
area of the sky.  This will be difficult with the \polar\ data, but
not impossible as we do make measurements at three different (albeit
closely spaced) frequencies.  Preliminary work on this front was done
by Keating \etal\ (1998) based on the work of Dodelson (1997)
\cite{bk98,dodelson97}.  More advanced subtraction techniques that
take into account the spatial structure of foregrounds were developed
by Tegmark and others \cite{te96, t98, bg99, tehd00}. Further
discussion of this topic as it relates to \polar\ will await
inspection of our data (for the reader that cannot wait, skip to
\chap{likes}).

\section{Free-free Emission} Bremsstrahlung (free-free) emission is
an important process at radio frequencies, though it is often
dominated by synchrotron radiation.  It is the well-understood
process of an electron passing close to a proton in an ionized
medium; the particles are accelerated as they pass near each other,
and hence radiate.  Bremsstrahlung radiation is well-described as a
power law, with the intensity of radiation proportional to
$\nu^{-\alpha}$, with $\alpha = 2.15 \pm 0.02$ in most regions of the
sky \cite{tehd00}.  Free-free emission is intrinsically unpolarized
\cite{rl79},
 but can become polarized by its parent HII region (where it
is produced) via Thomson scattering off the electrons in the cloud;
recall that this is the same process by which the CMB is polarized at
last scattering. It is expected that in the most extreme
circumstances, this would lead to a 10\% polarization fraction
\cite{bk98,dw99}.  Because synchrotron emission dominates
bremsstrahlung at almost all frequencies and locations on the sky,
and because it exhibits much greater polarization, we will not
consider free-free emission further.

\section{Dust Emission}
\subsection{Anomalous Emission at Low Frequencies}
Dust emission is complicated.  Until recently, the emission of dust
was assumed to be exclusively thermal (vibrational).
Cross-correlations of the COBE DIRBE data with IRAS support a dust
emission model where most dust is at a stable temperature of
approximately 20 K \cite{kogut96}; in the Rayleigh-Jeans portion of
its spectrum, this corresponds to a rapidly falling spectrum, such
that below 90 GHz dust emission was historically considered
unimportant as compared to synchrotron (see \fig{forspectra}).

However, multiple experiments at frequencies between 10 and 36 GHz
now support an additional mechanism of dust emission that is less
well understood. In 1995, Leitch \etal\ reported an excess emission
(inconsistent with synchrotron) at 14.5 and 32 GHz, and a large
correlation with IRAS far-infrared data at 14.5 GHz \cite{leitch97}.
The excess emission was initially attributed to free-free emission, but
that hypothesis was later ruled out on energetic grounds \cite{dl98a}.
The large correlation with IRAS (which is completely dominated by
dust emission) suggests a common emission source.

More recently, a strong correlation with IRAS data for both Tenerife
data at 10 and 15 GHz \cite{doc99}, as well the 19 GHz maps of
Cottingham \cite{cot87} has been noted.  In the former case, it was
compared to maps of $H\alpha$ (a good tracer of free-free) to test
the free-free hypothesis, and it was found that free-free emission is
a factor of ten too small to explain the emission \cite{doc98}.  The
conclusion drawn is that the correlation with IRAS data (dominated by
dust emission) is due to some previously unknown dust emission
mechanism \cite{doc00}.

Recently, new data by Finkbeiner \etal\ show a direct detection of
excess emission at low frequencies in two point sources
\cite{fink01}. This is in contrast to all previous sources of
evidence, which were statistical in nature.  The Finkbeiner team
surveyed ten fields at 5, 8, and 10 GHz with the 140-foot Greenbank
telescope.  In most cases the observed emission was consistent with
free-free or bremsstrahlung, and had a falling spectrum with
frequency. However, two fields show a \emph{rising} spectrum with
frequency (consistent with the dust emission mechanisms described
below), and one of these two fields is inconsistent with free-free or
synchrotron at the \app\ $10\sigma$ level.  What could this new
emission mechanism be?

\subsection{Dust Emission Mechanisms at Low Frequencies}
\subsubsection{Electric Dipole Emission from Spinning Grains}
To explain the anomalous dust emission at low frequencies, Draine and
Lazarian have proposed two new emission mechanisms for dust: electric
dipole emission from small, spinning grains \cite{dl98a} (hereafter
DL98), and magnetic dipole emission by magnetic grains \cite{dl99}.
In the former hypothesis, ultra-small grains emit via electric dipole
radiation because they are spinning rapidly (at frequencies of 10--30
GHz!).  A dust particle with electric dipole moment $\mu$ and angular
rotational frequency $\omega$ will exhibit emission proportional to
$\mu^2 \omega^4 / c^3$; the total emission is this factor times the
number of emitting grains $n_g$ along the line of sight.

There are many mechanisms that contribute to the rotational velocity
of dust grains, but in general small ($N < 150$ atoms) dust grains
can be spinning rapidly due to collisions with fast-moving neutrals
and ions in the ISM. Grains can acquire a net dipole moment through
two means. First, grains are expected to have an inherent dipole
moment due to the dipole moments of their chemical bonds, if they are
not completely symmetric molecules. Secondly, if the grain is charged
and the center-of-charge does not coincide with the center-of-mass,
this will induce a dipole moment.  DL98 estimated the parameters
$\omega$, $\mu$ and $n_g$ for the dust particles that would dominate
the overall emission.  The resulting spectrum is a blackbody-like
curve that peaks somewhere from $15-25$ GHz; a typical spectrum is
shown in \fig{forspectra} as the ``spinning dust'' model.

\begin{figure}[tb]
\begin{center}
\includegraphics[height=3.5in]{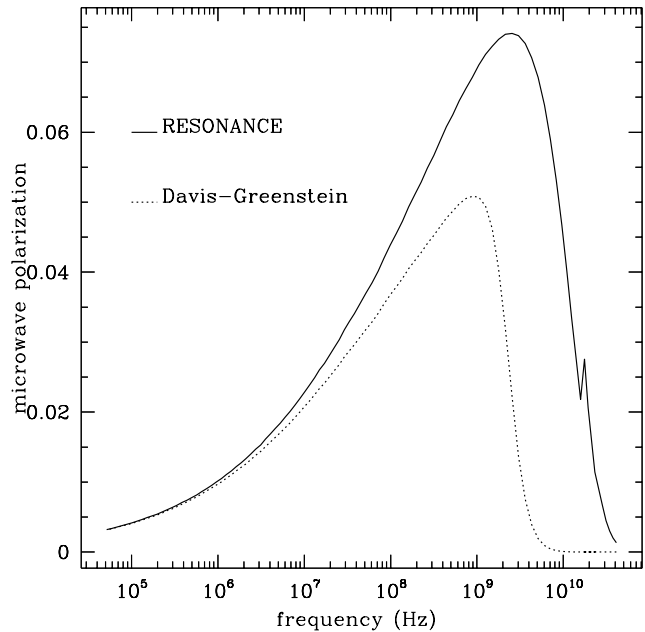}
\caption[Polarization Level of Spinning Dust vs. Frequency]
{\label{dgfig} \fixspacing Polarization for both the resonance
relaxation theory of Draine and Lazarian and Davis-Greenstein
relaxation for dust grains as a function of frequency (from
\cite{ld00}). For resonance relaxation, saturation effects are
neglected, which means that the upper curves correspond to the {\it
maximal} values allowed by the paramagnetic mechanism.  Even in the
worst case, polarization from spinning dust grain emission is
expected to be very small. }
\end{center}
\end{figure}

In terms of polarization, microwave emission from the spinning grains
is expected if the grains are aligned.  Lazarian and Prunet reviewed
this process recently in \cite{lp01}.  Grains in the galactic
magnetic field will experience a torque that tends to align their
angular momenta with the magnetic field via the paramagnetic
dissipation mechanism of Davis-Greenstein \cite{dg51}.  It turns out
this leads to essentially no polarization at frequencies of 10 GHz,
as shown in \fig{dgfig}.

However, Lazarian and Draine (2000) point out that this picture is
incomplete, and propose a modification to the theory to account for
an extra alignment mechanism, called ``Barnett Magnification''
\cite{ld00}. This effect involves spontaneous magnification of the
rotating grain in a local magnetic field, and leads to paramagnetic
relaxation, which the authors called ``resonance relaxation''.
However, this mechanism only marginally increases the polarization of
spinning grains. As shown in \fig{dgfig}, in the \polar\ frequency
range of 26--36 GHz, the polarization fraction is less than 1\% for
resonance relaxation, and is almost exactly zero for the
Davis-Greenstein theory. Thus, electric dipole emission from spinning
dust grains is not likely to be a problem for polarization
measurements, although until accurate measurements of polarization
are made at the frequency where this emission peaks (\app\ 15 GHz),
the picture will not be complete.

\subsection{Magnetic Dipole Emission from Dust}
Lazarian and Prunet (2001) point out that there is another,
often-overlooked dust emission mechanism also relevant at low
frequencies: \emph{thermal fluctuations of magnetization} in
individual dust grains.  This leads to magnetic dipole emission.
Often the magnetic characteristics of dust are neglected at higher
frequencies when calculating their emission properties; typically
electric dipole emission leads to the usual ``thermal'' (vibrational)
emission of dust that is so strong at higher frequencies.

Draine and Lazarian (1999) calculated the expected emission from
magnetic grains as a function of frequency; the emission is strongly
dependent on what magnetic species make up the grains \cite{dl99}.
The materials they considered were metallic iron and nickel, and
various iron-containing compounds.  The main result was that this
emission could be important for frequencies between 10 and 100 GHz,
although there is strong variability in the emission curves depending
on what types and amounts of magnetic materials are included in the
grains.  They note, however, that the anomalous emission reported by
Kogut \etal\ (1996) \cite{kogut96} at 90 GHz reported by the DMR can
be explained by magnetic emission if 5\% of interstellar iron is
locked up in dust.

These arguments then point to a potentially complicated picture of
dust, where at some frequencies, three different emission mechanisms
could be important.  Regarding polarization, it is possible that the
magnetic mechanism can lead to emission that is up to 30--40\%
polarized, depending again on the magnetic species dominating dust
grains.  This polarization results from alignment with the local
magnetic field, and due to their magnetization it is possible that a
large fraction of grains are aligned.  Thus, it is conceivable that
at frequencies around 30 GHz, dust emission could be dominated by
rotational emission in intensity, but by magneto-dipole emission in
polarization \cite{lp01}.

\subsection{Dust Templates and Conclusions}
\begin{figure}
\begin{center}
\includegraphics[width=6.2in]{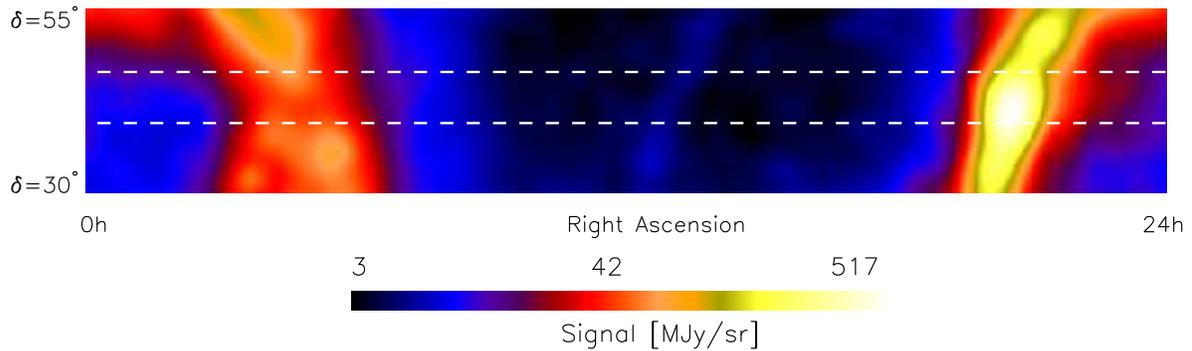}
\caption[DIRBE 100 $\mu m$ emission, along the \polar\ observing strip.]
{\label{dirbe} \fixspacing
DIRBE 100 $\mu m$ emission map, convolved with a 7 \deg\ beam, between
declinations 30\deg\ and 55\deg.  The strip covered by \polars\ FWHM
is shown between the dashed lines.  Intensity scaling is logarithmic.
Data from \emph{http://skyview.gsfc.nasa.gov}.
}
\end{center}
\end{figure}

It is interesting
to look at a template of dust emission in our observation region;
this is shown in \fig{dirbe} for the DIRBE 100 $\mu$m data
\cite{DIRBE}. We can see the same strong emission features due to the
galaxy as in the synchrotron map (\fig{has_strip}), and in between
the galactic features there is much less emission.

We can conclude from this map that the only
form of dust emission capable of affecting \polar\ is magneto-dipole
emission, unless the theories regarding polarized spinning grain
emission are inaccurate or incomplete.  This is because the DIRBE 100
$\mu m$ data show a level of 3-5 MJy/sr in our beam path at high
galactic latitudes ($|b| > 25\deg$).  This corresponds to roughly
10-20 \uK\ of emission at 100
$\mu m$.  Using the
Rayleigh-Jeans law, this \emph{thermal} emission can be extrapolated
to $\lambda = 1 \ cm$, giving roughly 0.1--0.2 \uK\ of thermal
emission.  A typical spinning dust scenario implies as much as 50
times more emission at 30 GHz from spinning dust as from thermal emission
in certain regions of the sky
\cite{doc98}.  This implies perhaps 5--10 \uK\ of emission from
rapidly rotating dust grains in \emph{intensity}.  Assuming a
fairly worst-case scenario of 1\% polarization, this leads to around
0.1 \uK\ polarized emission.  Compare this to the results of
\fig{has_strip}, which shows typical intensities in the same region
of the sky of 20--50 \uK, and could be as much as 75\% polarized!
Even if the excess emission reported in \cite{doc98} is purely
magneto-dipole, this is expected to be at most 20--40\% polarized (and
is probably much less).  This would put polarized dust emission at a level on par with
weakly (10\%) polarized synchrotron with a steep spectral index.

Thus for most scenarios, we expect the dominant foreground for
\polar\ to be synchrotron.  This is not necessarily bad, because it
is still interesting to map out \emph{any} polarized galactic
emission at our relatively under-studied frequency.  In addition to
any CMB knowledge we reap, it will still be useful to the community
to have an enhanced understanding of polarized emission from these
various galactic foregrounds.

\renewcommand\arraystretch{0.8} 
\chapter{Polarimeter Analysis Formalism}\label{PAF}

\section{Description of the Formalism}
    There are many possible radiometric schemes to measure
microwave polarization.  In order to achieve the low noise levels
necessary to see CMB polarization, however, the system must not
only be extremely sensitive, but also extremely stable.  It should
be insensitive to gain drifts in the front-end amplifiers and have
1/f noise that is as low as possible.

    A radiometer can in principle be designed to measure all four
Stokes parameters.  For CMB polarization, it is typically
desirable to be sensitive to both Stokes Q and U simultaneously;
as discussed in Chapter 2, the CMB is expected to exhibit only
linear polarization (although this certainly isn't a strong argument
against measuring V).  Thus, the polarimeter must be designed with
the issue in mind of which Stokes parameters are to be measured.
In this chapter, I describe a useful formalism for analyzing the
output of polarimeters in terms of the Stokes parameters.  The
formalism is based upon the Jones' matrix formalism of
polarization, but is a little bit different as we shall see.

\subsection{Description of Polarized Radiation}
The general situation for a microwave polarimeter is that
electromagnetic radiation is incident upon a feedhorn antenna.  This radiation can be
described by its electric field; for a single frequency of this
radiation, as in Chapter 1, the electric field can be written as
\begin{eqnarray}
\label{e:Efield}
\vec{E} & = E_x \hat{x} + E_y \hat{y} & \ , \ \text{with} \\
E_x & = E_{x_0} e^{i (kz - \omega t + \phi_x)} \notag & \\
E_y & = E_{y_0} e^{i (kz - \omega t + \phi_y)} \notag & \ .
\end{eqnarray}
It is implicit that one takes the real part of $\vec{E}$ to obtain the physical
field.  We can equally well describe this radiation as a complex,
2-element Jones' vector, or a real, 4-element Stokes vector
\cite{bornandwolf,azzam}:
\beq{}
|E\rangle \; = \; e^{i(kz - \omega t)}
\matrixb{ E_{x_0} e^{i\phi_x} \\ E_{y_0}e^{i\phi_y} }
\qquad \qquad \text{or} \qquad \qquad
|E\rangle \; = \;
\matrixb{I \\ Q \\ U \\ V }
\eeq
where the Stokes parameters for monochromatic radiation
are defined in the usual way \cite{rohlfs96} as
\begin{subequations}
\label{e:stokesdef}
\begin{align}
I &= \langle E_{x_0}^2 + E_{y_0}^2 \rangle \\
Q &= \langle E_{x_0}^2 - E_{y_0}^2 \rangle \\
U &= 2 \langle E_{x_0} E_{y_0} \cos(\phi_x - \phi_y) \rangle \\
V &= 2 \langle E_{x_0} E_{y_0} \sin(\phi_x - \phi_y) \rangle \ ,
\end{align}
\end{subequations}
where $\langle ... \rangle$ denotes a time average.
For quasimonochromatic light, each component of
\eqn{e:stokesdef} is understood to be averaged over the entire frequency band.
However, in this formalism we need only consider one frequency at
a time; the final results can then be averaged over frequency
space.

\subsection{The Stokes Operators}
It is useful to notice that we can think of the Jones'
vector of a field as a \emph{state} and describe each Stokes
parameter as an $2\times 2$ \emph{matrix operator}.
To evaluate the Stokes parameter of a state \ket{E}, we simply follow
the procedure from quantum mechanics: $\eval{\bf{O}} =
\bra{E}\bf{O}\ket{E}$, where $\bf{O}$ is the operator in question,
and \bra{E} is the hermitian conjugate of \ket{E}.  The correct
matrix operators for the Stokes parameters are given by
\begin{eqnarray}\label{e:paulistokes}
\I \;= \Im && \Q \ = \Qm \notag \\ \notag \\[1mm]
\U \; = \Um && \V \ = \Vm \; \; \text{.}
\end{eqnarray}
The astute reader will notice that these are none other than the
Pauli spin matrices, and hence have the following useful
properties:
\begin{subequations}
\begin{align}
\I \; \; \text{is the identity} \\
\Q^2 \;=\; \U^2 \;=\; \V^2 \;=\; \I  \\[1 ex]
\Q\U \;=\; -\U\Q \;=\; i\V  \\
\U\V \;=\; -\V\U \;=\; i\Q \notag \\
\V\Q \;=\; -\Q\V \;=\; i\U \notag
\end{align}
\end{subequations}

Let us verify \eqn{e:paulistokes} for U as an example.
\begin{align*}
\eval{\U} & = \bra{E}\U \ket{E} \\
\; & = [E_x^*\text{,} E_y^*] \ \Um \ \matrixb{E_x \\ E_y} \\
\; & = E_x^* E_y + E_y* E_x = 2 \ \text{\bf{Re}}(E_x^* E_y) \\
\; & =  2 \ \text{\bf{Re}}(E_{x_0} E_{y_0} e^{i(\phi_y-\phi_x)} )
= 2 E_{x_0} E_{y_0} \cos(\phi_y -\phi_x) \\
\; & = U \; \checkmark
\end{align*}
as promised.  The other Stokes parameter operators likewise yield
the desired results from \eqn{e:stokesdef}.

It is worthwhile to evaluate the Stokes parameters in a rotated
reference frame.  This is a straightforward calculation, but when done in the
usual notation of \eqn{e:Efield}, the mathematics is cumbersome.
The Jones-matrix formalism makes it particularly simple.
Let us work out the rotation of \Q\ to demonstrate this.
If the rotation matrix
corresponding to a rotation by the angle $\phi$ is given by
\beq{}
\mathbf{R}(\phi) = \mmb{\cphi}{\sphi}{-\sphi}{\cphi} \text{,}
\eeq
then the operator $\Q'$ as seen in the rotated frame is given by
$\Q' = \mathbf{R}^\dagger(\phi) \Q
\mathbf{R}(\phi)$.  Thus,
\begin{align*}
\Q' & = \mmb{\cphi}{-\sphi}{\sphi}{\cphi} \Qm
\mmb{\cphi}{\sphi}{-\sphi}{\cphi} \\
& =
\mmb{\cphi}{-\sphi}{\sphi}{\cphi}\mmb{\cphi}{\sphi}{\sphi}{-\cphi} \\
& = \mmb{\cos^2{\phi} - \sin^2{\phi}}{2\sphi\cphi}
{2\sphi\cphi}{\sin^2{\phi}-\cos^2{\phi}} \\
& = \mmb{\cos{2\phi}}{\sin{2\phi}}{\sin{2\phi}}{-\cos{2\phi}} \\
& = \cos{2\phi}\Qm + \sin{2\phi}\Um \\
& = \Q\cos{2\phi} + \U\sin{2\phi} \; \; \text{.}
\end{align*}
Working these rotations out for each of
the Stokes parameters, one obtains the usual result that
\vspace{-5mm}
\begin{subequations}
\label{eq:stokesrotate}
\begin{align}
\I' & = \I \\
\Q' & = \Q \cos{2\phi} + \U \sin{2\phi} \\
\U' & = - \Q \sin{2\phi} + \U \cos{2\phi} \\
\V' & = \V \; \text{.}
\end{align}
\end{subequations}

\subsection{Radiometric Operators}

Thus far we have described generic electric fields as Jones
vectors, and Stokes parameters as operators that can tell us what
each Stokes parameter for a given state is.  Next reconsider the
initial electric field state entering a polarimeter.  After passing through
the feedhorn antenna, an orthomode transducer (OMT) separates
the two polarizations and sends them down different rectangular
waveguides, which we refer to as \emph{arms} of the polarimeter
(see \fig{simplepol}).
However, we can still consider these two polarizations together as
part of the same state, described by the same Jones vector as
before OMT traversal.

The next piece of the formalism, then, is to construct \emph{operators} ($2
\times 2$ matrices) representing the action of various radiometric components
(such as amplifiers, magic tees, phase shifters, etc.) on the initial state.
Each successive component will then act on the incoming state, and produce
a new output.  We can describe the action of each component
as a matrix scrambling
together these two fields, and creating two outputs every time.
We can then describe the entire polarimeter as a
single $2\times2$ matrix, itself a sum of the four Stokes
parameters (in operator form).  It is these Stokes parameters that the polarimeter
will be capable of detecting.

\begin{table}
\begin{center}
\begin{tabular}{>{\centering}m{4cm} >{\centering}m{4cm} m{3cm}}
{\large Component} &
{\large Schematic/Symbol} &
\multicolumn{1}{c}{\large Operator}
\\
{\large Name} &
 &
\multicolumn{1}{c}{\large Equivalent}
\\ \hline
& & \\[-1.5 ex]
Orthomode Transducer
& \includegraphics[height=1.5cm]{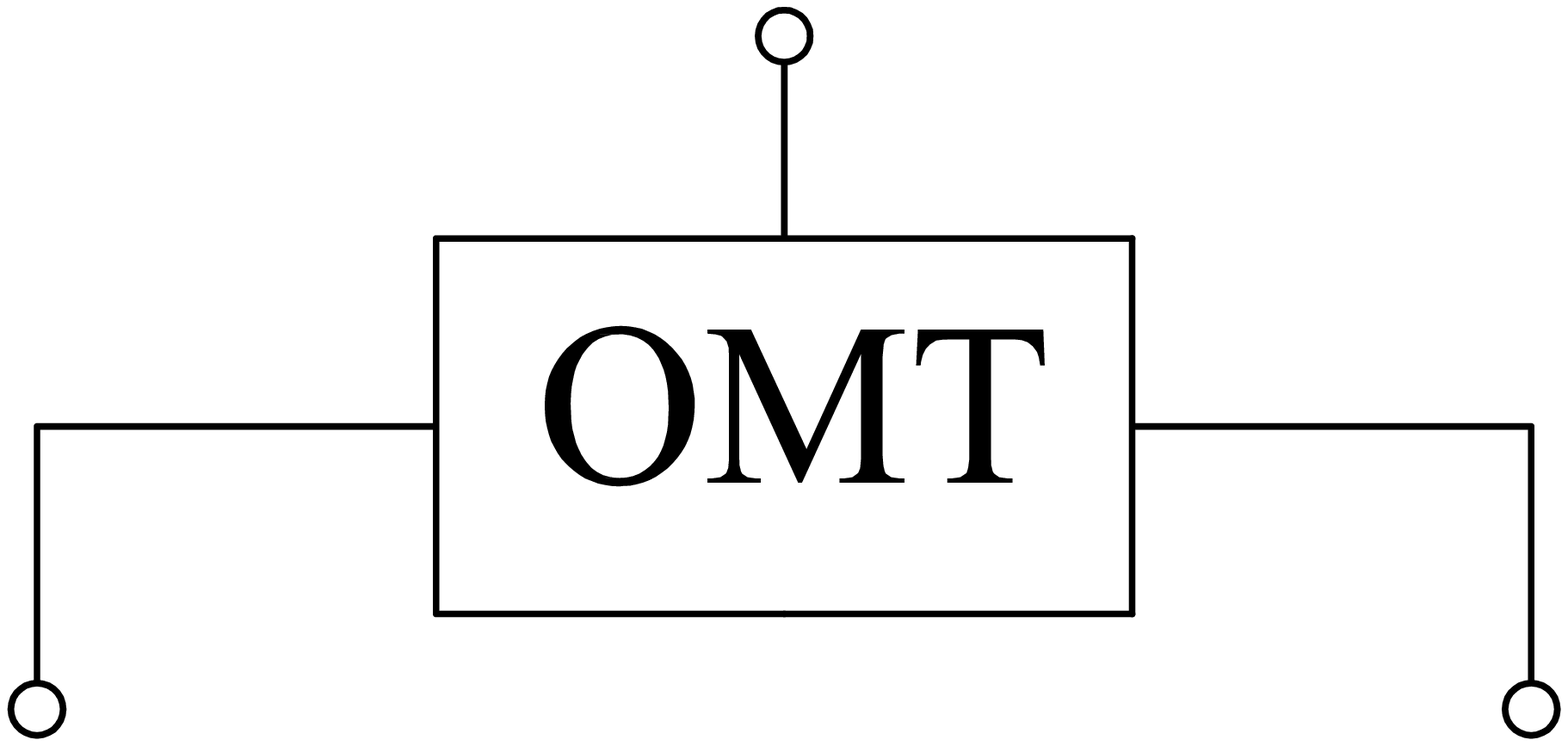}
& \multicolumn{1}{c}{ \mmbt{1}{0}{0}{1} }
\\
Septum Polarizer ({\footnotesize Circular Hybrid Polarizer})
& \includegraphics[height=1.5cm]{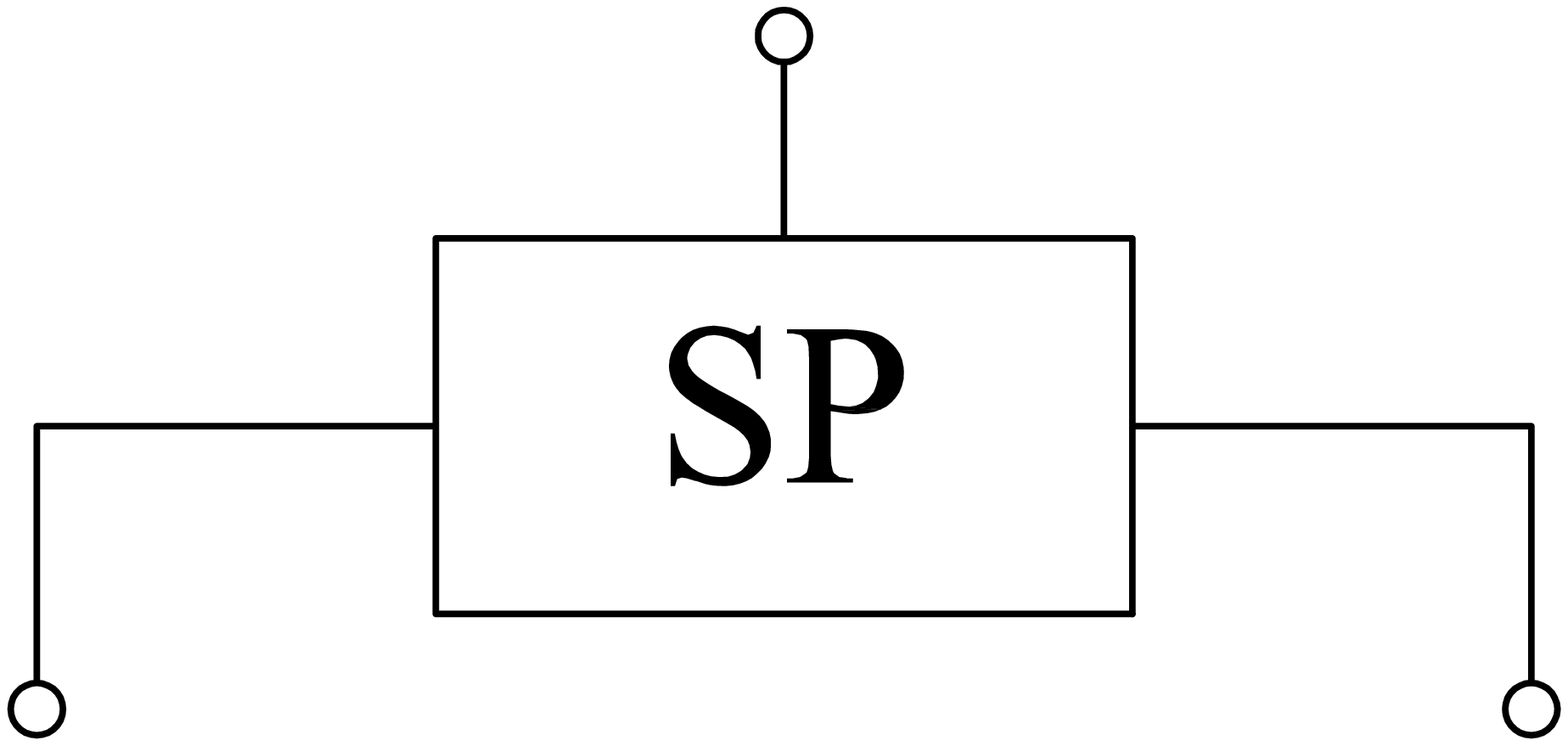}
& \multicolumn{1}{c}{ \sqinv \mmbt{1}{i}{i}{1} } \\
Amplifiers
& \includegraphics[height=2.5cm]{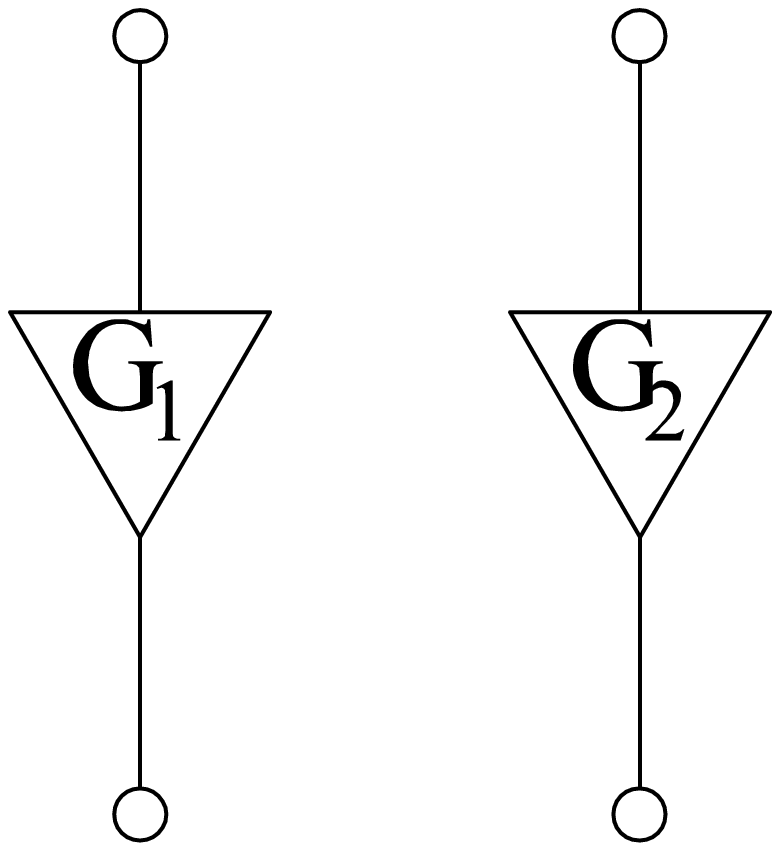}
& \multicolumn{1}{c}{ \mmbt{G_1}{0}{0}{G_2} } \\
Phase Shifters
& \includegraphics[height=2.5cm]{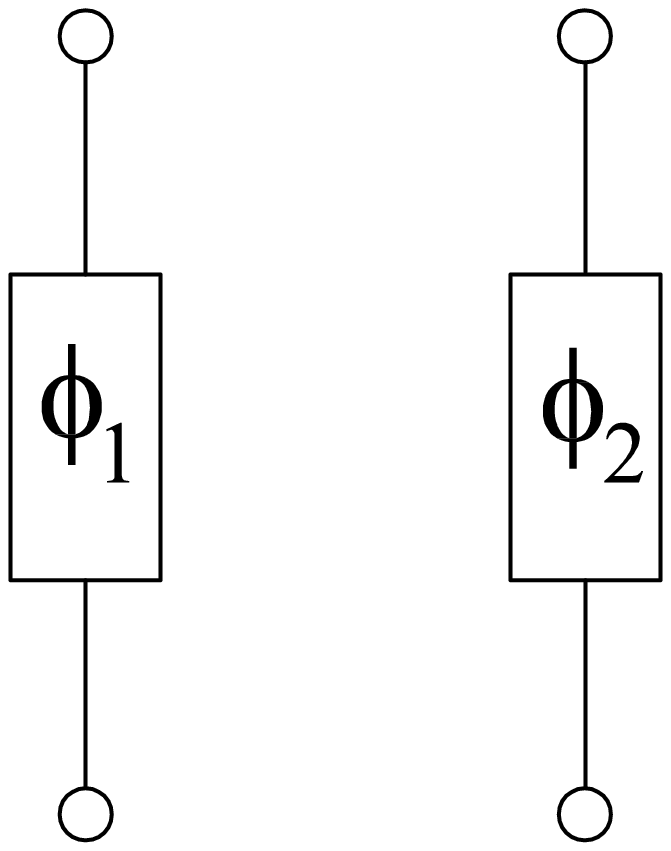}
& \multicolumn{1}{c}{\mmbt{e^{i\phi_1}}{0}{0}{e^{i\phi_2}}}
\\
Magic Tee \mbox{(180\deg\ Hybrid)}&
\includegraphics[height=2.8cm]{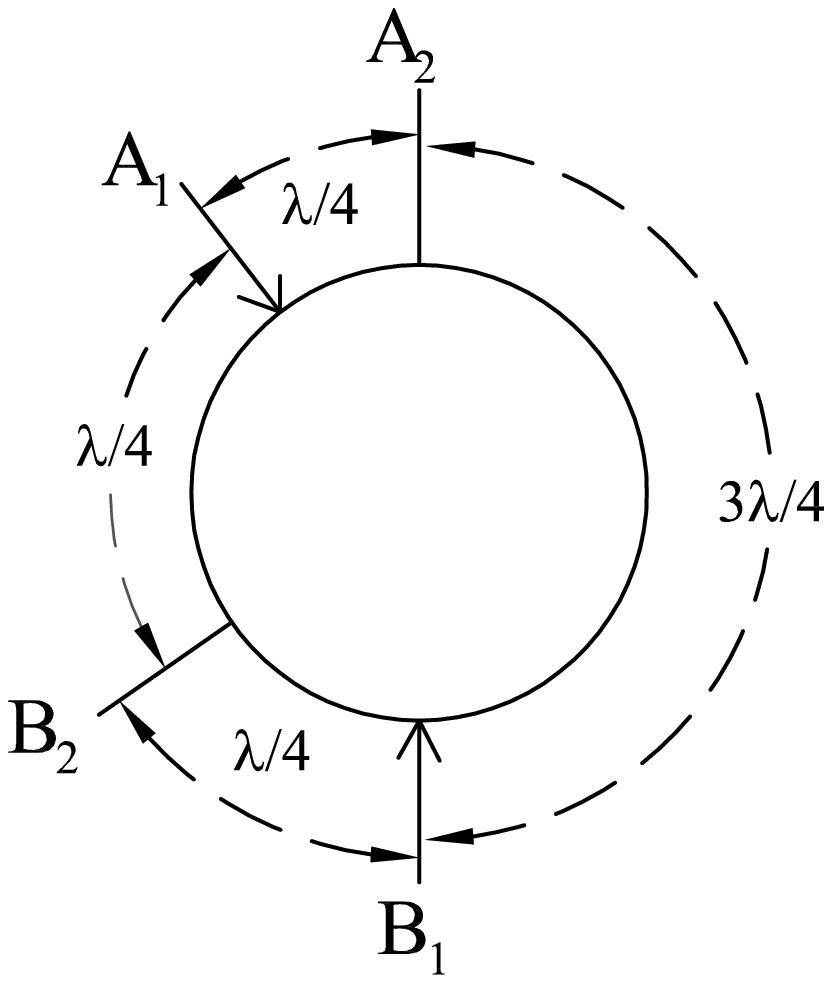} &
\multicolumn{1}{c}{ \sqinv \mmbt{1}{1}{1}{-1} }
\\
Short-Slot Hybrid Coupler \mbox{(90\deg\ Hybrid)}&
\includegraphics[height=1.5cm]{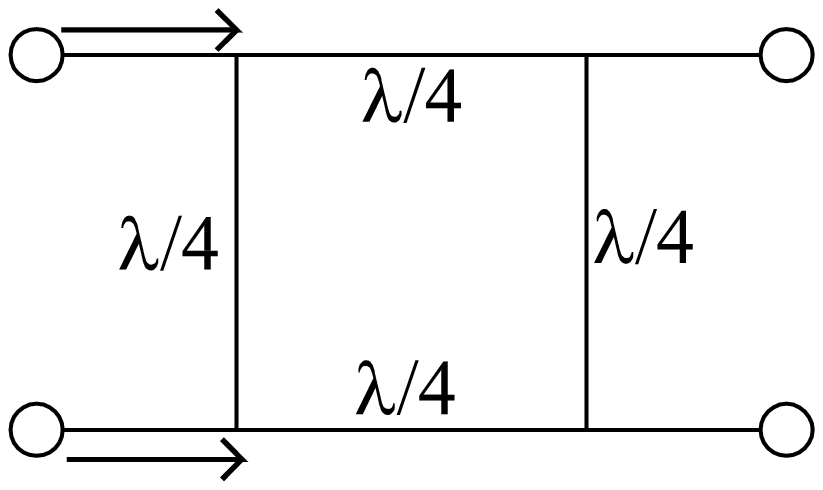} &
\multicolumn{1}{c}{ \sqinv \mmbt{1}{i}{i}{1} }
\\
& & \\[-1.5 ex]
Quarter-Wave Plate at Angle $\theta$ &
N/A &
\multicolumn{1}{c}{$ \sqinv
\mmb{-\cos{2\theta-i}}{\sin{2\theta}}{\sin{2\theta}}{\cos{2\theta}-i}$}
\\[0.5 ex]
& & \\[-1 ex]
\multicolumn{3}{c}{ {\bf Detection Devices}}
\\ \hline
& & \\[-1.5 ex]
Diode
& \includegraphics[width=2.5cm]{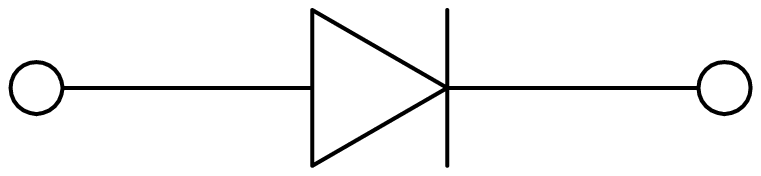}
& \multicolumn{1}{c}{\mmbt{1}{0}{0}{0} or \mmbt{0}{0}{0}{1}}
\\
Multiplier
& \includegraphics[width=2.5cm]{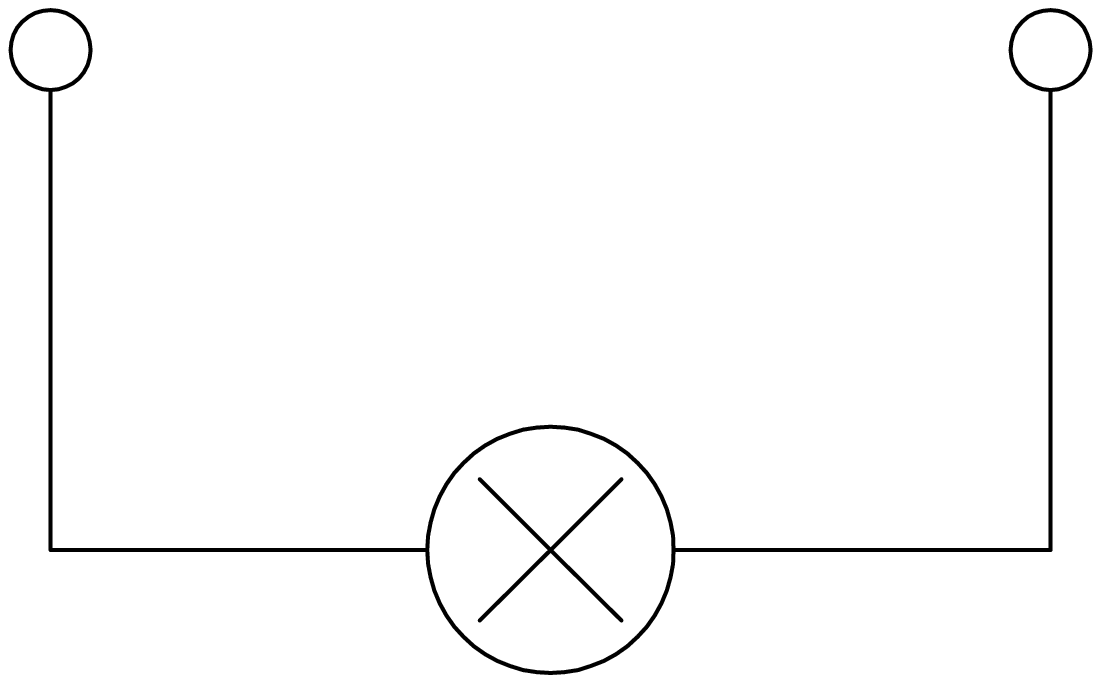}
& \multicolumn{1}{c}{\mmbt{0}{1}{1}{0} \ = U}
\end{tabular}
\end{center}
\caption[Matrix Operators for Selected Radiometric Components]
{\label{t:comp_matrices}
Schematics and matrix operators for selected radiometer components.}
\end{table}

\tbl{t:comp_matrices} shows various radiometric components,
their electrical representation, and their corresponding operator
matrix.  From the table
we see that the basic operators
take in each of the two arms of the radiometer, and produce some
linear combination of those fields in each output arm:
$\ket{E_{out}}=\M \ket{E_{in}}$, where $\M$ is the matrix operator
for that component.
``Detector Devices'' work the same but have only one (scalar)
output, so their formalism is $ \voltage = \bra{E_{in}} \D \ket{E_{in}} $,
where $\mathbf{D}$ is the matrix operator for that detector, and
\voltage\ represents the scalar output from the detector.  Hence, the
formalism can output what one actually measures, \ie\ a voltage.
The entire procedure can be described as follows:
\newpage
\begin{enumerate}
\item OMT creates the \emph{state} \ket{E_0}
\item Radiometric elements act to produce the final state before
detection, \\ $\ket{E_f} = \ldots \mathbf{M_3 M_2 M_1}
\ket{E_0} \; \equiv \; \PP\ket{E_0} $
\item Detector \D\ produces a scalar output \voltage\
$\propto \bra{E_f} \D \ket{E_f}$ = $\bra{E_0} \PP^\dagger\D\PP\ket{E_0}$ = \\
$\bra{E_0}
\mathbf{M_1}^\dagger \mathbf{M_2}^\dagger \mathbf{M_3}^\dagger
\ldots \D \ldots
\mathbf{M_3}\mathbf{M_2}\mathbf{M_1} \ket{E_0}$
\item The full gambit of  matrix operations for a given polarimeter
can be written as a linear combination of Stokes parameters,
$\PP^\dagger \D \PP = c_I \I + c_Q \Q + c_U \U + c_V \V$.
\item The $c_i$ coefficients then describe the polarimeter's
sensitivity to each of the Stokes parameters.
\end{enumerate}

\subsection{Comments on the Operators}
There are several other things to note from \tbl{t:comp_matrices}.
First, phase shifts common to both polarimeter arms are not shown,
as they are immaterial; only phase \emph{differences} between the two
arms matter.  Second, devices that neither amplify nor attenuate
have \emph{unitary} matrices, representing the conservation of
power.  Diagonal matrices do not mix the signals from the two
arms; this occurs when the signal arms do not cross (such is in
amplifier chains in the front-end).

It is interesting to see how some of these components behave.  A
septum polarizer is mathematically equivalent to a 90\deg\ hybrid
junction; the septum polarizer acts on a combined wave in square waveguide,
while the 90\deg\ hybrid acts on the already-split polarizations,
but the outputs are identical.  Thus, we conclude a septum
polarizer is the same as an OMT followed by a 90\deg\ hybrid.

Another comment is on the quarter-wave plate; when mounted with
its axis at 45\deg\ or 135\deg\ to the OMT, its effect is
equivalent to that of a 90\deg\ hybrid.  When mounted at 0\deg\ or
180\deg\, though, its effect is to add a 90\deg\ relative phase
shift between the two polarimeter arms (which can be useful in
certain situations).

\subsection{Phase Chopping}
\label{PhaseChopSec}
Many coherent radiometers employ some type of chopping
scheme within the radiometer to reduce offsets.  This is often
accomplished by electronically chopping a phase shifter in one or
both arms of the signal chain; examples of radiometers that
employ this technique are MAP and PIQUE, as well as \polar\ \cite{maphome,hed00}.
Electronically switching phase shifters are now available in
waveguide at frequencies up to 100 GHz \cite{hislop}.
Typically the relative phase difference between two arms is
chopped between 0\deg\ and 180\deg.

\renewcommand\arraystretch{0.6} 
\section{A Correlation Polarimeter}

\begin{figure}[tb]
\begin{center}
\includegraphics[height=4.3in]{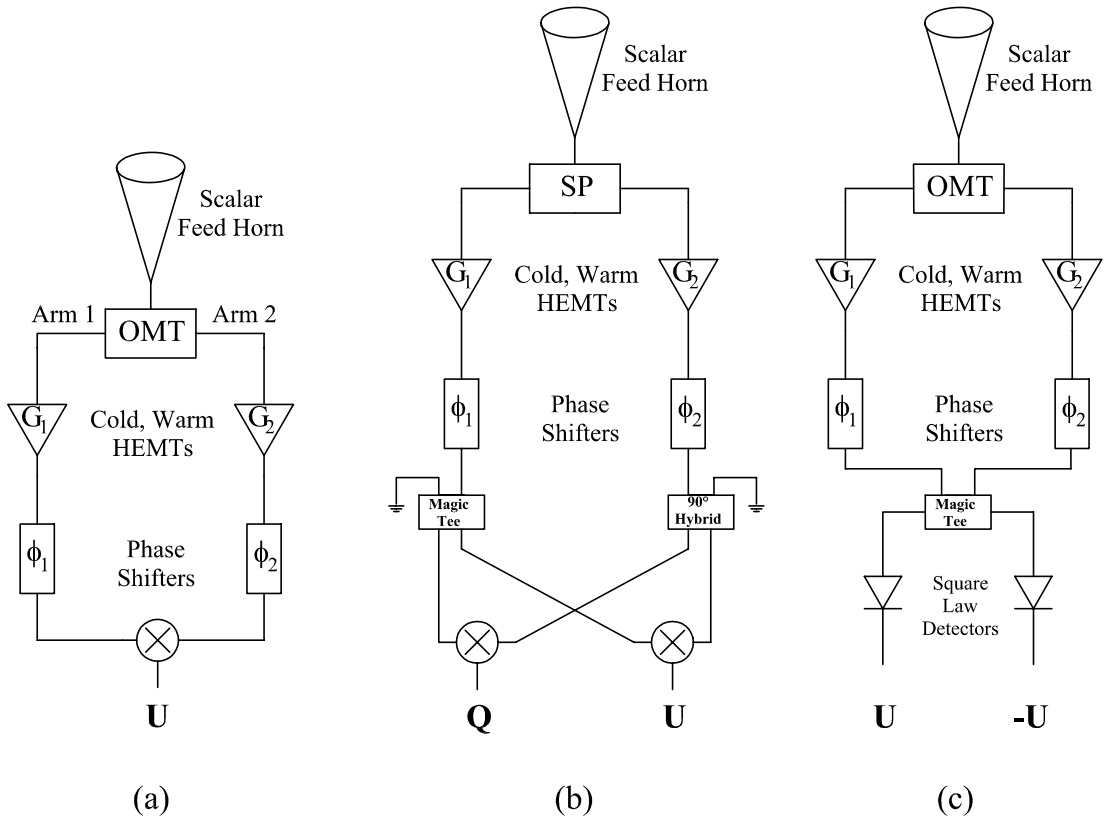}
\caption[Sample Polarimeter Configurations]
{\fixspacing\label{simplepol} Sample polarimeter configurations.  (a) depicts the basic
correlation polarimeter, which is sensitive only to U; (b) shows a
correlation polarimeter
that simultaneously detects both Q and U.
The Septum Polarizer is equivalent to
a quarter-wave plate mounted at 45\deg\ followed by an OMT, or
an OMT followed by a 90\deg\ hybrid tee; (c) is a basic
pseudo-correlation polarimeter, which also is only sensitive to U.
In all cases we assume phase-sensitive detection with a
phase chop of $\phi_2-\phi_1 = 0,180\deg$, which drastically reduces 1/f noise
introduced by the front-end amplifiers.}
\end{center}
\end{figure}

\subsection{Description}
Let us now apply this formalism to several different
polarimeter configurations,
to see which Stokes parameters these configurations actually detect.
Our first example is the basic correlation polarimeter, as shown
in \fig{simplepol}(a).  Radiation enters an axially symmetric
feed, and passes through an OMT.  The OMT sends the x-component of
the input field down one arm, and the y-component down the other.
Each signal is then amplified and then phase-shifted, although as
usual only the relative phase shift between the two arms is
important; in this idealized approach, we ignore the differential
phase shifts associated with different frequencies across the
band.  Finally, the two signals are multiplied with some type of
correlating element.
\subsection{Analysis}
Let us now analyze the behavior of this radiometer.
In our formalism, the OMT does
nothing to the Jones vector \ket{E_{in}}.  The
action of the amplifiers and phase shifters produces a state
\ket{E_f} entering the correlator given by
\begin{align} \label{e:Efsimpcor}
\ket{E_f}\; & = \; \mathbf{\Phi} \; \mathbf{G} \; \ket{E_{in}} \notag \\
& = \; \mmb{1}{0}{0}{e^{i\dphi}}\mmb{G_1}{0}{0}{G_2}\ket{E_{in}} \notag \\
& = \; \mmb{G_1}{0}{0}{G_2 e^{i\dphi}} \ket{E_{in}} \notag \\
& \equiv \; \PP \; \ket{E_{in}} \;\; \text{.}
\end{align}
where \PP\ simply encodes the action of all the components of the
radiometer preceding detection.
Now this state enters the correlator to produce
a voltage \voltage.  Referring back to \tbl{t:comp_matrices}, we
see that a correlator has the action of the \U\ operator, so:
\begin{eqnarray}\label{e:efcor}
\ & \voltage \; & \propto \bra{E_f}\; \U \;\ket{E_f}
= \bra{E_{in}} \;\PP^\dagger \U \PP \;\ket{E_{in}} \notag \\[1 ex]
\ & & = \bra{E_{in}} \mmb{G_1}{0}{0}{G_2 e^{-i\dphi}}\mmb{0}{1}{1}{0}
\mmb{G_1}{0}{0}{G_2 e^{i\dphi}} \ket{E_{in}} \notag \\[1 ex]
\ & & = \bra{E_{in}} \mmb{0}{G_1G_2e^{i\dphi}}{G_1G_2e^{-i\dphi}}{0} \ket{E_{in}}
\notag\\[1 ex]
\ & & = \; G_1 G_2 \; \bra{E_{in}}\; \U \cos{\dphi}\; -
\;\V\sin{\dphi} \;\ket{E_{in}} \notag \\
\text{or finally} \notag \\
& \voltage \; & \propto \; G_1 G_2 (U \cos{\dphi}\; - \;
V\sin{\dphi}) \quad \text{.}
\end{eqnarray}

\eqn{e:efcor} gives us a good deal of insight into the typical
OMT-based correlation polarimeter (of which \polar\ is a member).
The polarimeter is in general sensitive to a combination of both
$U$ and $V$, but is insensitive to both I and Q.
\polar\ itself has its phase difference \dphi\ chopped
from 0\deg\ to 180\deg\ (the usual case); after lock-in, it is sensitive only to
$U$.

\subsection{Possible Modifications}
Armed with this formalism, we can also answer questions pertaining
to simple changes in the correlation polarimeter configuration, such as by
adding a magic tee after the OMT or
a quarter-wave plate in front of the horn, replacing the
OMT with a Septum Polarizer, etc.  Just insert the appropriate
matrix into \eqn{e:efcor} and re-calculate.  In the case of adding
an element \M\ at the beginning, you can simply calculate
$\M^\dagger\PP^\dagger\U\PP\M$, without having to re-calculate \PP.

The effects of certain modifications to the simple correlation polarimeter
are listed in \tbl{t:spmods}.  If we phase chop from 0\deg\ to
180\deg, you can see we will be sensitive to $U$ in the simple case;
adding a magic tee really changes nothing.  Adding a quarter-wave
plate at 0\deg\ makes us sensitive to $V$.  Replacing the OMT with
a septum polarizer changes the $V$-sensitivity to $Q$ (this is equivalent
to adding a quarter-wave plat at 45\deg\ in the front-end optics,
either before or after the horn); coupling
this with another QWP at 0\deg\ would make the $\cos\dphi$
sensitive to $Q$.

\renewcommand\arraystretch{1.0}
\begin{table}\label{t:spmods}
\begin{center}
\begin{tabular}{|l|c|}
\hline
\multicolumn{1}{|c|}{{\bf Configuration}} & {\bf Output} \\ \hline
Simple Correlation Polarimeter &
$U\cos{\dphi} \;-\; V\sin{\dphi}$
\\ \hline
Insert Magic Tee after OMT &
$U\cos{\dphi}\; + \;V\sin{\dphi}$
\\ \hline
Add quarter-wave plate ($\theta$=0\deg) &
$U\sin{\dphi} \;+\; V\cos{\dphi}$
\\ \hline
Replace OMT with Septum Polarizer &
$U\cos{\dphi} \;-\; Q\sin{\dphi}$
\\ \hline
\end{tabular}
\caption[Effects of Modifications to the Simple Correlation Polarimeter]
{\fixspacing Effects of select modifications to the simple correlation polarimeter.}
\end{center}
\end{table}
\renewcommand\arraystretch{0.8}

\subsection{Simultaneous Detection of Two Stokes
Parameters}\label{QandUsimul}
It is relatively straightforward to detect both $Q$ and $U$
simultaneously (our original goal).
From \eqn{e:efcor}, we see that by introducing an
additional 90\deg\ phase shift between the two polarimeter arms,
we can detect $V$ instead of $U$.  By splitting each polarimeter arm,
introducing a 90\deg\ phase shift in one of the four resulting arms, and
correlating these two sets of arms, we can detect $U$ and $V$
simultaneously.  Finally, if we either replace the OMT with a
Septum Polarizer, or add a quarter-wave plate at 45\deg\ in the
front-end optics, we detect $Q$ and $U$ simultaneously.  The
configuration for this is shown in \fig{simplepol}(b).

Thus, it is evident from this simple example that there are many
games one can play to obtain sensitivities to different Stokes
parameters.  In the case of \polar\ we chose the simple scheme
of $U$-sensitivity only, but by rotating the instrument we
obtained both $Q$ and $U$ via \eqn{eq:stokesrotate},
at the expense only of spending 1/2 our time on each.

\subsection{Effect of OMT Cross-Polarization}\label{OMTcp}
We can extend the utility of this formalism by using it to
determine the effects of non-ideal radiometric components.
No component is perfect, and there is a specialized
lingo to discuss the degrees of these ``non-idealities'', with
motley terms such is Return Loss, VSWR, Isolation, and
Cross-Polarization.  For polarimeters, a common problem is
spurious polarization generated by cross-polarization in either
the horn or OMT (or both).  Let us now consider this effect in the
case of the simple correlation polarimeter.

Cross-Polarization refers to a portion of one of the input
polarization states being transferred down the wrong port
of the OMT; this effect can be
easily taken into account in the OMT operator as follows:
\beq{OMToperator}
\M_{omt} \; \; = \; \; \mmb{1}{\eta_{1}}{\eta_{2}}{1}
\eeq
where typically $\eta_{1} \approx \eta_{2}$, and each are of the order
of $0.001$ for a good OMT.
As the OMT is the first element in the polarimeter operator \PP\
(that is, the \emph{right-most} element), the overall operator
for the simple correlation polarimeter goes from $\PP^\dagger \U \PP$ to
$\M^\dagger_{omt} \PP^\dagger \U \PP \M_{omt}$.  Carrying out this
transformation of \eqn{e:efcor} yields
\begin{eqnarray}
\label{efcor_cp}
& \voltage \; & \propto \; G_1 G_2
[\;\{\;(\eta_1+\eta_2) \I \;+\; (1+\eta_1\eta_2) \U
\;+\; (\eta_2-\eta_1) \Q )\}\cos{\dphi} \notag \\
& \; & \; \; \;\;\;\;\;\;\;\;\; \;-\; (1 - \eta_1\eta_2) \V \sin{\dphi} \;\;] \notag \\
&  \; & \propto G_1 G_2 \left[\;(\eta_1+\eta_2) \I \;+\; \U
\; \right]
\end{eqnarray}
where the simplifications of the second step occurred because 1)
I only kept terms to first order in $\eta$,
2) I recognized that $\I\ \gg$ (\Q\ or \U\ or \V) for the CMB, thus I
kept only the \I\ terms, and 3) I assumed the usual
$\cos{\dphi}$ phase lock-in.  Thus, in actuality we detect the sum
of \U\ and $(\eta_1+\eta_2) \I$.  \U\ is
roughly $10^{-6}$ (or less) of \I\ for the CMB, and $\eta_1+\eta_2
\sim 2\cdot 10^{-3}$.  Does this mean we are defeated?  Do not
abandon hope, for as the polarimeter angle is rotated, \Q\ and \U\
are modulated, but \I\ and \V\ are not; this is how \polar\ can
discriminate the \I\ term from the \U\ term in \eqn{efcor_cp}.
Beam chopping would also be effective; as a beam is chopped
rapidly between two fairly close points in the sky, \I\ will be
roughly constant (at $T_{atm}+T_{cmb})$, but \Q\ and \U\ will
vary\footnote{\fixspacing CMB polarization is expected to vary most rapidly
on angular scales of a few arcminutes.}.  The \I\ term will of course lead to
an offset on the order of tens of mK, which is important to bear
in mind, and which indeed happened for \polar, but it is fairly
usual to have such an offset.

\section{The Pseudo-Correlation Polarimeter}
A pseudo-correlation polarimeter is the primary alternative to a
full correlation polarimeter.  The term ``pseudo-correlation''
typically refers to a radiometer that performs a multiplication of
two signals, say A and B, as $A \cdot B = \frac{1}{4}[(A+B)^2 -
(A-B)^2]$ \cite{rohlfs96}.  MAP employs this trick in its radiometer to compare
the temperatures of two different signals, however the trick can be
equally effective in the study of polarization. Although this has
never been done before in the field, it is planned on a couple
of upcoming experiments, including the Planck LFI receiver
\cite{planckhome}.

The primary advantage of pseudo-correlation is that no heterodyning
to IF frequencies is necessary.  Signals can be
detected with standard (phase-insensitive) square-law diode
detectors, which are readily available at RF.  This is in contrast to
true correlation radiometers which require a correlating device, which
generally only exist at IF (or lower) frequencies.
Another advantage to pseudo-correlation is that the diode detectors
have much lower power requirements (on the order of 30 dB
lower) than most types of
correlators.  Let us now examine what Stokes parameters this type of
system can detect.

\subsection{Analysis}
A simple pseudo-correlation polarimeter is displayed in
\fig{simplepol}(c).  As in the full correlation polarimeter,
radiation enters via a scalar feed and OMT, and is amplified in
each polarimeter arm.
Phase shifters are placed in each signal chain to give us the
opportunity to chop the relative phases of the two arms by
180\deg, as explained in \sct{PhaseChopSec}.  Next, the signals
traverse a magic tee, and the output signals are detected by
square-law detectors.

Using the matrices from \tbl{t:comp_matrices}, the operation
matrix \PP\ of the polarimeter immediately preceding detection is given
by:
\begin{eqnarray}\label{P_pcp}
\PP \; & = \; \sqinv \mmb{1}{1}{1}{-1}
\mmb{1}{0}{0}{e^{i\dphi}}\mmb{G_1}{0}{0}{G_2} \notag \\
\; & = \; \sqinv \mmb{G_1}{G_2 e^{i\dphi}}{G_1}{-G_2e^{i\dphi}}
\end{eqnarray}
Next we can determine the output of each of the two detectors,
denoted $V^{+}$ and $V^{-}$, as before:
\begin{subequations}
\begin{eqnarray}\label{V_pcp}
\mathbf{V^{+}} \; & \propto \; \PP^\dagger \mmb{1}{0}{0}{0} \PP
\\
\mathbf{V^{-}} \; & \propto \; \PP^\dagger \mmb{0}{0}{0}{1} \PP
\end{eqnarray}
\end{subequations}
which yields
\begin{equation}
\mathbf{V^\pm} \; = \; \frac{G_1^2 + G_2^2}{2}\, \I \;+\; \frac{G_1^2 -
G_2^2}{2}\, \Q \;\pm\; G_1 G_2 \left[\, \U \cos{\dphi} \,-\, \V \sin{\dphi}
\, \right] \; \; \text{.}
\end{equation}
Thus, we see in principle that our basic pseudo-correlation
polarimeter will produce a linear combination of all four Stokes
parameters, with a large offset proportional to \I.  This is of
course unwanted in a polarimeter, but if we now chop \dphi\
between the usual 0\deg\ and 180\deg, we will lock into the
$\cos{\dphi}$ term, or simply \U!  Thus, we obtain \U\ both with the
pseudo and true correlation polarimeter.

\subsection{Modifications to the pseudo-correlation polarimeter}

\begin{figure}[tb]
\begin{center}
\label{alanspol}
\includegraphics[width = 6in]{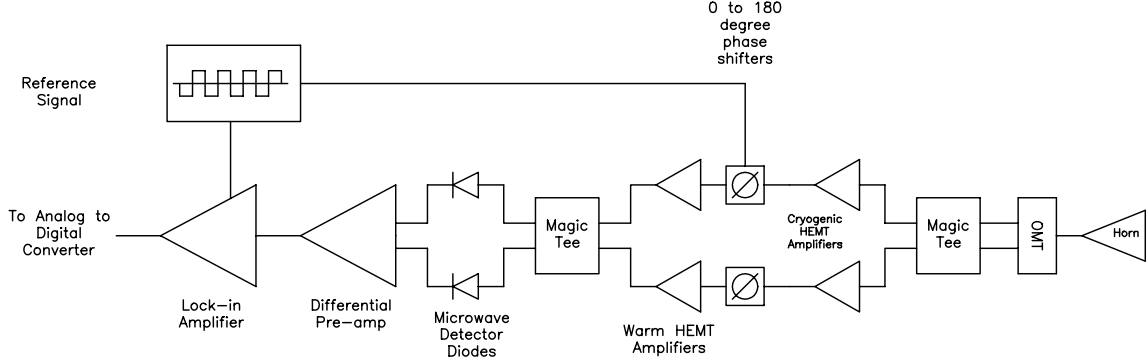}
\caption[A Complicated Pseudo-Correlation Polarimeter]
{\fixspacing A more complicated pseudo-correlation polarimeter,
employing two magic tees instead of one.  Courtesy of Alan Levy.}
\end{center}
\end{figure}

It is possible to add some bells and whistles to this
configuration.  By running the output through a differential
pre-amp, the $I$ and $Q$ offset terms are eliminated automatically
(without phase chopping), leaving only sensitivity to $U$ and $V$.
You can change this sensitivity from ($U$, $V$) to ($U$, $Q$) by
replacing the OMT with a Septum Polarizer, or by using a
quarter-wave plate.  You can
also employ the tricks described in \sct{QandUsimul} to detect
both $Q$ and $U$ simultaneously.
As a final demonstration of the power of this technique, let us
analyze the pseudo-correlation polarimeter shown in \fig{alanspol}.

Let the action of the magic tees, amplifiers, and phase shifters
be given by the matrix \PP, where
\begin{align}
\PP & = \mathbf{M G_B \Phi G_A M} \notag \\
   & = \frac{1}{2}\mmb{1}{1}{1}{-1}
    \mmb{G_{A_1}G_{B_1}}{0}{0}{G_{A_2}G_{B_2}e^{i\dphi}} \notag \\
& = \frac{1}{2}\mmb{G_1\pm G_2}{G_1\mp G_2}{G_1\mp G_2}{G_1\pm G_2}
\end{align}
where $G_1 \equiv G_{A_1}G_{B_1}$, \M\ is the matrix for a magic
tee, and the phase difference \dphi\ is chopped such that
$e^{i\dphi} = \pm 1$.

The output from the differential pre-amplifier, $\Delta\voltage$,
is the difference of the voltages from each diode detector:
\begin{align}
\Delta\voltage & = \voltage_1 - \voltage_2 \notag \\
& = \bra{E_0} \PP^\dagger \mmb{D_1}{0}{0}{0}\PP\ket{E_0} -
\bra{E_0} \PP^\dagger \mmb{0}{0}{0}{D_2}\PP\ket{E_0} \notag \\
& = \bra{E_0}\PP^\dagger \mmb{D_1}{0}{0}{-D_2}\PP\ket{E_0} \notag \\
& \equiv \bra{E_0}\PP^\dagger \D \PP\ket{E_0}
\end{align}
Solving for this in terms of Stokes parameters, we find
\begin{eqnarray}
(\PP^\dagger \D \PP)_{\pm} & \; = & \; \frac{1}{4}[
(D_1-D_2)(G_1^2+G_2^2)\: \I \;+\; (D_1-D_2)(G_1^2-G_2^2)\: \U \notag \\
& & \; \pm \; 2G_1G_2(D_1+D_2)\: \Q ] \; \; \text{.} \label{alan1}
\end{eqnarray}
Finally, upon lock-in, the constant part of \eqn{alan1} is lost
and the output is proportional to simply
$\frac{1}{2}(D_1+D_2)\:G_1 G_2 \cdot \Q$.  This is particularly nice, in
that it is insensitive to relative gain changes in the amplifiers
as well as relative gain changes in the diode detectors, and
there is no sensitivity to \I\ at all.

However, it is sometimes said that simplicity is a virtue, and it
was in that spirit that \polar\ was designed -- we opted to
construct the basic correlation polarimeter, described in full
detail in the next chapter.
\renewcommand\arraystretch{1.0} 



%
%
%
%
%

\chapter{The POLAR Instrument}\label{instrument}
The term \polar\ represents the entire experiment, including the
receiver, data acquisition system, grounds screens, rotation
mount, and housing. In this section I will give an overview of the
\polar\ instrument; the inner workings of \polar\ have been
described in detail in several previous works \cite{bkthesis,
kea01}, so I will present only a brief review of the instrument
here.  The basic specifications of \polar\ are given in
\tbl{polarspecs}.

\begin{table}
\begin{center}
\begin{tabular}{||c|c|c|c||} \hline \hline
Radiometer Type & \multicolumn{3}{c||}{Rotating Correlation Polarimeter} \\
\hline Beam Size & \multicolumn{3}{c||}{7\deg} \\ \hline Location &
\multicolumn{3}{c||}{Lat: 43.08\deg N, Lon: 89.69\deg W}\\ \hline
Scan Strategy & \multicolumn{3}{c||}{Zenith Drift Scan}\\ \hline
Rotation Frequency & \multicolumn{3}{c||}{0.0325 Hz} \\ \hline
Acquisition Frequency & \multicolumn{3}{c||}{20 Hz} \\ \hline
Frequency Bands [GHz]& \hs 32--36 \hs & \hs 29--32 \hs & 26--29 \\
\hline Bandwidths [GHz] & 3.25 & 3.0 & 2.45 \\ \hline Receiver
Temperatures & 32 K & 34 K & 45 K \\ \hline NETs [$mK s^{1/2}$] & 1.1
& 1.2 & 1.8 \\ \hline Amt. of Good Data & \multicolumn{3}{c||}{\app\
100 hours} \\ \hline \hline
\end{tabular}
\caption[POLAR specifications]
{\fixspacing \label{polarspecs}
POLAR specifications.
}
\end{center}
\end{table}

\begin{figure}
\begin{center}
\includegraphics[height=6in]{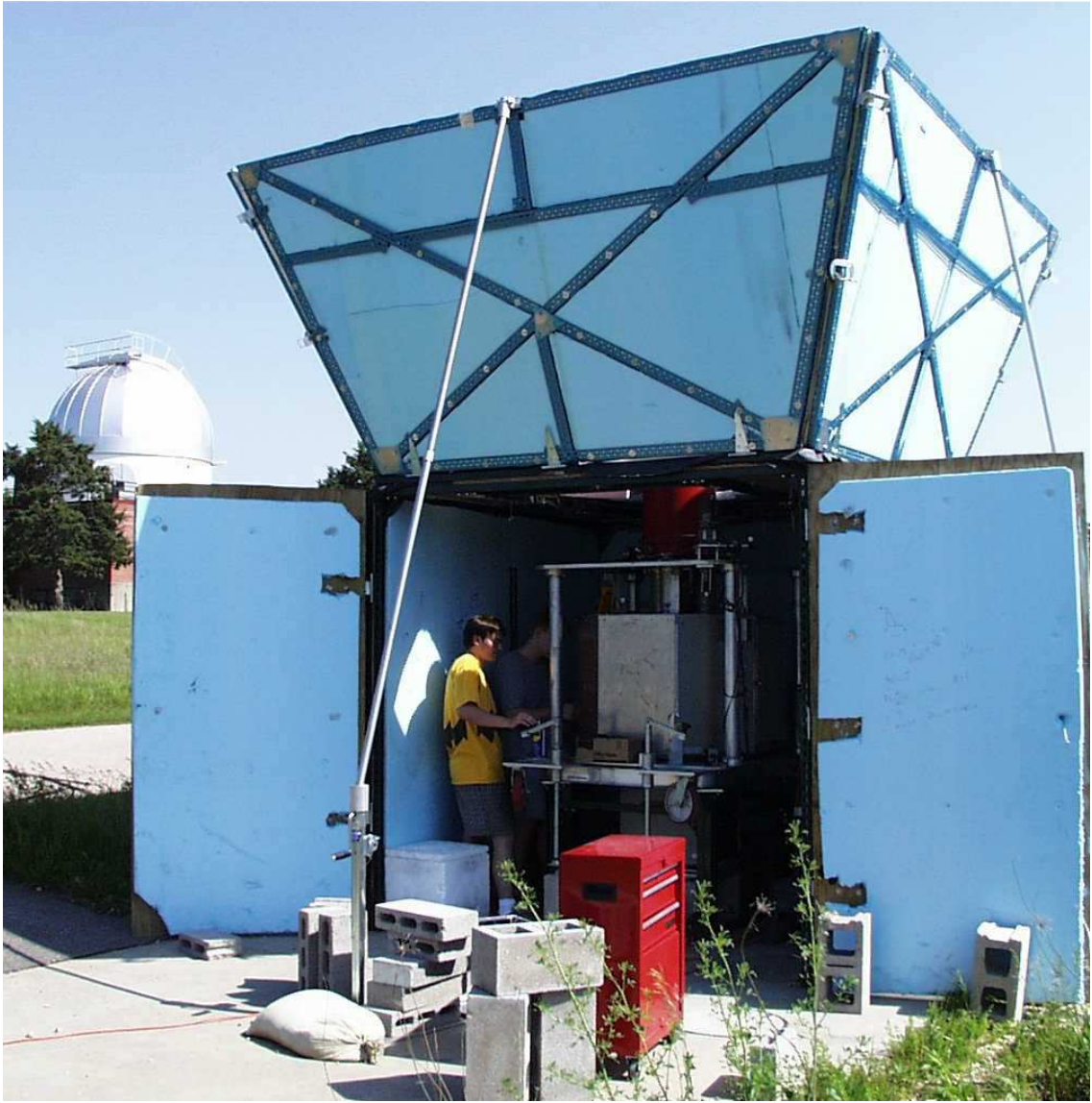}
\caption[A View of \polar\ From Outside its Enclosure]
{\fixspacing \label{dan_cube} A view of \polar\ from outside its
enclosure, in Pine Bluff, WI.}
\end{center}
\end{figure}

As mentioned in the previous chapter, \polar\ is sensitive
primarily to U, but as it rotates at \app\ 2 rpm, it can recover
both Q and U.  The \polar\ receiver itself is a superheterodyne correlation
polarimeter, and consists of three main parts: the cooled
front-end, the room temperature back-end receiver and electronics,
and post-detection electronics. A photograph of the instrument taken at the end of
the 2000 observing season is shown in \fig{dan_cube}.

\section{The Front End}
\begin{figure}[tb]
\begin{center}
\includegraphics[height=4in]{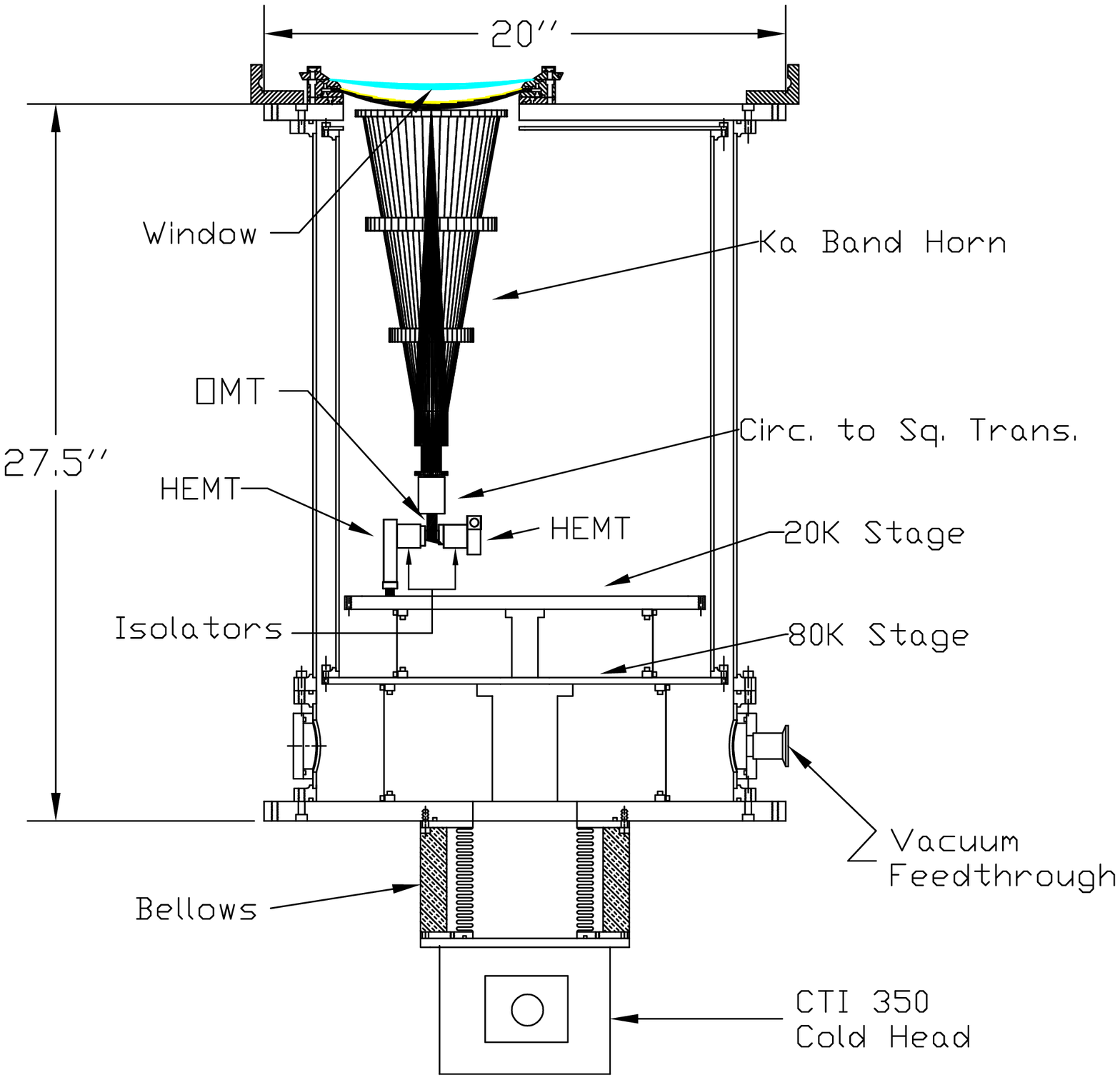}
\caption[The \polar\ Dewar]{\fixspacing \label{dewar} The \polar\ dewar.  The dewar is
overly-large due to allowance for future receivers to be added at
Q and W bands.}
\end{center}
\end{figure}

The front end of the receiver is shown schematically in figure
\fig{dewar}. The cold components reside in a cylindrical dewar
\footnote{\fixspacing Constructed by Precision Cryogenic Systems,
Inc., Indianapolis, IN}. The dewar has both a 20K and 80K stage, and
is cooled by a CTI-350 cold head in conjunction with an air-cooled,
mechanical compressor. The dewar pressure was $1\cdot 10^{-6}$ torr
for months at a time, and the cold head temperature was typically 22
K, although it was somewhat coupled with the ambient air temperature
and in practice fluctuated between 21 and 23 K daily.

\begin{figure}[tb]
\begin{center}
\includegraphics[width=6in]{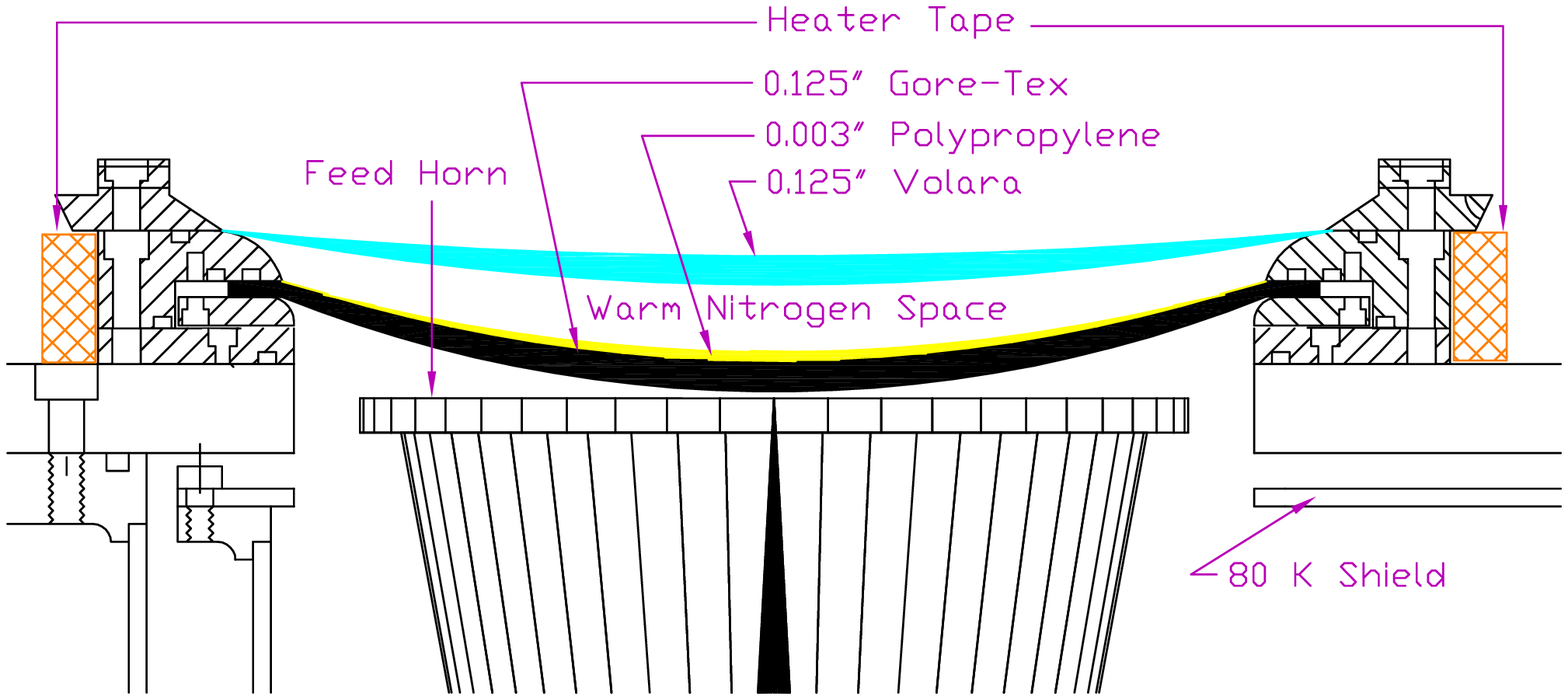}
\caption[The \polar\ Vacuum Window] {\fixspacing \label{window} The
\polar\ vacuum window.  The window consists of three layers: a 3-mil
polypropylene layer to hold the vacuum, a layer of gore-tex for
support against atmospheric pressure, and a layer of Volara above a
dry nitrogen space to prevent water condensation on the window.
Heater tape wrapped around the window kept the nitrogen space warm.
The top window ring was also designed to accept the calibrator, which
could be rotated above the dewar.}
\end{center}
\end{figure}
A brief summary of the signal chain is as follows:
microwave signals pass through a virtually transparent vacuum
window on the dewar face and into a cooled, corrugated microwave feedhorn. The
signal is then split by an orthomode transducer (OMT), which
separates the x- and y- components of the incident electric field.
Both signals then pass through cryogenic, low-loss isolators and
into HEMT amplifiers.  After amplification, the joint signals
travel out of the dewar and into the warm radiometer box, which
houses the back-end microwave components.  The signals are further
amplified, mixed down to IF frequencies, multiplexed into three
\ka\ sub-bands, and finally correlated together.  After the
post-detection electronics, the signals are recorded to computer.
Each of these steps will now be described in more detail.

\subsection{The Vacuum Window}
Our vacuum window is a true testament to human persistence. It
evolved over a period of years\footnote{\fixspacing Thanks to Karen
Lewis and Jodi Supanich, both of whom invested much of their lives in
the design and testing of our vacuum window technology.}, starting
from the window design of the MSAM-II radiometer \cite{gwthesis}.
Initially, a 20-mil thick polypropylene window was used, as this had
both the necessary strength and composition to hold the vacuum;
however, it had some severe drawbacks.  A minor drawback was the
transmissivity of the window to infrared; as most of the IR got
through, it increased the radiative loading on the dewar.  More
importantly, though, was the fact that the reflection coefficient is
relatively high for a 20-mil layer of polypropylene, \app\ 10\% for
30 GHz. For an anisotropy experiment only interested in intensity,
this would not be a big problem, because the roughly 2 K reflected
back into the dewar is fairly minor loading.  For a polarization
experiment however, this can be disastrous; a significant fraction of
that reflected radiation can be polarized, leading to a large offset
and much more stringent stability requirements.

Our team opted to avoid these potentially polarized offsets by
constructing a window with an explicitly low reflection coefficient
for microwaves. The vacuum window we arrived at is shown in
\fig{window}.  The window holds vacuum using a 3-mil (0.003") thick
polypropylene window. This thin layer of plastic has a reflection
coefficient of only $\sim$ 0.2\% (compared with 10\% for the 20-mil
case). By itself, the thin polypropylene layer was not strong enough
to hold against atmospheric pressure, and necessitated using a thin
(1/8'') layer of Gore-Tex, which has the strength to hold the
pressure across the 6.75''-diameter window (about 500 lbs). In
addition, we found the Gore-Tex was almost completely opaque to
infrared; this helped reduce the IR loading on the dewar, as the
Gore-Tex layer was somewhat cooler than ambient air.

When this window was used in the field, significant condensation
formed on the window, and proved an irksome issue throughout our
campaign.  The main defense against this condensation was use of a
layer of Volara\footnote{\fixspacing Voltek Corporation}, which is an
airtight, expanded polyethylene foam material.  The Volara created a
space which we filled with dry nitrogen and kept closed off from the
ambient air. In the mornings, dew often condensed on the Volara,
which was a separate problem.  To combat this, heater tape was
wrapped around the window structure; this kept the dry nitrogen (and
hence the Volara layer) warm enough to usually hold off the dew.  The
loss of the entire window structure was found to be very low
(emission less than 1 K).

\subsection{Corrugated Feed Horn}
After passing through the window, radiation enters our conical
corrugated feedhorn. Our feedhorn\footnote{\fixspacing designed by
Josh Gundersen} was loosely based on the COBE 7\deg\ corrugated
feedhorn \cite{cobehorn}. The recipe used to design the horn is given
in Grant Wilson's thesis \cite{gwthesis}, who himself used the
classic book by Clarricoats and Olver \cite{clarricoatsandolver84} to
design the feed, and the excellent paper by Zhang to design the
wide-band mode launcher \cite{zhang93}.

\begin{figure}[tb]
\begin{center}
\includegraphics[height=3.5in]{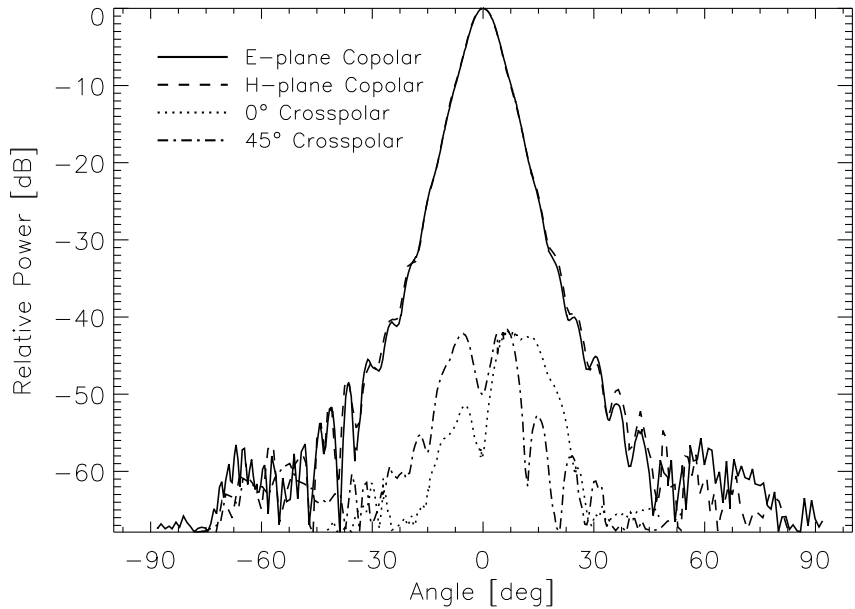}
\caption[Measured Feedhorn Beam Patterns at 29 GHz]
{\fixspacing \label{beam29meas} The measured beam patterns at 29 GHz for the
\polar\ corrugated feedhorn. The solid line is the E-plane copolar pattern,
the dashed line is the H-plane copolar pattern, the dotted line is the E-plane
cross-polarization pattern, and the dot-dashed line is the cross-polarization
pattern measured at 45\deg to the E-plane.  Note the close agreement
of the E- and H-plane patterns, the low sidelobes of the copolar pattern, and
the low level of cross-polarization in the main beam.}
\end{center}
\end{figure}

The measured beam pattern of the horn at 29 GHz is shown in \fig{beam29meas}.
Notice that the sidelobes are very low, near -70 dB at $\pm
90\deg$ off-axis.  Also, the cross-polarization response is quite
low, better than -40 dB down from the main lobe response.

Of course, the beam pattern will vary with
frequency, and this does not follow a strict 1/$\lambda$ behavior.
We performed beam pattern measurements at 26, 29,
and 36 GHz, but we would ideally like to know them at the \polar\
sub-band center frequencies of 27.5, 30.5, and 34 GHz. Luckily,
one can successfully model the beam response of a corrugated feed
using a simple model that expands the electric field in
Gauss-Laguerre modes \cite{wylde93}.
\begin{figure}[tb]
\begin{center}
\includegraphics[height=3.5in]{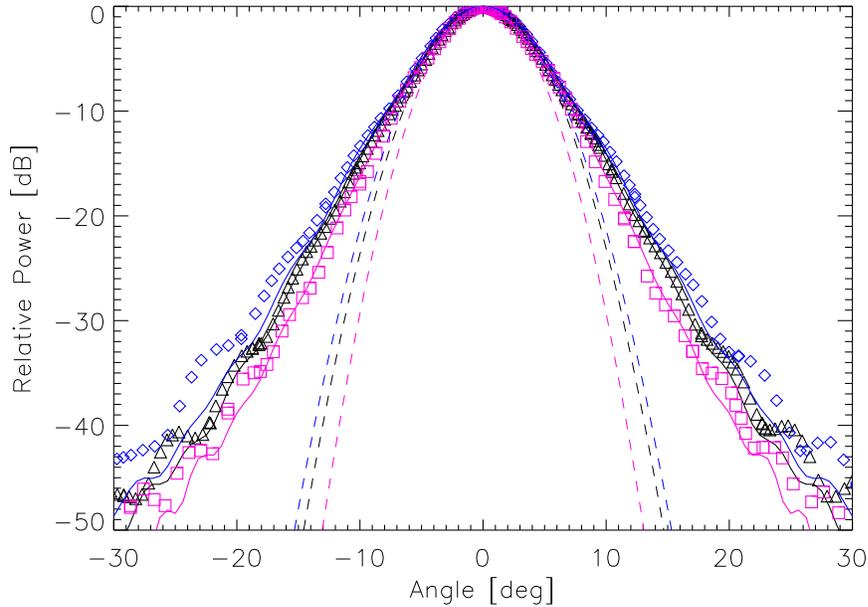}
\caption[Gauss-Laguerre and Gaussian beam models compared to measured
beam patterns]{\fixspacing \label{GLmodel} Gauss-Laguerre and Gaussian beam models compared to measured
beam patterns.  The diamonds (26 GHz), triangles (29 GHz), and squares (36 GHz)
are the measured beam patterns of the \ka\ horn in the E-plane.  The solid lines
represent the corresponding Gauss-Laguerre approximations, while the dashed lines
are the best-fit Gaussians to the main beam of the data.  The Gauss-Laguerre theory
involves no fitting -- it has zero free parameters.}
\end{center}
\end{figure}
As shown in \fig{GLmodel}, the Gauss-Laguerre model is an excellent
approximation to the beam out to \app\ -40 dB. The beam pattern is
Gaussian only over $\sim$ 1 FWHM, but later we will see that in the
simplest data analysis, modelling our beam as a Gaussian is
sufficient.  Using this model, we can predict the approximate
beamwidth at any frequency. The results for \polars\ frequencies are
shown in \tbl{t:fwhm}.

\begin{table}
\begin{center}
\begin{tabular}{|| c | c ||}
\hline\hline
Frequency [GHz] & $\theta_{FWHM} \: \pm \: 0.1\deg$ \\
\hline \hline
27.5 (J3)& 7.5\deg \\
30.5 (J2) & 7.0\deg \\
34.0 (J1) & 6.4\deg \\
\hline\hline
\end{tabular}
\end{center}
\caption[Beam FWHM's for \polars\ three \ka\ sub-bands using
Gauss-Laguerre theory]{\fixspacing \label{t:fwhm} Beam FWHM's for
\polars\ three \ka\ sub-bands, using Gauss-Laguerre theory.}
\end{table}

The principle information to take away from the beam maps is that
1) we accurately measured the main beam response, including the
FWHM's in the Gaussian approximation, and 2) the sidelobes are
very low, better than -50 dB at more than 35\deg\ off-axis.

\subsection{Cold Microwave Components: Isolators, OMT, HEMTs and Waveguides}
\begin{figure}[tb]
\begin{center}
\includegraphics[height=3in]{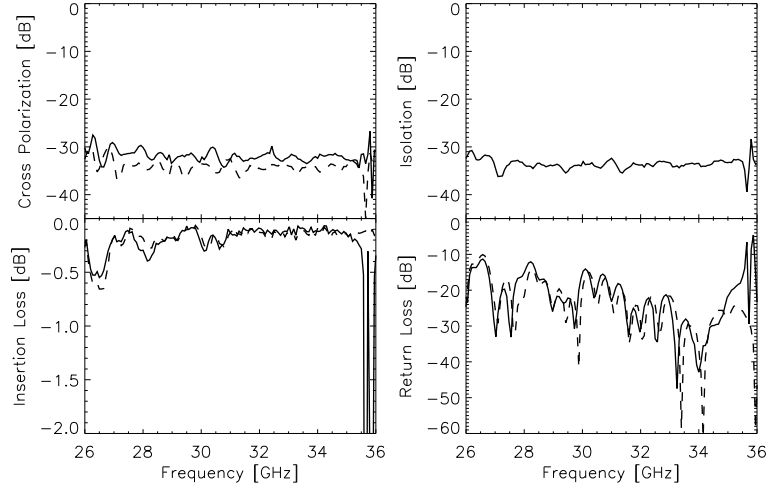}
\caption[Measured OMT Properties]
{\fixspacing \label{f:OMT}
Measured OMT Properties.
Solid lines correspond to the E-port properties, and dashed lines
to the H-port properties.  Isolation is symmetric between the two
ports.}
\end{center}
\end{figure}

The output signal from the corrugated feedhorn ends up in a circular
\ka\ band waveguide section, propagating the $TEM_{11}$ mode.  This
mode contains both x- and y-polarizations.  Then the signal traverses
a circular-to-square waveguide converter (maintaining both
polarizations) and enters an Orthomode Transducer
(OMT)\footnote{Fabricated by Atlantic Microwave, Bolton MA.}, which
separates the x- and y- components of the incident electric field.
The OMT is a critical component in our system, as our offset is
directly based on how well the OMT does its job of completely
separating the two polarization states; this effect was derived in
\sct{OMTcp}.  Graphs of some measured properties of our OMT are shown
in \fig{f:OMT}.  No OMT is ever perfect, and we measured our OMT to
have a roughly -30 dB cross-polarization (that is, roughly 0.1\% of
the power in the x-polarization state is transmitted to the y-output
port, and vice-versa).  Notice also that as much as 10\% of the input
power is reflected off the front port of the OMT; this effect serves
to make our system less efficient, and can lead to standing waves
through additional reflections from the window.

Both signals then pass through cryogenic, low-loss
isolators\footnote{PAMTech Inc.,  Camarillo, CA} and into HEMT
amplifiers. The isolators had better than -22 dB isolation. The
purpose of the isolators is to reject any backward propagating
radiation that can lead to false signals, such as happens when some
of the signal entering a HEMT is rejected off the input port and
returns through the OMT, and then propagates down the other arm of
the system.  This renders the OMT isolation rather unimportant; OMT
cross-polarization will be the dominant effect in offset generation.

The High Electron Mobility Transistor (HEMT) amplifiers
\footnote{\fixspacing Serial Numbers A31 and A32, loaned to us
graciously by John Carlstrom of the University of Chicago.} were
manufactured by Marion Pospieszalski at NRAO in 1994, and later
retrofitted with low-noise, first stage Indium Phosphide (InP)
transistors \cite{hemt95}.  These gave them noise temperatures of
\app\ 20 K.  It is currently not possible to obtain lower noise
amplifiers from any commercial vendor at these frequencies; they are
truly ``the most sensitive detectors in the world''.

Despite their amazing noise characteristics, the HEMTs are not
perfect.  As mentioned before, they reflect a non-negligible fraction
of the incident power back towards the OMT, and necessitate the use of isolators.
More importantly perhaps, their gain is not particularly flat; the
roughly 25 dB of gain varies by as much as 5 dB across our nominal 10
GHz-wide band.  These gain variations lower the effective
bandwidth of the instrument, although the effect is partially
compensated for by multiplexing into three sub-bands, as the
sub-bands have much smaller relative gain variations than does the
overall band; this effect is described in more detail in \S\ref{s:if}.

After this critical first stage of amplification the signals in
each arm of the polarimeter (now averaging about
$10^{-9}$ Watts) exit the dewar via a complicated,
three-dimensional bending waveguide path,
made of coin-silver waveguide.  The last six inches of this waveguide are made of
stainless steel, which provides the necessary thermal break
between the cold and warm components of the system.

\subsection{Dewar Temperature Observations and Control}
\begin{figure}
\begin{center}
\includegraphics[height=2in]{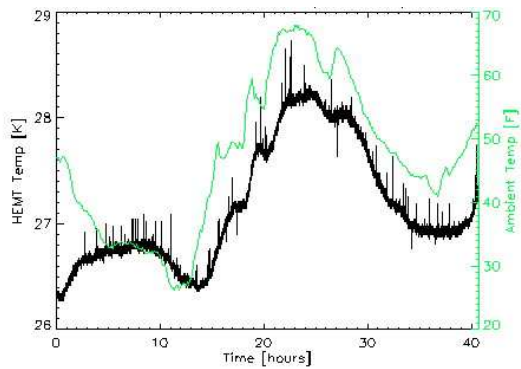}
\caption[Dewar Temperature Variations]
{\fixspacing \label{hemttemp} Dewar and ambient (air) temperature changes
over a two-day period.  The dewar changed by about 1 K for every 10 K change in air temperature.}
\end{center}
\end{figure}

The \polar\ dewar had no system of temperature control, perhaps to
the detriment of the experiment.  We recorded temperatures at three
points in the dewar using silicon diodes
\footnote{\fixspacing Serial Number DT-470, Lakeshore Cryotronics, Inc.,
Westerville, OH.}; on the feedhorn, cold
plate, and one HEMT body.  These temperatures turned out to be quite
coupled to the ambient air temperature; a 10\deg C change in air
temperature resulted in an approximately 1\deg K change in
temperature inside the dewar.  A sample two-day period is shown in
\fig{hemttemp}; the coupling between HEMT and ambient temperature is
quite striking.

Thus, we are faced with a diurnal temperature variation of \app 3
K.  Our signal was measured on the timescales of a single rotation
($\sim$ 30 sec).  Over this timescale a believable HEMT
temperature drift might be 10 mK.  This could lead to a change in
total power output by the HEMTs, but there is really no mechanism
for this to lead to an apparent polarized signal, as the HEMT noises are
uncorrelated.

\section{The Back End}
\begin{figure}
\begin{center}
\includegraphics[width=6in]{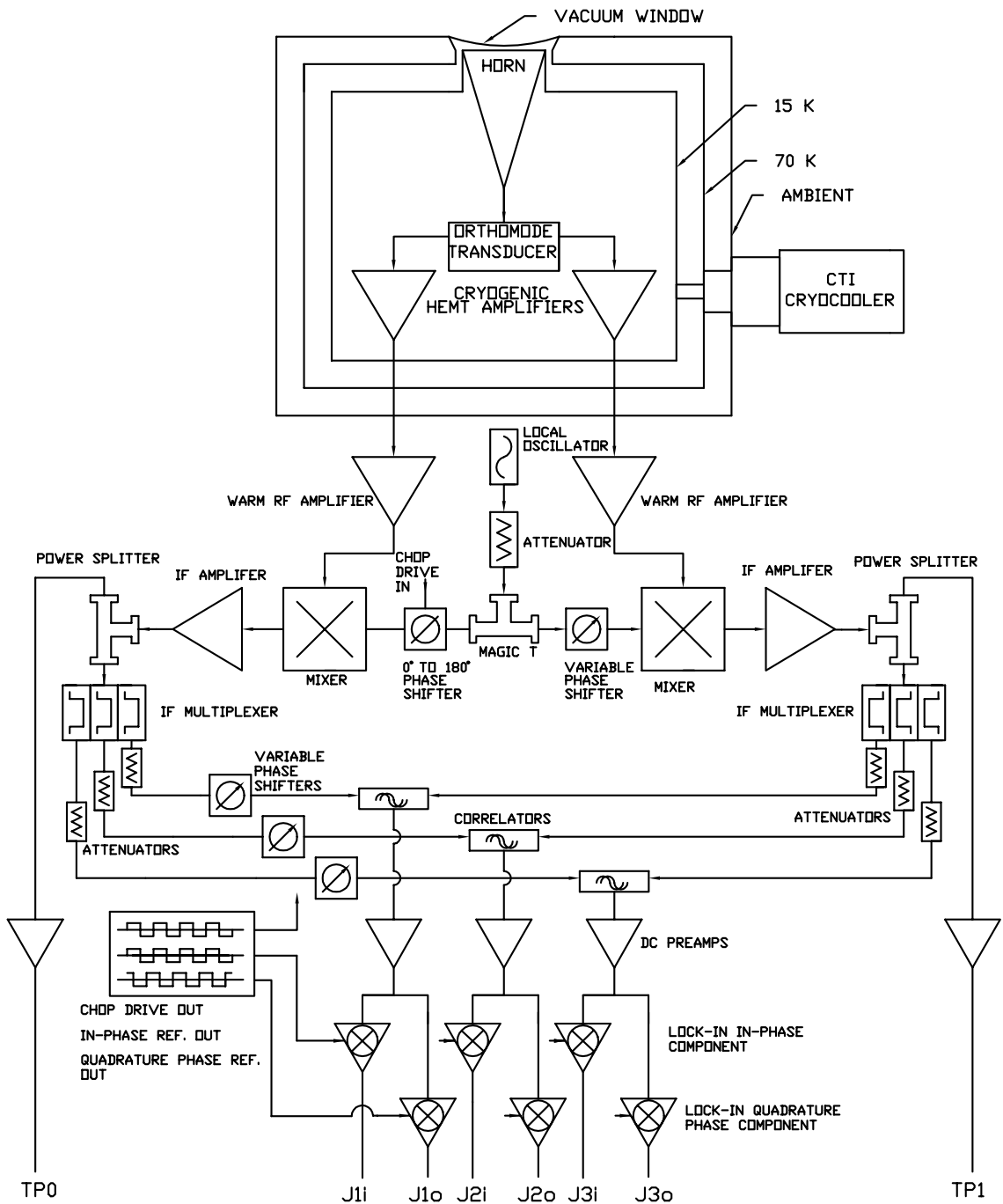}
\caption[The Signal Chain]
{\fixspacing \label{signalchain} The \polar\ signal chain.}
\end{center}
\end{figure}
\subsection{RF Components}
After leaving the dewar, the two signals traverse a section of copper
waveguide and then pass into the Room-Temperature Radiometer Box
(affectionately called the \textbf{RadBox}), which houses the warm
radiometric components. \fig{signalchain} shows the entire signal
chain. First, the nanowatt signals are amplified again by commercial
amplifiers
\footnote{\fixspacing Miteq Corp.} of approximately 20 dB of gain, and $\sim$
300 K noise temperatures. This may seem large, but the relative
amount of noise added goes like one over the gain already experienced
by the system, namely \app 25 dB in the cold HEMTs, so the relative
noise added in this second stage of gain is a mere $300/10^{2.5} \;
\sim \; 1$ K.
\subsection{Frequency Downconversion}
The signals are then mixed down from RF (radiofrequency, 26-36 GHz)
to IF (intermediate frequency, 2-12 GHz) via commercial mixers. The
mixers are fed with a pure 38 GHz signal from a local oscillator
(LO), which is subsequently split by a magic tee. One of these LO
signals has a manually adjustable phase-shifter, so that one can tune
the relative phase difference between the two LO signals, and thus
between the resulting IF signals. The other LO signal contains an
electronically-chopped phase switcher, which allows lock-in detection
at the end of the signal chain; the phase switcher is square-wave
chopped at 967 Hz, between 0\deg\ and 180\deg (this was previously
discussed in \sct{PhaseChopSec}).

\subsection{IF Amplification and Multiplexers}\label{s:if}
Following downconversion, the signals are further amplified by
commercial IF amplifiers, and then split by commercial power
splitters.  One of each of these signals is then sent through a diode
detector to monitor the total power level in that arm.  These total
power channels (referred to hereafter as TP0 and TP1) are very
powerful monitors of atmospheric fluctuations,
 HEMT gain variations, and other systematic effects.  Note that the TP signals
are not chopped at the LO chop frequency, as their diode detectors are not phase sensitive.

The remainder of the split signals are themselves
further amplified to approximately +12 dBm of power (approximately 15 mW),
which will be required to fulfill the correlator power requirements. The
signals are then \emph{multiplexed} into the three $K_a$ sub-bands
J1 (2--6 GHz), J2 (6--9 GHz), and J3 (9--12 GHz).  The purpose of
multiplexing is three-fold.  First, one gains information on the
frequency dependence of observed signals.  Second, as we have no
true filters, they are the primary band-defining element.  The
HEMTs and the feedhorn themselves have of course reasonably
well-defined bands which minimize the pickup of out-of-band
radiation, but the multiplexers ensure that \emph{no} out-of-band
signals propagate to the correlators.  And finally, as alluded to before,
the multiplexers serve to \emph{flatten} the frequency response of
the system, and hence increase the overall effective bandwidth.
This is because effective bandwidth is given by \cite{rohlfs96}
\begin{equation}\label{nu_eff}
\Delta \nu \; = \; \frac{\left(\int_{0}^\infty G(\nu) d\nu \right)^2}
{\int_{0}^\infty|G(\nu)|^2 d\nu}
\end{equation}
Notice if $G(\nu) = 1$ across a finite band (say from $f_a$ to
$f_b$) and is zero elsewhere, then the effective bandwidth is
$\Delta\nu = \frac{(f_b-f_a)^2}{f_b-f_a} = f_b - f_a$,
which is the maximum possible value.
The more the band deviates from flatness, the smaller will be the
effective bandwidth.  The bandwidth measurements are discussed at the
end of this chapter.

\subsection{The Correlators}
Once multiplexed, the six paths (three channels, two polarizations)
pass through attenuators to bring them all to approximately 0 dBm of
power (when looking at the sky).  We found that the correlators
needed at least -1 dBm of bias power into each arm to function
properly, but more than +5 dBm caused them to saturate.  This means
that the total bias power can vary by only a factor of four and still
ensure a linear response. Based on the radiometer equation,
\eqn{radeq}, it is possible to see the maximum antenna temperature
that the receiver can handle and still stay linear.  If there is +0
dBm of power when looking at the \app\ 10 K sky, then the radiometer
can accept at most 130 K of antenna temperature\footnote{\fixspacing
The real limiting factor for linearity turned out to be the final
stage of amplifiers, which only stayed linear up to $\sim$ + 16 dBm
output, which limited them to $\sim$ 93 K of signal temperature).}.

After this attenuation balancing act, each of the three channels in
the ``x'' signal chain passes through a tunable IF phase adjuster
before being correlated with its mate from the ``y'' signal chain.
This is highly desirable, because if the two lines have picked up
different phase shifts since they separated in the OMT, then their
correlation will decrease (even if they were perfectly correlated to
begin with), to a minimum of zero correlation if there was a 90\deg\
relative phase shift between them.  Finally, the two polarizations
from each of the three channels are brought together in the
correlators, which themselves are double-balanced mixers
\footnote{\fixspacing Miteq Corp.}.
The output from the correlators is DC, but only correlated
input signals will produce an non-zero output.  Because of the
chopped phase in the LO signal, a correlated input signal to the
polarimeter will produce, at the output of the correlators, a chopped
signal at 967 Hz of $\pm V$, where $V$\ is proportional to the
correlated power of the two input polarization states.

\subsection{Radbox Temperature Control}\label{radboxtemp}
The fact that many of the components in the Radbox were temperature
sensitive necessitated a robust system of temperature control.
Without temperature control, gains from amplifiers, and power output
by the mixers and LO would all drift due to their inherent
temperature dependence, and add to the 1/f noise of the system.  To
minimize temperature drifts, all components inside the Radbox were
heat sunk to a 0.25'' aluminum plate.  The plate itself was
temperature-controlled via a standard PID circuit\footnote{Manufacted
by OMEGA.}.  The PID circuit keyed off a temperature sensor mounted
directly to the local oscillator, which we found to be the most
critical component in terms of temperature dependence. The PID
circuit controlled the temperature via heaterfoil pads
\footnote{\fixspacing MINCO.}
mounted underneath the aluminum plate. We found the temperature
inside the RadBox to be stable to $\sim$ 100 mK per day.

\section{Post-Detection Electronics and Data Acquisition}
Five channels thus leave the radiometer box: the two total power
channels, TP0 and TP1, corresponding to the power in the x- and y-
polarizations of the signal, and the three correlator channels labelled J1,
J2, and J3.  In voltage units, all these signals are small (a few
$\mu V$ for the correlators), and are further amplified by a
home-made differential pre-amplification circuit.

The correlator channels then pass into a lock-in amplifier that
correlates the channels with the square-wave reference signal feeding
the LO phase chopper. They are also correlated with a
``quadrature-phase'' signal, which is the input chop signal shifted
in phase by 90\deg.  This results in six correlator signals, three
from the in-phase correlation and three from the quadrature-phase
correlation.  The quadrature-phase channels (hereafter QPC) are not
sensitive to signal and thus serve as a powerful check of systematics
and problems with the electronics.

Following lock-in detection, all channels pass through 5 Hz,
24 dB/octave low-pass filters to prevent
aliasing.  These eight signals are then recorded at 20 Hz via a
Labview-controlled, National Instruments data acquisition system running on a
Windows-98 laptop\footnote{\fixspacing This laptop did not crash a single time
during the March-May 2000 observing season, in spite of its operating
system.}, mounted inside the rotation stage.  Finally, data files were
written every 7.5 minutes over an ethernet connection to a remote
computer, where they awaited later analysis.

\begin{table}
\begin{center}
\begin{threeparttable}
\begin{tabular}{|c|c|}
\hline
\multicolumn{2}{|c|}{ {\bf\large Data Channels} } \\
\hline
\multicolumn{1}{|c|}{{\bf Channel}} & {\bf Description} \\ \hline
TP0 & Total Power Channel 0 (HEMT A31), ``$I_x$'' \\
TP1 & Total Power Channel 1 (HEMT A32), ``$I_y$'' \\
J1i & In-Phase Correlator Channel, 32-36 GHz \\
J2i & In-Phase Correlator Channel, 29-32 GHz \\
J3i & In-Phase Correlator Channel, 26-29 GHz \\
J1o & Quad-Phase Correlator Channel, 32-36 GHz \\
J2o & Quad-Phase Correlator Channel, 29-32 GHz \\
J3o & Quad-Phase Correlator Channel, 26-29 GHz \\
\hline
\hline
\multicolumn{2}{|c|}{ {\bf\large House-Keeping Channels} } \\
\hline
$T_{cp}$ & Temperature of Cold Plate (typ. 24 K) \\
$T_{hemt}$ & Temperature of HEMTs (typ. 27 K) \\
$T_{horn}$\tnote{1}
& Temperature of Horn (typ. 40 K) \\
$T_{LO}$ & Temperature of Local Oscillator (typ. 39\deg\ C) \\
AOE & Absolute One-Bit Encoder \\
Encoder\tnote{2} & 16-Bit Relative Encoder \\
Radmon & Current monitor for RadBox Temperature Control Unit \\
Octomon & Current monitor for Octagon Temperature Control Unit \\
\hline
\end{tabular}
\caption[Description of DAQ Channels]
{\fixspacing \label{daqchannels}
Data Acquisition Channels for \polar.}
\begin{tablenotes}
\fixspacing
\item [1] {\footnotesize Failed very early in season.}
\item [2] {\footnotesize Failed after approximately one month of data
taking.}
\end{tablenotes}
\end{threeparttable}
\end{center}
\end{table}
In addition to the eight primary signal channels, the data
acquisition system was capable of recording eight addition
channels of data (all were in non-referenced, single-ended mode).
\tbl{daqchannels} provides a list of all channels and their
descriptions.
These included three temperature sensors in the dewar, one
temperature sensor in the radiometer box, the AOE and encoder signals (described
below), and the current being used by the two temperature control
units.

\subsection{A Note on the Lock-In Technique}
As stated previously, the primary advantage of phase chopping the
signal and subsequent lock-in detection is to drastically reduce 1/f
noise, due primarily to gain fluctuations in the HEMTs. However, one
might naively assume that there is a price associated with phase
chopping, but this is not so.  Because we are chopping from 0\deg\ to
180\deg, a correlated signal will be chopped between some $\pm V$,
where V is proportional to the level of polarization in the signal.
As we are square-wave chopping, no integration time is actually lost,
and thus our sensitivity is not degraded by any factor of
$\sqrt{2}\;$ as might be expected.  This would \emph{not} be the case
if we were sine-wave chopping, where some integration time is lost,
and one would lose a factor of $\sqrt{2}\;$ in that case.

\section{\textsc{POLAR} Housing and Rotation Mount}
\begin{figure}
\begin{center}
\includegraphics[height=5in]{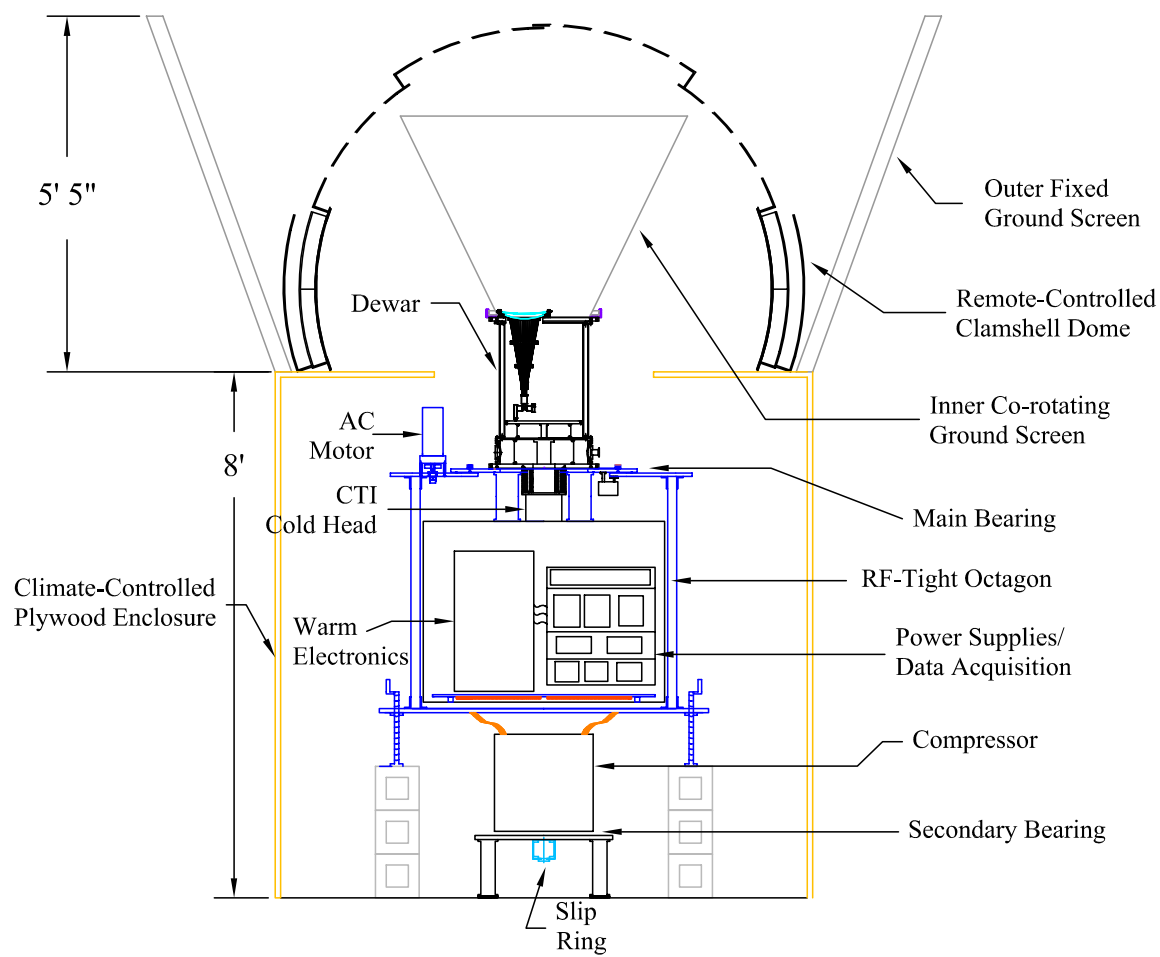}
\caption[Schematic of \polar\ inside the cubical plywood
housing]{\fixspacing \label{polarcube} Schematic of \polar\ inside the cubical plywood
housing.}
\end{center}
\end{figure}
\subsection{The POLAR cube}\label{s:cube}
\fig{polarcube} shows the \polar\ instrument inside its housing.
\polar\ lived at the Pine Bluff Observatory \footnote{Latitude
43\deg\ 4.7' N, Longitude 89\deg\ 41.1' W.} in Pine Bluff, WI, inside
a small housing built of plywood and unistrut, insulated with 2-inch
thick styrofoam.  This enclosure was temperature controlled by a
heater/air conditioner system which kept the cube temperature at
roughly 60\deg F, although this temperature varied somewhat over the
course of the season, where the outside temperature ranged from
10\deg F to 85\deg F.

\polar\ was guarded against the elements by a small fiberglass
clamshell dome made by AstroHaven Inc.  It used AC motors to
control the opening and closing of the dome.  During inclement
weather, the dome could be closed either locally using a small
control pad, or remotely via an Internet interface on a
private web page\footnote{\fixspacing
This wonderful remote-control system was designed and built by Kip Hyatt}.

\subsection{Weather Station}\label{s:weatherstation}
A small commercial weather station\footnote{PeetBros. Company Weather
Station Model ...} was mounted on the control building near the
\polar\ cube.  This station measured outside temperature, wind speed
and direction, and humidity, every five seconds.  This data was
recorded in our time stream every 7.5 minutes.

\subsection{Rotation Mount and AOE}
\polars\ rotation was driven by an AC motor, connected to a large
concentric gear at the base of the dewar.  The aluminum gear plate
sat on a groove filled with stainless steel bearings; we found the
rotation to be quite smooth.  The rotation was also robust; \polar\
underwent approximately 120,000 rotations in the 2000 observing
season with almost no problems\footnote{\fixspacing A problem did
occur later in the season when the bearings had ground away some of
the aluminum groove itself. This caused a viscous black paste to
form; the paste permeated the rotation groove and occasionally caused
the rotation to be jittery.  The problem was quickly identified and
solved by laborious cleaning of the groove.}  The rotation rate was
measured to be $0.0332 \ \pm \ 0.0003 \ Hz$.

At the beginning of the season, we recorded the angular position
of \polar\ with a relative 12-bit encoder.  To make it absolute,
we constructed an ``Absolute One-Bit Encoder'' (hereafter AOE).  This
simple optical device was mounted on the rotation stage, and
output a short voltage pulse whenever a certain point on the
non-rotating stage was passed.  The AOE contained an
opto-coupler that output 5 volts whenever a reflective surface
was close to its sensor (within $\sim$ 0.5'').  At due north on
the non-rotating stage was mounted a small piece of absorptive
tape, so the signal from the AOE dropped to zero volts whenever
the rotating stage passed due north.  This simple system, coupled
with the smooth rotation of \polar, obviated the need for a more
accurate encoder, which was ultimately abandoned when it failed in
the field.

\subsubsection{A brief note on POLAR's coordinate system}
\polars\ coordinate system is defined in actuality by the
OMT/circular-square transition; everything above this point is
axially symmetric.  It should be noted that the axis of symmetry
defined by this coordinate system was not aligned with local
North-South when the AOE pulse occurred.  However, we measured this
angle to be $23\deg \pm 1\deg$ at the end of the observing season.
This measurement ultimately enables presentation of results in the
standard IAU system (see, for example, reference \cite{kks}).

\subsection{The Octagon}
The warm radiometer box was mounted inside an RF-tight cage
that co-rotated with the dewar.  This cage, called the ``octagon''
(due to its octagonal shape), housed power supplies, the HEMT bias
cards, temperature control circuits, and the data acquisition
system, in addition to the Radbox.  It was temperature controlled
via a similar circuit as for the Radbox (\S\ref{radboxtemp}),
except with much larger heater pads.  The octagon was typically kept at
28\deg C, and varied by less than 0.5\deg C per day.  This
temperature control also made life easier for the temperature
control circuit inside the Radbox; as the air external to the
Radbox was at a fairly stable temperature, the intrinsic
temperature variations inside the Radbox were minimized, so only a
small amount of power was necessary to temperature stabilize the
Radbox internal components.

\subsection{Compressor and Slip Ring}
Beneath the RF cage sat the cryocooler compressor, which also rotated
with the rest of the system but on a separate bearing.  Underneath
the compressor sat an 8-line slip ring, through which data and power
were transferred between the fixed and rotating stages.  Four lines
were used for the $\sim$ 2 kW of power that \polar\ required for
operation, and two lines were used for data.  The data stream was
sent from the rotating laptop through the slip ring to an ethernet
cable, to the computer in a nearby control room which recorded the
data.

\subsection{Ground Screens and Spritz Tests}\label{groundscreens}
Although the sidelobe response of our feedhorn was very good, we
sought further sidelobe rejection through the use of two sets of
ground screens.  The inner ground screen was purely conical, and was
mounted to the top surface of the dewar.  In order to minimize
possible polarized reflections, we coated this aluminum surface with
Eccosorb; with the opening angle of 26\deg\ of this shield, it should
introduce less than 30 mK of unpolarized antenna temperature to our
system. Even if it is slightly polarized, it's immaterial as this
shield co-rotated with the dewar, and thus would not be modulated
like a true sky signal.

In addition to the inner ground screen, we built a non-rotating,
scoop-type outer ground screen, composed of four large, trapezoidal
aluminum panels.  In times of high winds, these panels were easily
lowered to the sides of the cube via hinges.  A simple scalar
diffraction calculation estimates an extra 40 dB of sidelobe
rejection from this ground screen; including the feed horns gives an
estimate of better than 100 dB of sidelobe rejection from the 300 K
earth. However, we discovered that the true rejection of side lobes
was not nearly as good.

\begin{figure}
\begin{center}
\includegraphics[height=2.5in]{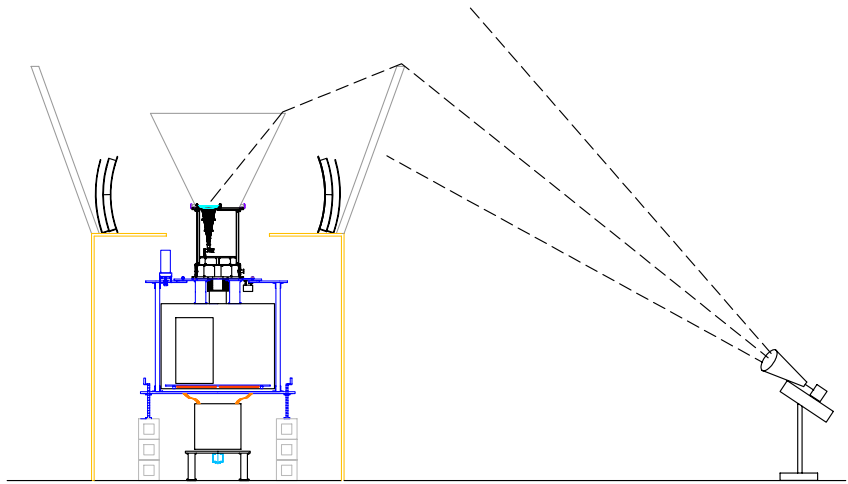}
\caption[Spritz Test Set-Up]{\fixspacing \label{spritz} Side view of the Spritz test set-up. A double refraction
is generally necessary for signals deep in the side lobes to be detected by the system.}
\end{center}
\end{figure}

``Spritz tests'' enable one to trace out the side
lobes of a telescope, by using a small, hand-held microwave
source.  The source is moved to different places in the side
lobes, and the system response is monitored.  As shown in
\fig{spritz}, a double refraction is in principle required for the
signal to reach the feedhorn. We conducted a series of these tests
at the end of the 2000 observing season \footnote{\fixspacing The
Spritz tests were carried out by Josh Friess and Dan Swetz .}, and
discovered that the signal was attenuated only by \app\ -50 dB when
placed further than 90\deg\ into the sidelobes.  However, the
signal was generally largest when aimed \emph{directly at the cube
housing}, meaning that somehow the signal was getting into the
system from within the cube, although it is difficult to construct
a mechanism for this.  In retrospect, we should have coated the
plywood cube (plywood is of course transparent to microwaves) with
something reflective to prevent such paths into the system.

Although this situation is definitely undesirable, it
is not perhaps as terrible as it seems: the 300 K earth is probably less than
1\% polarized, leading to a $\sim 30 \mu K$ earth-based signal,
modulated at the rotation frequency.  This is not consistent with
true polarized sky signals, which are modulated at twice the
rotation frequency; only the first harmonic would survive the
lock-in to twice the rotation frequency, which should be
significantly less than 30 $\mu K$.  However, this is still an
important systematic which may haunt us later in \chap{mapmaking}.

\section{Basic Radiometer Performance: Bandpasses and Power Spectra}
Now that the structure of the radiometer has been described,
let us discuss somewhat how well it worked.  In the lab, we attempted
to characterize the calibration, noise figures, bandwidths
of the radiometer.  We delay the discussion
of calibration and system temperature until \chap{calibration}.
\subsection{Bandpass}
\begin{figure}[tb]
\begin{center}
\includegraphics[height=4in]{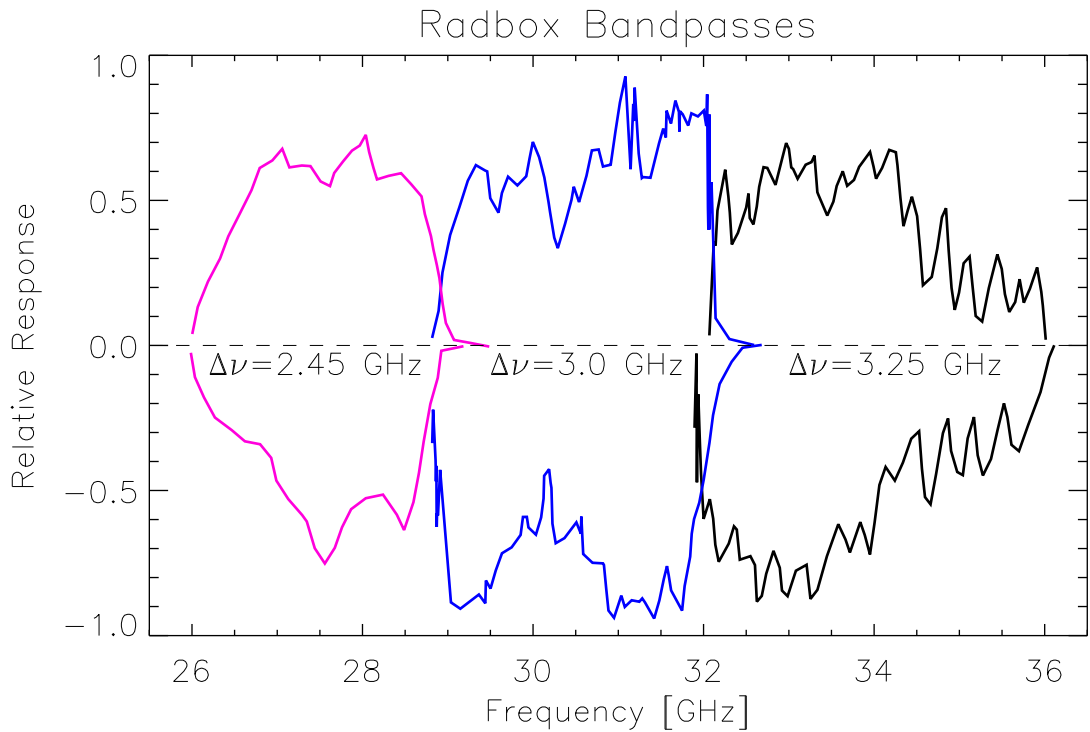}
\caption[RadBox Bandpasses for the Three Sub-bands]
{\label{bandpass} \fixspacing
Bandpasses through the radiometer box for the three in-phase (IPC)
correlator channels. From left to right, the channels are J3, J2,
and J1.  The upper curves represent the bandpasses when the
chopped phase shifter is the 0\deg\ (+) state, whereas the lower
curves correspond to the 180\deg\ (-) state.
Also shown are the effective bandwidths of each
sub-band in the Radbox, calculated using \eqn{nu_eff}.  This leads
to a total effective bandwidth of about 9.5 GHz for the Radbox,
but this does not include the HEMT amplifiers or any other
dewar or optical components.
}
\end{center}
\end{figure}

We never performed a full bandpass measurement on the system, but
we did do a test of the bandpass for the warm radiometer box.  These
tests did not include the cold radiometric components, such as the
HEMTs and OMT.
A 26-36 GHz signal was generated with an RF sweeper. This signal
was attenuated by about 50 dB to compensate for the \app\ 60 dB of gain
of the Radbox, and subsequently split
with a magic tee into two equal pieces.  These components were
fed into the two Radbox input ports, and we subsequently measured
the output from each correlator channel (after lock-in).  Because
the signals came from the same source, they were perfectly
correlated.  The resulting bandpasses are shown in \fig{bandpass}.

The bandpasses shown in the figure are quite acceptable.
Since the measurements did not include the cold radiometer
components, we cannot take this as the
true bandpass of the system, but these measurements are
suggestive of the generally good performance of the components we employed.

\subsection{Power Spectra}

\begin{figure}[tb]
\begin{center}
\includegraphics[height=4.5in]{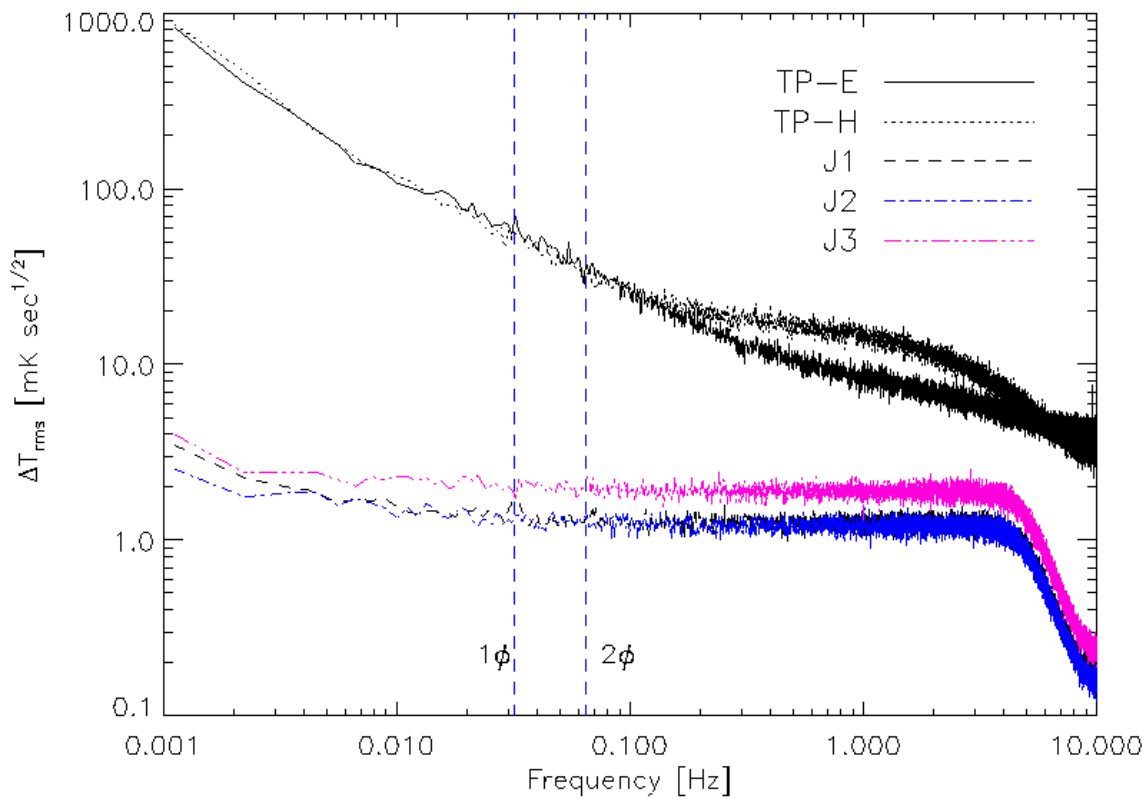}
\caption[Sample Power Spectra from the Radiometer]
{\label{powerspectra} \fixspacing Sample power spectra from the
\polar\ radiometer, in noise equivalent temperature (NET) units.
These spectra were taken under good weather conditions, while
observing the zenith. The total power channels display their
characteristic \emph{1/f} noise, which is noticeably absent in the
three correlator channels. Also seen is the effect above a few Hz of
the anti-aliasing filters. The vertical dashed lines correspond to
once and twice the instrument rotation frequency (labelled \onephi\
and \twophi); polarization signals induce a spike at the \twophi\
frequency.}
\end{center}
\end{figure}

Some sample power spectra for the system are displayed in
\fig{powerspectra}.  These were taken \emph{outside}, under good
weather conditions while staring at the zenith.  Notice how flat the
low-frequency power spectra are for the correlator channels; this is
quite good, and is a testament to the power of both the correlation
radiometer, and the phase-switching lock-in technique.  The power
spectra from the total power channels are also shown, and they
display the characteristic \emph{``1/f  noise''} intrinsic to the
HEMT amplifiers. At the frequency of interest where polarization
signals can be seen (namely twice the \polar\ rotation frequency),
the total power channels are roughly thirty times less sensitive than
the correlator channels! Note also the presence of the 5-Hz low-pass,
anti-aliasing filters in all of the channels.

In this chapter, we have seen that the \polar\ instrument is indeed a
stable, sensitive polarimeter, capable of viewing both Stokes $Q$ and
$U$ parameters, and hopefully finding the CMB polarization.  However,
before observations can begin, we must complete the final
instrumental task, calibration of the instrument.  This non-trivial
task is described in detail in the next chapter.



%




\chapter{Calibration}
\label{calibration}

\section{Background}
    Calibration is an ever-present challenge to all CMB
experiments.  However, a typical CMB anisotropy experiment
requires calibration only in the sense of \emph{total power}; that
is, phase information is irrelevant, and of the four Stokes
parameters, only \I\ is important.  This is not the case with a
polarization-sensitive receiver.  In that case, one should
calibrate all the Stokes parameters that one plans to measure.

    In addition to determining the calibration
proportionality constants (that is, the voltage one expects for a
given signal strength in Kelvins) for each Stokes parameter, the system calibration also
serves to tell us about the noise characteristics of our system.
Specifically, we would like to know how each of the following
quantities changes over time:
\begin{itemize}
\item $T_{rec}$ -- The equivalent antenna temperature due to
the noise of the radiometer itself.
\item $T_{sky}$ -- The antenna temperature input to the
radiometer (a sum of radiation from the sky, atmosphere, and CMB).
\item $\Delta\nu$ -- The RF bandwidth of each
radiometer channel.
\item $NET$ -- The ``noise equivalent temperature'' of the system, in
$mK \sqrt{s}$, which tells how long it takes to integrate to given
signal level.
\item System offsets for each channel.
\item Cross-talk between channels.
\end{itemize}

Ideally, we would not only measure these quantities accurately,
but we would measure them \emph{often}, to understand how our
system changes with time.  This chapter describes how (and how
well) this was done for \polar.

\section{Radiometry Basics}
Before going any further, it will be useful to lay out some of the
basics of microwave radiometry, as these techniques and formulas
are most useful in understanding the calibration and assessing the
noise of the instrument. These are only the absolute basics; for a
more detailed review of this material, some excellent sources are
\cite{partridge95, kraus86, rohlfs96}.

\subsection{Temperature in Radio Astronomy}
To understand radiometry, we must know the \emph{units} of
radiometry, which confusingly include no fewer than three different
types of temperature (in addition to many other quantities)! The
\emph{brightness temperature} of a source as a function of
frequency is a way of characterizing the power emitted by it at a
certain frequency, and is given by
\beq{tbright} T_B(\nu) \equiv
\frac{\lambda^2 B(\nu)}{2 k_B} \eeq
where $k_B$ is Boltzmann's
constant, and $B(\nu)$ is the blackbody emission spectrum, given
by
\beq{planck} B_{\nu} = \frac{2 h \nu^3}{c^2}
\frac{1}{e^{h\nu/kT}-1} \;\;\;\;\; [W m^{-2} Hz^{-1} sr^{-1}]
\;\;\;\text{;} \eeq
note that brightness temperature is only the same as
\emph{thermodynamic temperature} when in the Rayleigh--Jeans limit of
$h\nu \ll k_B T$.

However, radio telescopes measure a quantity called \emph{antenna
temperature}, which itself depends on the beam pattern of the telescope,
while brightness temperature is a function of the source only.  In the special case
that the source completely fills the beam, the antenna temperature
equals the object brightness temperature.  Otherwise, a brightness
distribution $T_B(\theta,\phi)$, when measured
by a system with a normalized beam power response of $G(\theta,\phi)$, viewing
in a direction $(\theta_0, \phi_0)$ yields an antenna temperature of
\beq{T_ant}
T_{ant} = \int_{4\pi} T_B(\theta,\phi)
G(\theta-\theta_0,\phi-\phi_0)d\Omega
\eeq
which is just the convolution of the source brightness distribution
with the beam pattern.

Thermodynamic temperature is the quantity typically reported;
based on \eqn{tbright}, we can see these quantities are related by
\beq{T2Tant}
T = T_{ant} \frac{e^x-1}{x}
\eeq
where $x \equiv \frac{h\nu}{k_B T}$.  In typical anisotropy
experiments, we measure not absolute temperatures, but rather
temperature \emph{differences}.  In polarization measurements it is the
same; recall for instance that $Q = I_x - I_y$.  That is,
polarization temperatures are really temperature differences, hence the same
temperature relationships hold for polarization as for temperature
anisotropy, and by differentiating \eqn{T2Tant}, one arrives at
\beq{dT2dTant}
\Delta T = \Delta T_{ant} \frac{(e^x-1)^2}{x^2 e^x}
\eeq
For the \ka\ band, this factor is about 1.025.
Unless otherwise noted, all temperatures in this work will be
reported in units of thermodynamic temperature.

\subsection{The Radiometer Equation}
By far, the most important formula in radiometry is known
as the ``Radiometer Equation'' \cite{rohlfs96}, which relates the
smallest detectable signal to the properties of the radiometer:
\beq{radeq} \Delta T \; = \; \kappa \frac{T_{rec} +
T_{load}}{\sqrt{\Delta\nu \; \tau}} \eeq where
\begin{description}
\item{$\Delta T$} is defined as
the smallest possible signal that can
be detected at the 1-$\sigma$ level with the system in question.
\item{$T_{rec}$ and $T_{load}$} are both antenna temperatures\footnote{\fixspacing
Often quoted is the \emph{system temperature}, $T_{sys}$, which is the sum
of $T_{rec}$ and $T_{load}$.}.
$T_{rec}$ is the equivalent antenna temperature due to
the noise of the radiometer itself, and $T_{load}$ is the antenna
temperature as seen by the feedhorn (not including our window or
other auxiliary optics).
\item{$\Delta\nu$} is the RF bandwidth of the
radiometer, and is given by \eqn{nu_eff}.  Note that it depends
only on the \emph{shape} of the bandpass; the flatter, the better.
\item{$\tau$} is the integration time.
\item{$\kappa$} is a constant of order unity that is associated with the type of
radiometer being used\footnote{\fixspacing $\kappa = 1$ for a total power radiometer and $\kappa =
2$ for a Dicke-switched radiometer.}.  For
correlation radiometers like \polar, $\kappa = \sqrt{2}$.
\end{description}

\eqn{radeq} can be 'compactified' by defining the \emph{Noise
Equivalent Temperature} (NET) of the system, which is the temperature limit
obtained after one second of integration:
\beq{NET}
NET \; = \; \kappa \frac{T_{rec} + T_{load}}{\sqrt{\Delta\nu}} \;\;\;\;
[K \; s^{-1}]
\eeq
Then the radiometer equation becomes simply $\Delta T = {NET}/{\sqrt{\tau}}$.
The \emph{Noise Equivalent Voltage} (NEV) is
the uncalibrated version of the NET.  It is measured in
$V \cdot \sqrt{s}$, and is the \emph{voltage} limit obtained
after one second of integration.
\eqn{NET} is only approximately correct for a correlation polarimeter.  A
more correct expression takes into account the fact that the two
different arms of the polarimeter can have different receiver
temperatures (as they contribute potentially different amounts of
noise to the system).  Then the NET becomes
\beq{NETcor}
NET \; = \; \kappa \sqrt{\frac{(T^L_{rec} + T_{load})(T^R_{rec} +
T_{load})}{\Delta\nu}}
\eeq
where $T^{L,R}_{rec}$ are the receiver temperatures for the two
polarimeter arms.  Even though this equation appears quite non-linear,
it is deceptive; if the two receiver temperatures are fairly
close, then the NET is approximately
\beq{NETapp}
NET \; \simeq \; \kappa\frac{(\sqrt{T^L_{rec} T^R_{rec}}+T_{load})}{\Delta\nu}
\qquad .
\eeq

\subsection{System Response and Y-factor}\label{sysresponse}
The general voltage response of our radiometer to an input antenna temperature
$T$ is given by
\beq{response}
\voltage = k (T_{rec} + T_{load}) + \voltage_{0}
\eeq
where k is the calibration coefficient, and
$\voltage_0$ is the offset due to the post-detection
electronics (such as a pre-amplifier offset), and is assumed
to be known.  Here, \voltage\
represents only our total power channels.  If \voltage\ is a correlator
channel, then we keep only the polarized fraction of $T_{rec}+T_{load}$.
Notice that \voltage\ is \emph{linear} in the input
temperature; this may surprise those who recall that the total power
emitted by a blackbody goes roughly like $T^4$.  The problem is
solved when we realize that the power collected by our radiometer
is
\beq{Peq}
P \; = \; k_B T \Delta\nu
\eeq
if we are in the Rayleigh-Jeans part of the emitting object's
spectrum. This is a simple consequence of integrating the Planck distribution
under the approximation $h\nu \ll k_B T$.  For a single-mode receiver and a single
polarization,
\beq{rjapprox}
P \ = \ \int_{\nu_1}^{\nu_2} \frac{h\nu}{e^{h\nu/k_B T}-1} d\nu
 \simeq \int_{\nu_1}^{\nu_2} k_B T d\nu \ = \ k_B T \Delta\nu \ .
\eeq
As the
radiometer response is linear in the power collected, we see it is
also linear in the antenna temperature.

A ``y-factor'' measurement is a simple technique to determine the
calibration coefficient $k$ as well as $T_{rec}$ from measurements of the system response
to two known temperatures, say $T_1$ and $T_2$ (typically 77K and room
temperature).  If these yield responses of $\voltage_1$ and
$\voltage_2$, respectively, then
\begin{subequations}
\begin{align}
k \; = &  \;\;\;\;\;\;\;\;\;\;\; \frac{\voltage_2-\voltage_1}{T_2 - T_1} \\
T_{rec} \; = & \;\;\; \frac{ T_2 (V_1-V_0) - T_1 (V_2-V_0) }{V_2 - V_1}
\end{align}
\end{subequations}
However, this simple technique only works for total power
channels; channels responsive to a polarized signal must be
calibrated by varying a polarized input signal, and tracking the
system response.  Simply varying the power in an input unpolarized
signal will only vary the noise in the channel; it will not (in
principle) generate a signal response.  The subject of calibrating a
polarization channel is treated in \sct{s:polarcal}.

\subsection{Noise Response and Y-factor}\label{s:noiseyfactor}
The signal is not the end of the story, however.  As a senior
graduate student
told me as I began to work on this project, ``Our signal \emph{is}
noise.''  While this was not quite true, understanding one's noise is
of critical importance.

It is assumed the reader is familiar with the basics of
\emph{auto-correlation}, \emph{cross-correlation}, and the \emph{Power
Spectral Density (PSD)}.  If not, an excellent introduction for
experimentalists is given in Numerical Recipes, reference
\cite{nrc}.
All channels have a power spectral density, or PSD, measured in
$V^2/Hz$ that is proportional to the system temperature \emph{squared}
(where for correlators we use the approximation of given in \eqn{NETapp},
where the receiver temperature is taken to be the geometric mean
of the two arms' receiver temperatures).
A more convenient quantity is the PSD amplitude, which is simply the
square root of the PSD, and has units of $V / \sqrt{Hz}$.

We measured the PSD amplitude of the system several times in the lab
before the instrument was deployed, and then several times during
the season (see \fig{powerspectra}).
For the correlator
channels, this provides the means to determine their receiver
temperatures.  Imagine we view a sky temperature $T_1$ and a
second temperature $T_2$
\footnote{\fixspacing For instance, a 300 K Eccosorb load}, then use the
following algorithm:
\begin{enumerate}
\item Record several minutes of data for each load condition.
\item For a given channel, compute the power spectral densities
$P_1(\nu)$ and $P_2(\nu)$ under the two load conditions.
\item Where the PSD is flat, or at the signal frequency, evaluate
each PSD.  Multiplying by $\sqrt{2}$ will yield the NEVs, $\eta_1$
and $\eta_2$, under each load condition
\footnote{\fixspacing It is a curious fact that the amplitude of the PSD has units
of $V Hz^{-1/2}$, while NEV has units of $V s^{1/2}$.  To go from
the PSD units to the NEV units, one must divide by $\sqrt{2}$.
An explanation of this phenomenon is given in \cite{bkthesis}.}
.
\item The receiver temperature in that channel is then given by
\beq{Noiseyfactor}
T_{rec} \; = \; \frac{ T_2 \eta_1 - T_1 \eta_2 }{\eta_2-\eta_1}
\eeq
\end{enumerate}
Once we know the calibration coefficient of a given
channel, we can calculate the NET of that channel by multiplying
the NEV by the calibration coefficient.  At this point, we can calculate
our channel bandwidth $\Delta\nu$ by using \eqn{radeq}.  If
$\Delta\nu$ is known by other means (such as by direct measurement
of the bandpass and then using \eqn{nu_eff}), then other quantities in
\eqn{radeq} can be verified.

\section{The POLAR Calibration}\label{s:polarcal}
\emph{The following section was originally published as a separate
paper, see \cite{odell01}}.

At this point, we must determine a method with which to calibrate our
polarization channels.
In general, there are three types of calibration we could appeal to:
internal, external (local), and astrophysical.  For internal-type
calibrations, some type of noise source, mounted within the
front-end, is used to inject a known signal into the radiometer.
However, to make a well-understood polarized signal requires an
additional OMT, and this was not within our budget.  Astrophysical
calibration is perhaps the easiest;  by pointing the radiometer at
sources of known flux (at your particular frequency), calibration
is achieved by comparing the response to the known flux.  To
determine the effectiveness of a point source calibration, first we
must compute our sensitivity to point sources, given by the
Rayleigh-Jeans relation \cite{rohlfs96}; for a Gaussian beam, this
relation is
\beq{RJ-relation}
\frac{T(\nu)}{S(\nu)} = 1.053
\;\frac{\lambda^2}{\theta^2_{1/2}} \;\;\;\; [K/Jy]
\eeq
where $S(\nu)$ is the flux density of the point source in Jy,
$\lambda$ is in meters, and $\theta_{1/2}$ is the beam FWHM in degrees. For
\polars\ 7\deg\ beamsize, this works out to only 2 \uK/J, which is
more than an order of magnitude too small to calibrate from even
the most powerful radio sources \cite{bkthesis}.

Thus, we abandon both internal and astrophysical calibration
techniques; instead, we must rely on some type of external
approach.  The standard way to generate a known
polarized external signal is by placing a wire grid above the radiometer at
a 45\deg\ angle to the antenna axis\cite{ls81,gas93}.
Wire grids are simple, reliable, and have well-understood
properties \cite{wait54,larsen62,lesurf90,houde01}.
If a hot load of temperature $T_H$ is placed
to the side (such that it will reflect in one polarization), and a
cold load of temperature $T_C$ is placed above the grid, both in such
a way that they completely fill the beam, the
radiometer will receive a polarized signal of temperature
$T_H-T_C$.  Using liquid nitrogen for the cold source results in a
\app\ 200 K polarized signal.  This is troublesome for two reasons.
First, this signal is \app\ 8 orders of magnitude larger than the
signal we are actually trying to measure; indeed, the signal
reaching the correlators will be almost 100\% correlated (provided
the radiometer is rotated to the appropriate angle).  Second,
and perhaps more important, we remember that our calibrators can
only handle \app\ 90 K of antenna temperature before compressing.
Clearly, we need a smaller polarized signal, and ideally one that
is only very slightly polarized, to more closely mimic the tiny CMB signal
that we seek.

Recently, Hedman \etal\ \cite{hed00} used reflection of a
known (unpolarized) source from a metal surface to create a
well-characterized but small polarization signal for their
calibration\cite{cor94}.  For \polar, we used yet another
approach -- reflection of thermal radiation from a thin dielectric sheet.
The rest of this section will describe what we learned in trying
to perform and understand this calibration technique; in order to
make the technique useful to as wide as possible an audience, the
discussion is kept relatively general.

\subsection{The Technique} 
\begin{figure}[tb]
\begin{center}
\includegraphics[height=8.5cm]{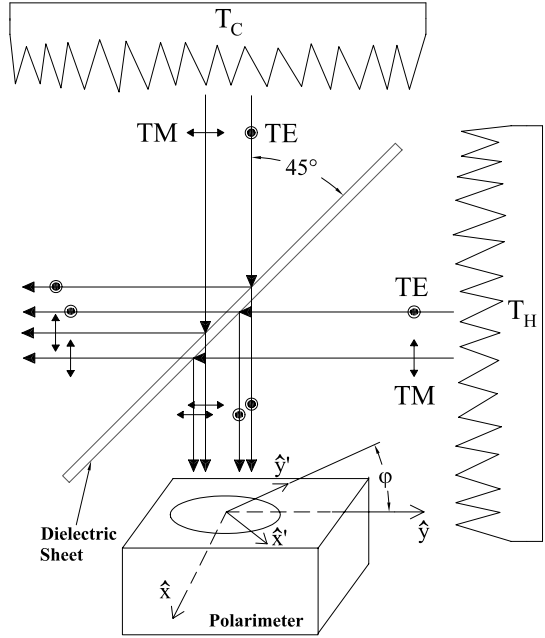}
\caption[Calibration Set-Up using the Thin Dielectric Sheet]
{\fixspacing\label{calsetup} Calibration Set-Up using the thin dielectric sheet.  Unpolarized
radiation from both a hot load (side) and cold load (top) is
partially polarized due to the slight difference in $R_{TE}$ and
$R_{TM}$ of the sheet, thus causing the polarimeter to see a slightly polarized
signal.  The angle between the polarimeter $x$-axis and sheet plane
of incidence is $\phi$.  The Stokes parameters can be modulated by variation
of the angle $\phi$.}
\end{center}
\end{figure}

In order to calibrate
\polar,  we replaced the wire grid in the
conventional set-up with a thin dielectric sheet (see \fig{calsetup}). The
sheet's composition
and thickness were chosen carefully according to criteria discussed
later in this section.
If the reflection and emission properties of the sheet can be
ascertained, through either direct measurement or calculation,
then it is straightforward to calculate the expected signal from
the dielectric.  Both the hot and cold loads emit blackbody
radiation at their physical temperatures, $T_H$ and $T_C$
respectively.  These unpolarized sources emit an equal amount of
radiation polarized both perpendicular (TE) and parallel (TM) to
the plane of incidence on the dielectric sheet.  Note that the TE
and TM radiation fields are \emph{uncorrelated} with each other.
Upon traversal of the sheet, a certain amount of each of these
four fields arrive at the aperture of the polarimeter, along with
the oblique emission from the sheet itself (which has a physical
temperature $T_S$).

In order to perform the calibration, we must determine the
intensity of fields at the aperture of the polarimeter from the calibrator.
We use the standard Stokes parameters $\{I,
Q, U, V\}$ to characterize field intensity.
The Stokes parameters are
additive quantities and hence simplify the following mathematics.

In \cite{gas93}, the Stokes parameters from a wire grid calibrator
are calculated.  For the dielectric sheet the derivation is
similar, but we must also take into account the emissivity of the
sheet, which may not be negligible.  We will make the simplifying
assumption that the microwave absorbers ($T_C$ and $T_H$) are
perfect blackbodies; this assumption will be discussed later in
detail.

First let us calculate the Stokes parameters in the reference
frame of the calibrator; once we have these, it is straightforward
to ``rotate'' them into the frame of the polarimeter.  We give the
Stokes parameters in units of brightness temperature; for a
single-mode antenna in the Rayleigh-Jeans limit, the brightness
temperature $T_B$ is related to power $P$ through $P = \Delta\nu
k_B T_B$, where $\Delta\nu$ is the frequency bandwidth and $k_B$
is Boltzmann`s constant. Let $\hat{x}$-$\hat{y}$ be the coordinate
system of the calibrator, and $\hat{x}'$-$\hat{y}'$ be the
coordinate system of the polarimeter; $\phi$ denotes the rotation
angle between these reference frames. Further, let $I_x$ and $I_y$
correspond to the brightness temperature of the total power
polarized along $\hat{x}$ and $\hat{y}$, respectively\footnote{\fixspacing We
work here with the more-convenient $I_x$ and $I_y$, rather than
their sum, $I$, because it is these quantities that polarimeters
usually measure.  Some polarimeters
also directly measure $Q$, $U$, and/or $V$, typically via
correlation or pseudo-correlation techniques.}. The Q Stokes
parameter is given by $Q = I_x - I_y$. The brightness temperatures
$I_x$, $I_y$, $Q$, and $U$ in the (unprimed)
calibrator coordinate system will be (see Appendix A):
\begin{subequations}
\label{unprimedstokes}
\begin{eqnarray}
I_{x} & = & T_C + \left(T_H - T_C\right) R_{TE} +
            \left(T_S - T_C\right)\epsilon_{TE} \\
I_{y} & = & T_C + \left(T_H - T_C\right) R_{TM} +
            \left(T_S - T_C\right)\epsilon_{TM} \\
\label{Qsignal}
Q & = & \left(T_H - T_C\right)\left(R_{TE} -  R_{TM}\right)
\nonumber \\
& & + \left(T_S - T_C\right)\left(\epsilon_{TE} - \epsilon_{TM}\right) \\
U & = & 0
\end{eqnarray}
\end{subequations}
If the angle between the
polarimeter $\hat{x'}$-axis and the sheet plane of incidence ($\hat{x}$-axis)
is $\phi$, then the Stokes parameters as seen by the polarimeter are
given by
\begin{subequations}
\label{primedstokes}
\begin{eqnarray}
I_{x'} & = & I_{x} \cos^2{\phi} + I_{y} \sin^2{\phi} \\
I_{y'} & = & I_{x} \sin^2{\phi} + I_{y} \cos^2{\phi}
\\
Q' & = & Q \cos{2\phi} \\
\label{rotatedQU}
U' & = & -Q \sin{2\phi} .
\end{eqnarray}
\end{subequations}
We note here that including the small reflectance $R_l$ of the
unpolarized loads would have the effect of increasing $T_C$ to
$T_C + R_l(T_H-T_C)$ in the reflection term in \eq{unprimedstokes},
assuming the environment has a temperature
$T_H$.  Typically the loads can be chosen such that
the overall effect can be neglected.  If this is not possible,
$R_l$ must be measured at the frequencies of interest, so that its
effect on \eq{unprimedstokes} can be included.

It is then a simple matter to calibrate the polarimeter by varying
the angle $\phi$, either by rotating the calibrator or the
polarimeter.  As we are primarily interested in calibrating
polarization channels, we will focus on the $Q$- and $U$-calibration
signals; each of these changes by a full 100\% over a complete
$\phi$ cycle.  In contrast, varying $\phi$ produced very
low signal-to-noise variations in $I_{x'}$ and $I_{y'}$
for the dielectric sheets we used, making a ``total power''
calibration with the sheet impractical.  However, this is
inconsequential because those channels are easily calibrated with simple
unpolarized loads through a conventional y-factor measurement.

The accuracy of the $Q$- or $U$-calibration depends on several factors.
First, one must know or determine the relevant
material properties of the sheet, namely the
reflection coefficient and emissivity both for the two polarization
states and as a function of incidence angle.
The angle of incidence $\theta$ must be known
to reasonable accuracy.   The sidelobes of the receiving horn should be low,
the sheet and loads should be large enough to completely fill the main beam
of the receiver, and the loads
should be near-perfect absorbers, else stray radiation from the surroundings will
enter the system.  All these conditions must be satisfied in the wire grid
approach as well,
with the exception that instead of understanding the grid properties,
now it is
the reflection and emission properties of the dielectric sheet
that we seek to understand. It is on these issues that we will now focus.

\section{Dielectric Reflection and Emission Properties}
\label{re_gen}
The general situation we wish to consider is as follows: an
electromagnetic wave
of wavelength $\lambda$ is incident upon
an infinite dielectric sheet of thickness $d$ and index of
refraction $n$.  Part of this wave will be reflected, part will be
transmitted, and part will be absorbed.  All these quantities will
depend upon the polarization state of the incident wave, which in
general will be a combination of $TE$- and $TM$-polarized radiation.
Thus, for the radiation incident upon the sheet (to be
distinguished from its own thermal emission), we have
\beq{RTA}
|r|^2 + |t|^2 + A \ = \ 1
\eeq
where $|r|^2$, $|t|^2$, and $A$ represent the fractional power reflected,
transmitted, and absorbed, respectively; $r$ and $t$ are the usual
Fresnel reflection and transmission coefficients, and are complex quantities.

If the sheet is in
thermal equilibrium, emission will equal absorption (\ie, $\epsilon = A$).
In general,
a material has a complex index of refraction
$N = n - j\kappa$ where $n$ corresponds to the real index of
refraction, and $\kappa$ is the extinction
coefficient and determines the loss of the material.  If $\kappa \ll
n$, then the \emph{loss tangent} of the material, the ratio of the
imaginary component
to the real component of the dielectric constant, is given approximately
by \footnote{\fixspacing The loss tangent,
$\tan{\delta}$, is not to be confused with the
unrelated quantity $\delta$, the phase change due to the
dielectric given in \eqn{deltadef}.}
\beq{losstan}
\tan{\delta} \approx \frac{2\kappa}{n} \ .
\eeq
Given N, it is possible to calculate both $r$ and $t$ for a lossy
dielectric slab \cite{azzam}.
Then the emissivity $\epsilon$ will be $1-|r|^2-|t|^2$, and in general will
be polarized.  However, for this treatment we assume that the
\emph{total} loss in the dielectric is negligible;
\sct{emission} deals with the conditions under which
this assumption is valid.

\subsection{The Reflection Term - Theory}
It is straightforward to derive the reflection coefficients for
our smooth dielectric sheet using the Fresnel equations,
under the assumptions that
the dielectric is homogeneous, optically isotropic,
non-amplifying, and the wavelength is on the order of or larger
than the film thickness, such that all the multiply-reflected
beams combine coherently (see \eg \cite{bornandwolf,pedrotti}).
Assuming the sheet is placed in air with a refractive
index of $\sim 1$, and absorption by the sheet is neglected,
the reflection coefficient can be shown to be:
\beq{Rgeneral}
R_i = \frac{\left[\cos^2{\theta} - \gamma_i^2\right]^2 \ \sin^2{\delta}}
        {4 \gamma_i^2 \cos^2{\theta} \cos^2{\delta} \ + \
\left[\cos^2{\theta}+\gamma_i^2\right]^2 \sin^2{\delta} }
\eeq
where $i \ \in \ \{TE, TM\}$ represents the incident field polarization direction, and
\begin{subequations}
\label{gamma}
\begin{eqnarray}\label{gammaTE}
\gamma_{TE} & \equiv & \sqrt{\ n^2 - \sin^2{\theta}} \\
\label{gammaTM}
\gamma_{TM} & \equiv & \frac{1}{n^2} \sqrt{n^2 -
\sin^2{\theta}}
\end{eqnarray}
\end{subequations}
and where
\beq{deltadef}
\delta \ = \ k d \sqrt{n^2 - \sin^2{\theta}}
\eeq
is the phase change that the wave undergoes upon traversal of the
sheet; $k = \frac{2 \pi}{\lambda}$ is the wavenumber of the wave in free space,
$d$ is the thickness of the sheet, $n$ is the (real) refractive
index of the dielectric, and $\theta$ is the angle
of incidence of the wave upon the sheet.

For this technique we are primarily interested in the Q and U
calibration; from \eqn{Qsignal} we see that the quantity of interest
here is $R_{TE} - R_{TM}$, the difference in the reflection coefficients
of the sheet.  The coefficients are only the same at normal and
grazing incidence; at all other angles a polarization signal will be
produced.  A useful formula can be
derived for the case of $\lambda \gg d$ and $\theta = 45 \deg$,
conditions which were satisfied by \polar\ (see Appendix B):
\beq{deltaRsmall}
R_{TE} - R_{TM} \ \simeq \  \left(
\frac{\pi f d}{c} \right)^2 \frac{(n^4-1)(n^2-1)(3 n^2 -1)}{2 n^4}
\eeq
This formula is informative as it shows how the calibration
signal behaves with varying frequency, sheet thickness, and
index of refraction.  Notice the signal varies quadratically in
both $f$ and $d$, and even faster with index of refraction.  This
implies that all these variables must be known with considerable
precision to result in an accurate calibration.

\subsubsection{The Reflection Term - Experimental Verification}
We devised a simple system to test the reflection equations
presented above,
in order to verify they worked on real-world
materials, and to ensure that we had not neglected other
potentially important effects.  We tested 0.003" (0.076 mm) and
0.020" (0.51 mm) thick polypropylene, for this material has a
well-characterized refractive index of $1.488-1.502$ in the useful
range of $30-890$ GHz \cite{gold98}. We also tested $0.030"$-thick teflon.
Other materials, such as polyethylene, TPX, or mylar could of course
be useful too,
and our results are directly applicable to those materials
assuming one knows the pertinent material properties.

\begin{figure}[tb]
\begin{center}
\includegraphics[height=7cm]{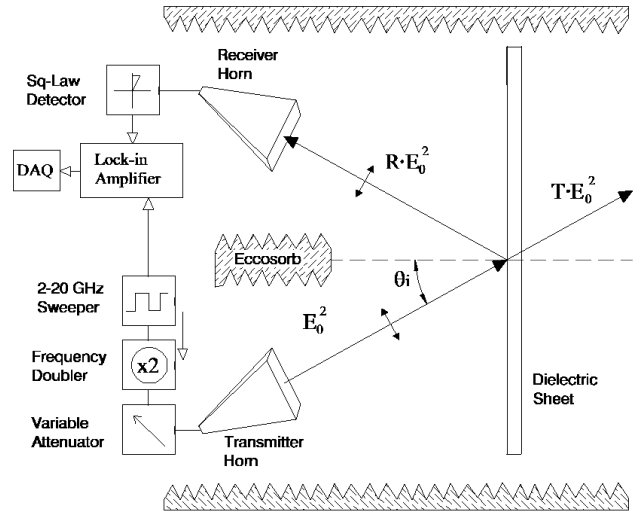}
\caption[Experimental configuration
used to test the reflectance of various materials]
{\fixspacing\label{setup} Experimental configuration used to test the reflectance of
various materials.  The incidence angle $\theta_i$ was kept fixed
at $45\deg$.  The frequencies used were the microwave Ka-band, 26--36 GHz.
The input signal was chopped at 1 KHz to
help eliminate 1/f noise.  The horns are shown here in the TM configuration;
for the TE configuration, the horns were rotated $90\deg$.}
\end{center}
\end{figure}
The experiments \footnote{\fixspacing These experiments were constructed and
performed excellently by Dan Swetz.}
were performed in a small homemade anechoic
chamber (see \fig{setup}), made of commercially available
Eccosorb$^{\circledR}$ CV-3 \cite{emersoncuming}.  Eccosorb CV-3
has a quoted reflectivity of less than -50 dB at frequencies
up to 25 GHz, and a reflectivity of -34 dB at 107 GHz \cite{lehto}, which
was adequate for our purposes.
We fixed the incidence angle at $45\deg$,
which was the primary angle of interest to us\footnote{\fixspacing Other angles
could potentially be used, but the calculations
for the calibration signal would be more complicated than those presented
above.}.  A standard-gain (25 dB) pyramidal feedhorn
transmitted a
signal of known frequency to a dielectric sheet approximately
$20" \times 20"$ in area.  The signal was generated by a
commercial 2--20 GHz microwave sweeper coupled to a frequency doubler
to obtain the $K_a$-band
frequencies of 26-36 GHz.  An identical horn was
placed symmetrically about
the sheet's normal in order to receive the reflected waves.
Reflected radiation from the room was found to be minimal.
A thin piece of Eccosorb was
placed between the two horns to minimize direct coupling between them.  The
transmitting source was swept through the $K_a$-band
over a period of $100$ seconds, and the amplitude square-wave chopped at 1
kHz (this frequency was well above the 1/f knee of the system).
The received signal was then sent to a lock-in amplifier and
recorded by a computer using a simple data acquisition system.
The reflected signal was quite small, and the lock-in
technique enabled us to significantly reduce our sensitivity to
1/f noise in the system.
A baseline reading was obtained using an aluminum flat instead of
the dielectric sheet; the flat had near-perfect reflectivity and
provided our normalization.

It was important to control systematic effects well; in particular,
the imperfect absorption of microwaves by the
Eccosorb walls of the anechoic chamber.  By varying the Eccosorb
configuration, we were able to virtually eliminate all spurious
signals related to imperfect Eccosorb absorption.
In the optimal configuration, tests with no reflector showed our
system was capable of measuring reflection coefficients as low as
a few $\times 10^{-5}$.  The primary systematic effect was
standing waves in the system, propagating between the source and reflecting
surface.  These were controlled (but not
eliminated) by placing an attenuator between the sweeper
and the transmitting horn.

\begin{figure}[tb]
\begin{center}
\includegraphics[height=7cm]{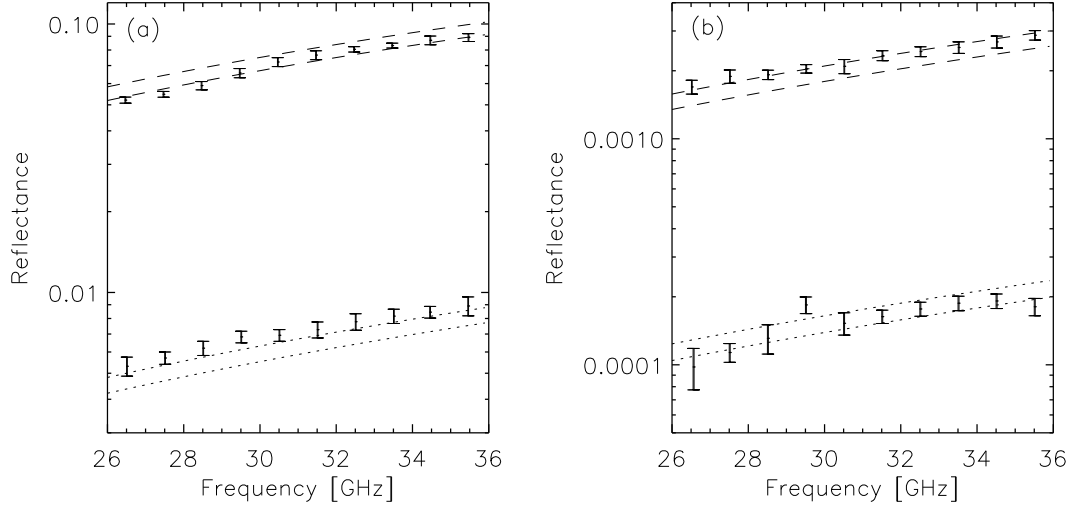}
\caption[Laboratory reflectivity measurements for polypropylene compared with theory]
{\fixspacing \label{danresults} Comparison between laboratory reflectivity measurements and theory
on polypropylene. The displayed $1\sigma$ errors
in the data are mostly systematic,
arising from standing waves in the system.  The uncertainty in
theory is due to both thickness variations and uncertainties in
the index of refraction.  $R_{TE}$ corresponds to the upper set of curves
(dashed), and
$R_{TM}$ to the lower set of curves (dotted). Measurements were
averaged into 1 GHz bins for convenience. (a) Results for 0.020" (0.51 mm) thick
polypropylene; (b) Results for 0.003" (0.076 mm) thick polypropylene.}
\end{center}
\end{figure}

Figs. \ref{danresults}(a) and (b) show the results for the 0.020" and 0.003" sheets,
respectively.  The errors bars shown on the measured points
are primarily due to standing
waves in the system.  The theoretical error contours drawn
represent the thickness variations in our plastic sheets.  We
found that both of these commercial sheets had thickness
variations on the order of 5\%;
because the reflection signal is roughly proportional to
$d^2$, the resulting uncertainty in the calibration is $\sim$ 10\%.
Uncertainty in the index of refraction of the dielectric is even
more important.  Luckily, for our chosen material of polypropylene in the
$K_a$ frequency band, the refractive index is known to an accuracy of
at worst $\sim 2\cdot 10^{-3}$ \cite{shim88,lync82}, which
contributes negligibly to our errors.
As \fig{danresults} shows, the measured
curves match the theory quite well for the displayed polypropylene
data.  Teflon (not shown) worked equally well, having an $R_{TE}-R_{TM}$ of
approximately $0.10$ for the $K_a$-band frequencies we tested.

In this section we sought to verify \eqn{Rgeneral} with laboratory
experiments.  The reader should note that we did not include any off-axis
beam effects when calculating the theoretical predictions for
these experiments.  The general calculation would involve
integrating over the antenna pattern of the transmitting and
receiving horns, for each polarization state. Off-axis rays,
reflecting from the dielectric at slightly
different angles from the on-axis rays, will then slightly affect the
measured reflection coefficients, due to the variation of the
reflection coefficients of the dielectric as a function of angle.
However, the remarkable agreement between the predicted (on-axis)
and measured reflection coefficients indicates that this was a small
effect.

\subsection{The Emission Term}\label{emission}
Oblique emission from a dielectric will in general be polarized
(for a review, see for example \cite{sandus}).
For this calibration technique to work, either
the emission must be known accurately (in both polarizations), or
it must be negligible. The emission of a material is determined by both its
thickness and loss tangent (or alternatively, its extinction
coefficient), will vary as a function of viewing angle, and
will generally be polarized
(that is, $\epsilon_{TE} \neq \epsilon_{TM}$).

As discussed in \sct{re_gen}, the complete way to determine
emission involves calculating both $R$ and $T$ using the complex
refractive index, and then using \eq{RTA} to find the absorption
(which equals the emission in thermodynamic equilibrium); then the calibration
signal can be calculated using \eq{unprimedstokes}.  For this
approach to work, the complex index of refraction (and hence the
loss tangent) must be known to reasonably good accuracy, and the
surface must be \emph{smooth}; if the surface roughness is
too high, the emission polarization will be less than theory
predicts \cite{jordan96}.
Typically, the loss tangent is known only poorly.  Luckily,
the \emph{total} emission can often be
made small compared to the reflection/transmission terms by
appropriate choice of dielectric material and thickness for the
frequencies of interest; then one can simply ignore the emission
terms in \eqn{unprimedstokes}.

An approximation for the total emission is \cite{gold98}
\beq{totemission}
\epsilon \ \approx \ \frac{2\pi n \tan{\delta}}{\lambda} \cdot \ d
\eeq
where $d$ denotes the thickness of the emitter,
and $\epsilon$ denotes the fraction of its
thermodynamic temperature that is emitted; hence, it produces a
brightness
temperature of $T_\epsilon = \epsilon \cdot T_S$.
As an example, the \polar\ calibration used a 0.003" thick polypropylene
sheet which had a loss tangent of $\sim 5 \times 10^{-4}$, leading to
$\sim$ 12 mK of total emission; this turned out to be small in
comparison with the calibration signal and hence was neglected.

\begin{figure}[tb]
\begin{center}
\includegraphics[height=7cm]{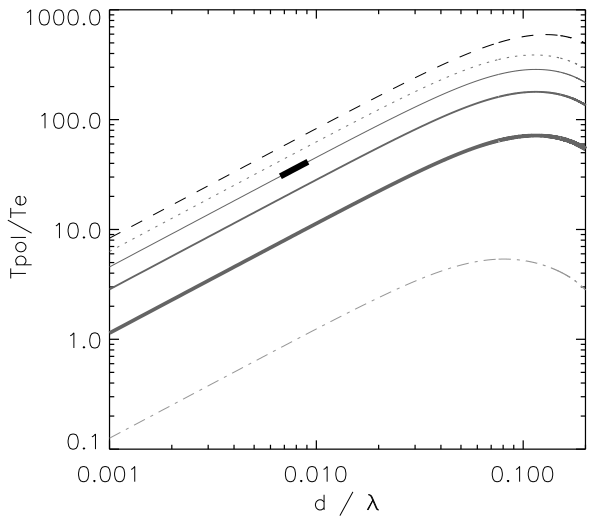}
\caption[Ratio of Polarization to Emission vs. $\frac{t}{\lambda}$ for various
materials]
{\fixspacing\label{tlfig} Ratio of the polarized radiation, $T_{pol}$, due entirely to reflection, to the
brightness temperature in emission, $T_e$, of the dielectric sheet, for various
materials, vs. $\frac{d}{\lambda}$, the ratio of sheet thickness to
free-space wavelength.  The six curves are for different
materials and/or frequency ranges; dashed: teflon ($30-300$ GHz),
dotted: TPX (30--270 GHz), solid-thin: polypropylene ($20-40$ GHz),
solid-medium: polypropylene (40--270 GHz), solid-thick:
polypropylene (270--900 GHz), dot-dashed: mylar (120--1000 GHz).
The darkened box shows \polars\ region in this parameter space.
Loss tangents were adopted from \cite{gold98}.}
\end{center}
\end{figure}

\fig{tlfig} shows
the ratio of the polarized reflection signal, $T_{pol}$,
to the emission signal, $T_{e}$, as a function of $\frac{d}{\lambda}$.
The higher this ratio is, the more safely emission can be
neglected in calculating the calibration Stokes parameters.
Notice that at higher frequencies and material thicknesses,
emission matters \emph{less} than at lower frequencies and
thicknesses.  This means that the smaller the desired polarization
signal, the \emph{more} emission will matter.  This result may seem
counter-intuitive, but it is directly evident from the reflection and
emission equations; emission goes like $\frac{d}{\lambda}$, while
typically the reflection portion of the signal goes like
$\left[\frac{d}{\lambda}\right]^2$.
In terms of absolute emission, polypropylene,
polyethylene, TPX, and teflon are all useful.  However, mylar's
high loss makes it non-ideal for this technique, unless one has
good data on the directional emissivity of the material at the
frequencies of interest.

\subsection{Pitfalls}


We discovered several pitfalls during the development of this calibration
that should be avoided if possible.  The first is to make sure the
dielectric sheet is kept as taut and flat as possible.  In our
first version of the calibrator, we didn't pay much attention to
this and the plastic sheet had a slight bow in it. Laboratory
results found this bowing to have a significant impact on the
resulting calibration signal, causing it to deviate from theory by
as much as $20\%$ for a barely-visible bowing.  Reducing the
bowing by increasing the tension in the sheet resulted in the
signal matching theoretical predictions.

A second source of error was variation in material thickness. We
found that, in practice, some of the materials we tested varied by
as much as $10\%$ in thickness across a sheet;  this is rather
large and leads to a high uncertainty in the calibration signal,
due to its approximate $d^2$--dependence.  Sheets with
manufacturing processes that lead to a more uniform thickness
should be used if possible.

\section{Daily Calibration Observations}\label{s:dailycal}
\begin{figure}[tb]
\begin{center}
\includegraphics[height=3.5in]{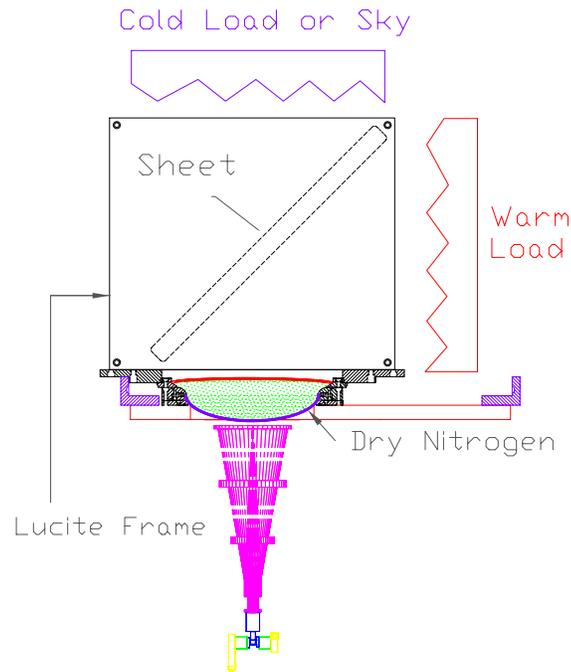}
\caption[The \polar\ Dielectric
Sheet Calibrator (DSC)]
{\label{calibrator}\fixspacing The \polar\ Dielectric
Sheet Calibrator (DSC). The DSC could be rotated manually about the horn, and locked
into any of 16 evenly-spaced positions.}
\end{center}
\end{figure}
    Now that we understand the theoretical and experimental basis
of the calibration technique, let us discuss how it worked in
practice.  Calibrations were performed roughly once/day during
periods of good observing weather.  The dielectric sheet calibrator
(DSC, see \fig{calibrator}) we built
for observations had a lucite frame, and had several dielectric sheets
that we could easily switch in and out: 3-mil polypropylene,
20-mil polypropylene, and a wire grid made of 8-mil wide copper strips deposited
on 2-mil mylar.  The DSC sat snugly on top of the vacuum
window, and was rotated among eight different positions (one position every
45\deg); ideally, the polarimetric response would then be
($0$, $V_{max}$, $0$, $-V_{max}$, $0$, $V_{max}$, $0$, $-V_{max}$)
for each of the eight positions, as the response varies as $\sin{2\theta}$.
The resulting calibration coefficient is then given by:
\beq{calsignal}
\frac{1}{k_i} = \frac{(T_H-T_C)(R_{TE}-R_{TM})}{V_{max}} \ \ \ \ \ \
[\frac{K}{V}] \ ,
\eeq
since for these for frequencies in polypropylene, emission can be
neglected in \eqn{Qsignal}.

A full calibration consisted of techniques to calibrate all signal
channels (including the total powers) and used these measurements
to extract the calibration coefficients, system temperatures and
NETs for each channel, as well as the overall \ka-band sky
temperature.
In order to accomplish this, the following procedure
was repeated daily (in good weather).  We placed each of the
following "loads" in front of the radiometer for the duration
indicated:

\begin{itemize}
\item Pure Sky Load (1 min)
\item Pure 77 K Eccosorb Load (1 min)
\item Pure Ambient Eccosorb Load (1 min)
\item DSC with sky, 300K loads (2 rev, 10sec/position)
\item DSC with sky, 300K loads (2 rev, 10sec/position)
\item Pure Ambient Eccosorb Load (1 min)
\item Pure 77 K Eccosorb Load (1 min)
\item Pure Sky Load (1 min)
\end{itemize}

The total-power channels
were calibrated with a simple y-factor between 77 K and 300 K Eccosorb
loads. Sky temperature was then extracted via the total power (now
calibrated) sky measurements.  System temperatures were determined
from noise y-factor measurements (discussed in \sct{s:noiseyfactor}),
and NETs were determined from
the PSD of each channel viewing pure sky (applying the appropriate calibration).
The entire calibration procedure took about fifteen minutes.

\subsection{A Sample Calibration}
\begin{figure}[tb]
\begin{center}
\includegraphics[height=3.5in]{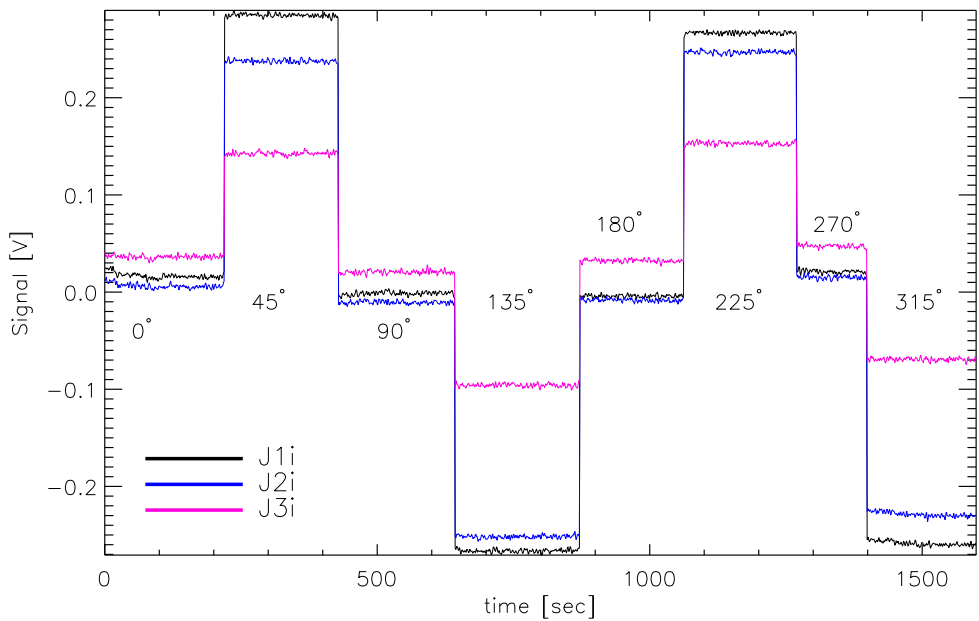}
\caption[A Sample DSC Calibration Signal]
{\label{samplecal}\fixspacing A sample calibration signal,
showing the response of the three correlator channels to the
spinning dielectric sheet viewing a 300 K load to the side and a 10
K (sky) load above.  The raw calibration signal is a function of frequency,
leading to the different responses of the three channels.
The angular labels correspond to the eight evenly-spaced positions of the DSC
referred to in the text.}
\end{center}
\end{figure}
A sample calibration of the DSC viewing (sky ,300 K)
is shown in \fig{samplecal}.  You can see the
signal-to-noise ratio is quite high; indeed, it is $\sim$ 100 for J1.
Remember, these calibration signals are of order 500 mK!
Note also that the strength of the calibration signal varies as a
function of frequency, roughly according to \eqn{deltaRsmall}, as
$f^2$.  Our center frequencies of (27.5, 30.5, 34.0) GHz
correspond to a calibration ratio of 1.0:1.23:1.53, for
$J3$:$J2$:$J1$.

\subsubsection{The Calibrator Background}\label{calback}
Of course, a problem is immediately evident in the calibration
plot.  Besides slight overall offsets, which are expected as
explained in \sct{OMTcp}, the signals are not quite the same when
rotate 180\deg\ in angle, nor are the signals all equal
at multiples of 90\deg, which should correspond to no
polarization.  Although the
exact source of this effect is unknown, we believe it to be
related to beam-spillover to the calibrator frame, which was made
entire of lucite.  It is possible that via both reflection and
emission, sidelobe pick-up contributed additional polarized signals
to the radiometer during calibration; if these were not constant as
a function of angle (say, due to the presence of the person
calibrating), this could lead to the observed asymmetry, and have
a non-negligible effect.

Thus, we believe there was a ``background'' associated with the
calibrator that only manifested itself when the polarization signal
generated by the dielectric sheet was $\lesssim$ 100 $mK$, as in the
case of the 0.003" sheet we used for most of the season.
Calibrations done at the beginning of the season with a 0.020"
sheet, which itself produces close to 50 times as much signal, do
not exhibit this effect.  Therefore, we presume the effect to be
only important in the case of the 0.003" sheet calibration.

Presumably, there is some small spurious polarization generated by
the lucite calibrator frame, scattering radiation
that is some mixture of sky and 300 K, and this mixture is
dependent upon the orientation of the calibrator.
In order to correct for the effect, calibrations done with the two
sheet thicknesses were compared, and correction factors were
derived for each channel.  The correction factor was $\lesssim
2\%$
for J1 and J2, but was about $1.25$ for J3, such that the
derived NET was lower than without performing the correction.

The errors on these corrections factors are based simply on the
statistical variation in the correction factors between the \app\ 3
calibrations where correction factors could be derived.
Unfortunately, this effect was not realized until well after the
experiment had been dismantled
\footnote{\fixspacing Thanks very much to Phil Farese, who noticed this effect
while trying to use this calibration scheme in order to calibrate
the COMPASS telescope.}, so we could not fully characterize
it; the lesson is to always check your background, even if you
think it is non-existent!

\section{Atmospheric Correction}\label{s:at}
There is one final consideration in our calibration to be
discussed.  In addition to the correction factor for converting
from brightness to thermodynamic temperature, there is also a
correction factor due to atmospheric absorption.
The actual antenna temperature we see on the ground is given by:
\beq{atmos1}
T_{ant} = T_{ant}^{cmb} e^{-\tau} \ + \ T_{phys}^{atm} (1 -
e^{-\tau})
\eeq
where $\tau$ is the optical depth of the atmosphere, and is a
function of both frequency and atmospheric conditions.
$T_{ant}^{cmb}$ is the CMB antenna temperature, and $T_{phys}^{atm}$
is the effective physical temperature of the atmosphere.  The
atmospheric condition that most strongly affects $\tau$ in the
\ka\ frequency band is the precipitable water vapor content.

When we calibrate on the ground, the calibration signal is
obviously not travelling through the atmosphere, and so receives
no attenuation, but any astrophysical signal we receive will be
attenuated by the factor $e^{-\tau}$.  Hence, we must multiply
all our calibration coefficients by $1/e^{-\tau} \simeq 1 + \tau$,
as $\tau$ will typically be small, less than 0.1 in our band.

The main question is then how to determine the optical depth at
any observation time, for each of our three channels.  If we had
simply one large band spanning 26-36 GHz, we could use our total
power channels as atmospheric monitors; then the optical depth is
obtained via solving \eq{atmos1}.
The physical temperature of the atmosphere, $T_{phys}^{atm}$, is typically
\app\ 250 K, and varies very little.  Thus, this method works over a
whole band, but is not very helpful in determining the individual
sub-band corrections.

\begin{figure}[tb]
\begin{center}
\includegraphics[width=6in]{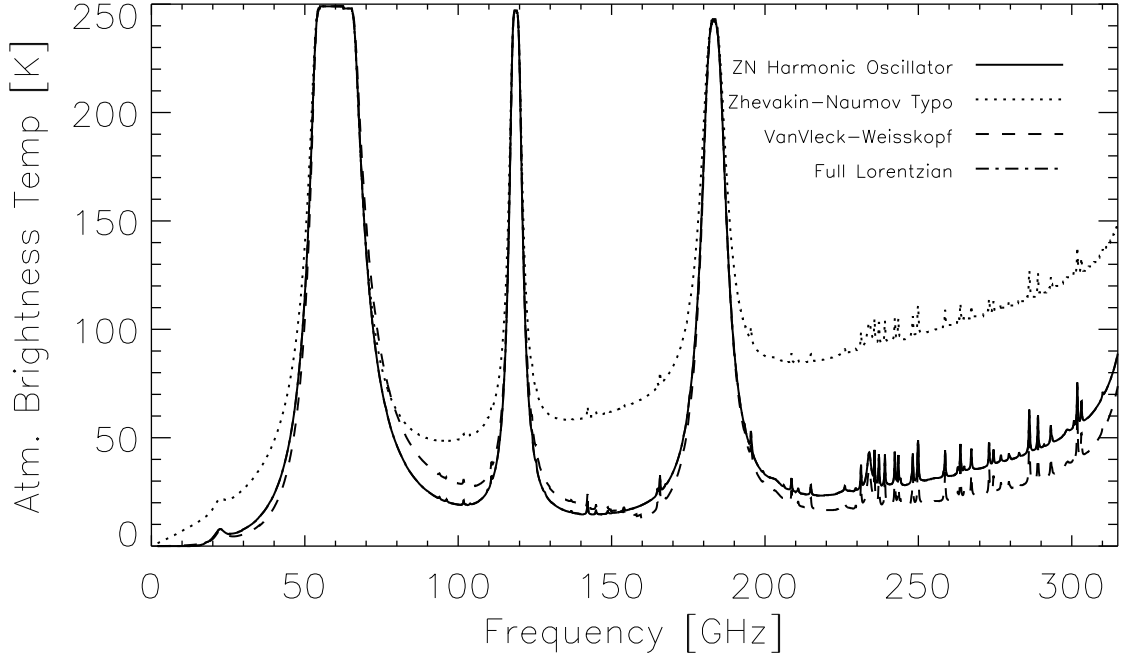}
\caption[Atmospheric Emission Profiles for Different Lineshape Types]
{\label{lineshapes} \fixspacing
Atmospheric emission profiles for different lineshape types,
assuming 3 mm precipitable water vapor.
The emission, $1-e^{-\tau}$, depends strongly on the lineshape
assumption.  This graph contains several prominent features:
the atmospheric ``windows'' centered about 30, 100, and 150 GHz;
the water lines at 22 and 182 GHz, and the oxygen lines at 60
and 119 GHz.
}
\end{center}
\end{figure}
At this point, we utilize the atmospheric model by Erich
Grossman, as implemented in his commercial $AT^\copyright$ software
\cite{ATmanual}.  We fix the latitude and altitude to be those of
our observing site, and vary the precipitable water vapor (PWV) content.
However, the least well-constrained aspect of the model is the
lineshape profile of the various lines contributing to the
atmospheric absorption profile.  This is illustrated in
\fig{lineshapes}, which shows the atmospheric brightness
temperature from 0-300 GHz.  There are four line profile shapes
typically used: Full Lorentzian, VanVleck-Weisskopf,
Zhevakin-Naumov (ZN) Harmonic Oscillator, and Zhevakin-Naumov
Typo.  The last of these was actually derived from a typographic
error in the original paper on the ZN-Harmonic Oscillator
lineshape, but was found to give reasonable, if somewhat
pessimistic (large optical depth) values.  All four profiles
give similar results near the line peak; it is how they treat the ``wings''
of the line that is different. The Full Lorentzian and
ZN-Harmonic Oscillator give virtually identical atmospheric
profiles.  In this work, we assume a ZN-Harmonic Oscillator
lineshape profile, as it is a fairly "middle-of-the-road" case,
is recommended by the AT model author, and yields atmospheric
antenna temperatures similar to those we measured in our total
power channels.

\begin{figure}[tb]
\begin{center}
\includegraphics[height=3.6in]{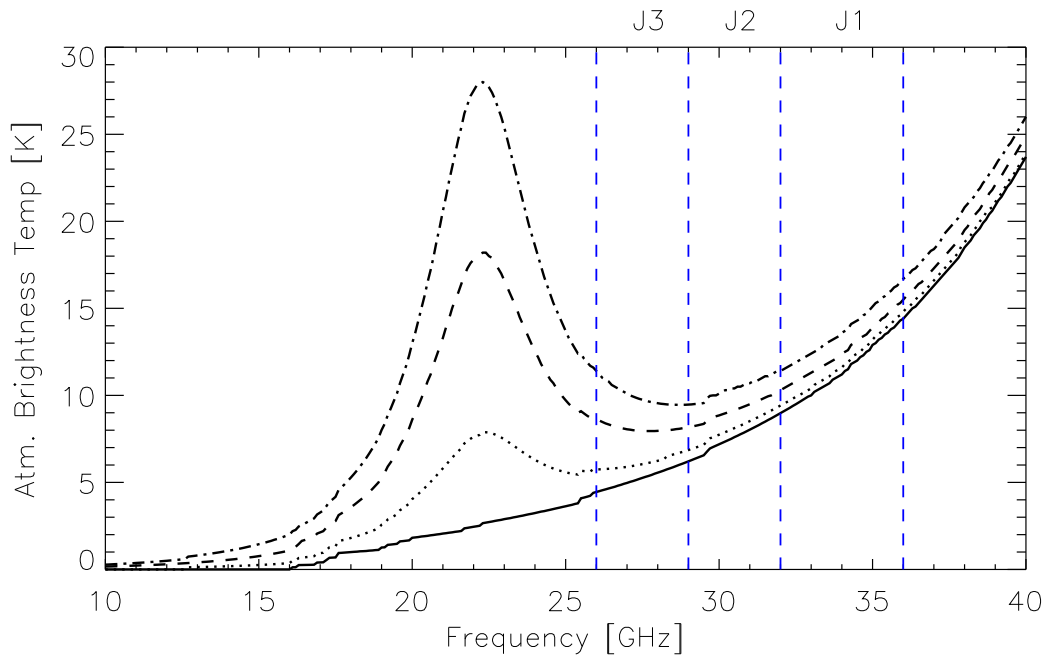}
\caption[Atmospheric Emission with Varying PWV]
{\label{Ka_pwv} \fixspacing
Atmospheric emission with varying PWV.
The PWV values are 0, 3, 9, and 15 mm, corresponding
to the solid, dotted, dashed, and dot-dashed curves, respectively.
Notice the rapid increase of the 22 GHz ``water line'' with
increasing PWV.  The vertical dashed lines correspond to our
sub-band boundaries, labelled as shown.
}
\end{center}
\end{figure}

\begin{figure}[tb]
\begin{center}
\includegraphics[height=3.2in, width=6in]{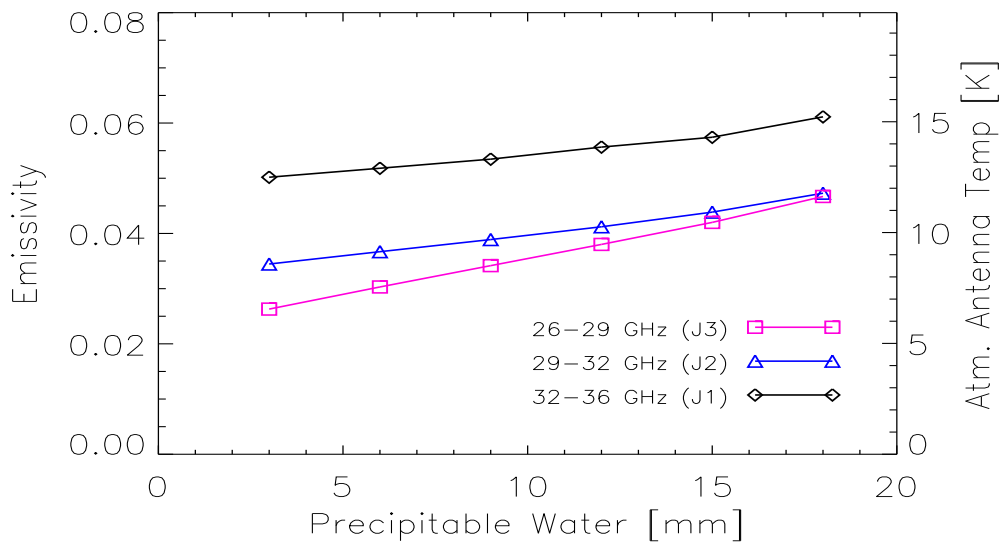}
\caption[Atmospheric Emission in the \polar\ Sub-Bands]
{\label{at_cfactor} \fixspacing
Atmospheric emission in the \polar\ sub-bands, as a function
of atmospheric water vapor.
}
\end{center}
\end{figure}

\fig{Ka_pwv} shows the effect of varying PWV across the \ka\
band, given our lineshape assumption. We chose a sea-level
temperature of 280 K, although changing this by $\pm \ 10K$ has only
a small effect on the results.  The 22-GHz ``water line'' is very
striking as the PWV is increased.  The large emission at higher
frequencies is due to the 60 GHz oxygen line.
By using this model, we can derive the atmospheric correction
factor $e^\tau$ simply by knowing the water vapor content of the
atmosphere.  \fig{at_cfactor} plots the emission, $1-e^{-\tau}$,
for each of the \ka\ sub-bands, where flat sub-bands are assumed
for each channel.  Note that the emission and hence the
correction factor is highest in the \emph{J1} sub-band.

We combined the above model with water vapor measurements made
by the Geostationary Operational Environmental Satellite (GOES)
\cite{goes01}, which was publicly available for download hourly,
in order to derive hourly atmospheric correction
factors for each channel.  The mean correction factors are given in
\tbl{t:at_cfactor}, for \emph{good} data only; that is, data
that survived the quality cuts given in \chap{cuts}.
\begin{table}[tb]
\begin{center}
\begin{tabular}{|c|c|c|} \hline
\textbf{\large Channel} & $\mathbf{e^{\tau}}$
\\ \hline
J1 & $1.0555 \ \pm \ 0.0020$ \\
J2 & $1.0391 \ \pm \ 0.0025$ \\
J3 & $1.0389 \ \pm \ 0.0019$ \\ \hline
\end{tabular}
\caption[Atmospheric Calibration Correction Factors]{
\label{t:at_cfactor}\fixspacing
Atmospheric calibration correction factors, $e^\tau$, for
each of \polars\ three sub-bands, during periods
of data that survive all quality cuts.  For each channel,
this results in a distribution of correction factors;
the numbers listed are the mean plus or minus the standard deviation
of those distributions.  Because of the relatively small variation with precipitable
water vapor content, a single correction factor was applied to each channel for
the entire season.}
\end{center}
\end{table}

\section{Putting It All Together}
Now that we have all the pieces of the calibration, let us put
them together, so that not only will we understand
where the full calibration comes from, but we'll also
get a better idea of the calibration uncertainty.
The full calibration coefficient for each channel is given by
\beq{Full_Cal_Eq}
k_i = \frac{(T_H - T_C)(\Delta R)}{\Delta V} \ \eta_{bright}
 \ \eta_{atm} \ \eta_{beam}
\eeq
where the quantities in the equation are as follows:
\begin{itemize}
\item $T_H$: Hot Eccosorb load temperature
\item $T_C$: Sky brightness temperature, sum of atmospheric + CMB
temperatures
\item $\Delta R$: $R_{TE} - R_{TM}$ for the channel, given by
\eqn{Rgeneral}
\item $\Delta V$: The voltage amplitude the channel passes
through as the calibrator spins through 360\deg
\item $\eta_{bright}$: The conversion factor from brightness
temperature to thermodynamic temperature
\item $\eta_{atm}$: The correction factor due to atmospheric
attenuation of celestial signals
\item $\eta_{beam}$: The correction factor due to beam spillover
off the hot and cold loads during calibration
\end{itemize}

It is possible to place uncertainties on each of these terms in
order to arrive at a final calibration uncertainty for each
channel, although for certain terms the uncertainties are not
particularly well known (\eg, the beam spill-over correction).
Estimates of all the above terms and their uncertainties have been
compiled into \tbl{finalcal}, as well as the calculated final
uncertainties in each channel's calibration.
Notice that the only uncertainties that really matter in the final
calibration error budget are due to $R_{TE} - R_{TM}$ and the beam
spill-over effect described in \sct{calback}.  The J1 and J2 final
uncertainties are both roughly 10\%, while the J3 uncertainty is
significantly higher due to the beam spill-over effect.

\begin{table}
\begin{center}
\begin{threeparttable}
\begin{tabular}{||c||r@{}l|r@{}l|r@{}l||} 
\hline\hline
\textbf{\large Term} & \multicolumn{2}{c|}{\textbf{\large J1}} & \multicolumn{2}{c|}{\textbf{\large J2}} &
\multicolumn{2}{c||}{\textbf{\large J3}} \\ \hline\hline
$T_H \ - \ T_C$ \tnote{1} & & $\pm\ 4 K$ (1.5\%) & & $\pm\ 4 K$ (1.5\%)&
& $\pm\ 4 K$ (1.5\%)
\\
$\Delta R$ \tnote{2} & $0.0024$ &  $\pm\ 8\%$ &
$0.0020$ & $\pm\ 8\%$ & $0.0017$ & $\pm\ 9\%$
\\
$\Delta V$ \tnote{3} & &$\pm\ \sim 1\%$& &$\pm\ \sim 1\%$& &$\pm\ \sim 1\%$
\\
$\eta_{bright}$ & \multicolumn{2}{c|}{1.0303} &
\multicolumn{2}{c|}{1.0243} & \multicolumn{2}{c||}{1.0197}
\\
$\eta{atm}$ \tnote{4}& $1.056$ & $\pm\ 0.5\%$ & $1.039$ & $
\pm\ 0.5\%$ &
$1.039$ & $ \pm \ 0.5\%$
\\
$\eta{beam}$ \tnote{5} & $0.98 $ & $\pm\ 2\%$ & $0.99$ & $\ \ ^{+1\%}_{-2\%}$
& $0.80$ & $\ \ ^{+5\%}_{-15\%}$
\\ \hline
Final Uncertainty&
\multicolumn{2}{c|}{$\pm \ 8.5\%$} &
\multicolumn{2}{c|}{$\pm \ 8.5\%$} &
\multicolumn{2}{c||}{$\ +11\%, \, -18\%$}
\\ \hline\hline
\end{tabular}
\caption[Calibration Uncertainties as a Function of Channel]
{\label{finalcal} \fixspacing
Values and uncertainties in all calibration terms from
\eqn{Full_Cal_Eq}, as a function of channel.  For those terms that
changed throughout the season, only the uncertainties are shown.
Standard statistical error propagation was used to arrive
at the final uncertainties.\\
}
\begin{tablenotes}
\fixspacing
\item [1] {\footnotesize Uncertainty due to rough TP measurement of sky
temperature, as well as interpolating to the given channel.}
\item [2] {\footnotesize Uncertainty due primarily to index of refraction
    and thickness of the dielectric.}
\item [3] {\footnotesize Uncertainty due to noise in channel, and offset
drifts during each calibration.}
\item [4] {\footnotesize Larger error adopted because of
uncertainty associated with lineshape profile.}
\item [5] {\footnotesize Uncertainty due to lack of knowledge of beam spillover
location, and in measurement from 20-mil dielectric to determine
correction factor.}
\end{tablenotes}
\end{threeparttable}
\end{center}
\end{table}
\clearpage
\subsection{Calibration Variations Throughout the Season}
\begin{figure}
\begin{center}
\includegraphics[height=3.5in]{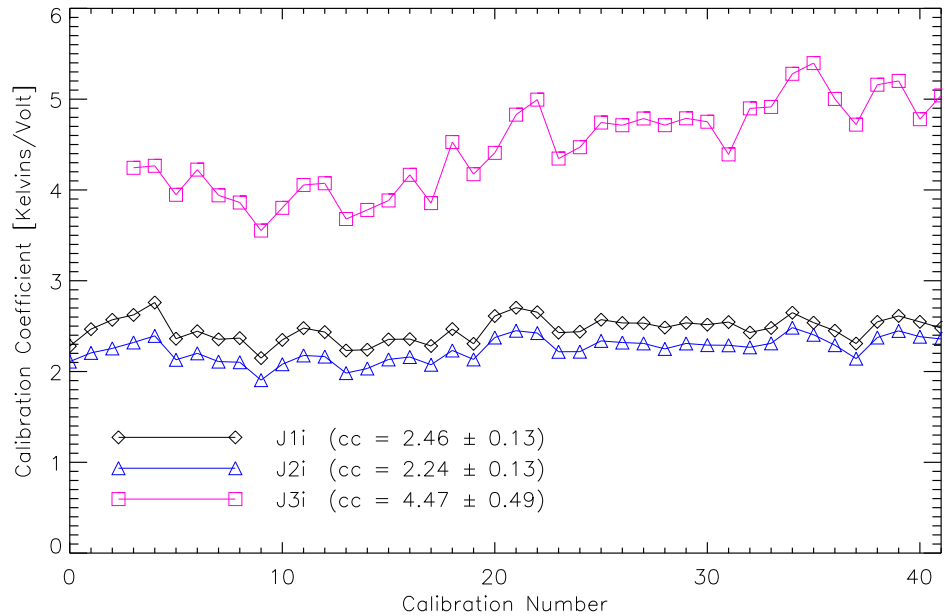}
\caption[Calibration Variations Throughout the Season]
{\label{calvariations}\fixspacing The calibration constants
of our three signal channels (J1i, J2i, J3i) measured throughout the observing
season, using the DSC with $T_C = sky$ and $T_H = 300 K$.
The term \emph{cc} denotes ``calibration coefficient''.  Some innate scatter, due to both system variations and human error, is evident
in the plots.  The J3 beam-correction factor has not been applied here.}
\end{center}
\end{figure}
\fig{calvariations} shows our \app\ 40 calibration measurements taken
throughout the season.  Overall, we were quite pleased with the
stability.  In the two best channels, J1i and J2i,
the 1-$\sigma$ variation is
about 5\% over the course of the season.
Computing a $\chi^2$ is uninformative, because the
inherent uncertainty in each calibration is mostly systematic.
There is some random error due to human error, such as not
stopping the calibrator at the correct point.  Also, as
certain system parameters varied over the course of the season,
such as the HEMT temperature, we do not expect these curves to
be precisely constant.
J3i's calibration varied somewhat more, by about 10\%, but
this is still not enough to worry about it for the very same
reasons.

For a section of data taken between two calibrations, we
used the mean calibration coefficients of those two
measurements.  We opted this over a more complicated
linear extraction because of its simplicity, and the fact that calibrations
did not vary much between two adjacent calibrations.

\clearpage
\begin{chapappendix}
\section{Derivation of Calibration Signal Stokes parameters}
Our goal in this appendix is to determine the Stokes parameters
due to the electric fields generated by the dielectric sheet
 calibrator (DSC) as shown in
\fig{calsetup}.  Specifically, we would like to know $I_x$, $I_y$,
$Q$, $U$, and $V$, where $I_x$ ($I_y$) is the intensity of the
electric field polarized along the $\hat{x}$ ($\hat{y}$) axis. The
Stokes parameter I is then simply given by $I = I_x + I_y$.

By looking at \fig{calsetup}, we see that the $\hat{x}$-axis
corresponds with TE-polarized electric fields, of which there are
essentially three: the TE-field from $T_C$ transmitted through the
sheet, the TE-field from $T_H$ reflected from the sheet, and the
TE-field emitted from the sheet itself.  Similarly, the
$\hat{y}$-axis corresponds with TM-polarized fields.  Thus, we
have
\be
I_x = \abs{t_{TE}}^2 T_C + \abs{r_{TE}}^2 T_H +
\epsilon_{TE} T_S \ ,
\ee
where $r_i$ is the ratio of the
reflected to incident electric field polarized along $\hat{i}$ due
to the sheet, and likewise $t_i$ is the ratio of transmitted to
incident electric field polarized along $\hat{i}$.  Using the fact
that $\abs{r_i}^2 = R_i$ and $\abs{t_i}^2 = 1 - R_i - \epsilon_i$
(the latter being due to \eqn{RTA}), we can recast Equation (A1) as
\be I_x = T_C + (T_H - T_C) R_{TE} + (T_S - T_C)
\epsilon_{TE} \ ,
\ee
which is the form given in
\eqn{unprimedstokes}.  The derivation for $I_y$ follows the same
format and yields
\be I_y = T_C + (T_H - T_C) R_{TM} + (T_S
- T_C) \epsilon_{TM} \ .
\ee

Now we can use the fact that $Q = I_x - I_y$, which immediately
leads to \eqn{Qsignal}.  Next, due to the rotation properties of Q
and U we can write U as \beq{Ueq} U = I_{x_{45}} - I_{y_{45}} \eeq
where ${\hat{x}}_{45}$ refers the the axis rotated +45$\deg$ from
$\hat{x}$, and ${\hat{y}}_{45}$ is the orthogonal axis. However,
as the $\hat{x}$ and $\hat{y}$ axes are exactly aligned with the
TE and TM states, the 45$\deg$-rotated axes will contain equal
amounts of TE and TM fields, and their intensity difference will
be zero.  Thus we have \be U = 0 \ . \ee

Finally, we must consider the possibility of the sheet
contributing a circular polarization signal V to our hypothetical
polarimeter.  We only expect this if there is some coherent phase
delay between TE and TM polarizations, to give the final
polarization state some ellipticity.  This cannot happen from the
unpolarized loads, but the emission from the sheet as seen at
oblique angles will in general be elliptically polarized due to
its imperfect transparency \cite{sandus}; however, this will be
proportional to the emissivity of the sheet and hence will be
small enough in comparison to the other Stokes parameters that it
can be ignored for the purposes of this paper, and we take
\be
V \approx 0 \ .
\ee

\section{Derivation of Simplified $R_{TE}-R_{TM}$}
The purpose of this appendix is to derive the quantity
$R_{TE}-R_{TM}$ under the simplifying assumptions that $\lambda
\gg t$ (which is equivalent to $\delta \ll 1$), and $\theta = 45
\deg$.  Applying the latter assumption to Equations (\ref{gamma})
and (\ref{deltadef}) yields
\begin{subequations}
\begin{eqnarray}
\gamma_{TE}^2 & = & n^2 - 0.5 ,\\
\gamma_{TM}^2 & = & \frac{1}{n^4}\left(n^2-0.5\right)
\end{eqnarray}
\end{subequations}
and
\be
\delta = k t \sqrt{n^2-0.5} \ .
\ee Substituting these
expressions into \eqn{Rgeneral}, and requiring that $\delta \ll
1$, we have
\begin{eqnarray}
R_{TE} & \simeq & (k t)^2 \frac{(n^2-1)^2}{2} \\
R_{TM} & \simeq & (k t)^2 \frac{(n^2-1)^4}{8 n^4}
\end{eqnarray}
Finally, solving for $R_{TE}-R_{TM}$ we find
\be
R_{TE}-R_{TM}
\simeq (k t)^2 \frac{(n^2-1)^2}{8 n^4}\left[4 n^4 - (n^2-1)^2
\right]
\ee
which factors into
\be R_{TE}-R_{TM} \ \simeq \
\left( \frac{\pi f t}{c} \right)^2 \frac{(n^4-1)(n^2-1)(3 n^2
-1)}{2 n^4}
\ee
as desired.
\end{chapappendix}



%
%
%
%
\chapter{Observations and Weather Data}\label{observations}
\polar\ was deployed into the field on March 1, 2000.  After a period
of setting up, calibrations, and systematics tests, observations were
begun in earnest on March 11, and continued through May 29, 2000.
\polar\ was situated at the Pine Bluff Observatory, in Pine Bluff,
Wisconsin (latitude = 43.08\deg, longitude = 89.69\deg W). This is a
small astronomical site that the UW-Madison Department of Astronomy
owns and operates; it sits on a small hill in the country,
circumscribed by trees that are all less than 10\deg\ above the
horizon.  Our observing philosophy was to always take data unless it
was raining or snowing.  This was possible through the internet-based
dome operation system described in \sct{s:cube}; this system allowed
for fast response times, sometimes enabling us to take data very
close in time to severe weather \footnote{\fixspacing This system was
not infallible, and more than once resulted in \polar\ being rained
upon.}.

\section{Observing Region}
\begin{figure}[tb]
\begin{center}
\includegraphics[height=4in]{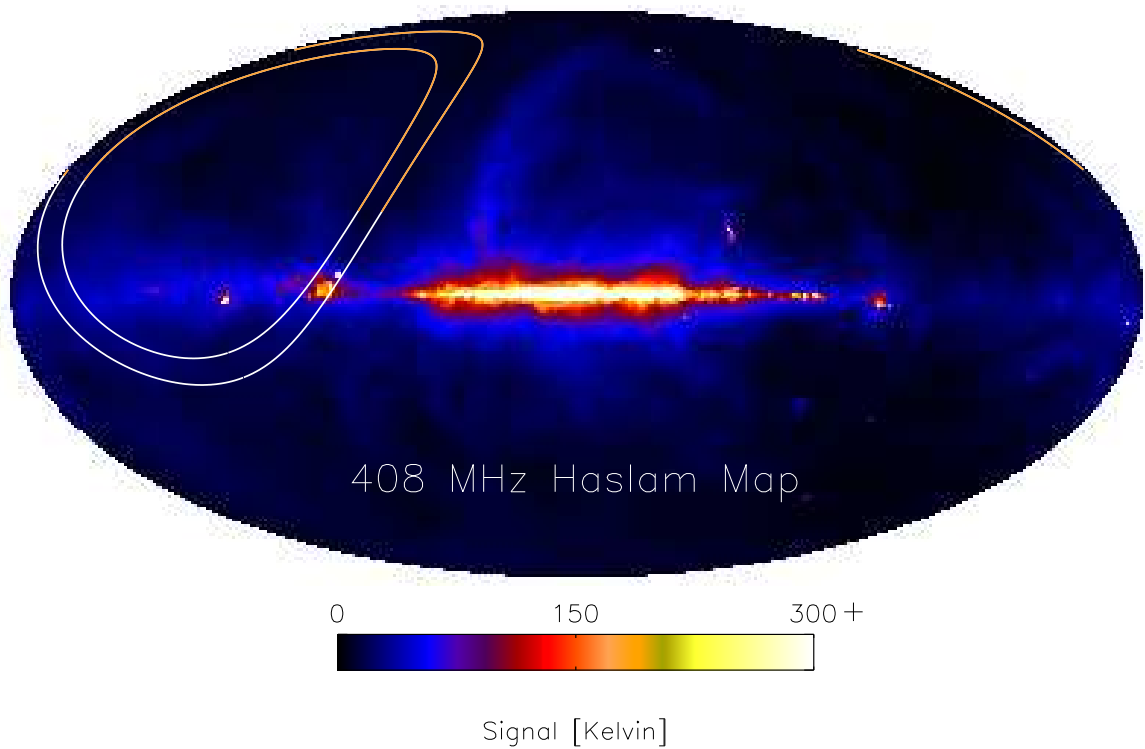}
\caption[Scan Strategy overlaid on Galactic Synchrotron map.]
{\label{obs_strategy}\fixspacing
The \polar\ scan strategy,
overlaid on the 408 MHz Haslam Synchrotron map, plotted in
galactic coordinates.  The off-colored portion of the scan
path denotes the area where galactic latitude $|b| > 25\deg$;
we ended up cutting all data taken below this point to avoid
galactic contamination.}
\end{center}
\end{figure}

\polar\ utilized a simple zenith drift scan.  The instrument was
initially levelled with the jacks on each of the four corners of the
mount, to within 0.2\deg\ of the actual zenith.The zenith scan
resulted in an observation region spanning right ascensions $0h$ to
$24h$, at a declination equal to our latitude of 43.08\deg. For our
beamwidth of \app\ 7\deg, this corresponds to approximately 38
uncorrelated pixels on the sky.  The path of our observing strategy
is shown in \fig{obs_strategy}.  Note that it passes through both
high-- and low-- galactic latitude regions.  The low galactic regions
are interesting as they are likely to house the strongest polarized
foregrounds, especially the region around right ascension 20\emph{h},
which corresponds to the very bright radio source Cygnus A, and is
very obvious in \fig{obs_strategy}.  In the actual CMB analysis, we
eliminated all regions with galactic latitude $|b| < 25\deg$.

\section{Structure of the Data}
\polar\ collected approximately 750 hours of data during the 2000
observing season.  There were three principle operators of the
instrument\footnote{The operators were Slade Klawikowksi, Chris
O'Dell, and Peter Timbie.}.  Every day one of the operators would
drive to the Pine Bluff site, check on \polar, perform a calibration
(described in \sct{s:dailycal}), and of course, fix things when they
broke.

\begin{table}[tb]
\renewcommand{\baselinestretch}{0.8}\small\normalsize
\setlength{\extrarowheight}{2mm}
\begin{center}
\begin{tabular}{||c|m{2.5in}|c||}
\hline \hline
\textbf{\large Name} &
\multicolumn{1}{c|}{\large \textbf{Description}}
& \textbf{\large Duration} \\
\hline
\emph{Section} &
One deployment, between calibrations, during which the
dome was open and data was recorded.  There were 49 total
sections in this data set. & 2-48 hours \\
& & \\[-1.5 ex]
\emph{Hour File (HF)} & One data file, containing 9000 samples/channel.
The Header of the file contains data/time, pwv, and weather
information. & 7.5 min \\
& & \\[-1.5 ex]
\emph{Rotation} & A single rotation's worth of data, defined between
consecutive AOE pulses.  Typically contained 630 samples. &
30.6 sec \\
& & \\[-1.5 ex]
\emph{Sample} & A single sample of data, containing one recording of each
of the sixteen data channels.  & 0.05 sec \\
& & \\[-1.5 ex]
\emph{Submap} & Sky map (signal vs. RA) derived from the \textit{good}
data within one section. & 2-15 hours \\
\hline \hline
\end{tabular}
\end{center}
\caption[Structure of the \polar\ Data]{\fixspacing
\label{t:datastructure} The various levels of \polar\ data
organization.}
\end{table}

The structure of the data is described in \tbl{t:datastructure}.
Periods between calibrations during which we were acquiring data were
called \textbf{\textit{Sections}}. During acquisition, data were
recorded to files 7.5 minutes in length, called \textbf{\textit{Hour
Files}} (hereafter HF)\footnote{\fixspacing The misnomer ``Hour
File'', labelling a 7.5 minute data file, can be traced to Brian
Keating.}.  As we took data at 20 Hz, there were precisely 9000
samples per channel recorded in one Hour File. Additionally, several
pieces of information were recorded in the header of each HF:
\begin{itemize}
\item Date and Time measured by a commercial time-stamping card.
\item Precipitable Water Vapor, as extracted by the GOES
satellite\footnote{\fixspacing PWV and cloud cover are available
hourly from the SSEC site at ftp://suomi.ssec.wisc.edu/pub/rtascii
.}\cite{goes01}, for the region over Pine Bluff, WI.
\item Latest weather information from the \polar\ weather station:
temperature, relative humidity, wind speed and direction.
\end{itemize}
When the instrument was spinning normally, it rotated between 13 and
14 times per HF.  Throughout the course of the season, \polar\
rotated more than 90,000 times.

\section{Weather Conditions}
Some members of our team had reservations about performing sensitive
CMB observations from a site that traditionally sees fairly wet and
unstable weather.  Most ground-based CMB experiments are performed at
high, dry sites, such as the Atacama desert in Chile, the South Pole,
or the beautiful Tenerife island in the Mediterranean.  However, as
described in \chap{foregrounds}, the atmosphere is not expected to be
polarized, and thus ground-based polarization observations are
theoretically possible. Of course, increased sky loading will add to
the \emph{noise} of our experiment, requiring longer integration
times to see a given signal.

\begin{figure}[tb]
\begin{center}
\subfigure[]{\label{weath-a}
\includegraphics[width=3in]{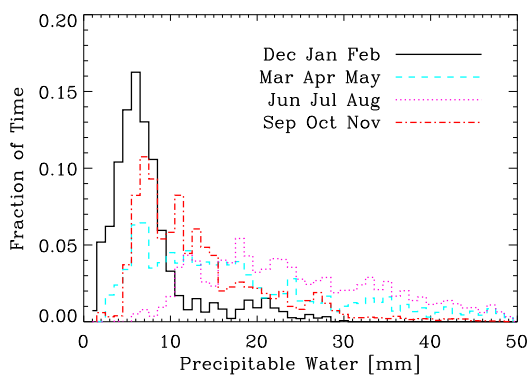}
}
\subfigure[]{\label{weath-b}
\includegraphics[width=3in]{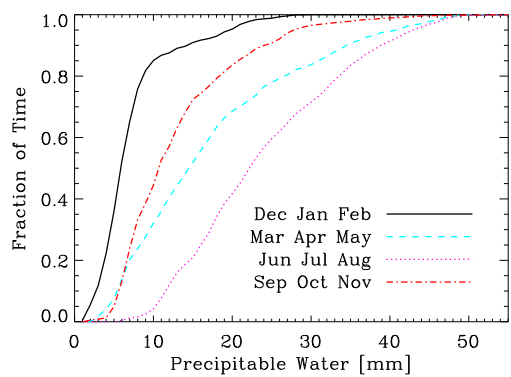}
}\\
\subfigure[]{\label{weath-c}
\includegraphics[width=3in]{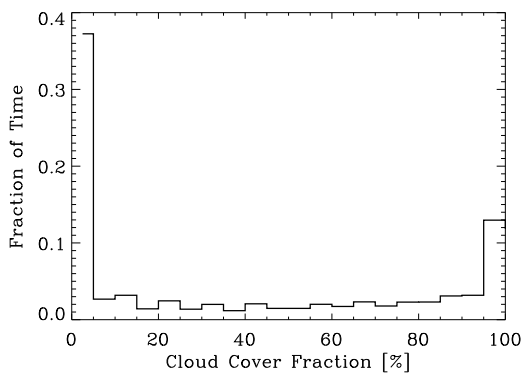}
}
\subfigure[]{\label{weath-d}
\includegraphics[width=3in]{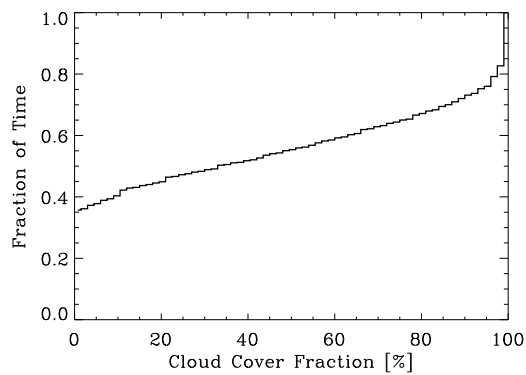}
} \caption[Selected Pine Bluff Weather
Data]{\label{f:weather}\fixspacing Precipitable Water Vapor (PWV) and
Cloud Cover for Pine Bluff, WI, as determined from hourly GOES
satellite data. (a): Shows the PWV histogram for the years 1997,
1998, and 2000, separated into seasons.  (b): Shows the same
information as (a), but in integrated form (fraction of time less
than a certain value). (c): Displays the cloud cover percentage
histogram for 2000. As cloud cover was not very seasonally dependent,
the full year is displayed as a single curve. (d): Same as (c), but
in integrated form. All data were provided by Gail Bayler and Gary
Wade at the UW-Madison Space Sciences and Engineering Center.}
\end{center}
\end{figure}

We compiled data on the Pine Bluff area from both the National
Weather Service and the GOES satellite data served by the Space
Sciences and Engineering Center at UW-Madison \cite{goes01}. The
GOES-8 data proved most useful as it was recorded hourly, and
typically hit a patch with area 5 km by 5 km, within 20 km of Pine
Bluff. It provided cloud cover fraction of the area, as well as
precipitable water vapor (PWV) column height (along with a host of
other weather variables). We acquired all the data for 1997, 1998,
and 2000 on precipitable water, but only the year 2000 for cloud
cover information. These data are displayed in \fig{f:weather}.
Figures \ref{weath-a} and (c) show PWV and cloud cover fraction
histograms, while Figures \ref{weath-b} and (d) show the same
information in ``integrated'' form; that is, they display the
fraction of time the variable was less than a certain value.  PWV
showed a strong seasonal dependence; in the figures, the three years
of data are averaged, but separated by season as shown.  Notice that
in the winter, more than 50\% of the time has PWV $<$ 5 mm, and
almost 90\% less than 10 mm. This is in stark contrast to summer,
where around 95\% of the time the PWV is greater than 10 mm! Clearly,
summer is a bad time for CMB observations in Wisconsin. Spring and
Autumn lie in between, but are still less than ideal.  It is perhaps
noteworthy that Autumn is markedly better than Spring in terms of
PWV.

Cloud cover fraction is not seasonally dependent, but does exhibit a
bimodal distribution, with more than 35\% of the time showing totally
clear weather.  However, about 15\% of the time is categorized as
completely overcast (and will likely be useless for CMB
observations).  The ``partially cloudy'' days account for the other
50\% of the distribution.

\begin{figure}
\begin{center}
\subfigure[]{\label{weath2-a}
\includegraphics[width=3in]{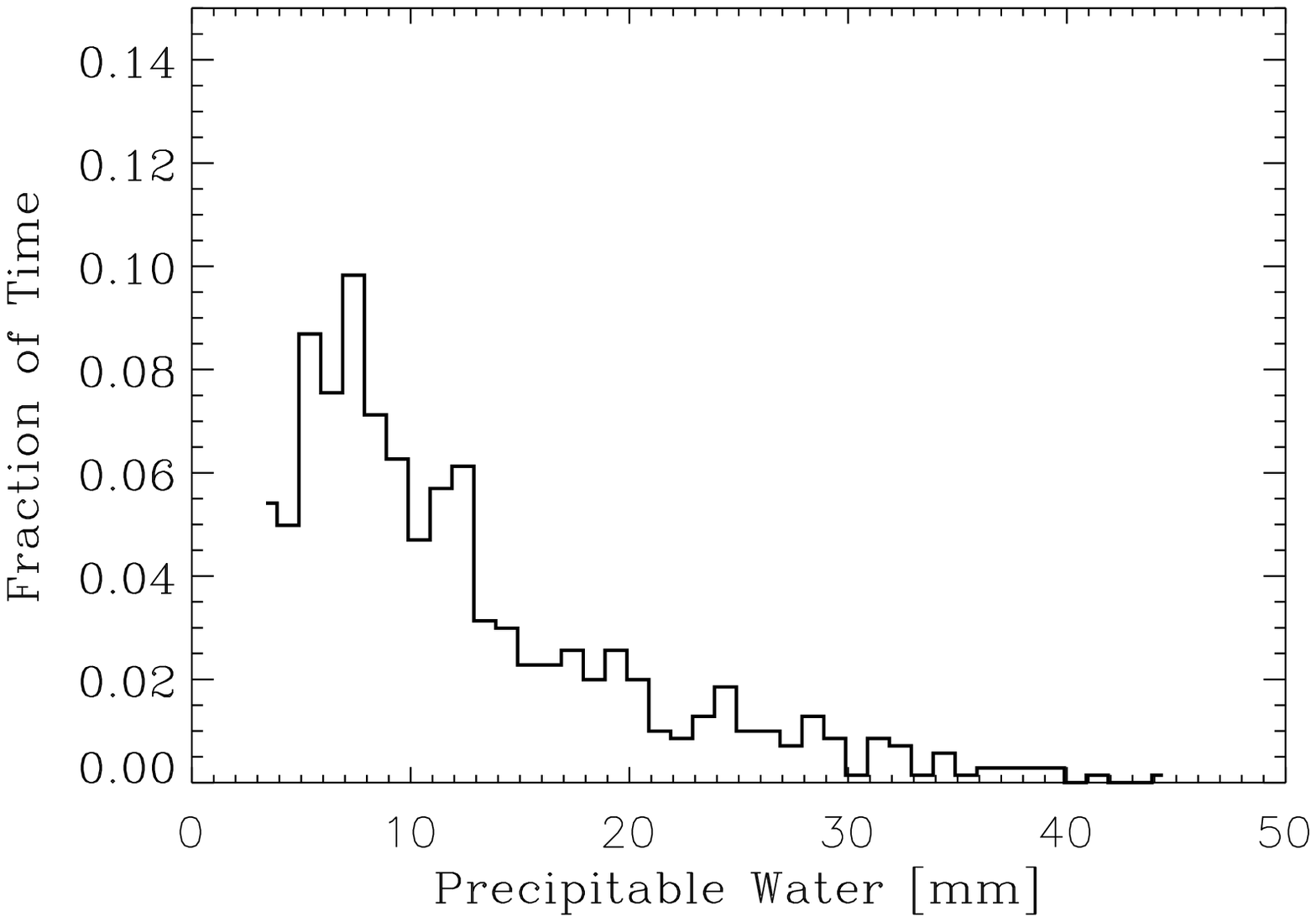}
} \subfigure[]{\label{weath2-b}
\includegraphics[width=3in]{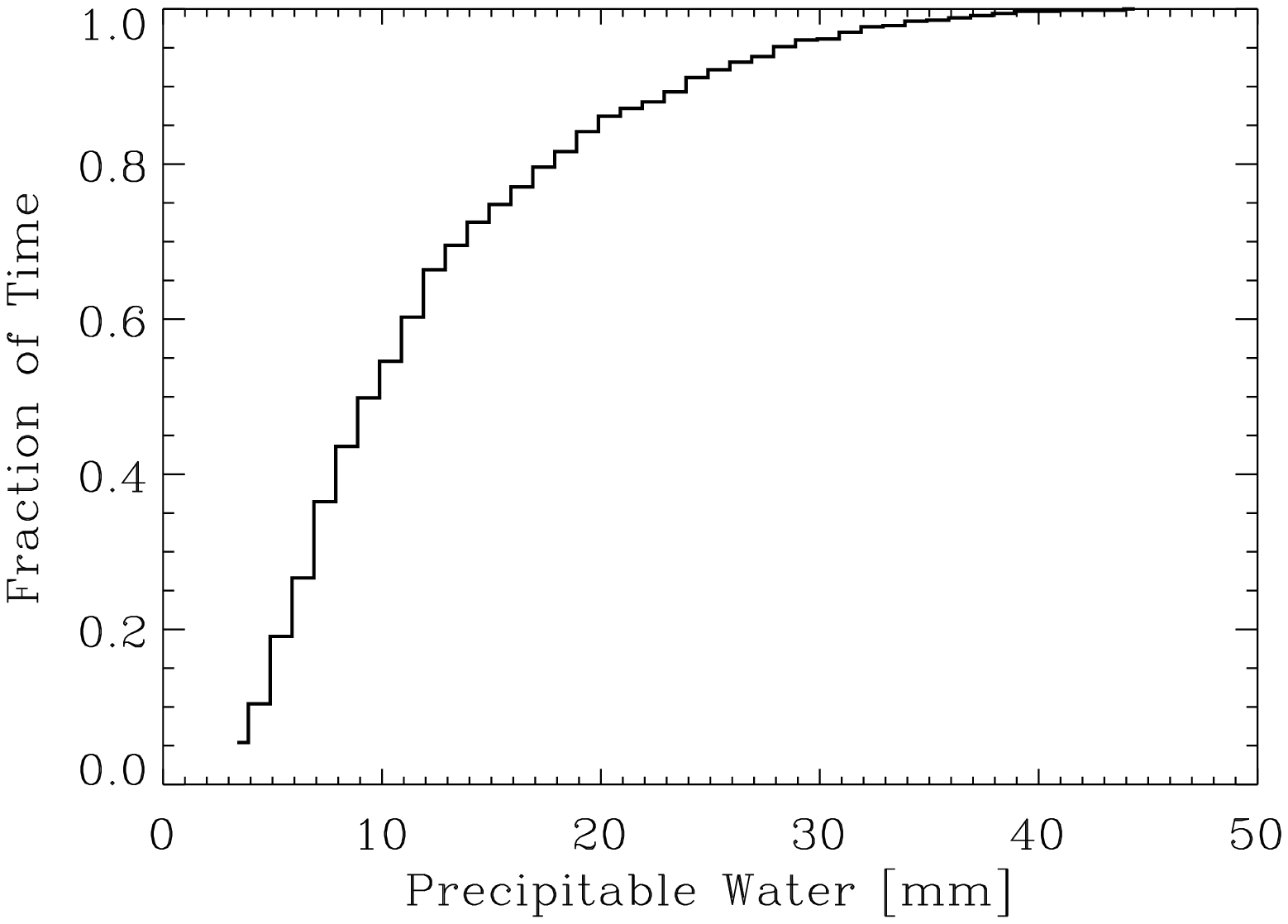}
} \caption[Precipitable Water Vapor during Observation Season]
{\label{f:weather2}\fixspacing Precipitable Water Vapor during
Observation Season.  (a) shows the PWV histogram during the observing
season, while (b) shows the same data in integrated form.  More than
50\% of this period has PWV less than 10 mm. }
\end{center}
\end{figure}

While the general Wisconsin weather displayed in \fig{f:weather} is
interesting in its own right, what were these values during actual
observations in the Spring of 2000?  The cloud cover distribution is
virtually unchanged, and the water vapor follows the general
Springtime trend, as shown in \fig{f:weather2}. They are consistent
with the previous distributions for Spring weather in Wisconsin.   We
leave it to the next chapter to discuss the effects of the weather
upon the data, but the general conclusion here is that there will be
a reasonable number of clear, dry days in Wisconsin, even in Autumn
and Spring.  Sensitive measurements of CMB polarization are clearly
possible from the ground, even from a relatively poor site such as
Wisconsin.


%
%

\chapter{Data Selection}\label{cuts}
The \polar\ observing season contained a large diversity of
weather conditions, and this led to a correspondingly large
diversity of data quality.  This is in stark contrast to a typical
balloon or satellite mission, where often most of the data set is
of the same quality, and various statistics of the data do not
change much over the data set.  For \polar, we observed through
cloudy, humid conditions as often as through clear, dry
conditions.

Thus, developing robust data selection techniques was one of the most
critical tasks in the data analysis pipeline. We used standard
techniques of time series analysis to look at the data in different
ways, in order to assess its quality and to develop criteria on which
to cut. Unfortunately, because of the data's diversity, we were not
able to arrive at a single selection criterion; rather, we developed a
battery of conditions that the data must pass before being accepted.
Our underlying philosophy was to cut ``more rather than less'', to
ensure a final data set of the highest possible quality.

\section{Data Analysis Overview}
There are many measures of data quality, but the critical thing for
this experiment is that \polar\ \emph{rotated}. It is then natural to
describe the output of \polar's correlator channels in the following
form:
\beq{polaresponse}
y(t) \;=\; I_0 + C \cos{\omega t} +
S\sin{\omega t} + Q \cos{2\omega t} + U\sin{2\omega t} + n(t)\qquad
\eeq
where $\omega = 2\pi f = 0.2055 \;rad\; s^{-1}$ was \polars\
(angular) rotation frequency. The constant offset $I_0$ is due
to coupling of the unpolarized total power signal into the correlators via the
nonzero cross-polarization of the OMT described in \sct{OMTcp}. This
offset term was typically 10-100 mK, depending on the channel; during
good weather its stability was better than 0.6 mK per hour.

$C$ and $S$ are signals modulated at the rotation frequency
(referred to hereafter as \onephi\ signals), and can be caused by
various types of ground pickup and other systematic effects.
During good weather, there is often no ``offset'' associated with
$C$ or $S$. An offset is quite conceivable, as any type of
ground-source picked up in the sidelobes of our beam would have a
strong component at \onephi. We often saw
quite strong signals at the \onephi\ frequency, but these signals
were not ever particularly stable, mainly being associated with
time-varying atmospheric conditions.

The signal we want to extract from our raw data is the astrophysical
polarization signal in terms of $Q$ and $U$, and this corresponds
only to signals modulated at \emph{twice the rotation frequency} (hereafter
\twophi). Thinking in frequency space, only signals modulated at this
particular frequency can produce true polarization signals, and all
other frequencies will be rejected in the analysis. The next chapter
will discuss how this extraction was done in practice.

Many effects can conspire to contaminate the $Q$ and $U$ signals,
be they instrumental, atmospheric, or celestial. In general it was
not possible for us to cut bad data based on a single quantity
that would apply to all causes of bad data. In fact, it was often
difficult to determine the precise cause of bad data. This last
statement begs the question, what precisely is ``bad data''? The
simple answer is any data with a non-cosmological contribution to
our signal, that mimics a cosmological signal in a way that we
cannot account for and remove. For instance, attenuation by the
atmosphere scatters CMB photons as described in \sct{s:at}; but
the effect is calculable, and does not (in principle) add a
spurious polarized component. On the other hand, observing a finite,
asymmetric cloud will lead to a spurious signal if we have an
asymmetric beam, which we do at some small level. This will lead
to an apparent signal in $Q$ and $U$, and this signal will change
as the cloud moves through the beam.  This is too difficult to
model for and subtract out; therefore, the data must be cut.

The rest of this chapter will describe the various criteria we
established, to separate the ``good'' data from the ``bad''. I use
quotes here with a purpose; ``good'' is a relative term, and in
some sense the argument is circular because we define
what is good data with the cuts will we establish. \tbl{cuttable}
lists all of the cut criteria used in \polar; the entries in the
table will be described throughout the rest of this chapter.

\begin{table}
\begin{center}
\begin{tabular}{|c|c|c|c|c|} \hline
\textbf{\large Cut Type} & \textbf{\large Remaining} & \textbf{\large Additional Cut} & \textbf{\large Individual Cut}
\\ \hline
No Cuts (full data set) & 746.5 (100\%) & 0 (0\%) & 0 (0\%) \\
Sun Elevation $< \  20\deg$ & 458.6 (61.4\%) & 287.9 (38.6\%) & 287.9 (38.6\%) \\
Moon Elevation $< \ 50\deg$ & 430.8 (57.7\%) & 27.9 (3.7\%) & 60.6 (8.1\%) \\
Dew Cut & 348.4 (46.7\%) & 82.4 (11\%) & 105.4 (14.1\%) \\
Proper Rotation & 346.8 (46.5\%) & 1.6 (0.2\%) & 13.8 (1.8\%) \\
$\onephir\ < \ 2.1$ & 238.1 (31.9\%) & 108.7 (14.6\%) & 387.7 (51.9\%) \\
$\Zeta \ < \ 4.0$ & 210.2 (28.2\%) & 27.9 (3.7\%) & 303.9 (40.7\%) \\
1/f Knee $< .060$ Hz & 204.8 (27.4\%) & 5.4 (0.7\%) & 294.9 (39.5\%) \\
Spike Two $< \ 5\sigma$ & 195.8 (26.2\%) & 9 (1.2\%) & 122 (16.3\%) \\
Galactic Lat $> \ 25\deg$ & 152 (20.4\%) & 43.9 (5.9\%) & 398.6 (53.4\%) \\
Nearest Neighbor Cut & 113.3 (15.2\%) & 38.7 (5.2\%) & N/A \\
8 consecutive HFs & 94.8 (12.7\%) & 18.5 (2.5\%) & N/A \\
Duration $\geq \ 3$ Hours & 78.1 (10.5\%) & 16.7 (2.2\%) & N/A \\
Remove Q,U Spikes & 77.3 (10.4\%) & 0.8 (0.1\%) & N/A \\
Executive cuts & 71.1 (9.5\%) & 6.2 (0.8\%) & N/A \\ \hline
\end{tabular}
\end{center}
\caption[Effect of Various Data Cuts]{\fixspacing \label{cuttable}
Effect of various data cuts.  The data set was viewed as three
independent data sets (one for each channel), and shown are the cuts
for this entire data set as an average of the individual channel cuts.
\textbf{Remaining} is the amount of data left after that cut and
all the cuts above it have been applied; \textbf{Additional Cut} is the amount of data
cut at that stage; \textbf{Individual Cut} is the amount of
data that would have been cut if that particular criterion were the
only one applied. Thus, many cuts are overlapping.  Of the 750
hours of data taken, only about 10\% survived our cut criteria.}
\end{table}

\section{Data Selection Techniques}
\subsection{The Fiducial Data Unit}
Before deciding those statistics upon which to cut, one must
ponder the question of the \emph{time scale} over which to cut.
That is, do we simply remove individual data points we do not
like?  Or variable-length segments of data?  If we cut one
channel, do we cut them all?

We chose to cut at the hour-file (HF) level of the data, and we did this
for several reasons.  The simple answer was convenience: that is how
our data were packaged.  Another reason was that an HF contained 9000
samples/channel, which is a large enough number to get a good
representation of the low-frequency power spectrum of the data during
that particular HF.  This was important because our mapmaking
technique (described in \chap{mapmaking}) needs to know the noise
properties of the data in order to turn the data into a map. We could
also have cut on individual rotations of the system, but the
mapmaking technique didn't work nearly as well on individual
rotations, so we stuck with full hour files.

Finally, we chose to perform the analysis on each of our three
sub-bands individually, and thus we ended up cutting somewhat
different portions of the data for each channel. Our channels
exhibited fairly different responses to spurious signals and
systematic effects, in particular channel J3, which often showed
contamination when J1 and J2 did not.  In order to save many
``good'' sections of data in J1 and J2, many of our data-based
cuts applied to each channel individually.

\subsection{Weather-Based Cuts}
The first level of cuts was weather-based.  Remembering that our
philosophy was to always take data if possible, it is not surprising
that much of our data were taken during periods of highly non-ideal
weather conditions: high water vapor or humidity, rapidly changing
temperature, strong winds, and diverse types of cloud cover, as well
as the sun or moon being close to our main beam. Most of these
conditions have a mechanism to contaminate our data,
and thus it is important to consider each of these mechanisms and
understand at what level they contaminate the data, and then remove
these periods when the contamination is unacceptable.

\subsubsection{Dew on the Window}
\begin{figure}[tb]
\begin{center}
\includegraphics[height=3.5in]{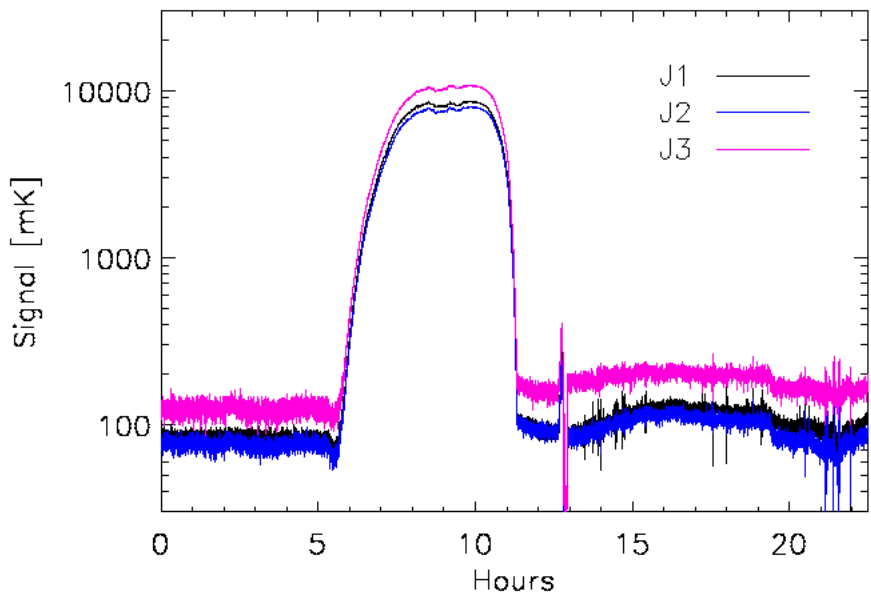}
\caption[Dew Effect on the Time Stream] {\label{dew} \fixspacing
The Effect of Dew on the time stream.  All three in-phase
correlator bands are shown.  Not only does dew on the
window increase loading, but it generates a strongly polarized
signal, presumably by scattering 300 K radiation into the system, and
partially polarizing it.  It is interesting to note that the
resultant polarized signal is largely independent of frequency.
}
\end{center}
\end{figure}

The first issue was dew formation on the optics, a common problem
in ground-based experiments. Specifically, when the relative
humidity reached 95\% or so for a long period of time (an hour or
more), moisture would condense on the top surface of the window
(namely, the layer of Volara).  This immediately led to a strong
polarization effect in the data, as shown in \fig{dew}.  Most
likely ambient (300 K) radiation was scattered off the layer of
water; differential scattering coefficients coupled with a beam or
ground asymmetry would lead to a spurious polarization signal.

This problem didn't manifest itself until later in the season,
when conditions became warmer and wetter, and we never came up
with a good moisture-prevention system for the window surface.
However, most characteristics of our data got worse during periods
of high humidity, and thus it is quite likely that even absent
this spurious polarization, data through a wet atmosphere would
not have been useable.  Of the 750 hours of recorded data, dew
contaminated approximately 105 hours, or about 14\% of the data
set.  As it had such a striking and characteristic effect on the
data, it was simply removed by hand.

\subsubsection{The Effect of Clouds}
Clouds had a large effect on the correlator channels that was
often visible directly in the time stream.  Clouds are essentially
non-uniformities in the three-dimensional water/ice field in the
sky through which we observe.  A patch in this field with a high
but asymmetric water/ice content can induce a false signal, as our
beam pattern is slightly asymmetric; thus, as we rotate, the
asymmetric beam convolved with the asymmetric cloud will produce
an offset that varies with rotation angle, and will in general
have components at both \onephi\ and \twophi.  This suggest that data
taken through thick clouds should be cut.

Unfortunately, we were unsuccessful in developing a good cloud
monitor.  Data containing some cloud cover information is available
from both the National Weather Service (NWS) and the GOES-8
satellite.  We found that neither correlated particularly well with
periods of bad data. The NWS data were taken from the Dane County
airport, roughly 35 km from our site.  The GOES data were in the form
of fields 5 km x 5 km wide, but typically the closest such field was
about 20 km from our site.   As cloud cover varies substantially over
these distances, our cloud cover data are of little use as a cut
statistic.  As our only alternative, we opted to cut directly on statistics of the data
themselves.  These statistics are described in \sct{deepercuts}.

\subsubsection{Sun Spikes} 
\begin{figure}
\begin{center}
\subfigure[Solar Contamination in the Time Stream]{\label{suncrazies}
\includegraphics[width=5in]{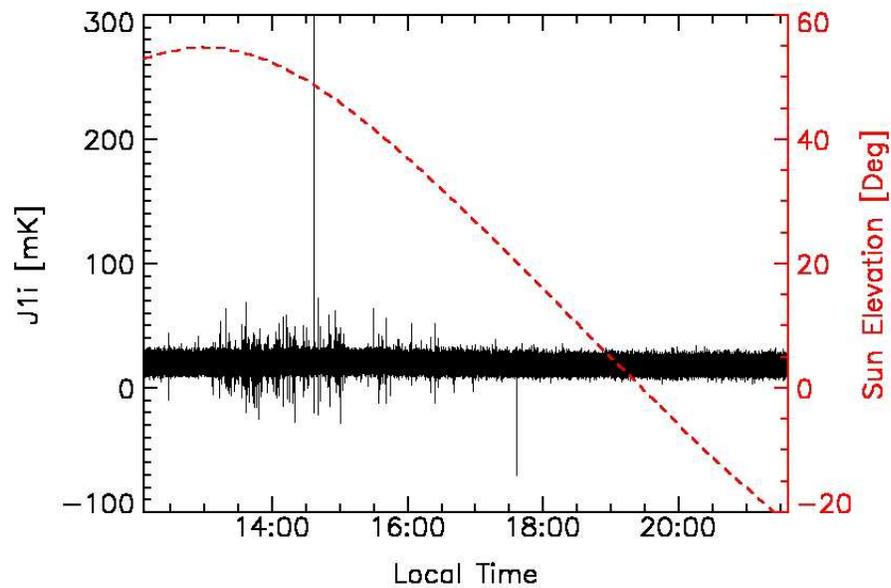}
}\\
\subfigure[Solar Contamination in the Power Spectrum]{\label{sunps}
\includegraphics[width=5in]{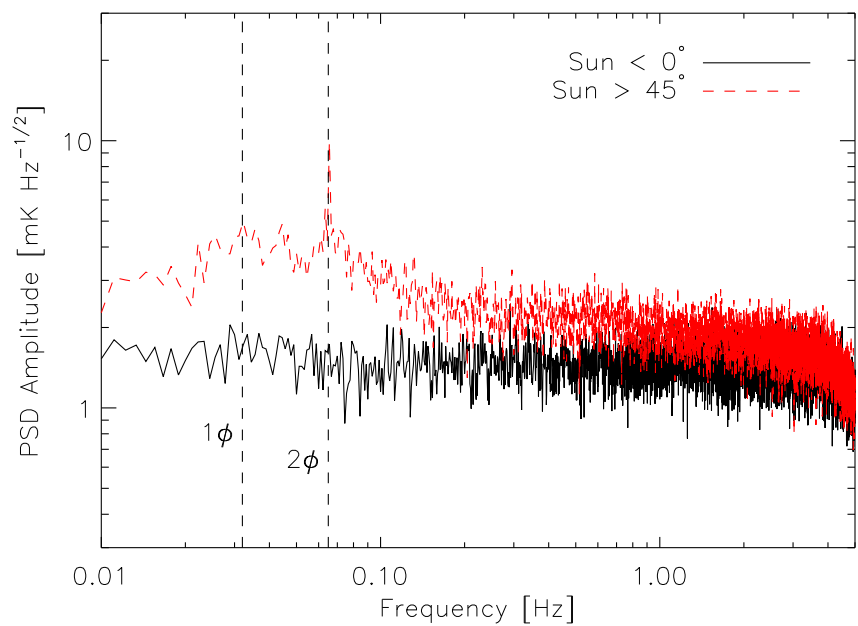}
}
\caption[The Effect of Solar Contamination on the Data]
{\label{sun}\fixspacing The Effect of Solar Radiation on the Data.
(a) shows the effect on the time stream.
When the sun climbs higher than about 20\deg\ in the sky,
some sun-related effect comes into play, increasing the noise and
generating a spurious polarization signal.
This is evident in (b), which displays the power spectra of a
two segments of data, when the sun elevation is low and high.
}
\end{center}
\end{figure}

\fig{suncrazies} shows an 18-hour
segment of time-ordered data, with the sun elevation overplotted. It
is evident that the noise characteristics of the data are radically
different when the sun is more than 30\deg\ above the horizon.
Plotting the power spectrum of two segments of the data
(\fig{sunps}), when the sun is high and low, confirms this, and shows
the noise properties get much worse at all frequencies, and
significant \twophi\ noise is added, which is unacceptable.

Based on the geometry of the cone, some solar radiation will enter
when the sunlight can just make it over the outer (fixed)
groundscreen and strike the inner ground screen.  This happens at
an elevation of \app 10\deg.  However, this is a very tiny effect,
because for this light to make it into the horn, it must scatter
many times off the inner ground screen as it ``runs around'' the
perimeter of the screen; it will then be mostly absorbed.  Below
this elevation of 10\deg, solar radiation must undergo a double
diffraction to enter the system. The amount of sunlight reaching
the horn then steeply increases as the sun climbs higher into the
sky, until the sun's elevation reaches \app 40\deg, at which time
radiation from the sun can directly enter the horn.  We found
that, in practice, sun contamination was undetectable below a
solar elevation of 30\deg.  To be conservative, we eliminated all
data taken with the sun more than 20\deg\ above the horizon.  As
we attempted to take data 24 hours per day, this represents a
sizeable 38.6\% of our data, or \app\ 288 hours.

Recent additional evidence shows that this sun-correlated effect
may actually not have been an optical effect caused by the sun,
but rather was related to an increase in temperature.  It appears that
some electronic effect was initiated when the system was too hot; this
could have been related to our heating temperature pads, but it is
unclear at this point.  What is clear is that this effect occurred
during the daytime and late afternoon, and thus all of these
periods had to be eliminated from the analysis.

One unfortunate consequence of our scan strategy was that much of
the data near the galaxy crossings had to be eliminated due to sun
and dew contamination, in particular the bright region near Cygnus A.
It is conceivable we could have seen this region in polarization with
reasonable integration time on it, but sun and dew contamination ruled out
this possibility.

\subsubsection{Moon Cut}
\begin{figure}[tb]
\begin{center}
\includegraphics*[height=4in]{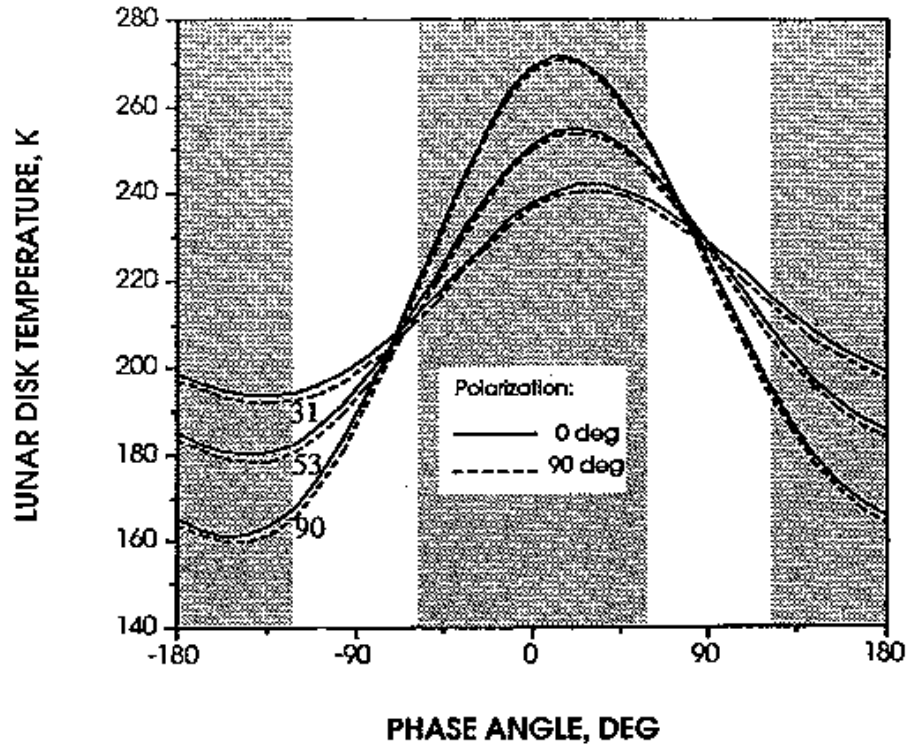}
\caption[Moon Emission at Three Frequencies versus Phase Angle]
{\label{moonmodel}\fixspacing Model of Lunar Emission at the
frequencies 31, 53, and 90 GHz.  The solid and dotted curves show
the two different emission polarizations.  Notice that they differ
by approximately 1 K on average, independent of frequency.
Reproduced from the COBE-DMR team, reference \cite{bennett92a}.}
\end{center}
\end{figure}

The moon is a bright microwave source, corresponding to an
emission temperature of roughly 220 K.  Its emission is dependent
upon frequency, phase, and polarization. The standard model of
lunar emission at microwave frequencies is Keihm's 1983 model
\cite{keihm83}. Using this model, the COBE team calculated the
lunar emission in both polarization states at the three COBE DMR
frequencies, and showed that the polarization temperature at 31 GHz
of the moon (viewed as a point source) is $\lesssim 1 K$
\cite{bennett92a}.  We have reproduced their model in
\fig{moonmodel}.

\begin{figure}[tb]
\begin{center}
\includegraphics[height=3.5in]{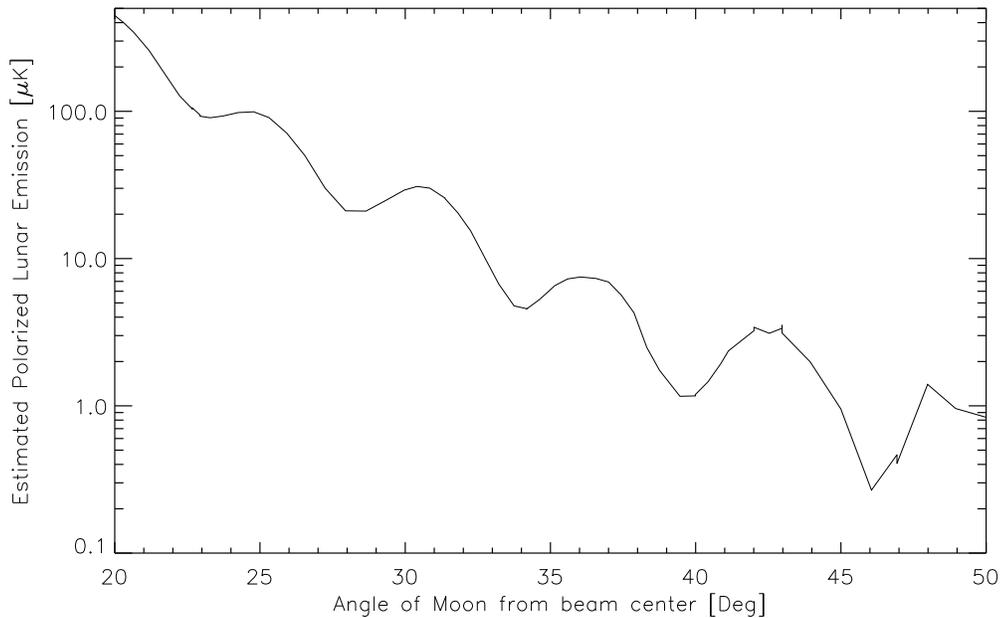}
\caption[Estimated Polarized Lunar Emission versus Lunar Elevation]
{\label{moonsignal}\fixspacing
Lunar Contamination.  This model assumes the moon has a 1 K polarized emission
temperature at 30 GHz, when viewed as a point source, and is based on
the assumptions of Bennett \etal, 1992 \cite{bennett92a}.  We further
assume attenuation only by the feedhorn beam pattern for this range of lunar angles.
The figure shows that it is desirable to have the moon more than 40\deg\
from the main beam.}
\end{center}
\end{figure}

Let us now attempt to estimate how \polar\ will view this emission.
If we see the fully polarized effect, it will be attenuated by our
beam function at the very least, and ground shields will only
increase the attenuation. Modelling the moon as a point source in our
beam (0.5\deg\ is much less than 7\deg, so this is indeed a valid
model), we can then estimate the polarization signal we will see as a
function of moon elevation (or alternatively, angle from the
boresight). \fig{moonsignal} shows this estimate, assuming an
azimuthally-symmetric beam and a 1 K polarized lunar brightness
temperature. When the moon is more than approximately 40\deg\ from
zenith, it cannot shine directly on the feed horn; a single
diffraction over the rotating inner ground screen is required.
Assuming this provides at least an extra 20 dB of shielding, we can
safely assume lunar signals are negligible at elevations below
50\deg.

The moon's highest elevation in Wisconsin is about 70\deg, and
during our observation season the highest lunar elevation was
68.4\deg, corresponding to an angle from the zenith of 21.6\deg.
We removed all data when the moon was more than 50\deg\ in
elevation; this corresponds to about 8\% of the data.  However,
taking into account its overlap with other cuts, it only cuts
about 3.7\% of the total data set (28 hours).

\subsubsection{Galaxy Cut}
At low galactic latitudes, there is the possibility of a serious
galactic foreground.  We hoped to bypass this ``demon'' by cutting
all data occurring below a certain galactic latitude.  Thus, we
elected to cut any data with galactic latitude $|b| < 25\deg$. This
did not cut a significant amount of data in the end, but we felt it
was important to have a well-defined section of sky where hopefully
foreground contamination is kept to a minimum. \fig{obs_strategy}
from the previous chapter shows this cut overlaid with our scan
strategy, on the 408 MHz Haslam synchrotron map. Clearly, this area
of sky appears to be of much lower radio emission than the low
galactic latitude regions.

\subsection{System-Based Cuts}
Although in principle most experiments have cuts based on equipment
failure, high system temperatures and the like, we only had to
perform two system based-cut because our systems generally operated
as expected throughout the season.  The first was cutting data
whenever the clamshell dome was closed, although these data were cut
right at the beginning of the analysis and was never included in what
we called the ``data set''.  The second system-based cut was removing
any HFs during which the system wasn't rotating for the entire file,
or the rotation was unnaturally slow or jittery.  These primarily
occurred at the beginnings of deployments, when we started taking
data and then began system rotation.  This cut accounts for about
1.8\% of the data, and is referred to as ``Proper Rotation'' in
\tbl{cuttable}.

\section{Deeper Cuts}\label{deepercuts}
The above three sections deal with data problems that were quite
simple to recognize and remove.  According to \tbl{cuttable}, the
sun and dew cuts have removed close to 50\% of the data we took
during 2000. However, many spurious signals remain lurking in the
data. Sometimes they are associated with clouds, or high humidity,
or other weather-related events, but sometimes they are not, and
it is our job to identify and remove them.

\subsection{The \onephi\ Cut}
The first powerful statistic that we learned correlated with spurious
polarization signals was a high \onephi\ signal (signals modulated at
precisely our rotation frequency). Recall that only signals modulated
at twice our rotation frequency can correspond to true polarization
signals; thus, a signal that has harmonics at both \onephi\ and
\twophi\ cannot correspond to a true celestial signal.

\begin{figure}[tb]
\begin{center}
\subfigure[]{\label{lowfreqpsd-a}
\includegraphics[width=3in]{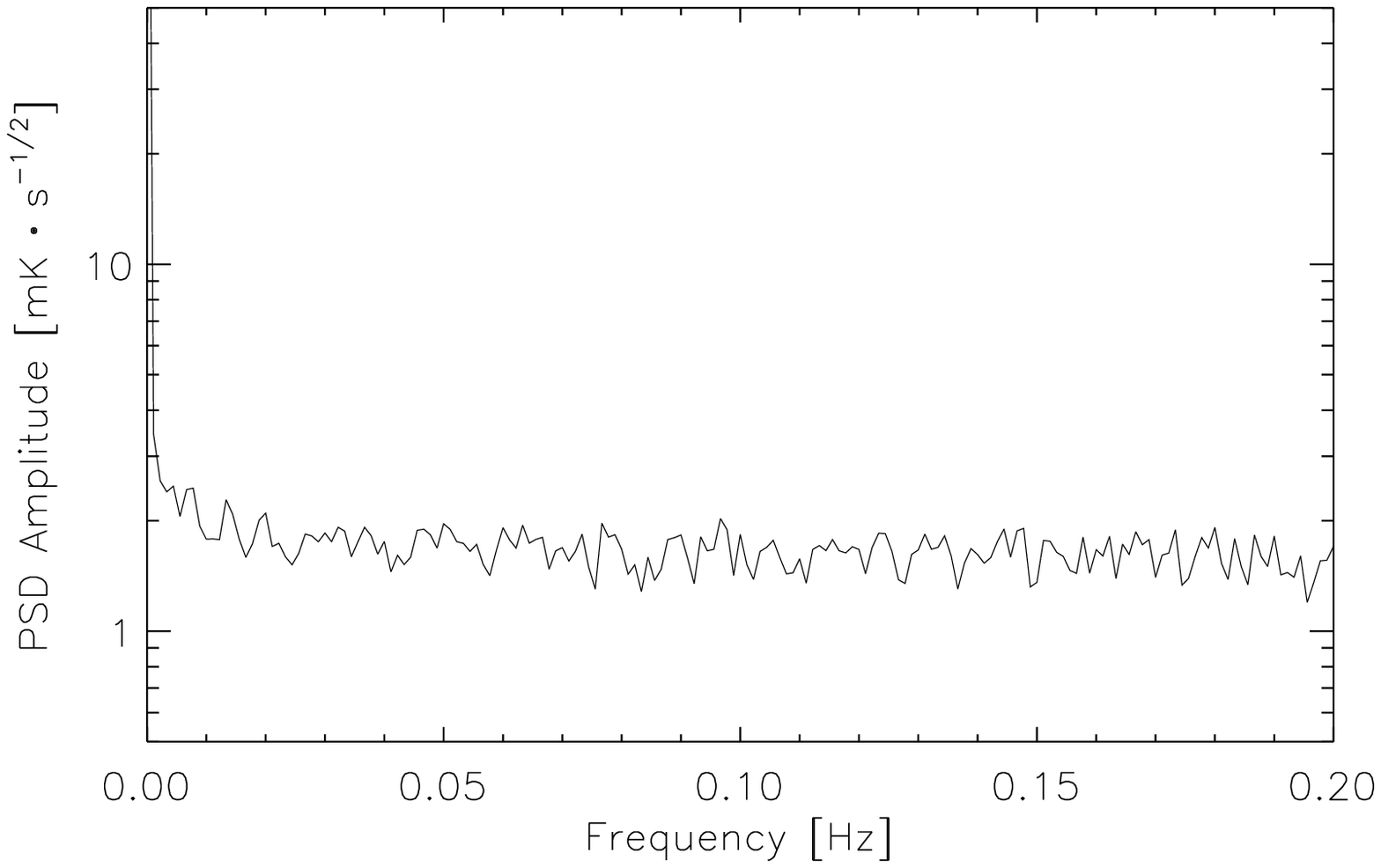}
} \subfigure[]{\label{lowfreqpsd-b}
\includegraphics[width=3in]{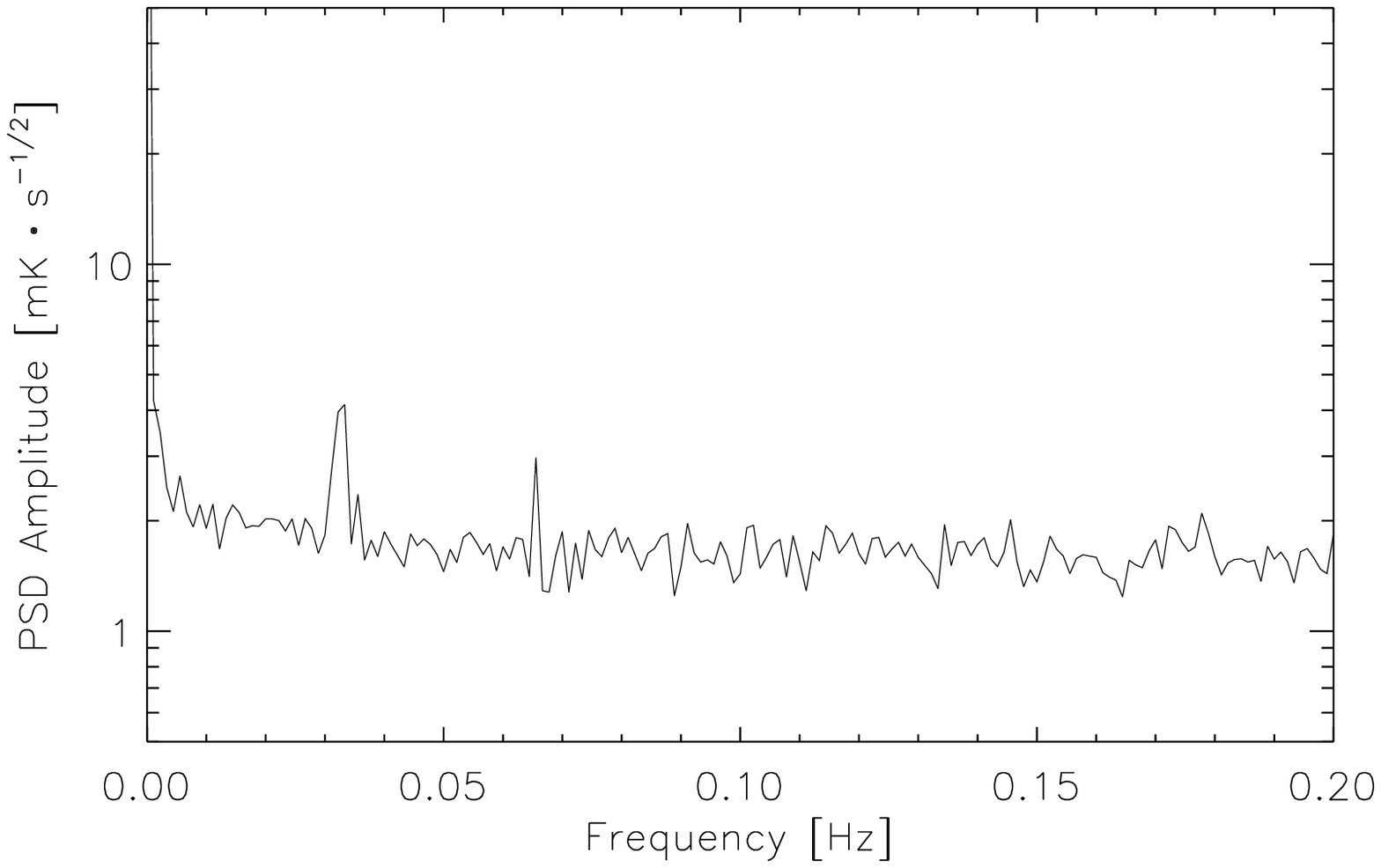}
} \caption[Low Frequency PSD of Two Sections of Data]
{\label{lowfreqpsd}\fixspacing Low frequency power spectrum of two
sections of data.  Panel (a) shows a featureless, white noise power
spectrum when the weather is good and systematic effects are low.
Panel (b) displays a period when clouds appeared: in addition to 1/f
noise increasing slightly, noise at the \onephi\ and \twophi\
frequencies appears as well.  This motivated our using the height of
the \onephi\ peak as a cut statistic. }
\end{center}
\end{figure}

As an example, consider \fig{lowfreqpsd}.  The left-hand (a) panel
shows a classic, featureless white noise power spectrum, taken
during a period of good weather for channel J2i.  But as the
weather gets bad, due to clouds, sun, or something else, features
at the \onephi\ and \twophi\ frequencies appear, in addition to
1/f noise, as shown in \fig{lowfreqpsd-b}.

We computed the heights of each of the \onephi\ peaks for every
hour file in the data set (and for each channel). This was
performed as follows: a given HF and data channel gives a
9000-element data set, for which we wish to calculate the
low-frequency power spectrum.   We use the standard, FFT-based
periodogram construction of the power spectrum. With this
algorithm, the highest frequency resolution given these parameters
is $2/9000 = 2.22$ mHz.  Recall that $f_{\onephi} = 32.5$ mHz.
However, a significantly better reconstruction of the power
spectrum can be performed, at the expense of some degradation in
frequency resolution, by splitting up the data into several
overlapping equal-length segments, and averaging the power
spectrum of each. A full discussion of these techniques is given
in the classic \textit{Numerical Recipes} \cite{nrc}, chapter 13.

\begin{figure}[tb]
\begin{center}
\includegraphics[height=3in]{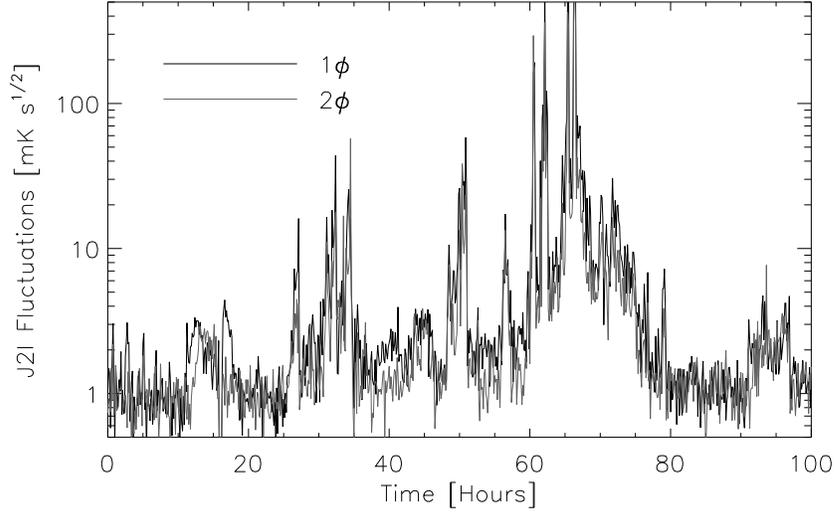}
\caption[\onephi\ and \twophi\ Fluctuations vs. Time]
{\label{onephi_and_twophi} \fixspacing
\onephi\ and \twophi\ Fluctuations vs. Time, for channel J2i.
The high degree of correlations between the two rotational harmonics
is evident.  The \onephi\ statistic cannot correspond to a true
sky signal, thus motivating its use as a cut statistic.
}
\end{center}
\end{figure}

We found that, in practice, this technique gave very similar
results to direct Fourier extraction of the \onephi\ and \twophi\
components.  \fig{onephi_and_twophi} displays the \onephi\ and
\twophi\ statistics through a 100--hour section of the observing
season; periods of sun, moon, and dew have all been removed.  The
correlation coefficient between the \onephi\ and \twophi\
statistics are 0.73, 0.89, and 0.93 for channels J1i, J2i, and
J3i, respectively. This strongly motivates a cut based on the
\onephi\ statistic. However, because we found the base noise level
to fluctuate somewhat, we defined statistics relative to the base
noise level, which will refer to as \onephir\ and \twophir, where
the 'r' stands for 'relative'.

\begin{figure}[tb]
\begin{center}
\includegraphics[height=3in]{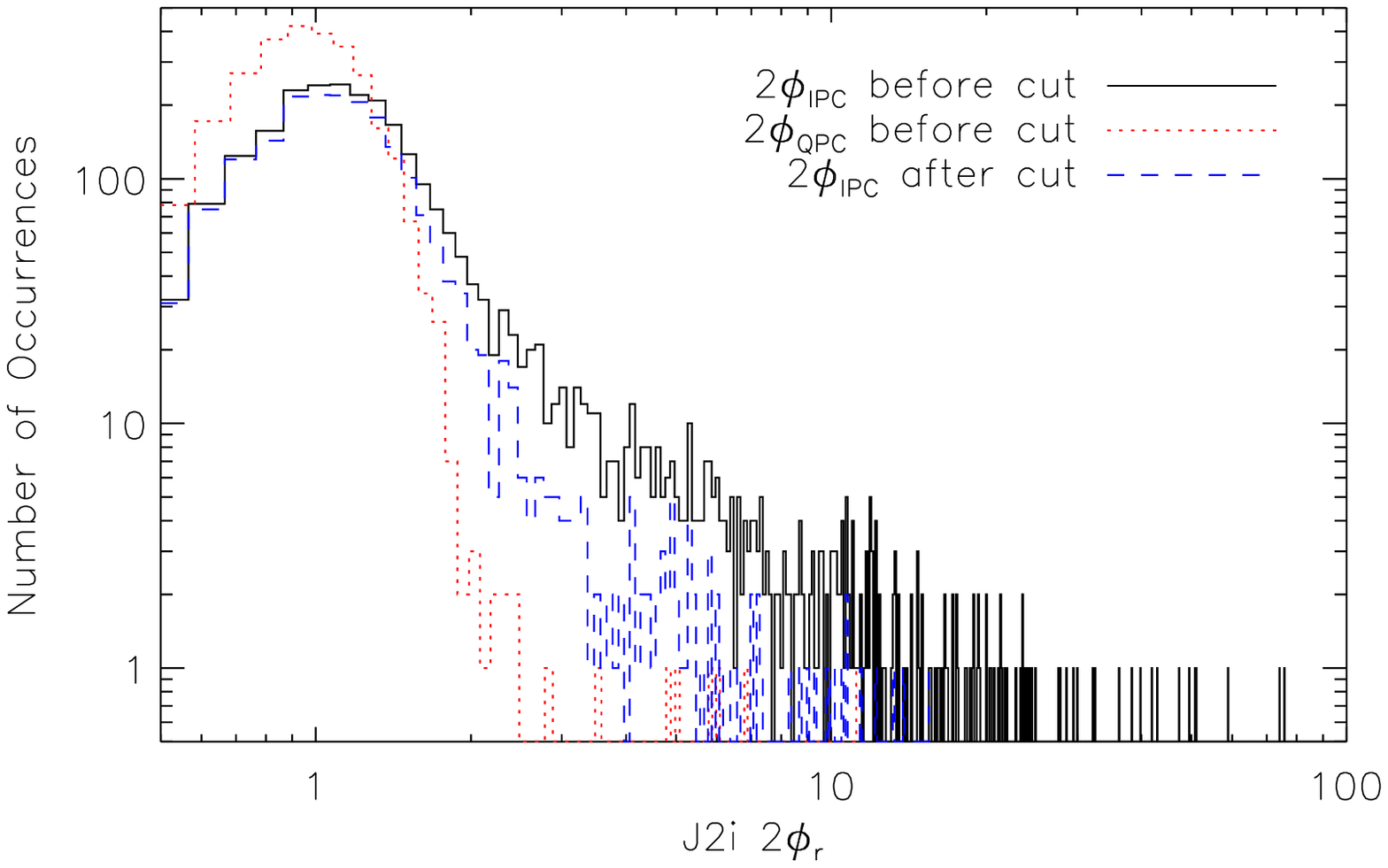}
\caption[Histogram of \twophir\ data for channel J2i]
{\label{onephi_hist} \fixspacing
Histogram of \twophir\ data for channel J2i.
The solid (black) line represents all the data, with sun, moon,
and dew-contaminated data removed. The dotted (red) line is the
same data set for the QPC channel J2o; it can be thought of as the
ideal \twophir\ distribution for white noise.
The dashed (blue) line is the \twophir\
data after the $\onephi_r < 2.1$ cut was applied.  Notice that a
large fraction of the tail of the distribution has been removed,
but a small portion still remains.
}
\end{center}
\end{figure}

\fig{onephi_hist} shows a histogram of the \onephi\ and \twophi\
data relative to the NET of each hour file (this is to remove the
effects of gain drifts).  A value of $1$ thus denotes a flat
(white) spectrum; Monte-Carlos of Gaussian white noise show that
this statistic should be $1.0 \pm 0.27$. We cut whenever
$\onephi_r > 2.1$; this is more than $4\sigma$ from the mean for
white noise.

\begin{figure}
\begin{center}
\subfigure[]{\label{one_v_two_j1i}
\includegraphics[width=3in]{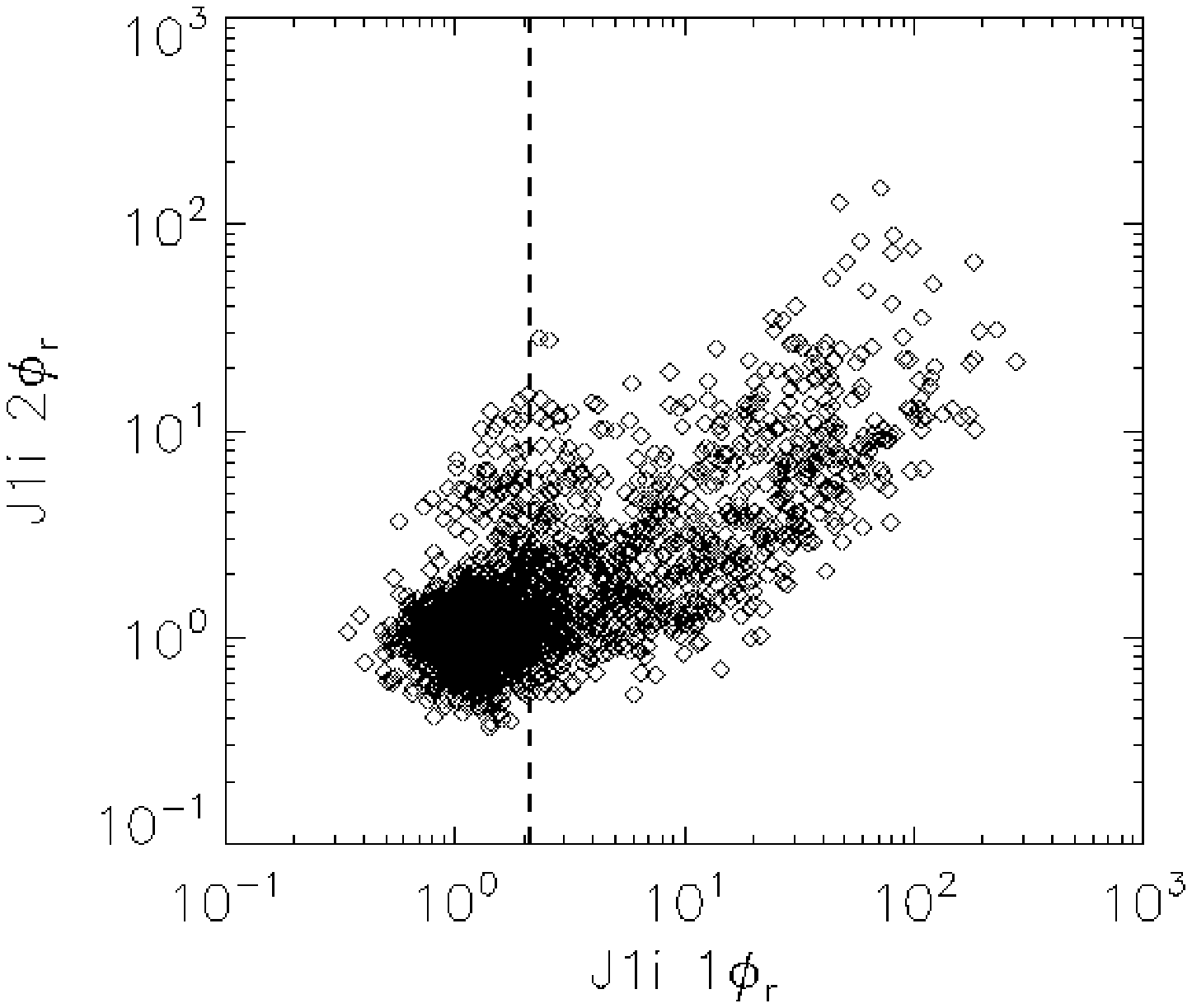}
} \subfigure[]{\label{one_v_two_j2i}
\includegraphics[width=3in]{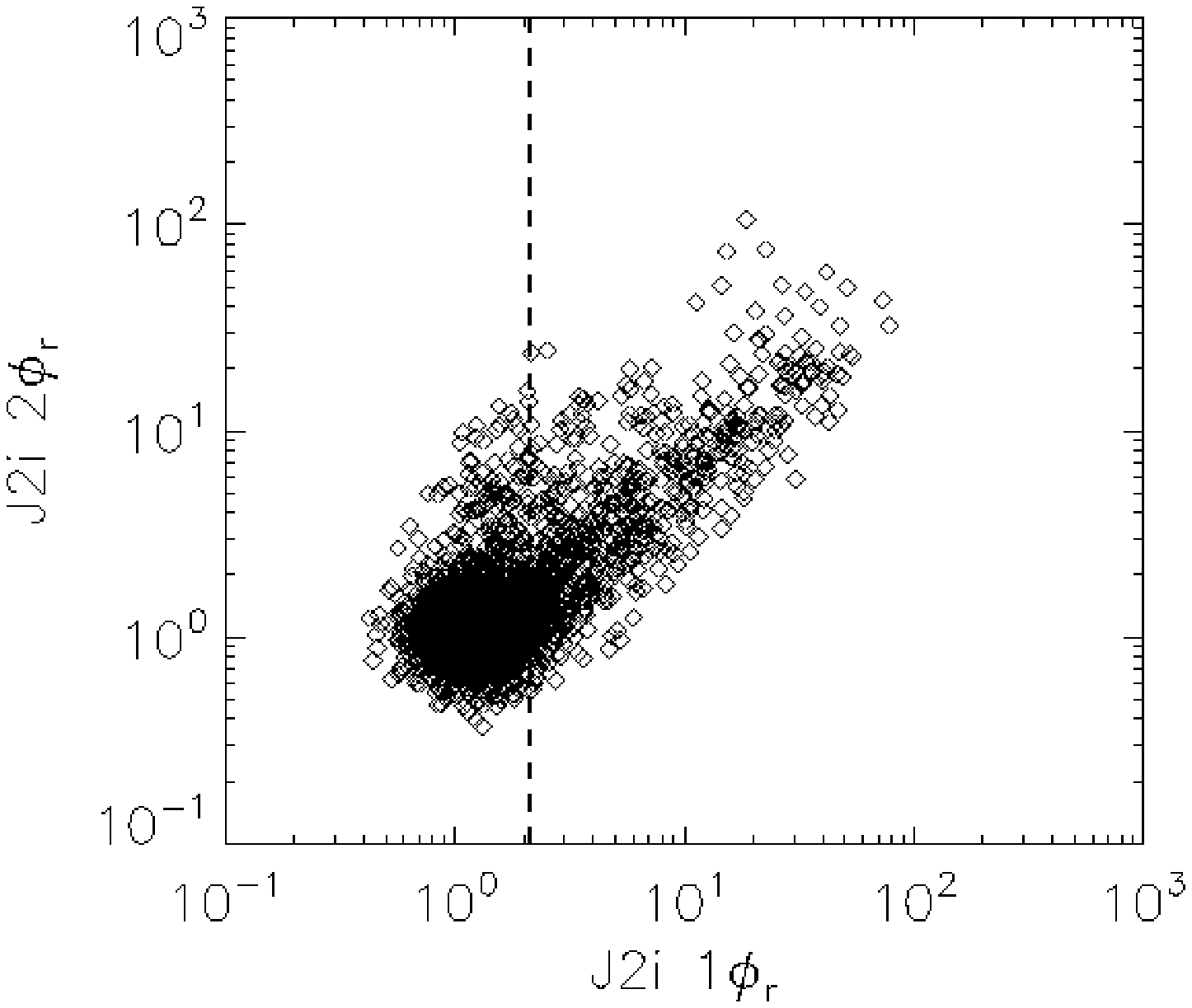}
}\\
\subfigure[]{\label{one_v_two_j3i}
\includegraphics[width=3in]{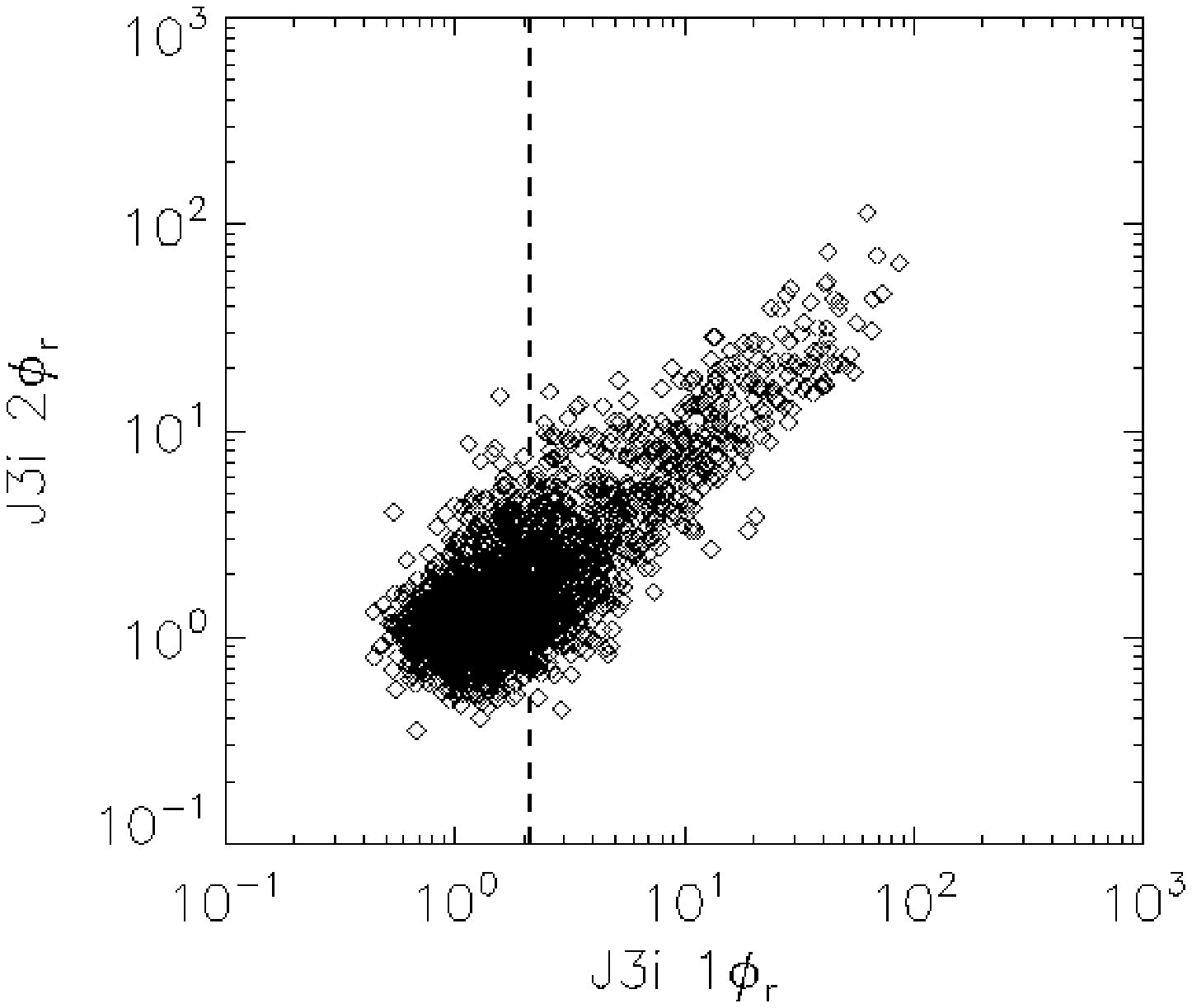}
} \caption[\twophir\ vs \onephir\ for each the data channels J1i, J2i, and
J3i]
{\label{one_v_two}\fixspacing
\twophir\ vs \onephir\ for each the data channels J1i, J2i, and
J3i.  Periods contaminated by sun, moon, and dew have been
removed.  The vertical dashed line shows the $\onephir \ < \ 2.1$
cut level.
}
\end{center}
\end{figure}

\fig{one_v_two} displays the same data, but plots \twophir\ vs.
\onephir; the cut level is shown as the dashed vertical line. Again,
the high degree of correlation between the two quantities is obvious
in all channels.  Using $\onephir < 2.1$, an additional 109 hours, or
about 15\% of the total data was cut.

\subsection{Additional Selection Criteria}
It is apparent from \fig{one_v_two} that there will still be some
residual data points with high \twophir, regardless of the
\onephir\ cut. This motivated us to examine additional statistics
that may also be correlated with the \twophi\ level, and therefore
may serve as useful data selection criteria. The next section will
briefly describe each of these statistics.

\subsubsection{The Zeta Cut}
Auto-correlation is a powerful and illuminating technique for
viewing data. The auto-correlation $C(y)$ of a data set $y$ is
intimately connected to the power spectrum $S(\nu)$ via the
Wiener-Khincin theorem \cite{nrc}:
\beq{WKtheorem} C(y)
\Longleftrightarrow |S(\nu)|^2 \ .
\eeq
That is, the
autocorrelation function $C(y)$ and power spectrum
$|S(\nu)|^2$ of any data set form a
Fourier transform pair, and hence the information in one is the
same as the information in the other. A rise in
1/f noise leads directly to a higher ``floor'' in the
autocorrelation.  We used that fact to our advantage and defined
the following statistic for each HF of data:
\beq{e:zeta}
\zeta \ \equiv \ \frac{\sum_{lag=1}^{1000}{C(y_{in})}^2}
{\sum_{lag=1}^{1000}{C(y_{quad})}^2}
\eeq
where $y_{in}$ denotes
data from an in-phase (IPC) channel and $y_{quad}$ denotes data
from the corresponding quad-phase (QPC) channel. \eqn{e:zeta} may
seem a rather arbitrary definition, but it is fairly easy to
analyze what it does.  Using the Wiener-Khincin theorem, we can
think of any timescale as corresponding to a certain frequency.
The autocorrelation at lag zero corresponds to the offset of the
data, which we do not care about.  Lag one corresponds roughly to
10 Hz, our Nyquist frequency, and lag 1000 roughly to 0.01 Hz.
\Zeta\ is roughly the integral of the power spectrum, weighted by
$1/f$, so low-frequency drifts cause \Zeta\ to increase rapidly.
Since we would like to remove periods with high 1/f noise
(presumably due to atmospheric fluctuations), this makes a \Zeta\
a sensible cut statistic.
\begin{figure}[tb]
\begin{center}
\includegraphics[height=3in]{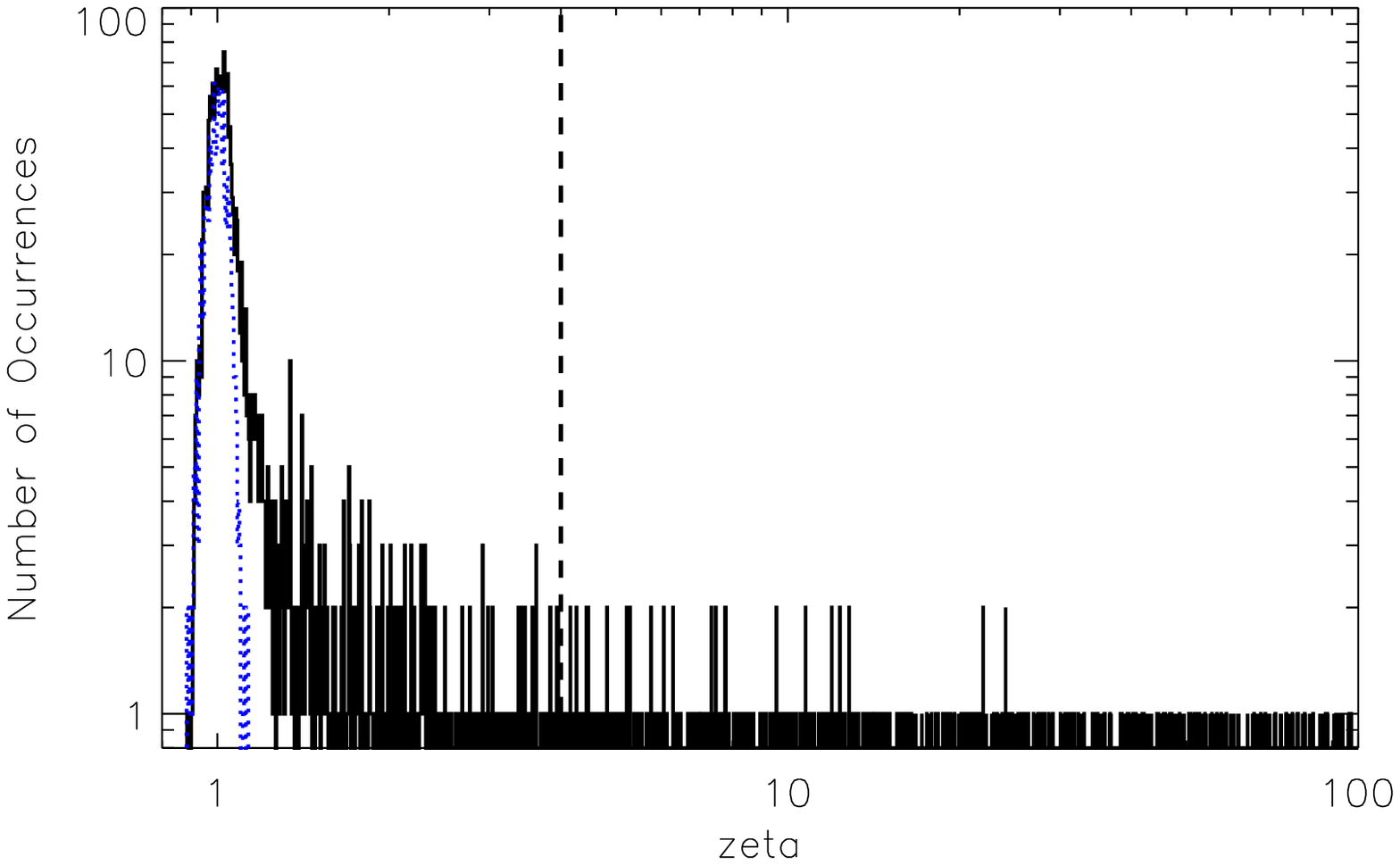}
\caption[Distribution of the \Zeta\ Variable]
{\label{zeta_hist} \fixspacing
Distribution of the \Zeta\ Variable.  The solid (black) curve shows the
distribution of \Zeta\ for channel J2, with the basic sun, moon,
and dew cuts applied.  The dotted (blue) curve shows the
\Zeta-distribution for simulated white noise run through our
anti-aliasing filter.  The vertical dashed curve is the cut level
of \Zeta=4.0.
}
\end{center}
\end{figure}

\fig{zeta_hist} shows a histogram of \Zeta\ values throughout the
season (with the basic sun, moon, and dew cuts applied).  As this
cut mainly correlates with 1/f noise, we only applied a very mild
cut on this parameter to the data, $\zeta < 4.0$.  Also shown in
the plot is a model of our data stream with no 1/f noise or signal
of any kind.  As can be seen, this distribution is sharply peaked
around 1.0, with more than 99.9\% of the data lying below 1.2.
We tried cuts around this level, but found the cut to simply be
far to sensitive at this level.  At levels above 4.0, however, the
cut typically correlates with our concept of bad weather, and
hence this is the \Zeta-cut level we chose.
If a further analysis of the data occurs, we
will most likely remove this cut in favor of something more
correlated with spurious systematic effects, but for now the
\Zeta-cut is in line with our highly-demanding approach to data
quality.

\subsubsection{Outliers in the TOD}
\begin{figure}
\begin{center}
\includegraphics[height=3in]{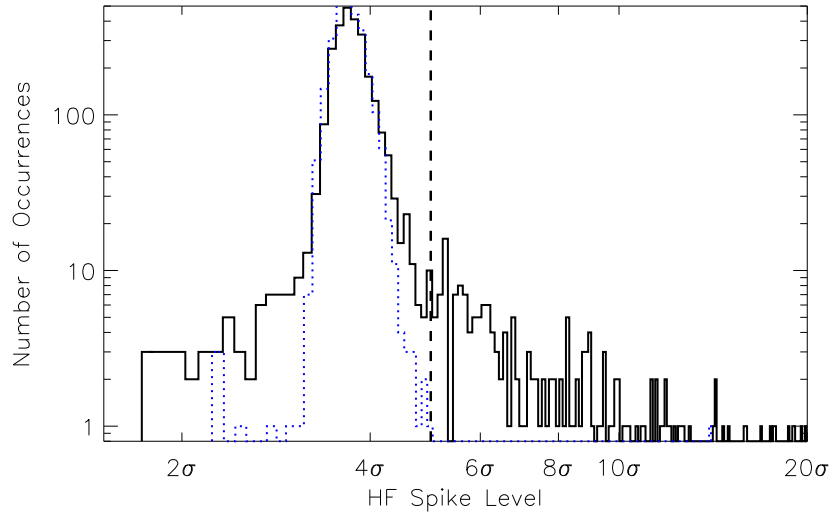}
\caption[Distribution of Second Worst Outlier in each Hour File]
{\label{todsigma} \fixspacing The distribution of second-worst
outliers for each hour file. Nominal sun, moon, and dew cuts have
been applied.  The solid (black) curve is the J2i channel, while
the dotted (blue) curve is the J2o channel.  The J2i channel has
large wings on both sides of the main distribution, representing
non-Gaussian behavior on either side.  The J2o behavior is
strictly Gaussian.  The vertical dashed line shows the applied cut
level of $5\sigma$. }
\end{center}
\end{figure}

Occasionally, birds, planes, etc. would cause large short-lived
spikes in the data stream.  Rather than trying to identify the
cause of these, we took a different tack.  For each hour file and
channel, we calculated the mean and standard deviation for that
file, and recorded how many standard deviations the first, second,
and third outliers were from the mean.  Our philosophy was that
any data file could have one non-Gaussian outlier, but not two.
Two strong outliers were evidence of non-Gaussian behavior which
would therefore not be consistent with our signal (which is at
such a low level that it would not cause significant deviations
from Guassinity in the data stream).  Thus, we kept track of the
number of standard deviations the \emph{second worst} outlier was from the
mean.  This distribution is shown in \fig{todsigma}. Any HF with a
more than $5\sigma$ second-worst outlier was removed.
This cut accounted for \app\ 1.2\% of the data, or about 9 hours
(in addition to all previous cuts).

\subsection{Duration-Based Cuts}
A final series of cutting measures was based upon the length of
surviving data segments.  First, we required that if an HF were to
survive, both its neighbors had to survive as well. Thus, if a
segment of say 10 HF's survived, the first and last of these would
be chopped off.  This is to ensure that the bad weather has truly
ended, so this provides a nice 7.5-minute buffer.  This we called
the ``Nearest Neighbor'' cut in \tbl{cuttable};
it is a fairly common criterion in CMB data analysis.

Secondly, we required a minimum of eight (8) consecutive data
files to survive in order to keep any piece of data.  This is to
make sure we don't get into a situation where we're keeping only a
bunch of very short segments of data where statistically they just
happened to survive the cuts, even though perhaps they were part
of a larger distribution of data that should have entirely failed
a given cut.

Lastly, we required a minimum of three hours total to survive a
\emph{section} of data.  This was primarily for the offset removal
step that will take place in the mapmaking part of the analysis
pipeline, to be described in the next chapter. This algorithm
takes place on a per-section basis, and effectively kills
all information from very short sections of data; in the end we found it
simpler to just remove these short sections of data.

\tbl{cuttable} shows the sizeable effect of these duration-based
cuts; together, they remove more than 10\% of the full data set.
However, they were quite critical; since many analysis techniques
only work for long, contiguous sections of data where the noise is
fairly stable, it is in our best interests to have relatively longs sections
of good data.  If things are changing quickly, the noise is not
stable and the analysis techniques will fail.

\subsection{Final Cuts: ROD Spikes and Executive Cuts}
After we had applied these cuts to the entire data set, we were left
with a set of HF's for each channel and section.  The cuts for a
given QPC channel were the same as for the corresponding IPC channel.
The mapmaking techniques described in the next chapter enabled us to
determine a mean $Q$ and $U$ from each Hour File; this was called the
\emph{ROD}, or Rotation-Ordered Data set (see \sct{s:ROD} for a
complete description of the ROD data).

For each section and channel, we examined the distribution of these
$Q$'s and $U$'s, and found there were still a fair number of
outliers. We chose to proceed as in the ROD case; for any section
and channel with surviving data, we found the mean and standard
deviation for the \twophi\ signal, which is $\sqrt{Q^2 + U^2}$. If
this quantity was more than three standard deviations from the
mean for that section, we cut it. Notice that this cut is NOT
independent of the other cuts; it requires a fairly clean segment
of data to get meaningful, clean distributions of $Q$ and $U$ in
the first place. It then finds sites of potential non-Gaussian
behavior. Notice also that this cut is essentially insurance; like
the \Zeta\ cut, it may easily be cutting good data, but our
philosophy was ``better safe than sorry'' in this analysis.
However, after instituting all the other cuts, this requirement
only removed an additional 0.1\% of the data (less than 1 hour).

At this point, there were on the order of 15 sections per channel
that had some surviving data, which amounted to about 100 hours
per channel. There were a couple of places where inexplicably ``bad
data'' had survived; \ie\ data with strong drifts, or a strong
\twophi\ component but no other obvious statistics indicating that
anything was wrong. This latter case is entirely consistent with
signal, except that the expected CMB polarization is predicted to
be so low as to not be recognizable at these levels (many tens of
\uK). This only amounted to about two sections per channel, and
again was consistent with our very restrictive cut hypothesis, so
we removed them.

\section{Final Remarks on the Cuts}
It is perhaps questionable that so much of the season's data were
eliminated through the cutting process.  Our aim was simply to take a
very restrictive approach to the data cutting, and then loosen up
these restrictions in the hope of teasing some signal from the data.
Unfortunately, time constraints have prevented us from going beyond
this initial step, although it is still possible in the future. The
reader should keep in mind that these cuts were quite strong, so in
principle there should be a minimum of spurious signals left in the
data. However, even with these harshest of criteria, a spurious
signal \emph{did} end up residing in the data, which was only
discovered through our mapmaking analysis described fully in the next
chapter.





\chapter{From Data to Maps}\label{mapmaking}
Now that we have our cleaned data set, our goals are fairly clear: extract as
much astrophysical and cosmological information from the data as
possible. However, the data set is relatively large, on the order
of $10^8$ numbers (about 400 MB). In the ``olden days'' of CMB
analysis, many standard algorithms would compute cosmological
parameters directly from the time-ordered data. These algorithms used
techniques that needed to invert an $n\times n$ matrix, where $n$
is the number of elements in the data set. Recalling that the time
it takes to invert an $n\times n$ matrix goes like $n^3$, we can
see that these techniques will rapidly become intractable once $n$
is too great.

Luckily, it is possible to reduce a large astrophysical data set
to a \emph{map} of the sky, with no loss of cosmological
information \cite{teg97a}. This perhaps surprising result lets us
kill two birds with one stone: we can make maps to visualize our
data, and we can radically compress our data set to make
extracting cosmology from it much more tractable. In this chapter,
I will describe the algorithms that take the \polar\ data and
transform it into map form. The next chapter will discuss what we
can learn about astrophysics and cosmology from the maps
themselves.

\section{The General Mapmaking Problem}
Many authors have written on the mapmaking problem and all the
nuances that can arise during its solution.  For several excellent
papers on the subject, see  \cite{teg97a,teg97b,qmap98c,bjk98}.
Most of this section is adopted directly from Max Tegmark's
excellent paper on mapmaking, reference \cite{teg97a}.

The most intuitive and simple mapmaking algorithm is by far the
``simple binning'' approach, in which all data belonging to the
same pixel are simply averaged together. However, drifts in the
data (as from 1/f noise) can lead to stripes in the map, and also
to nonsensical maps if the drifts are too strong. Thus, data
belonging to the same pixel must be averaged in such a way as to
remove the effect of 1/f noise. Overall offsets can also be a
problem, and subtracting them in a mathematically consistent way
leads to further complexities in the mapmaking algorithm. Luckily,
there are well-tested algorithms that treat all the problems we
will encounter.

\subsection{Mapmaking Notation}
Throughout the next sections, the mathematics of mapmaking will be
discussed in great detail. Whenever discussing a vector, I will
use a lowercase boldface letter (\eg, $\y$) to represent it.
Similarly, matrices will be represented by uppercase boldface
letters (\eg, $\W$). A vector such as $\xt$ denotes our
``best-guess'' of the true underlying vector $\x$.

\subsection{Definition of the Problem}
The mapmaking problem proceeds as follows \cite{teg97a}.
In general, if you
have a time ordered data vector $\y = \{y_1, \ldots , y_n
\}$, it can be written as a sum of signal plus noise. If the
signal comes from an underlying map on the sky $\x= \{x_1,
\ldots , x_m \}$ with $m$ pixels, then one can write
\beq{LinearProblemEq}
\y = \A\x+\n
\eeq
where $\n$ represents the system noise, and $\A$ is called the
``Pointing Matrix'', and describes how to move your data from
``data space'' to ``map space''. The matrix $\A$ is determined
entirely by the scan strategy; its evaluation is described in
detail in \sct{pointingmatrix}. Regarding the noise $\n$, the
assumption is usually made that the noise is \emph{stationary}
throughout the data set $\y$, which is equivalent to the power
spectral density of the noise being constant in time. Under this
assumption, the noise can be characterized by its \emph{noise
covariance matrix}, $\N$, which is defined as
\beq{NoiseCovEq}
\N\equiv \langle \n\n^t \rangle \ ;
\eeq
we assume that $\langle
\n \rangle = 0$ without loss of generality. For Gaussian noise,
the noise covariance matrix will reduce to a multiple of the
identity matrix.  More complicated noise will result in
\emph{correlating} different measurements $y_i$ with each other
(for instance, in the presence of 1/f noise).  It is
assumed that the instrumental noise is uncorrelated with the
signal.

\subsection{General Solutions to the Mapmaking Problem}
Given the pointing matrix $\A$ and the noise
statistics $\N$, it is straightforward to
solve \eqn{LinearProblemEq} for the underlying map $\x$. We
want to find an estimate of the map $\xt$, given our
noise (because we have incomplete, noisy data, we can't find the
exact underlying map).  All {\it linear} methods can be written in
the form
\beq{WdefEq}
\tilde{\x} = \W\y ,
\eeq

Ideally we would like to minimize the difference between $\xt$ and
$\x$; that is, minimize
$\langle | \epsilon |^2 \rangle$ where $\epsilon \equiv \xt -
\x$.  A good choice for $\W$ that essentially accomplishes this is
the COBE-style solution \cite{jansen92,wright96},
which gives
\beq{MaxMethod1}
\W = [\A^t\N^{-1}\A]^{-1}\A^t\N^{-1} \ .
\eeq
In general, the final map pixels in $\xt$ will be correlated;
this information is recorded in the \emph{noise covariance matrix
in the map}, $\NN$, which is defined as
\beq{Sigmadef}
\NN \ \equiv \ \langle (\xt - \x)(\xt - \x)^t \rangle
\ = \ \W \N \W^t \ .
\eeq
It can be shown that, for the COBE method, this covariance matrix
becomes
\beq{SigmaCOBE}
\NN  = [\A^t\N^{-1}\A]^{-1} \ .
\eeq
\eqn{MaxMethod1} is not the only $\W$ you can choose, but it does
have several nice properties \cite{teg97a}.
It minimizes $\langle | \epsilon |^2 \rangle$, subject to the
constraint $\W \A = \I$.
It is the maximum-likelihood estimate of $\x$, if the
underlying probability distribution for $\n$ is Gaussian
\cite{teg97a}.  And finally, the final reconstruction
error $\epsilon$ is independent of the underlying map $\x$.
We will hereafter refer to the mapmaking
solution with this particular choice of $\W$ as \emph{\mm}.

A couple other choices for $\W$ are noteworthy. Choosing $\W =
[\A^t \A]^{-1} \A^t$, which is equivalent to $\N = \sigma^2 \I$,
is just the case of simple data averaging. As stated previously,
it works well when there are no correlations between separate data
points.  It is easy to see physically how this works.  When $\W$
acts on $\y$, the first operation (proceeding right to left) is
$\A^t \y$, which carries your data vector into map space; it is
this step that averages all data from the same pixel, and yields
the unnormalized map. $[\A^t \A]^{-1}$ is simply the normalization
factor.

If the signal covariance matrix,
 $\Sc = \langle \x \x^t \rangle$,
is known, a Wiener-filtered version of the map can by made by
choosing $\W = [\Sc^{-1} + \A^t \N^{-1} \A]^{-1} \A^t \N^{-1}$
\cite{bunn94, zaroubi95}.
A Wiener-filtered map is typically used for visual presentation of
maps, as it is less noisy (at the price of introducing additional
pixel-pixel correlations).  This technique
also has the property of minimizing
$\langle | \epsilon |^2 \rangle$.
It can be shown that the Wiener-filtered map, taken
together with its noise covariance matrix $\NN$, contains identical
information to the \mm-derived map.

Overall, the linear method (based on \eqn{WdefEq})
is extremely general, and can be applied to
any type of problem where a linear combination of data is made in
order to determine some physical parameter. Whenever the noise of
said data isn't white, this is the best approach. For \polar, we
will exploit this technique no fewer than four different times
throughout the analysis pipeline. However, this method breaks down
when the noise is not stable, so the trick is to apply the
technique to short ``chunks'' of data, where the noise was
stable, and then piece all these submaps together at the end.
It will turn out that the same mapmaking trick will also
tell us how to do this.

\section{Mapmaking for POLAR}
\subsection{The POLAR Pixelization}\label{s:pixelization}
Let us now lay out the details of the mapmaking problem for the case
of \polar.  We would like to construct maps of both $Q$ and $U$;
because of our scan strategy, a zenith drift scan at
$\delta = 43\deg$, this will be a 1D map in right ascension.
We can choose to pixelize this however we like, but will lose
information if the pixels we choose are too big.  As we have a
7\deg\ beam (FWHM), the number of independent pixels
is $\frac{360}{7} \cos{43\deg} \sim 38$.  We lose nothing in
choosing smaller pixels, because we can always average them
together to form larger pixels later, but you can lose
information if your pixel size is too large.  The rule of thumb
is to pixelize at \app\ 40\% of your beam FWHM or smaller.

For \polar, we chose to use 180, 2\deg\ RA-pixels, with the first
pixel arbitrarily centered at RA = 0\deg.  These were large enough
so that there weren't a ridiculous number of them (the more pixels
there are, the harder the final likelihood analysis becomes), and
small enough not to lose any potential information.

\subsection{Overview of POLAR's Analysis Pipeline}
We would like
to obtain maps of $Q$ and $U$ for each of our three polarization
channels (26--29, 29--32, and 32--36 GHz).
If frequency-dependent foregrounds do not prove
to be a problem, we can then average
the maps of the three channels into final overall maps of $Q$ and
$U$, and proceed to do cosmology with those maps.

\begin{figure}
\begin{center}
\includegraphics[height=6.9in]{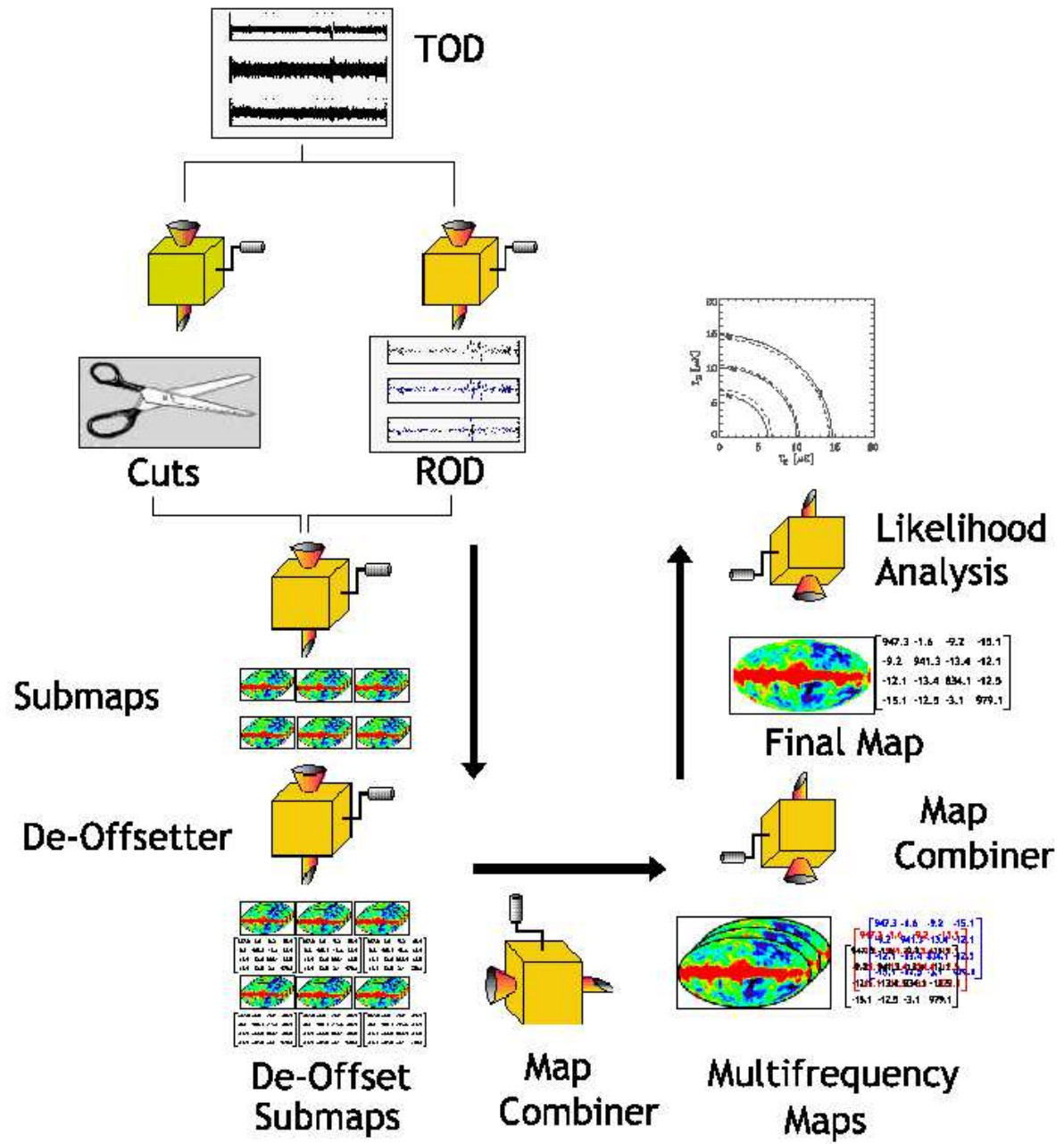}
\caption[Flow Chart of Data Analysis]
{\label{flowchart} \fixspacing
Flow Chart of the POLAR Data Analysis Pipeline
}
\end{center}
\end{figure}

\fig{flowchart} gives an overview of the entire analysis pipeline.
We begin with the calibrated data time stream, which we refer
to as the \emph{Time--Ordered Data}, or ``TOD''.
Instead of simply going straight from $TOD \rightarrow
Maps$, we will make a couple of intermediate ``rest stops'' to
simplify the problem.  The primary intermediary we call
the \emph{Rotation--Ordered Data}, or ``ROD''.  The ROD
is the projection of the \onephi\ and \twophi\ components from
the data set (effectively, a time stream of $Q$'s and $U$'s);
its construction is described in \sct{s:ROD}.
For all sections of data passing the cuts, we form
``submaps'' of $Q$ and $U$ (see \sct{s:addmaps}).
For example, if one day we observed
for seven hours of clear weather, we would obtain submaps for $Q$
and $U$, for each channel,
each \app\ 53 pixels long (given the above pixelization).
When we do this for all sections of data, we will arrive at some
number of submaps.  See the \emph{glossary}, Appendix A, for
additional clarification of these terms.

Unfortunately, during the analysis we discovered that the ROD data
itself had unforeseen offsets at the level of hundreds of \uK.  These
were offsets in the parameters $Q$ and $U$ themselves, to be
distinguished from the overall offsets $I_0$ in each channel;
these offsets are discussed in great detail in \sct{RODoffsets}.
This new problem
necessitated an additional step to remove the offsets(see \sct{offsetremoval}), forming
``de-offset submaps''.  We then combine these de-offset
submaps into final maps for $Q$ and $U$ for each channel (\sct{s:subfinal}),
and taking into account the inter-channel correlations, combine
these channel maps into final maps for $Q$ and $U$ (\sct{s:finalmaps}).

\section{Rotation-Ordered Data (ROD)}\label{s:ROD}

To form the ROD, we must robustly extract the coefficients
of the simple Fourier expansion of the time-ordered data.
If y(t) is some segment of the TOD, then
\beq{eq:spin}
y(t_i) \;=\; I_0 + c_i \cos{\phi_i} + s_i \sin{\phi_i} +
q_i \cos{2\phi_i} + u_i \sin{2\phi_i} + n(t) \quad ;
\eeq
this is the signal seen at time $t_i$ (the i$^{th}$ sample),
when the polarimeter is at a rotation angle of $\phi_i = \omega t_i$
with respect to geographic North, and $n(t)$ denotes the noise. The
coefficients $\{c,s,q,u\}_i$ are lower-case to denote they are
elements of vectors, and will hereafter be referred to as the
\emph{Rotational Coefficients}.  The term ``ROD'' will
hereafter refer to this time series of rotational coefficients.
$I_0$ is the overall DC level, and is not used in the analysis.

We choose to extract these coefficients once per hour file, due to
our stationarity assumption.  We could have chosen to extract the
coefficients for each individual rotation, but one rotation covers
only about 0.13\deg\ on the sky, whereas our beam size is
7\deg.  It is to our benefit to average \emph{more}, since our beam smears
out all information at scales much below our beam size. The
hour-file standard is a nice compromise between having a section
too short (less compression), and too long (difficult to compute
the rotational coefficients, and failure of noise stationarity assumption).

\subsection{The Simple Approach to ROD extraction}
The most straightforward way to form the ROD data is simply to
``project out''
the components we are interested in by using the orthogonality
of sines and cosines.  For a data vector $\y$ from
a single hour file with $N_f$ elements, $q$ and $u$ are given by
\beqa{FourierTrick}
q \ = \ \frac{2}{N_f}\sum_{i=1}^{N_f} y_i \cos{\phi_i} \\
u \ = \ \frac{2}{N_f}\sum_{i=1}^{N_f} y_i \sin{\phi_i}
\eeqa
where $\phi_i$ is the polarimeter rotation angle at sample $i$.
Although it is very simple and fast,
this approach has two important drawbacks.  First,
it ignores the effects of 1/f noise in our data.  Because
the noise from our polarization channels is almost always
flat, 1/f noise will play a small role, but in order to get
a good estimation of our errors, it is important to know the
contribution to the errors from this noise component.

The second drawback to this simple projection
method is that it ignores our pixelization scheme, described
in \sct{s:pixelization}.  An hour file spans 1.875\deg\ in Right
Ascension, while our chosen pixel size on the sky spans 2\deg;
these are fairly well-matched, but not perfectly, and they will not
always line up.  There could easily be an hour file that straddles
two sky pixels equally.  This is most likely a small effect, but
it is easy to treat \emph{exactly}, using the formalism of
\mm\ as discussed in the next section.

\subsection{ROD Extraction Using \mm}
To this end, we seek to discern which sky pixels a given data file
spans (for this pixelization it is 1 or 2 only); then we must
determine the rotational coefficients implied by the data file
for each of these pixels.  Couple this to the
1/f noise effect and it seems a pretty daunting task, but the amazing
thing is that \mm\ covers it all!

To use this technique, we need our data vector $\y$, our noise covariance
matrix $\N$, and the pointing matrix $\A$.  We obviously know \y, but keep in
mind that we do one of these operations for each channel
(including the QPC for completeness).  Next we show
how to calculate the $\N$ and $\A$ matrices.

\subsubsection{The Pointing Matrix}\label{pointingmatrix}
$\A$ is the transformation matrix that maps the sky map
into our data stream; it depends only on the scan strategy
and beam parameters.
$\A$ has a number of columns equal to the number of map
pixels and a number of rows equal to the number of data samples.

The case of assembling $\A$ for a one-horn total power experiment is particularly
easy.
For each row, you simply put a `1' in the column of the pixel that
this particular sample came from, and zeros everywhere else.
Thus, $\A$ is an exceedingly sparse
matrix made up of 1's and 0's in the total power (intensity)
case, such as most anisotropy experiments.  A good example of
this is given in \cite{wright96}, where the case of
multiple horns is also treated.

Because of \polars\ rotation and the fact that we are observing the
Stokes parameters $Q$ and $U$,
our situation is somewhat more
complicated.  Let us write the map vector as $\x = \{\cv,\sv,\q,\u\}$.
At some time $t_i$, the signal we observe is
given by \eqn{eq:spin}. Now it is straightforward
to form the pointing matrix $\A$, so that \eqn{LinearProblemEq}
holds.  Let us consider a toy model in which there are four
sky pixels spanned by the data segment we are analyzing.
Our map in this case is a 16-element vector,
containing the four rotational coefficients for each of the four
pixels
\footnote{\fixspacing This may seem strange to the beginner; our four maps
we treat as one big map, which can be thought of as
four vectors concatenated together.
This will treat inter-map correlations naturally, but we hope
those inter-map correlations will turn out to be small or zero.}
.
For the observations $\y$ within the data segment, a sample
pointing matrix would look like the following (I show $\A^t$ for simplicity):
\be
\A^t = \left(
\begin{array}{cccccccccc}
  0       &    0     &  \ldots  &   0      &    0     &   \ldots &   0      &    0     &  \ldots  &    0     \\
  \ct{1}  & \ct{2}   &  \ldots  &  \ct{a}  &    0     &   \ldots &   0      &    0     &  \ldots  &    0     \\
  0       &    0     &  \ldots  &   0      & \ct{a+1} &   \ldots & \ct{b}   &    0     &  \ldots  &    0     \\
   0      &    0     &  \ldots  &   0      &    0     &   \ldots &    0     & \ct{b+1} &  \ldots  & \ct{N_f} \\
\hline \vspace{-5mm} \\
 0      &    0     &  \ldots  &   0      &     0    &   \ldots &    0     &    0     &  \ldots  &   0      \\
\st{1}  &  \st{2}  &  \ldots  & \st{a}   &     0    &   \ldots &    0     &    0     &  \ldots  &   0      \\
  0     &    0     &  \ldots  &   0      &  \st{a+1}&   \ldots & \st{b}   &    0     &  \ldots  &  0       \\
  0     &     0    &  \ldots  &    0     &     0    &   \ldots &    0     & \st{b+1} &  \ldots  & \st{N_f} \\
\hline \vspace{-5mm} \\
 0       &    0     &  \ldots  &   0      &    0     &   \ldots &   0      &    0     &  \ldots  &    0     \\
 \ctt{1} & \ctt{2}  &  \ldots  &  \ctt{a} &    0     &   \ldots &   0      &    0     &  \ldots  &    0     \\
 0       &    0     &  \ldots  &   0      & \ctt{a+1}&   \ldots & \ctt{b}  &    0     &  \ldots  &    0     \\
  0      &    0     &  \ldots  &   0      &    0     &   \ldots &    0     & \ctt{b+1}&  \ldots  & \ctt{N_f}\\
\hline \vspace{-5mm} \\
 0      &    0     &  \ldots  &   0      &     0    &   \ldots &    0     &    0     &  \ldots  &   0      \\
\stt{1} &  \stt{2} &  \ldots  & \stt{a}  &     0    &   \ldots &    0     &    0     &  \ldots  &   0      \\
  0     &    0     &  \ldots  &   0      & \stt{a+1}&   \ldots & \stt{b}  &    0     &  \ldots  &  0       \\
  0     &     0    &  \ldots  &    0     &     0    &   \ldots &    0     & \stt{b+1}&  \ldots  & \stt{N_f}\\
\end{array}\right)
\vspace{3mm}
\ee
In this example, for observations $t_a$ through $t_b$, we were aimed
at pixel 2, and likewise pixel 3 corresponds to $t_b+1$--$t_c$,
and pixel 4 to $t_c+1$--$t_d$.  I included pixel 1 here to show
that we are not penalized for keeping extra pixels that don't have
any data; they simply get all 0's in $\A$, and will have infinite error
bars in the final map.
Then, by multiplying any row in $\A$ (any column in $\A^t$) by
the map vector $\x$, we recover \eqn{eq:spin} in the absence of noise (and with
no DC level), which was our original goal.

\subsubsection{Noise Covariance Matrix Construction for ROD extraction}
In order for this machinery to work we must adequately understand
our noise, in the form of its noise covariance matrix $\N$.  For
our data files (HF's), this will be a $9000\times 9000$ symmetric,
positive-definite matrix ($N_f=9000$ samples per data file).
It has the special property that each
row equals the row above it, right-shifted by one element (with
wrap-around boundary conditions).  Thus, instead of containing
$N_f^2$ pieces of information, in this special case of noise
stationarity, the data covariance matrix contains only $N_f$
independent numbers.

In order to characterize $\N$, as I will show below,
all we really need to know is the power spectrum of the
noise.
A reasonable estimate of the PSD is obtained by
using only the data from this file; however,
with this method the PSD comes from a mere
$N_f=9000$ numbers and is
very noisy at low frequencies, precisely
where we most need to know its shape.
We therefore perform a fit of the PSD to the
following model
\footnote{\fixspacing
The fit was performed in log-frequency space, so as not to weight
higher frequencies more (where there is more data).  The fitting
algorithm was provided by Craig Marquardt, see \cite{marquardt}.}
:
\beq{PSDmodel}
S(\nu) = \sigma^2 (1 + \frac{\nu_{knee}}{\nu}) \qquad [K^2 \ Hz^{-1}] \ ;
\eeq
this is simply the case of white plus 1/f noise, where the knee
frequency is $\nu_{knee}$.
There is also a 5 Hz low-pass filter on our data, which correlates
samples taken within about 0.2 seconds of each other (or
about every four samples). However, this is irrelevant because all our
information is contained at much lower frequencies, \app\ 0.067 Hz,
where the filter response is effectively unity, and so we leave
it out of our noise model.

\subsubsection{Algorithmic Tricks for ROD Extraction}
Let us return to the algorithm at hand, which to remind
the reader, is
\begin{subequations}
\label{mmROD}
\beqa{RODy}
\xt & = & [\A^t\N^{-1}\A]^{-1}\A^t\N^{-1} \y \\
\label{RODN}
\NN & = & [\A^t\N^{-1}\A]^{-1}
\eeqa
\end{subequations}
where $\xt$ are our rotational coefficients for a particular 7.5
minute data file, and the algorithm for $\A$ was given above.
As we will need to perform
this algorithm for each channel and each hour file, that is,
about $6\times 800 \simeq 5000$ times, we would like it to be as fast as possible.
$\N$ may not seem large, but inverting a $9000\times9000$ matrix
5000 times takes a while on any computer.  And we haven't
even fully evaluated $\N$ yet, only its power spectral density.

The algorithm becomes more numerically
palatable by making the following definitions:
\begin{subequations}
\label{RODeff}
\beqa{yeff}
\yeff \ = \  \N^{-1/2} \y \\
\Aeff \ = \ \N^{-1/2} \A
\eeqa
\end{subequations}
in which case $\xt$ and $\NN$ simplify to
\begin{subequations}
\label{mmRODeff}
\begin{eqnarray}
\xt & = & [\Aeff^t \Aeff]^{-1} \Aeff^t \yeff \\
\NN & = & [\Aeff^t \Aeff]^{-1}
\end{eqnarray}
\end{subequations}
The reason this is numerically better than \eqn{mmROD}
is that there is a fast way to calculate $\yeff$ and $\Aeff$.
Rather than work in the time domain, we can work
in the Fourier domain, where we can use the Fast Fourier Transform
(for example, see \emph{Numerical Recipes} Chapter 12 \cite{nrc})
to great advantage.

Instead of constructing $\N$, we simply work with
the PSD of N (this has units of $K/\sqrt{Hz}$),
which we denote as $S_N(\nu)$.  Using Fourier tricks,
\eqn{RODeff} becomes
\begin{subequations}
\label{RODeff_F}
\beqa{yeff_F}
\yeff \ = \  \fftinv{\frac{\fft{\y}}{S_N(\nu)}} \\
\label{Aeff_F}
\Aeff^i \ = \ \fftinv{\frac{\fft{\A^i}}{S_N(\nu)}} \ ,
\eeqa
\end{subequations}
where $\A^i$ is the $i^{th}$ row of $\A$, and ``$\ft$'' denotes
the Fourier Transform.  So we've reduced the algorithm to a bunch
of FFT's, which are quite fast and make the computation very
manageable on almost any workstation.

\begin{figure}[tb]
\begin{center}
\includegraphics[width=6.2in]{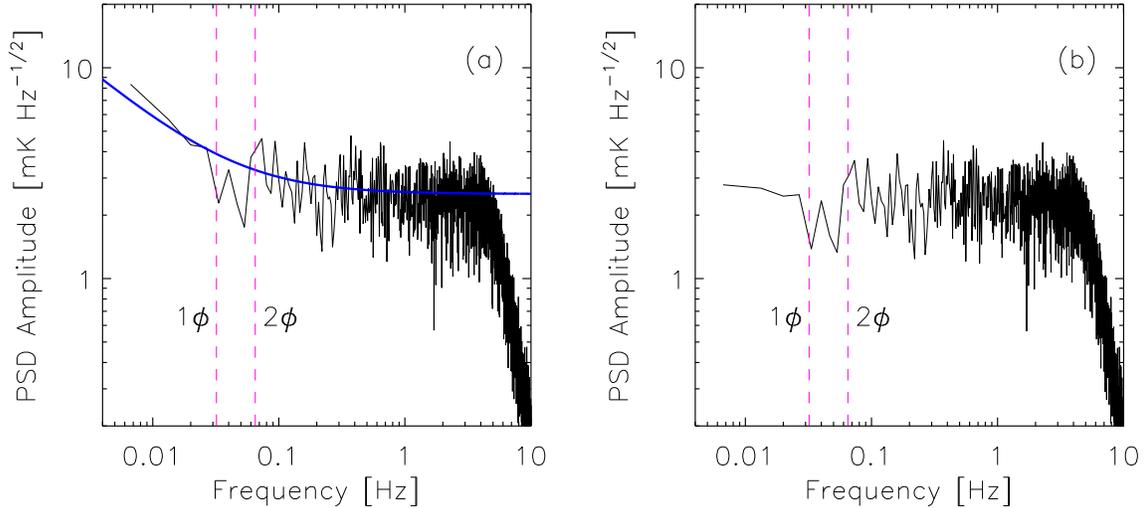}
\caption[Effect of Prewhitening on Sample Noise PSD]
{\label{psdfilter} \fixspacing
Effect of Prewhitening on sample noise PSD.
Panel (a): PSD for a sample Hour File, with a knee
frequency of \app\ 0.03 Hz, a high but not unreasonable value.
The solid (blue) curve is the logarithmic-weighted fit to the PSD.
The vertical dashed lines show the \onephi\ and \twophi\
rotational frequencies.  Panel (b): Same as (a), but for
the $\yeff$ version of the data.  The effect was to whiten
the low-frequency section of the PSD; it is equivalent
to dividing the PSD by the fitted curve.  Note that the roll-off
due to the anti-aliasing filter remains untouched.
}
\end{center}
\end{figure}

However, Equations \ref{RODeff} have physical meaning as well.
Turn your attention to \fig{psdfilter}; the figure
displays the PSD amplitude of one short section of data.
It shows the action of $\N^{-1/2}$ as a ``pre-whitening
filter'' on $\y$; that is, the effect is that
the low-frequency 1/f rise in the PSD is fitted for and removed.

\begin{figure}[tb]
\begin{center}
\includegraphics[height=4in]{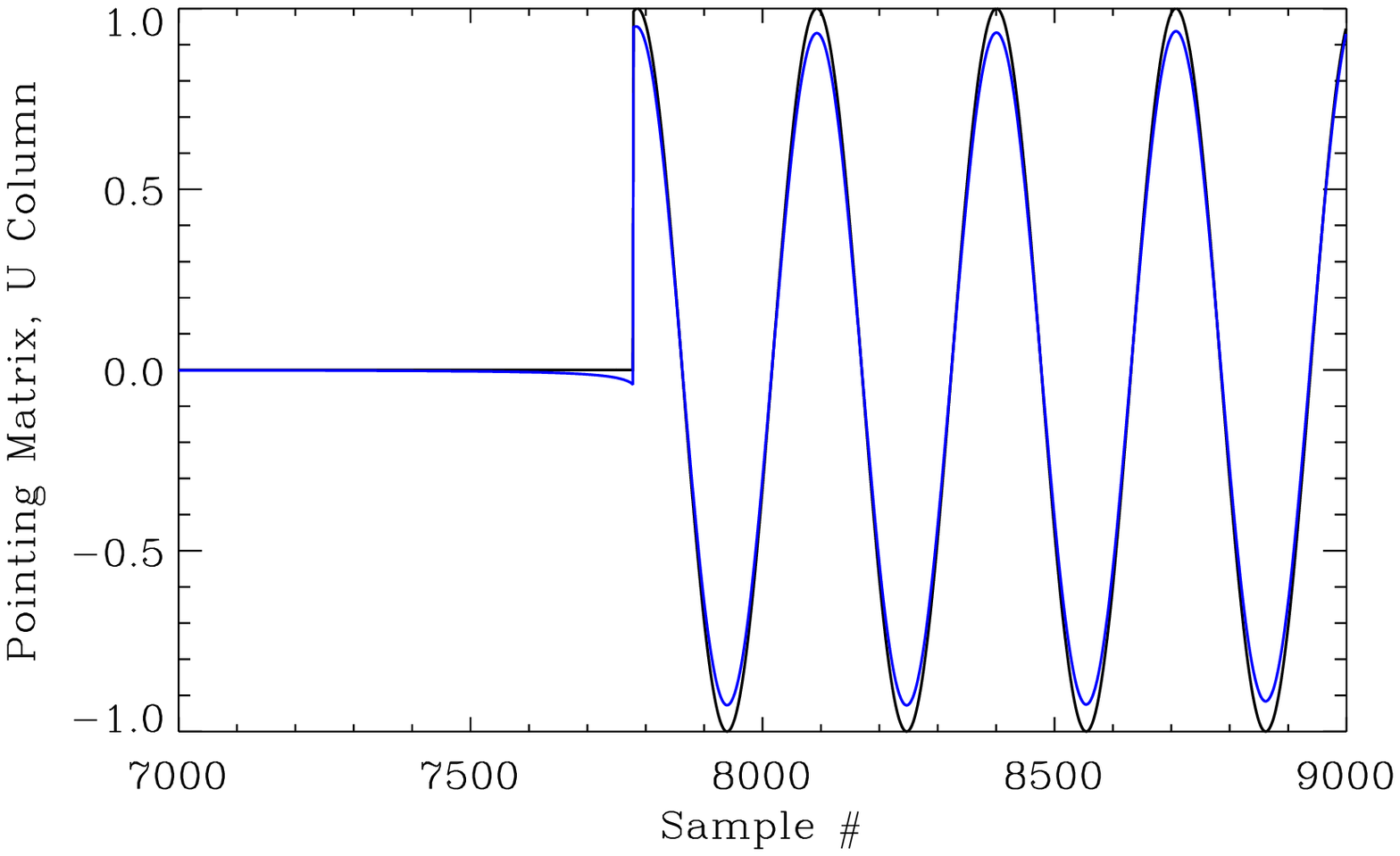}
\caption[Pointing Matrix Before and After Processing]
{\label{f:Aeff} \fixspacing
Effect of $\N^{-1/2}$ on a column of the
Pointing matrix $\A$.  The black (solid) curve shows a section of
the U-column of a sample pointing matrix.  The blue curve shows
the same column from the effective pointing matrix $\Aeff$,
processed according to \eqn{Aeff_F}, with a knee frequency of 0.01
Hz (about 1/6 of the \twophi\ frequency shown here). Notice that
some of the response has been smoothed out to the left, where the
original response was zero, at the price of lowering the primary
pixel response.}
\end{center}
\end{figure}

The effect on the pointing matrix $\A$ is also interesting.
\fig{f:Aeff} shows a section of one column of $\A$, a column
that extracts $U$ for a certain pixel.  With a knee frequency of
0.01 Hz (about 1/6 of the \twophi\ frequency), you can see the
slight difference between $\A$ and $\Aeff$.  The pixel we want sensitivity
to comes into view about sample number 7800.  However, some ``power'' is
removed from this main pixel, and there is some leakage from the
previously-viewed pixel to the desired one due to the presence of 1/f noise.

At this point, the reader may ask him/herself if such ``processing of
the data'' is warranted.
Let me be clear about this: \mm\ as I have defined
it yields the best map in the sense that it has minimum
variance while retaining all the cosmological information
\cite{teg97c}.
There will be some correlation between pixels
manifested in off-diagonal elements of the map covariance matrix.
Thus, there is no cheating involved; we are not prewhitening the data
and then constructing
maps.  Rather, in the process of constructing maps, a prewhitened
version of the data arises naturally, but because we also modify
the pointing matrix ($\A \rightarrow \Aeff$), we are not throwing
away any information.

\subsection{First Peek at the ROD: Offsets
revealed}\label{firstpeek}
We perform the above algorithm on all HF's, even those that
didn't survive data quality checks; we institute those in the next
step (see \fig{flowchart}). Thus, we now have a new data set, the
so-called ROD data, which again are the $Q,U$ rotational
coefficients on the sky for each HF.  However, another useful data
set to construct is one which is
identical to ROD except that precisely one set of rotational
coefficients (that is, \{c,s,q,u\}) is determined for each HF
(rather than tagging them to specific sky pixels).  Because data from
consecutive rotations are averaged together, this data set
lets us see the characteristics of the ROD data more directly.
In the following analysis of the timestream of rotational
coefficients and other related quantities, it is this latter data
set that was used.

\begin{figure}
\centering
\includegraphics[height=7.25in]{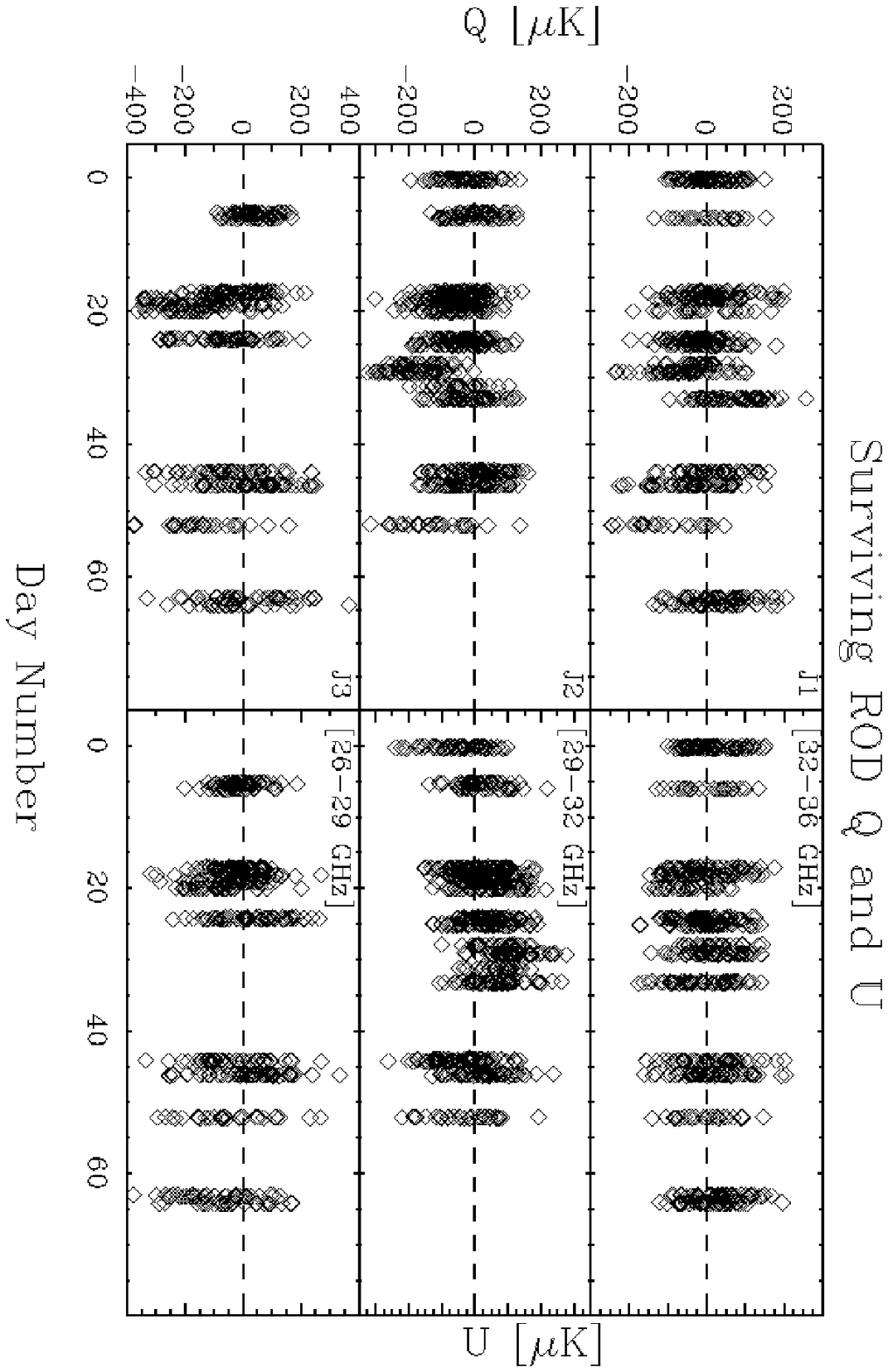}
\caption[Time Stream of $Q$ and $U$ for IPC Channels]{\fixspacing
\label{QUtimestream} The mean values of $Q$ and $U$ for the ROD
data set versus time; each data point represents one Hour File.
Non-Gaussian behavior is immediately apparent at the 50-100 \uK\
level. }
\end{figure}

\begin{figure}
\centering
\includegraphics[height=7.25in]{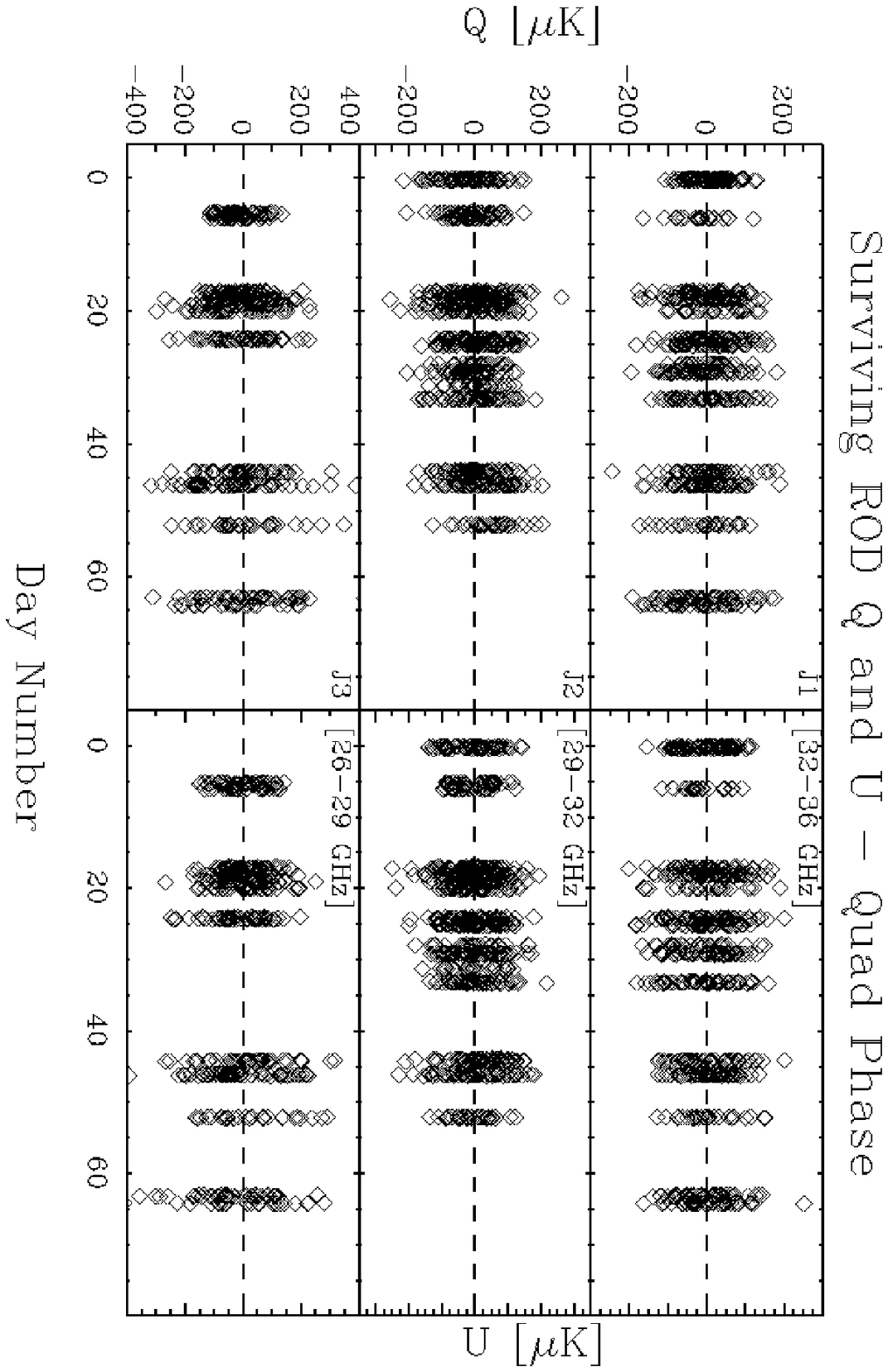}
\caption[Time Stream of $Q$ and $U$ for QPC channels]{\fixspacing \label{QUtimestream_quad}
Same as \fig{QUtimestream} but for the QPC channels.  There is
no evidence of any offsets from these distributions in the QPC
channels, eliminating any sort of electrical cause of the offsets.
}
\end{figure}

\fig{QUtimestream} shows the derived $Q$ and $U$ values for all
the surviving data for the three primary IPC polarization
channels, while \fig{QUtimestream_quad} shows this same
information for the QPC channels.  Notice that the data lie in
``chunks'' along the time axis; this is because the surviving data
were in sections 3-12 hours long each (there were about 20 of
them); we call these surviving chunks simply ``sections'', and the
maps on the sky corresponding to each one we call a ``submap''
(see glossary).

The primary feature of these graphs is the changing offset
level of each section of data,
present for the in-phase (IPC) channels, but not seen in the
quadrature (QPC) channels; \ie, the IPC
channels are obviously not consistent with purely Gaussian
noise. From these plots alone, it is difficult
to discern what the nature of the apparent signal is, but it is obvious
that it is a global issue in our data that may strongly affect
our results.  Hereafter we refer to these
apparent signals as ``offsets'', for lack of a better description.
In \sct{offset_character} we will examine the offset issue
in more detail.

\section{From ROD to Submaps: How to Add Maps}
\label{s:addmaps} At this point in the data pipeline we find
ourself with the ROD data: a time stream of rotational
coefficients, each tagged to a particular pixel on the sky.  For
each HF, we have a ``mini-map'' of just one or two sky pixels, and
determine what the Stokes' parameters are for each of these one
or two pixels, and also determine the covariance matrix between
these two pixels and all the calculated parameters.  We do things
on the level of the HF for calculational convenience, and to
ensure that our noise stationarity assumption is valid.

Now we work our way up the ``time-scale'' ladder: for each
\emph{section} of data, we combine all the little 1--2 pixel
HF maps together into one map for each channel, and calculate its
corresponding covariance matrix; these are our \emph{submaps}.
We find robustly that the ``inter-coefficient correlations'',
that is, offsets between $Q$ and $U$ and the like within a pixel,
were completely negligible. We forget about our $C$ and $S$
data at this point; these coefficients do contain information on
systematic effects, but making maps from them will be difficult to
interpret and hence not be particularly informative.

We are hence faced with the standard ``map combination'' problem:
given a collection of $m$ maps $\{\x_1 \ldots \x_m\}$ with
corresponding covariance matrices $\{\N_1 \ldots \N_m\}$, what is
the best estimate of the full map? The problem is complicated by
the fact that our maps are incomplete -- each only covers a part
of the total map.

Let us deal with the latter problem first.  The solution to
partial sky coverage is to ``expand'' each initial map to cover
the final map by making up values for the unmeasured map pixels,
but giving those values zero weight by assigning them infinite
uncertainty in the noise covariance matrix. As an example, let us
say one of our initial maps measures pixels 2 and 3 of a
four-pixel map.  We perform the following ``expansion of this
initial map'':
\beq{}
\x_i \ = \ \{x_2,x_3\} \ \rightarrow \
\{0,x_2,x_3,0\} \eeq and
\begin{subequations}
\label{expandN}
\beqa{}
\N_i & = & \left[ \begin{array}{cc}
N_{11} & N_{12} \\
N_{21} & N_{22}
\end{array} \right] \rightarrow \
\left[ \begin{array}{cccc}
\infty & 0 & 0 & 0 \\
0 & N_{11} & N_{12} & 0 \\
0 & N_{21} & N_{22} & 0 \\
0 & 0 & 0 & \infty
\end{array} \right] \\
\text{and} & & \qquad \qquad \N_i^{-1} \ \rightarrow \
\left[ \begin{array}{cccc}
0 & 0 & 0 & 0 \\
0 & N^{-1}_{11} & N^{-1}_{12} & 0 \\
0 & N^{-1}_{21} & N^{-1}_{22} & 0 \\
0 & 0 & 0 & 0
\end{array} \right]
\eeqa
\end{subequations}
We see it is a trivial task to deal with the partial sky
coverage problem.  I show the inverse covariance matrix because
that is what is actually used in the formalism.

Now let us assume that our maps $\{\x_1 \ldots \x_m\}$ and covariance matrices
$\{\N_1 \ldots \N_m\}$ have been expanded to all have equal
coverage.  How do we then find the most-likely/minimum noise map
containing all the information?  The answer is, of course,
\mm!  We simply take our data vector to be $\y \ = \ \{\x_1 \ldots
\x_m\}$, and we make a block-diagonal
``mega-covariance matrix'' $\N$ with all the little $\N_i$'s on the
diagonal.  Since we've expanded each $\x_i$ map to have the same
pixel coverage as the final map we're pointing to, each has a
``pointing matrix`` that is simply the $n$-element identity
(assuming the maps all have $n$ pixels).  The full
pointing matrix is then $\A = [\I,\ldots,\I]$, so \A\ is a $(nm)
\times n$ rectangular matrix.  Applying the standard \mm\ formalism,
Equations \ref{WdefEq}--\ref{MaxMethod1}, the final map is
\beqa{blah}
\x_f & = & [\A^t \N^{-1} \A]^{-1}\A^t \N^{-1} \y \notag \\
& = &\left[\sum_{i=1}^m \N_i^{-1}\right]^{-1}[\N_1^{-1},\ldots,
\N_m^{-1}]
\left[
\begin{array}{c} \x_1 \\ \vdots \\ \x_m \end{array}
\right] \notag
\eeqa
Then the final map $\x_f$ and final covariance matrix
$\N_f$ are given by
\begin{subequations}
\label{addmaps}
\begin{eqnarray}
\N_f \ = \ \left[ \sum_{i=1}^m \N_i^{-1} \right]^{-1} \\
\x_f \ = \ \N_f \ \left[ \sum_{i=1}^m \N_i^{-1} \ \x_i \right]
\end{eqnarray}
\end{subequations}
which is the standard result (\eg, \cite{qmap98c,maxima01b}).

We applied this technique to all the data passing our cuts, in
order to make a submap for each section, channel and \{Q or U\}
combination, for both the IPC and QPC channels: this generates
approximately $6 \times 20 \times 2 = 240$ submaps on the sky. As
will be discussed in the next section, the offset that was noticed
in the ROD data is even more apparent once these submaps are
formed; the offset problem and how we dealt with it are discussed
in the next two sections.

\section{Characterizing the ROD Offsets}\label{RODoffsets}
\label{offset_character}

\begin{figure}
\begin{center}
\includegraphics[height=6in]{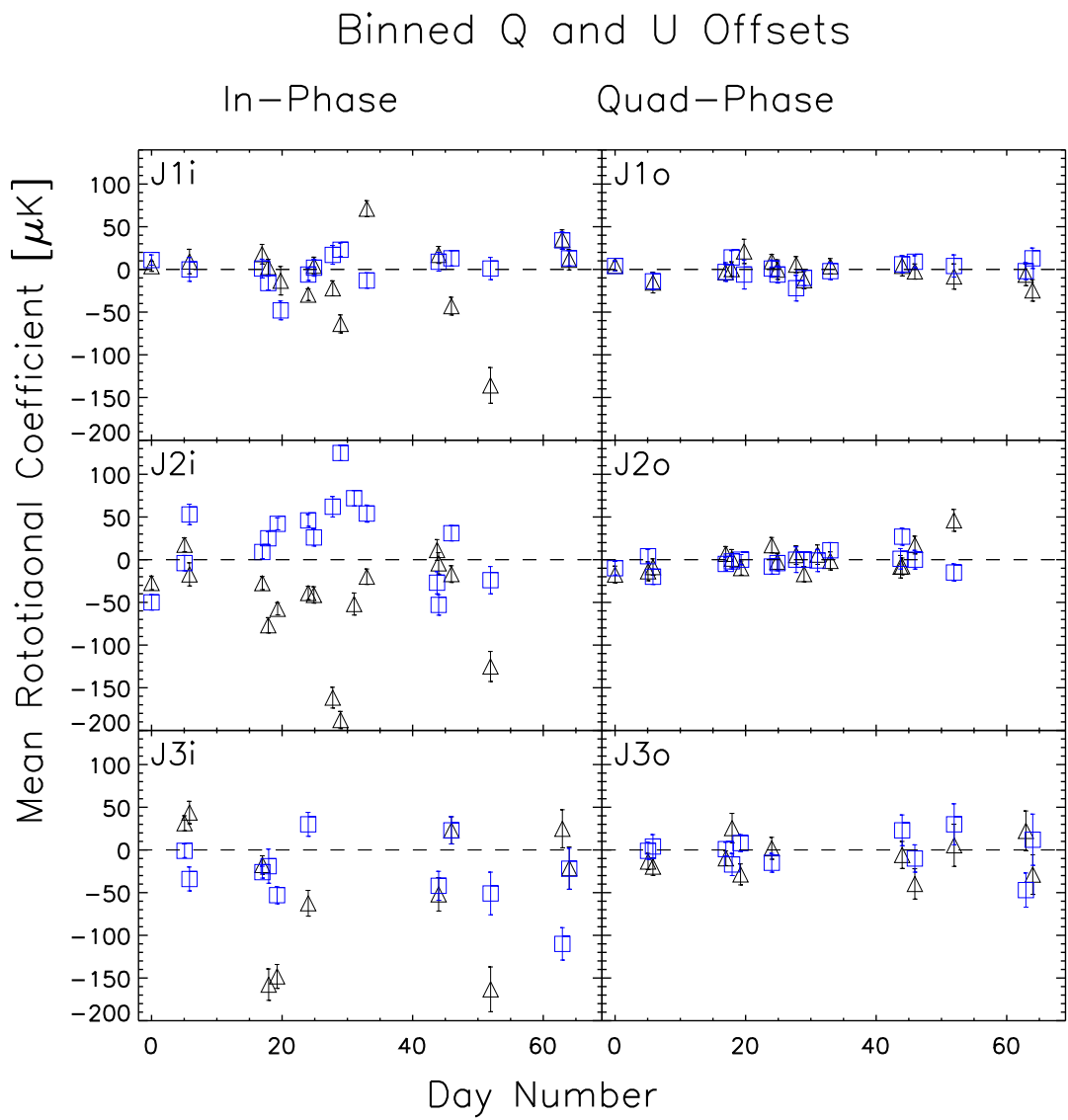}
\caption[Q and U offsets binned by section] {\label{offset_eplot}
\fixspacing Q and U offsets binned by section, for all six
polarization channels.  The (black) triangles correspond to the
$Q$ offsets, while the (blue) squares correspond to the $U$
offsets.  The channels are indicated in the upper left-hand corner
of each plot.  The displayed error bars are statistical.  Only
data that survived the quality cuts contributed to this plot. }
\end{center}
\end{figure}

Let us now return to the issue of the ROD ``offsets'' discussed
previously.
Perhaps the first logical question to ask is, just how
statistically significant are these offsets? \fig{offset_eplot}
shows the offsets in $Q$ and $U$ for all six channels as a
function of time, but averaged down to a single number per
section, with the corresponding derived statistical error. As you
can see the effect is strong (multi-sigma) and quite variable, an
experimentalist's worst nightmare.

Based on this plot, we can already give some qualitative
characteristics of the offsets.  First, there is no visible effect
for the QPC channels, eliminating some kind of electrical
explanation. It seems that the effect is sometimes correlated
among the three channels, but not always.  If there were always a
strong correlation among the channels, we could probably attribute
the effect to some specific optical phenomenon, such as sidelobe
pickup of the 300 K earth.  But its time-varying nature makes it
difficult to ascribe a particular cause to it.

%
%

\begin{figure}[tb]
\begin{center}
\includegraphics[height=3.7in]{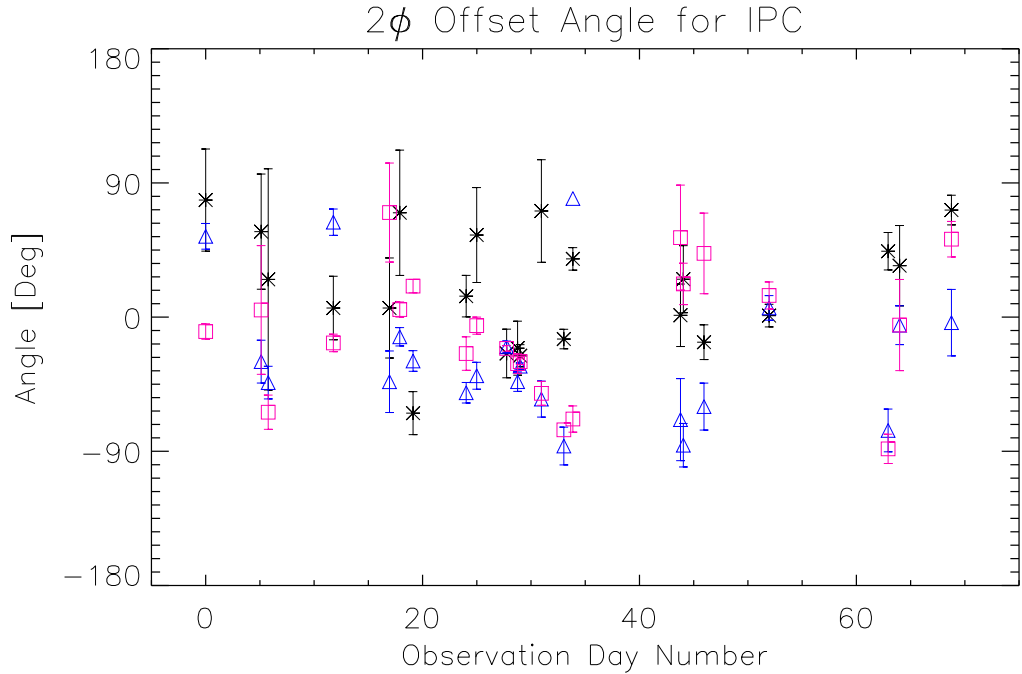}
\caption[\twophi\ Offset Angles for each of the three IPC channels
during the season] {\label{offset_angle2} \fixspacing \twophi\
Offset Angles for each of the three IPC channels during the
season.  For a given section, the \twophi\ offset angle is defined
as $\arctan{\frac{\langle U \rangle} {\langle Q \rangle}}$. The
stars, triangles, and squares, correspond to channels J1i, J2i and
J3i, respectively.  Standard error propagation is used to derive
errors on the angle from the errors on $\langle Q \rangle$ and
$\langle U \rangle$. Data cuts used are slightly less restrictive
than the standard cuts, in order to enable direct comparison
between the three IPC channels. }
\end{center}
\end{figure}

A final interesting quantity to look at is the relationship
between $Q$ and $U$.  Specifically, let us define the
\emph{\twophi\ offset angle} as $\arctan{\frac{\langle U \rangle}
{\langle Q \rangle}}$.  In \fig{offset_angle2}, this quantity is
calculated for each submap as in \fig{offset_eplot}, but a
slightly less restrictive cut has been used that cuts on our three
channels equally.  Corresponding error bars on the angles are
derived.  First, notice that the errors on J1i's angle are quite
high, indicating that it doesn't have a strong offset problem.
J2i's offset angle is relatively stable, especially for a long
period in the middle of the season. J3i's offset is less stable.
However, there is a telling correlation between all three channels
that implies a possible common origin. Similar results are
obtained for the \onephi\ coefficients $C$ and $S$ and the
resulting angle; J1i shows small offset angles, J2i shows a fairly
stable offset angle, and all three channels exhibit some
correlation.

\subsection{A Possible Sky Signal?}

\begin{figure}[tb]
\begin{center}
\includegraphics[height=3.6in]{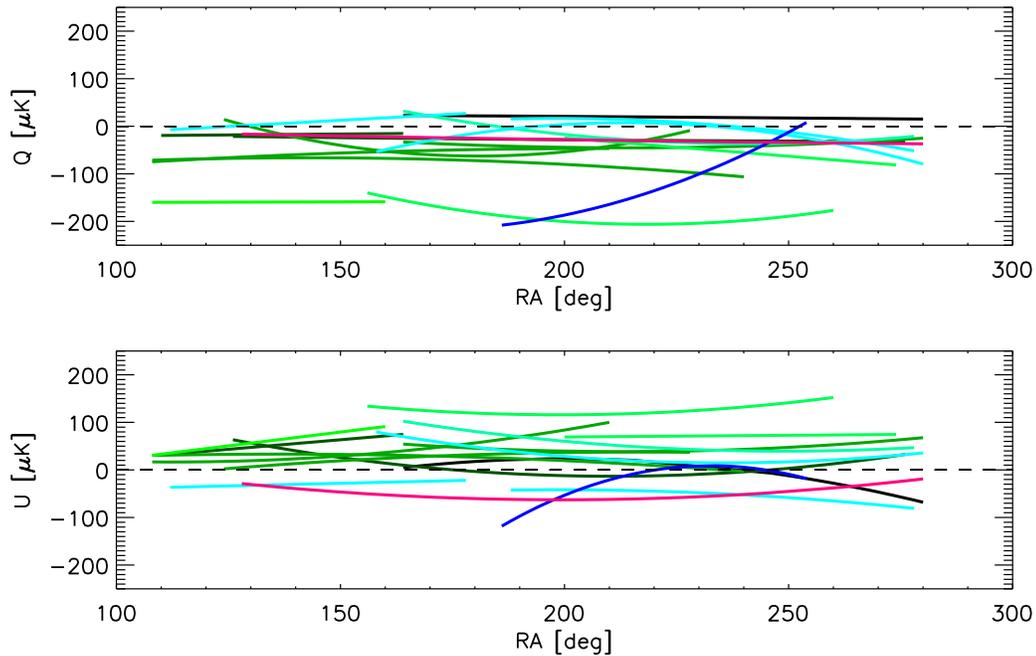}
\caption[Polynomial Fits to J2i Offsets, Plotted in Right
Ascension] {\label{skyoffsetplot} \fixspacing Polynomial fits to
J2i offsets, plotted in right ascension. Polynomials of varying
degree were fit to each submap; for submaps with less than four
hours of data, a line was fit to the data, while a quadratic was
fit to submaps with more than four hours of data. Different
sections are indicated by different colors. }
\end{center}
\end{figure}

Is the effect consistent with some type of signal attached to the
sky? It is far too strong to be CMB polarization, but it is
conceivable that the structure is due to strong synchrotron or
spinning dust.  \fig{skyoffsetplot} shows the same data again for
channel J2i (which exhibits particularly strong ``offsets''),
plotted in right ascension. Polynomials of varying degree were fit
to each submap, as described in the figure caption. Clearly, there
is nothing visibly consistent with a true sky signal.  Certain
submaps are totally inconsistent with other submaps covering the
same section of sky. The same conclusion holds true for the other
polarization channels as well.

\subsection{Rotation-Synchronous Analysis}
It is quite useful to take a step back at this point, and form a
``map'' straight from the time-ordered-data \emph{that is binned
into coordinates fixed to the ground rather than to the sky}.
Offsets in $Q$ and $U$ can only come from signals that are
synchronous with our rotation frequency (specifically, at its
first harmonic); these offsets are naturally due to a
rotation-synchronous noise component.

\begin{figure}[tb]
\begin{center}
\includegraphics[height=4.5in]{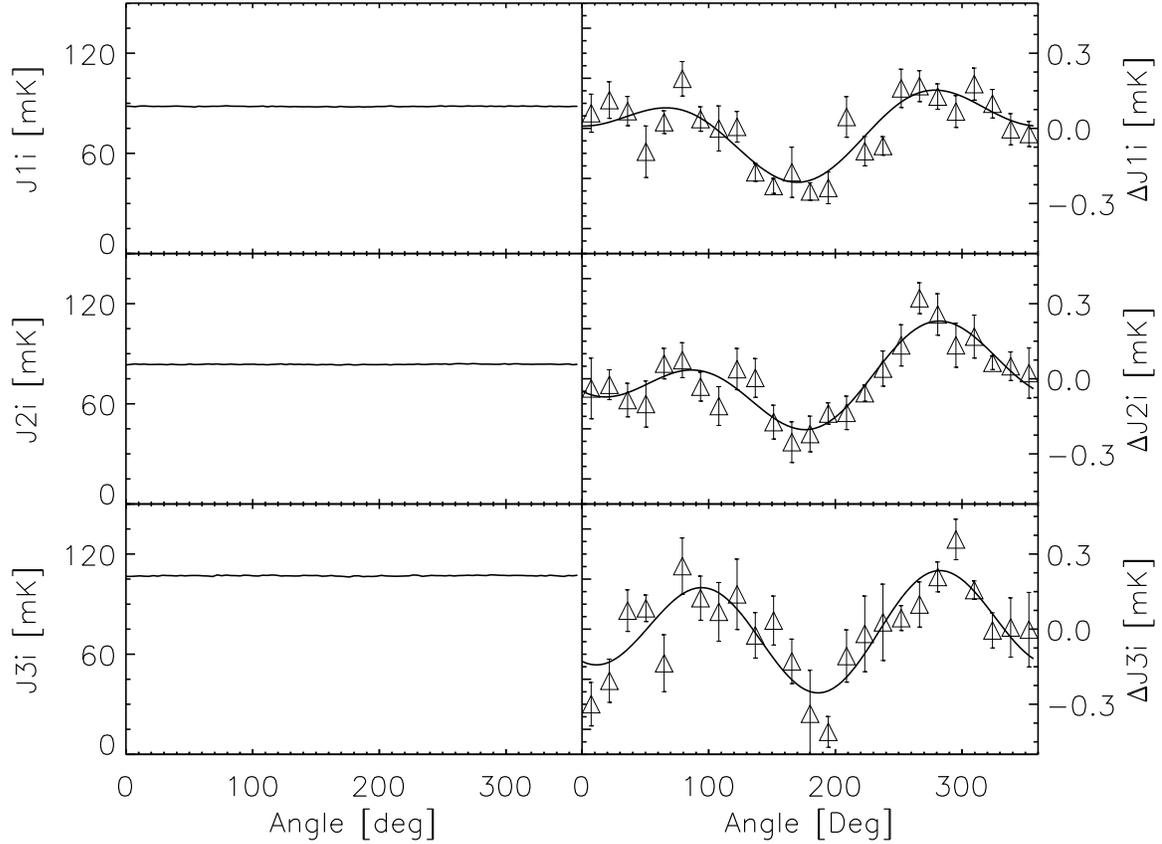}
\caption[Correlator time streams plotted in ground-based
coordinates, for May 5, 2000] {\label{gb31} \fixspacing Correlator
time streams plotted in ground-based coordinates, for the night of
May 5, 2000. Only data passing the quality cuts are shown.  The
left three panels show the overall polarimeter response as a
function of angle; it is quite stable considering the large (tens
of mK) offset on each channel. The right three panels show the
same information with the offsets removed; the solid line shows
the best-fit curve including both \onephi\ and \twophi\
frequencies. }
\end{center}
\end{figure}

\begin{figure}[tb]
\begin{center}
\includegraphics[height=4.5in]{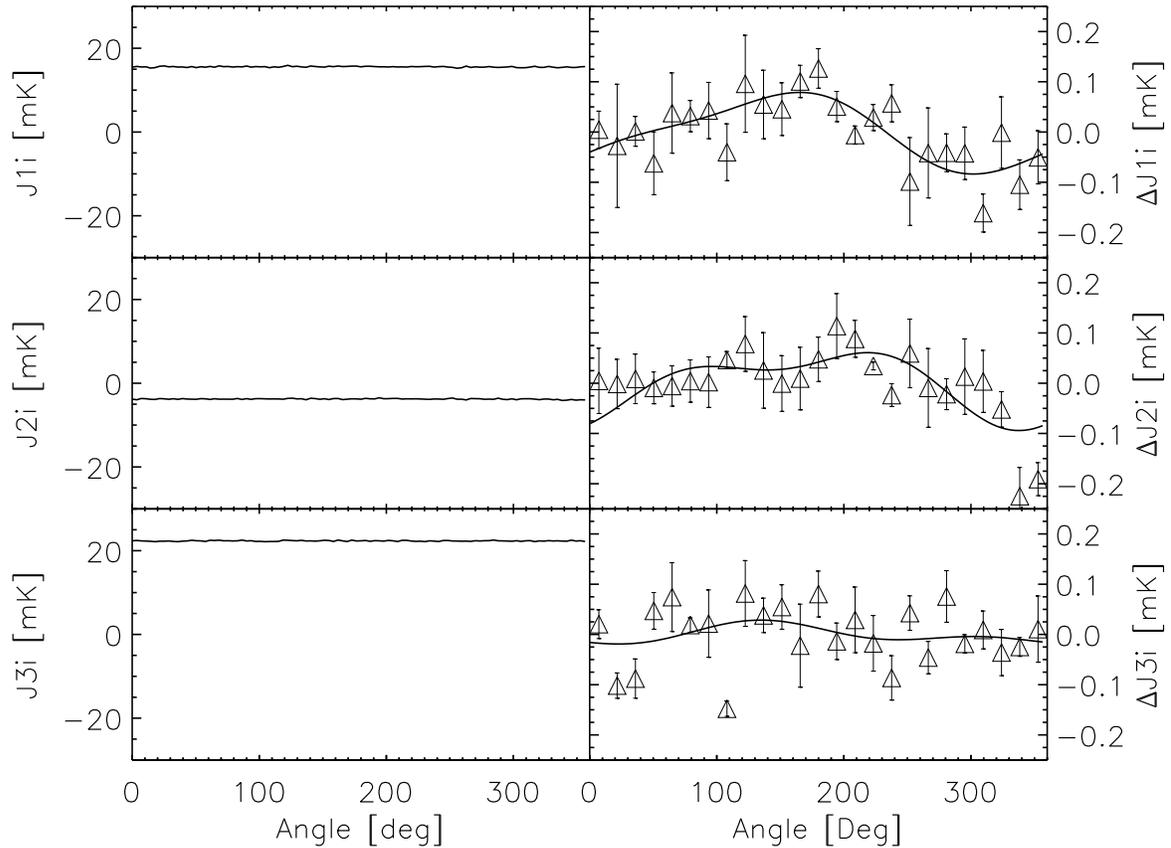}
\caption[Correlator time streams plotted in ground-based
coordinates for March 28, 2000] {\label{gb4} \fixspacing
Correlator time streams plotted in ground-based coordinates for
the night of March 28, 2000.  Plot conventions are the same as
\fig{gb31}. This plot shows a striking lack of rotation-modulated
systematics. In general, the $C$, $S$, $Q$, and $U$ offsets varied
quite a bit throughout the season, as discussed in the text. }
\end{center}
\end{figure}

Rotation-synchronous plots are made of two sections of data in
Figures \ref{gb31} and \ref{gb4}.  The first figure shows this
information for the three IPC channels from a particular night of
good data. The panels on the left of the figure are simply to
remind the reader of the large offsets upon which the
rotation-synchronous signals ride.  These ``$I_0$'' offsets are
typically 10--100 mK (see \fig{offset_eplot}), and are due to the
cross-polarization of the OMT.  The right-hand panels show the
same plots with $I_0$ removed; it is evident that there is a
strong rotation-synchronous effect (100-200 \uK), correlated among
the three channels.

However, it was not always this way.  \fig{gb4} shows a night of
data from earlier in the season where the synchronous effect was
significantly smaller, though not negligible.  In this example,
J1i and J2i appear somewhat correlated, but J3i doesn't exhibit
signs of visible contamination.  The fact the that overall offsets
were significantly lower in this case is somewhat misleading;
there are cases throughout the season when the TOD offsets were
high and the ROD offsets were low, and vice versa (and, indeed,
everything in between).

\subsection{Conjectures on the Rotation Synchronous Effect}
It is worthwhile to speculate on the cause of these
rotation-synchronous offsets, although let me be clear from the
outset that ultimately we never identified the culprit(s). For
reference, here are some specific facts on the rotation-synchronous
offsets:

\begin{itemize}
\item
The offsets were often strongly correlated among channels,
especially when the offsets were strong.
\item
There was marked anti-correlation between Q and U, especially in
channel J2i.
\item
J1i had the weakest offset, with almost no visible effect on U.
\item
Offsets were highly variable throughout the season, although for
J1i and J2i typically $Q$ was negative and $U$ was positive.
\item There was a degree of correlation in the offset angles
($\arctan{\frac{\langle U \rangle}
{\langle Q \rangle}}$) between the three IPC channels.
\item No offsets are visible in the QPC channels.
\end{itemize}

\subsection{Asymmetric Beam Shape+Anistropic Sky}
So, what could this mystery effect be? A real possibility is
atmospheric emission, coupled with an asymmetric beam pattern.
Recall that $I_0$, the TOD offset in the polarization channels,
can be coupled into the radiometer via non-zero cross-polarization
of the OMT. If we have a symmetric beam shape, as \polar\ rotates,
you should get the same offset.  However, if our beam has a
\twophi\ component, this will cause the offset to similarly be
modulated at this level.  As $I_0$ varies with atmospheric and
ground-based variables, such as temperature and precipitable water
vapor, it is easy to see how this could enter in.

\subsection{Outer Ground Screen Pick-Up}
Another possibility is associated with the outer ground screens.
While the inner conical ground screen rotates with the experiment,
the outer ground screens are fixed (see \sct{groundscreens}). The
outer screens have a natural quadrupolar component, being composed
of four large aluminum panels. The sky will reflect off of these
panels and enter our sidelobes, and at some level there will be a
natural quadupolar component to this contamination.  Again, it
will vary throughout the season as the atmospheric emission
varies. The frequency dependence will be a complicated interplay
of the frequency-dependent features of the horn, OMT,
ground-screens and atmosphere.

\section{Offset Removal Techniques}\label{offsetremoval}
Whatever the ultimate cause of the offsets, it is clear we must do
something about them in order to proceed with the data analysis.
If we had some
model of the effects that could make successful predictions about
its level, we could use it to safely subtract out the offsets
without strongly affecting our signal recovery.  However, without
such a model we must proceed along a different path, which
assumes no knowledge of the offset cause.

Most experiments have offsets that they must deal with, although
ours is particularly insidious because it is so variable. Luckily,
machinery has been developed to account for this contingency.  A
literature search reveals that the topic was apparently first
addressed by Rybicki and Press \cite{rb92,rbh92}. The QMAP group
introduced the method of ``virtual pixels'' (sometimes called
``extra pixels'') to estimate and remove offsets, see
\cite{qmap98c}.  Max Tegmark derived tricks for the removal of
unwanted modes in CMB maps in \cite{teg97b}, although the
techniques are quite general.  Bond, Jaffe, and Knox introduced
the method of \emph{marginalization}, where unwanted modes of a
time stream or map have infinite noise added to them, thus
removing them \cite{bjk98} (hereafter BJK). Recently, the DASI
group used the BJK technique to deal with their offset problems
\cite{dasi01b}, while the MAXIMA group explored both extra pixels
and marginalization, and provide a great review of these
techniques in their excellent mapmaking paper \cite{maxima01b}.

Although the mechanics of the two primary mode-removal techniques,
the Tegmark-QMAP technique of \emph{extra pixels} and the
BJK/Tegmark technique of \emph{marginalization}, are quite
different, they produce identical results.  Typically one or the
other is more convenient, depending on the situation. We will
briefly explore each of these methods below; our discussion draws
heavily upon \cite{maxima01b}, to which the reader is referred if
more details are required.

\subsection{Direct Offset Removal}
Let us first define our problem.  Consider a data vector $\y$ and
corresponding noise covariance matrix $\M$. $\y$ may correspond to
a map, submap, or to time-ordered data; it doesn't matter.  First,
what is the offset in $\y$ -- that is, what is its weighted mean?
The easiest way to calculate this is using \mm\ again, but
taking your output ``map'' to be a single number! Using this
approach it can be shown that for a piece of data $\y$ with noise
covariance $\N$, the weighted mean is given by
\beq{meaneq}
\expec{\y} = [\eO^t \N^{-1} \eO]^{-1}\eO^t\N^{-1} \, \y ]
\eeq
where $\eO$ is a column vector of all 1's.  Notice that $\eO$ is our
pointing matrix in this case -- it ``points'' to only one pixel, the
mean. Similarly, the data with the mean removed is given by
\begin{subequations}
\label{PiEq}
\begin{eqnarray}
\y - \expec{\y} & = & \ \PI \ \y \\
\PI & \equiv & \I - \eO[\eO^t \N^{-1}\eO]^{-1}\eO^t\N^{-1}
\end{eqnarray}
\end{subequations}
I have intentionally expressed this relation as some matrix $\PI$
operating on our data vector; it is inordinately handy
to use an \emph{operator} that extracts the weighted mean of
something, and have that operator be independent of the data itself
(it only depends on the noise properties of the data).
Note that $\PI$ is not in general symmetric.  Now, the covariance
matrix of our new data set $\y - \expec{\y}$ can be easily
calculated:
\beq{CorrectedCov}
\C \ \equiv \ (\y - \expec{\y})(\y - \expec{\y})^t
   \ = \ \PI\ \N\ \PI^t
   \ = \ \PI\ \N \ = \ \N\ \PI^t
\eeq
where the final two equalities are easily shown given the
definition of $\PI$ given in \eqn{PiEq}.
\footnote{\fixspacing It can also be shown that $\PI^2 \ = \
\PI$.}

It turns out that $\C$ is singular; this is a direct property
of the fact that we have lost sensitivity to the mean of $\y$,
which indirectly leads to a zero eigenvalue in $\C$ for this
technique.
However, for the purposes of a likelihood analysis, all we will
need is the ``pseudo-inverse'' of $\C$, which we denote as $\Cp$:
\beq{pseudoinverse}
\Cp \equiv \N^{-1} \, \PI = \PI^t \, \N^{-1} \quad .
\eeq
This cannot be the true inverse of $\C$, but it is true that $\PI \
[ \C \ \Cp - \I ] = {\bf 0}$, which means that $\Cp$ is the
inverse of $\C$ once you project out the unwanted mean \cite{teg97b}.

This method can be directly expanded to allow for elimination of
any unwanted modes in a data stream or map, such as a linear or
quadratic term.  In that case, we construct a $m \times n$
matrix, $\Z$, where $n$ is the number of data/map elements, and $m$ is the
number of modes.  Each column of $\Z$ contains the template for
that mode
\footnote{\fixspacing You need not normalize the modes; it will
happen automatically through this formalism.}
.  We then simply replace $\eO$ with $\Z$ in each of the
above equations, and everything still works.

There is one final interesting property of our new data vector
$\y - \expec{\y}$ and noise matrix $\C$ worth discussing.
When calculating the
likelihood function for some model, what you really care about is
the $\chi^2$ of the data, defined in general as $\y^t \ \N^{-1} \
\y$ for data vector $\y$ and covariance matrix $\N$.  After mode
removal, our new $\chi^2$ is given by:
\beqa{newchisq}
\chi^2 & = & (\y - \expec{\y})^t\ (\N^{-1} \PI) \ (\y -
\expec{\y}) \notag \\
 & = & \y^t \ \PI^t \ \N^{-1} \PI \ \PI \y \notag \\
 & = & \y^t \ \N^{-1} \ \PI \ \y  \notag \\
 & = & \y^t \ \Cp \ \y \quad .
\eeqa
The point here is that you only need to change the
covariance matrix to get the same $\chi^2$; you don't
need to mess with the data vector at all.  We can determine the
mean and subtract it off, but it doesn't matter --- any final
model predictions will be the same.  The same argument holds for
removal of multiple modes from the data.

\subsection{Marginalization and Constraint Matrices}
Our last discovery lends understanding to the BJK marginalization
technique for
removing unwanted modes (like an offset) from your data.  The
standard formulation of this technique is, again given data $\y$
and covariance matrix $\N$, one simply adds ``constraint matrices''
to the covariance matrix to \emph{remove sensitivity to unwanted
modes}
\cite{bjk98}:
\beq{constrmatrices}
\N_t \ = \ \N + \sigma_c^2 \ \Z \Z^t
\eeq
where $\Z$ is the template of the unwanted mode(s), and $\sigma_c^2$
represents the variance of the unknown amplitude of the modes.
Remember that in the case of removing a single offset, $\Z = \eO$.
One then takes $\sigma_c^2$ to be much larger than the instrument
noise, so these unwanted modes get zero weight, but not large
enough to cause matrix inversion problems.

Formally, you can also take the limit as $\sigma_c^2 \ \rightarrow \
\infty$.  In that case, $\N_t$ has an infinite eigenvalue, but its
inverse still exists and is given by
\beq{Ninv_marg}
\N_t^{-1} \ = \N^{-1} - (\N^{-1} \Z)[\Z^t \ \N^{-1} \Z]^{-1}
(\N^{-1} \ \Z)^t
\eeq
The astute reader will notice that this is \emph{identical} to
\eqn{pseudoinverse} (replacing $\eO$'s with $\Z$'s), which gave the pseudo-inverse of the
corrected covariance matrix in the technique that directly
subtracted off the mean (or other unwanted modes).
Thus, the trick is simply to add infinite noise to the unwanted
modes in your covariance matrix, and find the new \emph{inverse
covariance matrix}.

\section{From Submaps to Final Maps}\label{s:subfinal}
The application of the algorithms described in the previous
section for offset removal are quite straightforward, at least
to construct the ``de-offsetted'' submaps.  These are exactly how you
might anticipate them, each submap being now centered around $0$ in both
$Q$ and $U$, and the covariance matrix keeping track of
our information loss. An offset was subtracted in both $Q$
\emph{and} $U$, as these are independent variables, with no clear
systematic relationship between their offsets.  These offsets are
removed for each of the three IPC as well as QPC channels, as
described previously there was no clear relationship between the
offsets of the different channels that could be exploited.

\subsection{How to Combine Maps with Singular Inverse Covariance
Matrices}
We have done a great deal of data processing at this point, in our
march from TOD to final maps (see \fig{flowchart} for the
analysis pipeline), and we are almost there.  We
currently have a set of submaps and their corresponding
covariance matrices for each channel, a set for $Q$ and $U$ each.
With each of these sets, we shall combine the submaps
together into a single map and covariance matrix.  \sct{s:addmaps}
gives the machinery to perform this step:
we simply apply \eqn{addmaps} to our set of submaps
(for $Q$ and $U$ and for each channel, of course)
, with one complicating
factor -- the covariance matrices for each submap have singular
inverses!  That is OK, we know the inverse matrices from
\eqn{Ninv_marg}; however, we still have to perform a final inverse
to find the final noise covariance matrix, which from
\eqn{addmaps} is  $\left[ \sum_{i=1}^m \N_i^{-1} \right]^{-1}$.
We may hope that even though each $\N_i^{-1}$ is singular,
their sum might not be singular, but
our hopes will be dashed as we see the error messages flying
across our computer screen.

Do not fear.  What we have done is remove unwanted
modes from our maps; hence, the final map will also
have those unwanted modes removed, and will have infinite
eigenvalues in its covariance matrix -- \ie, its covariance matrix
does not formally exist.  But we don't care about these modes, and
they hold no information anyway, so we are free to set their
eigenvalues to whatever we want, as long as we take care to
remember what we have done in future processing.
If we denote the final covariance matrix as $\Sig$, such that
$\Sig^{-1} = \left[\sum_{i=1}^m \N_i^{-1} \right]$, then
the final covariance matrix is taken to be
\beq{final_cov}
\Sig \ = \ (\Sig^{-1} + \epsilon \Z \Z^t)^{-1} - \epsilon^{-1}
\Z \Z^t
\eeq
where $\epsilon$ is any small positive number.
In practice, it is
best to choose $\epsilon$ to be on the same order as the nonzero
eigenvalues of $\Sig^{-1}$.  The trick performed in
\eqn{final_cov} replaces the infinite eigenvalue(s) in $\Sig$
with zero eigenvalue(s); all the information in the other modes
of the matrix remains unchanged.
This will have consequences for any type of further analysis.
In performing any likelihood analysis one forms the full
covariance matrix ($\C$) by adding together the covariance matrices from
theory ($\Sc$) and data ($\Sig$):
\beq{full_cov}
\C \ = \ \Sc + \Sig \quad \text{,}
\eeq
but we have explicitly set the infinite eigenvalue of $\Sig$ to
zero, so it is quite possible that $\C$ will also have a zero
eigenvalue.  In any likelihood analysis we will have to invert $\C$,
which we can perform as follows:
\beq{full_cov_inv}
\C^{-1} \ = \ \lim_{\sigma^2 \rightarrow \infty}[ \Sc \ + \ \Sig \ +
\ \sigma^2 \ \Z \Z^t]^{-1} \ \ \text{.}
\eeq
All we're doing in this equation is adding back in the large
uncertainty to the modes which are supposed to have infinite
eigenvalues, which we had previously set to zero.  In practice, we
just take $\sigma^2$ to be much larger than any of the other
eigenvalues of $\Sig$, to ensure the unwanted modes get zero
weight.  Recall that in our simple case of offset removal, $\Z
\Z^t$ is just the matrix of all $1$'s, so all we're doing is
playing with the overall offset of the covariance matrix (which
itself is often inconsequential in the likelihood analysis anyway).

\subsection{A Brief Comment on Information Loss}
\begin{figure}[tb]
\begin{center}
\includegraphics[height=3.2in]{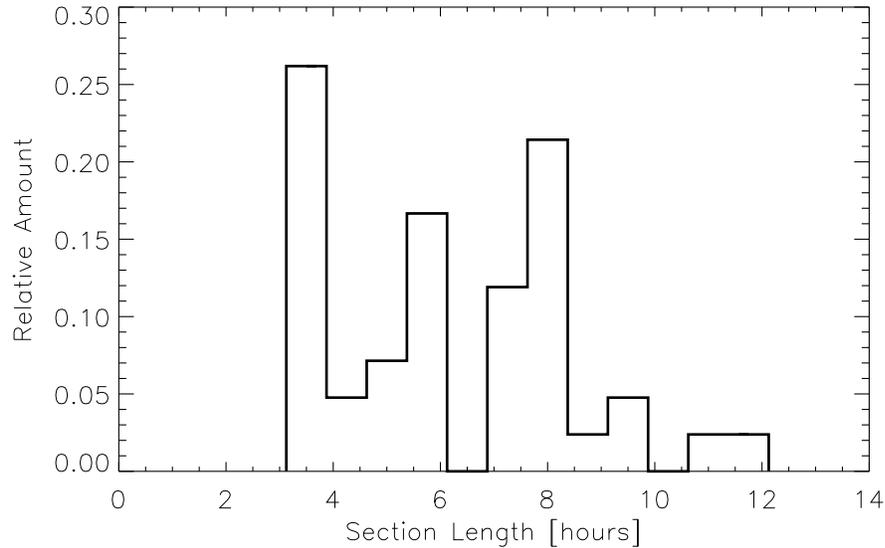}
\caption[Distribution of Surviving Section Lengths]
{\label{sldist} \fixspacing
Distribution of Surviving Section Lengths.
The distributions for each of the three channels have been
averaged, as they are slightly different due to the
different cuts each channel undergoes.  Clearly there are several
surviving sections with a mere 3-5 hours of data, resulting
in a significant information loss due to offset removal.
}
\end{center}
\end{figure}

Recall that, for our 7\deg\ beamsize, we
stare at one true 7\deg\ pixel roughly every 47 minutes.
The surviving data from each section varied in duration; we
required it to be at least 3 hours long, and the longest was about
8 hours long; the distribution of section lengths (averaged
between the three channels) is shown in \fig{sldist}.

For a short section containing, for instance, only four independent pixels,
we naively seem to
remove \app\ 25\% of its information when we remove its offset.
\footnote{\fixspacing As we will see in the next chapter, the
actual information loss can be even worse in the
a likelihood analysis, due to the
characteristics of the model we are trying to constrain.}
It is important to bear this penalty in mind.  If we instead had
further chopped up our data into smaller sections, the information
loss would have been correspondingly greater, and vice versa.
This emphasizes to the experimenter that having long, clean
sections of data is key to obtaining the best final noise
possible.

\subsection{Qualitative Analysis of Final Maps}

\begin{figure}[tb]
\begin{center}
\includegraphics[height=3.5in]{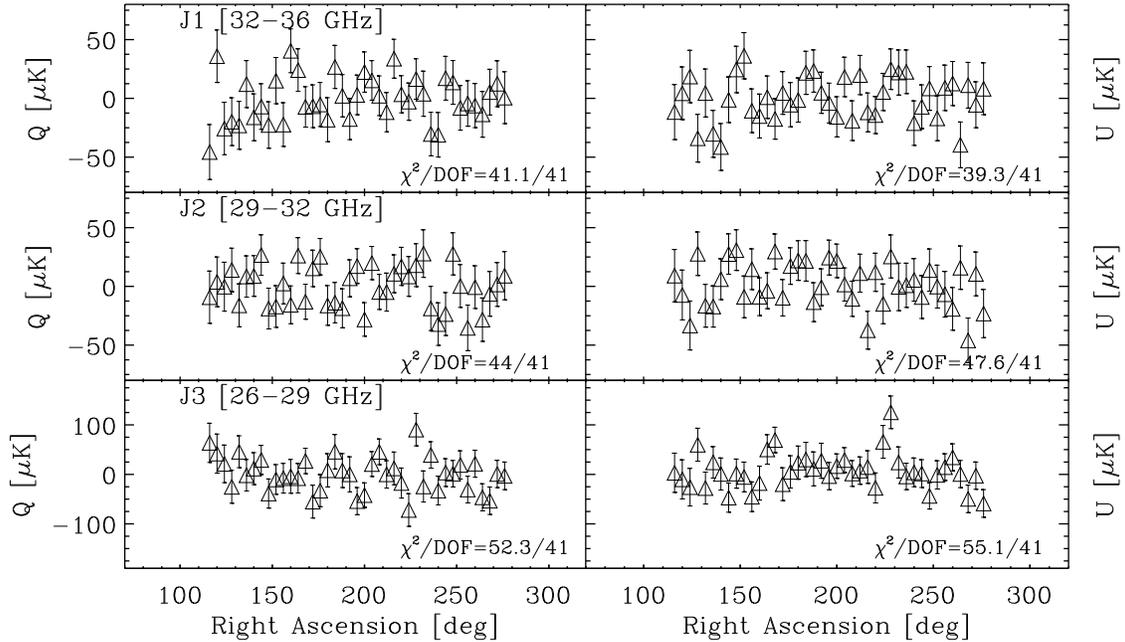}
\caption[Sky Maps for the IPC Channels]
{\label{maps_in} \fixspacing
The final maps for each in-phase channel, in both $Q$ and $U$.
The displayed error bars are simply the square-roots of the
diagonal elements of the covariance matrix.
The $\chi^2$ values displayed for each map include the off-diagonal
covariance for each map, using the full $\chi^2$-equation in the
case where the model is Gaussian noise distributed about zero,
 $\chi^2 \ = \ \y^t \N^{-1} \y$, where $\y$ is the map and $\N$ is
its noise covariance matrix.  Visually, there is no clear evidence
of signal correlated among the three channels.
}
\end{center}
\end{figure}

\begin{figure}[tb]
\begin{center}
\includegraphics[height=3.5in]{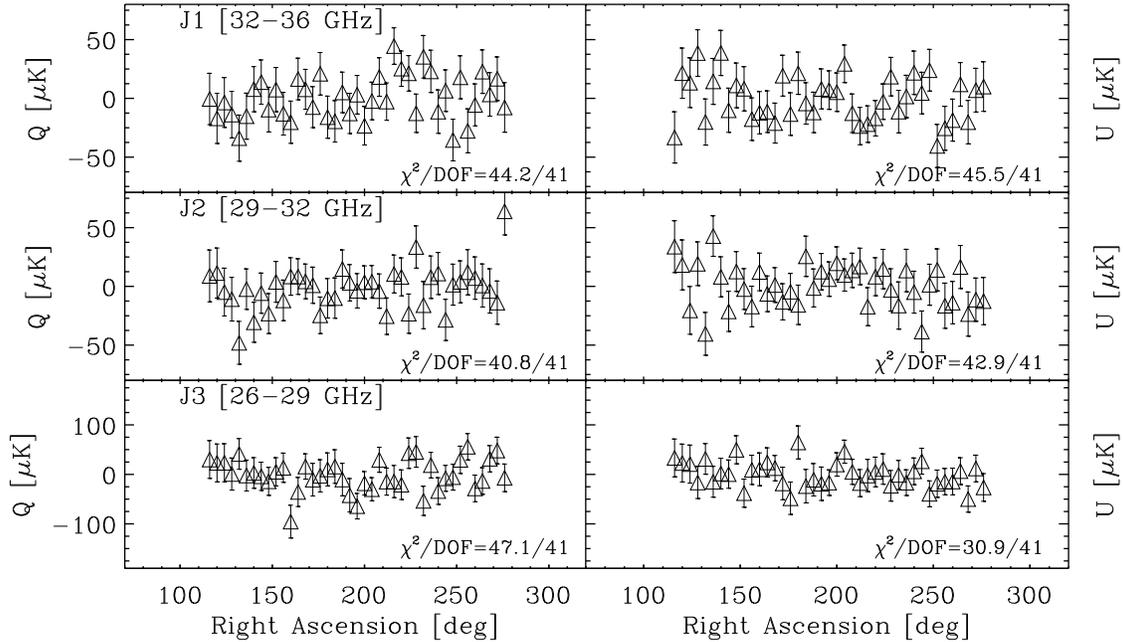}
\caption[Sky Maps for the QPC Channels]
{\label{maps_quad} \fixspacing
Same as \fig{maps_in}, but for the Quad-Phase Channels \{QPC\}.
No signal is expected in any of these maps, and clearly these maps
are all consistent with zero.  The QPC channels underwent
identical processing as the IPC channels, including offset
subtraction.
}
\end{center}
\end{figure}

The final maps for \polar\ are shown in \fig{maps_in}.
Qualitatively, there is not strong evidence of a common
signal among the three sub-bands, in either $Q$ or $U$.
The $\chi^2$ values from each map are also not consistent
with a statistically significant signal.  For comparison,
\fig{maps_quad} shows the corresponding maps made from the QPC
channels.  We do not expect these to contain signals either, but
rather they serve as useful litmus when viewing the IPC maps:
if the IPC maps
differ strongly from the QPC maps, that is evidence of either
signal or some type of contamination.  However, that is not the
case; none of the maps contain strong outliers, and
all exhibit $\chi^2$ values consistent with zero signal.

\subsection{Combining the Channel Maps}\label{s:finalmaps}
In order to perform our CMB analysis, we would like to find the
joint map of \emph{all the channels}, based on our three individual channel maps.
The reader can probably guess that we will employ the standard map
co-addition algorithm introduced in \sct{s:addmaps} to perform
this task.  However,
that algorithm assumes that the measurements made of each
individual map are \emph{independent}; if there was some
systematic effect that introduced correlations between the
measurements from different channels, then we would have less
information than we think we do, and we must take this into
account.

\subsubsection{Inter-Channel Cross-Correlation Coefficients}
\begin{figure}[tb]
\begin{center}
\subfigure[$\expec{J1 \ J2}$]{\label{todcc-a}
\includegraphics[width=2.8in]{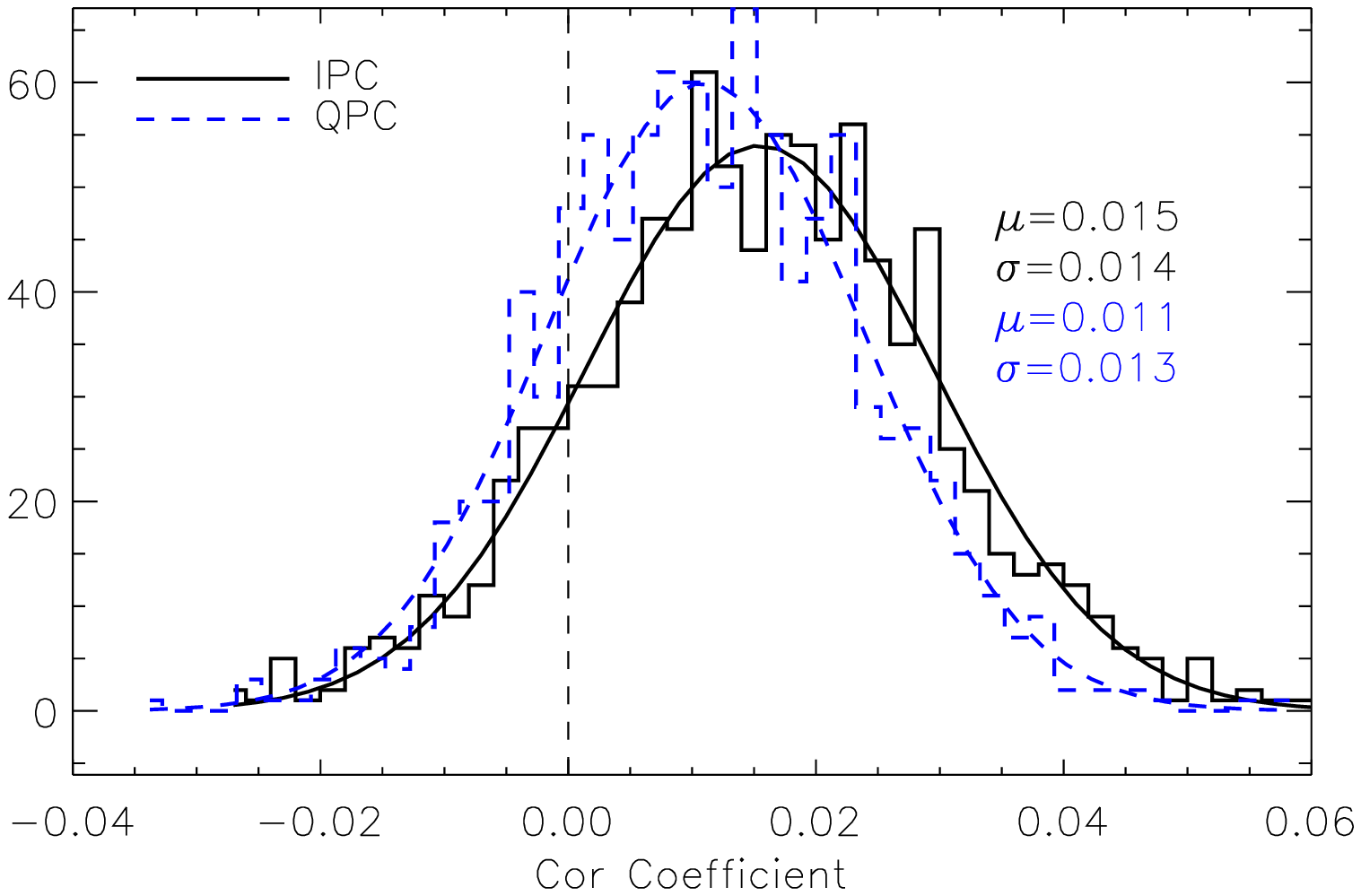}
}
\subfigure[$\expec{J1 \ J3}$]{\label{todcc-b}
\includegraphics[width=2.8in]{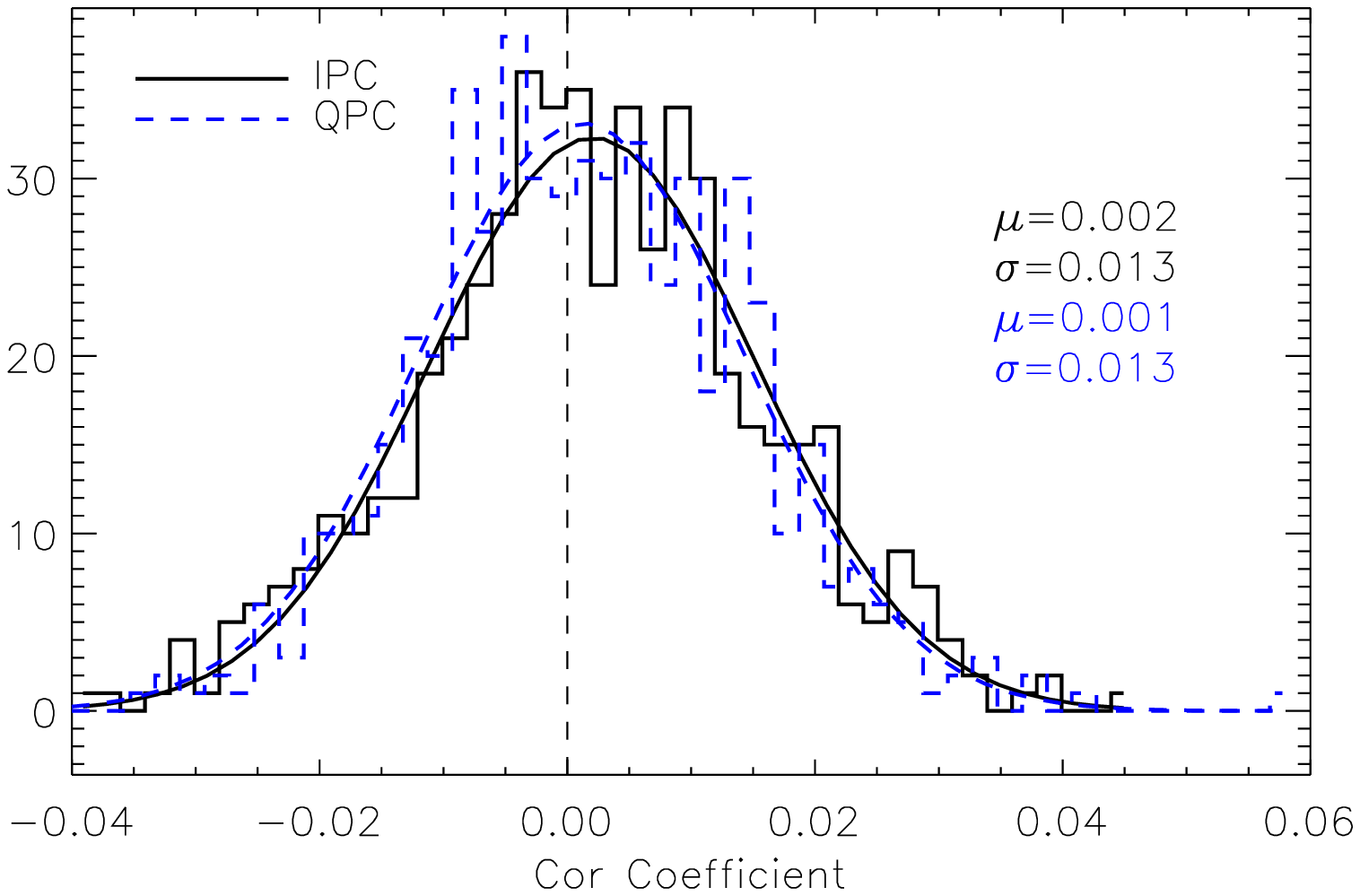}
 }\\
\subfigure[$\expec{J2 \ J3}$]{\label{todcc-c}
\includegraphics[width=2.8in]{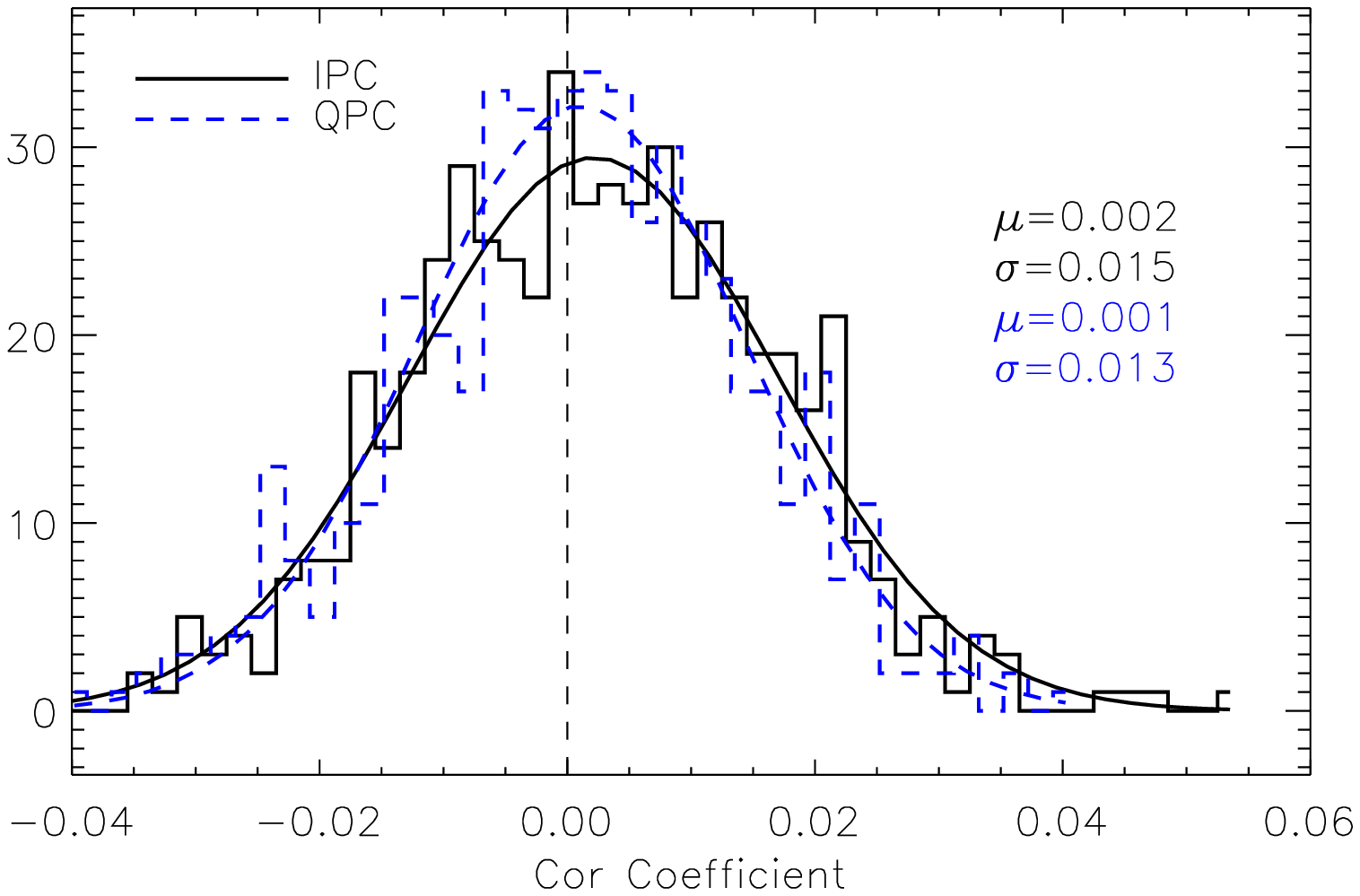}
}
\caption[Distribution of TOD Correlation Coefficients]
{\label{todcc}\fixspacing
Distribution of Correlation Coefficients between various
channels in the time-ordered data (TOD).  The correlations
were small, and the distributions were the same for both IPC
QPC inter-channel correlations, implying a common (electrical)
source.   This is most likely explained through the slight
overlap of the channel band passes.
}
\end{center}
\end{figure}

We can get a feel for the raw inter-channel cross-correlations by finding
the correlation coefficients between our three channels in the
time-ordered data.  \fig{todcc} shows a histogram of the
correlation coefficients between all three channels; one
correlation coefficient (Pearson's) was calculated for each
surviving file in the data set.  As you can see, the correlations in
the time stream are on the order of (or less than) 1\%.

However, it is not truly the time stream correlations that we so
much care about, it is the correlations between $Q$ or $U$ for the
channels.  For instance, if there were a 10\% correlation between
J1i-$Q$ and J2i-$Q$, it could be hidden in the smaller time stream
correlations.  We must therefore evaluate these correlations
directly.

\begin{table}
\begin{center}
\begin{tabular}{|c|c|c|c|c|} \hline
& $\mathbf{\expec{QQ}}$ & $\mathbf{\expec{UU}}$
& $\mathbf{\expec{QU}}$ & $\mathbf{\expec{UQ}}$ \\ \hline
$\mathbf{\expec{J1 \ J2}_{IPC}}$ & 0.144 $\pm$ 0.034 & 0.134
& -0.024 & 0.021 \\ \hline
$\mathbf{\expec{J1 \ J2}_{QPC}}$ & 0.005 $\pm$ 0.034 & 0.062
& -0.011 & -0.002 \\ \hline
$\mathbf{\expec{J1 \ J3}_{IPC}}$ & 0.074 $\pm$ 0.041 & 0.063
& -0.069 & -0.003 \\ \hline
$\mathbf{\expec{J1 \ J3}_{QPC}}$ & 0.023 $\pm$ 0.041 & 0.048
& -0.009 & -0.042 \\ \hline
$\mathbf{\expec{J2 \ J3}_{IPC}}$ & 0.104 $\pm$ 0.041 & 0.093
& -0.024 & -0.029 \\ \hline
$\mathbf{\expec{J2 \ J3}_{QPC}}$ & 0.066 $\pm$ 0.041 & 0.001
& -0.054 & -0.047 \\ \hline
\end{tabular}
\caption[ROD Inter-Channel Cross-Correlation Coefficients]
{\fixspacing \label{t:iccc}
ROD Inter-Channel Cross-Correlation Coefficients,
calculated from the surviving ROD data.
The errors are the same within each row, and
assume that the underlying distribution
of correlation coefficients is Gaussian.
}
\end{center}
\end{table}

In order to measure these correlations, we used the ROD
data set and found the Pearson's
correlation coefficient in
the same way as for the TOD, but because there is so much less
data, we found only one correlation coefficient for each surviving
section.  We then calculated means and errors by averaging from
the distribution of these sectional values.
\tbl{t:iccc} shows these values with their errors.
The numbers in this table are very suggestive.  For instance,
$\expec{Q \ Q}$ for all IPC channels is about the same as $\expec{U \
U}$, suggesting a common source.  All QPC correlation coefficients
are consistent with zero, as are all correlations of the
$\expec{Q \ U}$ variety
\footnote{\fixspacing
Except perhaps $\expec{Q_1 \ U_3}$, but because all the other
coefficients of this type are consistent with zero, we assume
it is an outlier.
}.
Luckily, as $Q$ and $U$ show no correlation between them
(either within a channel or between channels), we can
keep treating $Q$ and $U$ as completely independent measurements.
This is not too surprising, considering they are essentially
the $\sin{\twophi}$ and $\cos{\twophi}$ projections from each
rotation, which are orthogonal functions.  However, the
correlations between IPC channels (for the same Stokes parameter)
are \app\ 10\%, so we cannot ignore them in constructing a final
map.

\subsubsection{Combining non-independent maps : the machinery}
Now, how do we add together maps of $Q$ (or $U$) from the three
non-independent channels, armed with the knowledge of their mutual
correlations?
The algorithms in \sct{s:addmaps} did not deal with adding
non-independent maps together, but it is relatively easy to expand
them.  We will treat $Q$ and $U$ separately, as they are completely
uncorrelated.
Let us consider our situation for $Q$; $U$ will follow an identical format.
We have three channel
maps, call them $\q_1$, $\q_2$, and $\q_3$, with their
corresponding covariance matrices $\Sig_{Q1}$, $\Sig_{Q2}$, and
$\Sig_{Q3}$.  Let the correlation coefficient between $\q_i$ and
$\q_j$ be $\rho_{ij}$. We appeal to \mm\ to form
the best possible map.

First, we form our ``mega-map'', which is the three maps
concatenated together, and a corresponding ``mega-covariance
matrix'':
\begin{subequations}
\label{megamap}
\begin{eqnarray}
\q_{mega} & = & \{\q_1, \q_2, \ \q_3\} \\
\Sig_{mega} & = & \left[
\begin{array}{ccc}
\Sig_{Q1} & \rho_{12}\sqrt{\Sig_{Q1}}\sqrt{\Sig_{Q2}}
& \rho_{13}\sqrt{\Sig_{Q1}}\sqrt{\Sig_{Q3}} \\
\rho_{12}\sqrt{\Sig_{Q1}}\sqrt{\Sig_{Q2}} & \Sig_{Q2}
& \rho_{23}\sqrt{\Sig_{Q2}}\sqrt{\Sig_{Q3}} \\
\rho_{13}\sqrt{\Sig_{Q1}}\sqrt{\Sig_{Q3}} &
\rho_{23}\sqrt{\Sig_{Q2}}\sqrt{\Sig_{Q3}} &
\Sig_{Q3}
\end{array} \right]
\end{eqnarray}
\end{subequations}
We can take the square roots of the $\Sig$-matrices since they are all
positive definite, as long as we add a large offset to each matrix
(corresponding to the uncertainty in the offset, which is formally
infinite)
\footnote{\fixspacing
For a symmetric, positive-definite, $n\times n$ matrix $\M$, its
square root is given by $\PP^t \D^{1/2} \PP$, where $\PP$  is an $n
\times n$ matrix such that the $i^{\text{th}}$ row of $\PP$ contains the
$i^{\text{th}}$ eigenvector of $\M$, and $\D^{1/2}$ is a diagonal
matrix with the square-roots of the eigenvalues of $\M$ along its
diagonal. The eigenvectors must be normalized, such that $\PP \PP^t = \I$.}
.  The final full covariance matrix, $\Sig_q$, will then
also have a large offset, but because of the arguments already
discussed, this will not affect the final CMB likelihood analysis.
We simply must remember that the very large eigenvalue of the final
map covariance matrix represents our infinite uncertainty in the
overall map offset.

Now that we have re-expressed our individual maps in the
``mega-map'' and ``mega-covariance matrix'' format, we apply \mm.  Our
pointing matrix is given by
\beq{jointpointingmatrix}
\A_{mega} = \left[ \begin{array}{c} \I_n \\ \I_n \\ \I_n
\end{array}\right] \ .
\eeq
This points our three individual maps to the same final map; each
$\I$ is the $n\times n$ identity matrix, where $n$ is the
number of pixels in our maps.
Explicitly, the final joint map and covariance matrix are
given by:
\begin{subequations}
\label{addchannelmaps}
\begin{eqnarray}
\q & = & \Sig_q \
\A_{mega}^t \Sig_{mega}^{-1} \q_{mega} \\
\Sig_q & = & [\A_{mega}^t \ \Sig_{mega}^{-1} \A_{mega}]^{-1} \quad
\text{.}
\end{eqnarray}
\end{subequations}
Because of the large offset each covariance matrix possesses,
the final map $\q$ may have some random offset to it, but
it is meaningless, and can be safely subtracted out.

\begin{figure}
\begin{center}
\subfigure[]{\label{finalmap_in}
\includegraphics[width=5in]{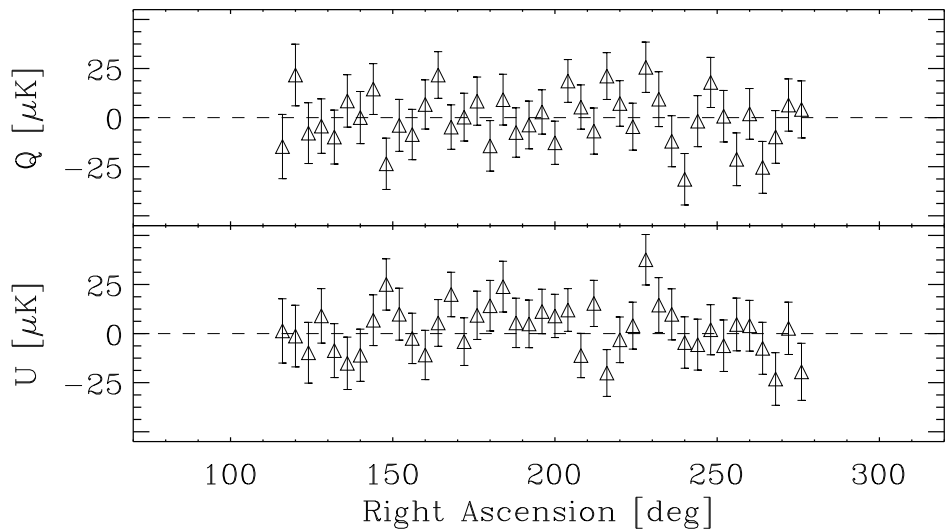}
} \\
\subfigure[]{\label{finalmap_quad}
\includegraphics[width=5in]{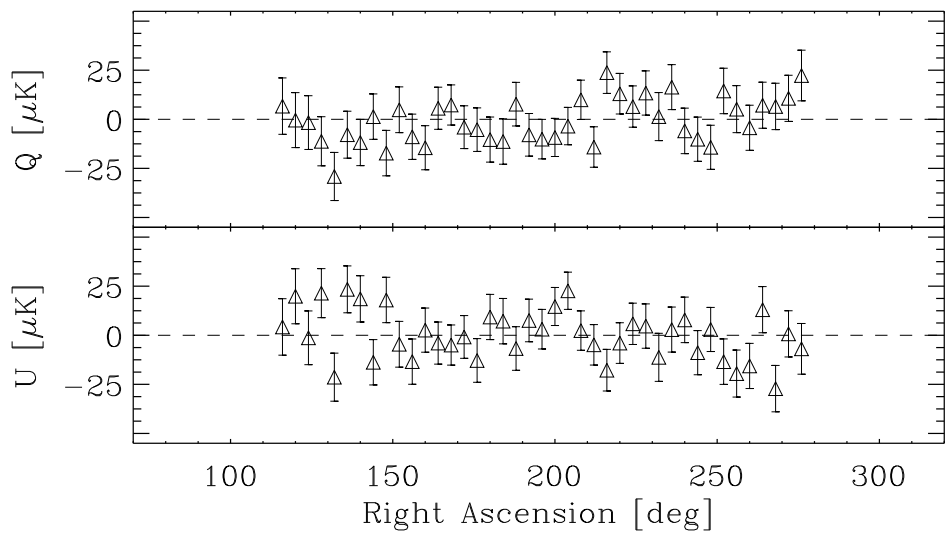}
}
\caption[Final Joint-Channel Sky Maps]
{\label{finalmap}\fixspacing
Final joint-channel sky maps.  The joint IPC maps for $Q$ and $U$
are displayed in (a), while (b) shows the QPC maps.
}
\end{center}
\end{figure}

The final joint-maps for the IPC and QPC are shown in
\fig{finalmap}.  Again, we see no evidence of an underlying sky
signal.  Instead of calculating $\chi^2$'s and the like for these
maps, in order to determine if they show evidence of a sky signal,
the route along which we will proceed is the full \emph{Likelihood
Analysis} in order to constrain a specific model of CMB
polarization.  This process is described fully in the next
chapter.

\section{A Simple Simulation}
The number of steps that the data passed through in
the route from time-ordered data to final maps was large, and
several of steps were not exactly simple to understand.  The
chance for errors in at least one of these steps is high, so we
found it very useful to generate model data streams for
which we knew the underlying signal, and run these ``Fake \polar\
Signals'' through the full machinery of the data pipeline, to
check for errors and make sure that everything worked as
anticipated.

\subsection{Parameters of the Simulation}
There were three primary steps involved in the simulation process: build the
underlying map, let \polar\ ``observe'' and generate data based on
these ``observations'', and then run this data through the full
mapmaking analysis, from $TOD \rightarrow Maps$.

\subsubsection{Underlying Maps}
We built the underlying sky maps out of simple sine and cosine
modes.  We assumed a basic flat band-power model with \app\ 10
\uK\ per band, from $\ell \sim 0 \ to \ 100$, and then convolved
these signals with a 7\deg\ Gaussian beam.
We assumed the same underlying map for
all three channels (thus, only CMB, no foregrounds).  We did not
need a complicated map because this procedure is not to test
foreground removal or parameter extraction, but rather to ensure
our map reconstruction algorithms performed well.

\subsubsection{Simulated POLAR Observations}
\begin{figure}[tb]
\begin{center}
\includegraphics[height=3.2in]{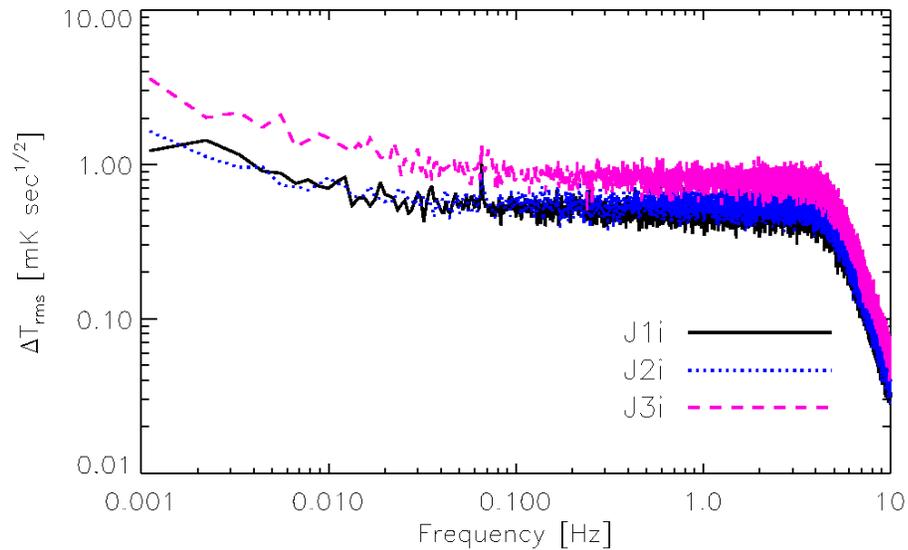}
\caption[Power Spectra of Simulated Correlator Channels]
{\label{simpsd} \fixspacing
Power spectra of simulated IPC data.  Notice the presence of \emph{1/f}
noise, our 5 Hz anti-aliasing low-pass filter, and also a
small peak at the \twophi\ frequency.  The noise-equivalent
temperatures for these channels are somewhat better than for
actual \polar, in order to speed up the processing time (less
data were required to reach the same signal level).  Otherwise,
these power spectra qualitatively match those of \polar\ (when
the weather was good) quite well.
}
\end{center}
\end{figure}

To shorten processing time, we made \emph{the simulated \polar}
about a factor of two
more sensitive than the real \polar,
with a full sensitivity of $340 \uK \sqrt{sec}$.
We included all IPC and QPC channels in the analysis, but no total
power channels.  We assumed the noise was almost white, with
a small amount of 1/f noise in each channel.  We convolved
each data stream with our 5 Hz anti-aliasing filter.  The
resulting power spectra from the three IPC channels are shown in
\fig{simpsd}; notice the strong similarity to \polars\ true power
spectra, such as shown in \fig{powerspectra}.

We included the effect of the ``AOE'' (absolute one-bit encoder),
which fired every time \polar\ completed a rotation;
the analysis techniques require this data in order to
form the \emph{ROD} data.  We even included the short gaps
in data taking that occurred after each 7.5 minute data file was
written (these gaps were 2-5 seconds long), just in case this might
have led to an observable effect in the maps.
We also added random offsets in $I_0$, $Q$, and $U$ for each section
and channel, of levels consistent with those experienced by \polar.
We generated five sections of data, with a total of 31 hours of
observation time.  The true \polar\ data set had 49 sections of
data, of which about 15 contributed to the final analysis
(about 100 hours of data).  Additionally,
we made the sky signals much larger than theory expects,
so the simulation could be run in a reasonable amount of time.
The simulation parameters for each channel are shown in \tbl{simparams}.

\begin{table}
\begin{center}
\begin{tabular}{|c|cccccc|} \hline
Channel & NET [$mK \ s^{1/2}$] & \emph{1/f} Knee & 5 Hz Filter &
$I_0$ [mK] & $Q_{off}$ [\uK] & $U_{off}$ [\uK] \\
\hline
J1i & 0.50 & 0.01 & Yes & $3r$ & $30r$ & $20r$ \\
J2i & 0.55 & 0.005 & Yes & $10r$ & $60r$ & $40r$ \\
J3i & 0.80 & 0.02 & Yes & $50r$ & $90r$ & $60r$ \\
J1o & 0.50 & 0.01 & Yes & $0.03r$ & $0$ & $0$ \\
J2o & 0.55 & 0.005 & Yes & $0.1r$ & $0$ & $0$ \\
J3o & 0.80 & 0.02 & Yes & $0.5r$ & $0$ & $0$ \\ \hline
\end{tabular}
\caption[Simulation Parameters for Each Correlation Channel]
{\fixspacing \label{simparams}
Simulation parameters for each correlator channel.
The ``$r$'' in several columns indicates a random number between
$-1$ and $1$, drawn from a flat distribution.
Thus, for example, the $Q$ offsets for
channel J1i were between $-20$--$20$ \uK, and were different for each
section.  Five sections were generated with a total of 31 hours of
data.
}
\end{center}
\end{table}

\subsubsection{Simulation Map Reconstruction}
\begin{figure}[tb]
\begin{center}
\includegraphics[height=5.5in]{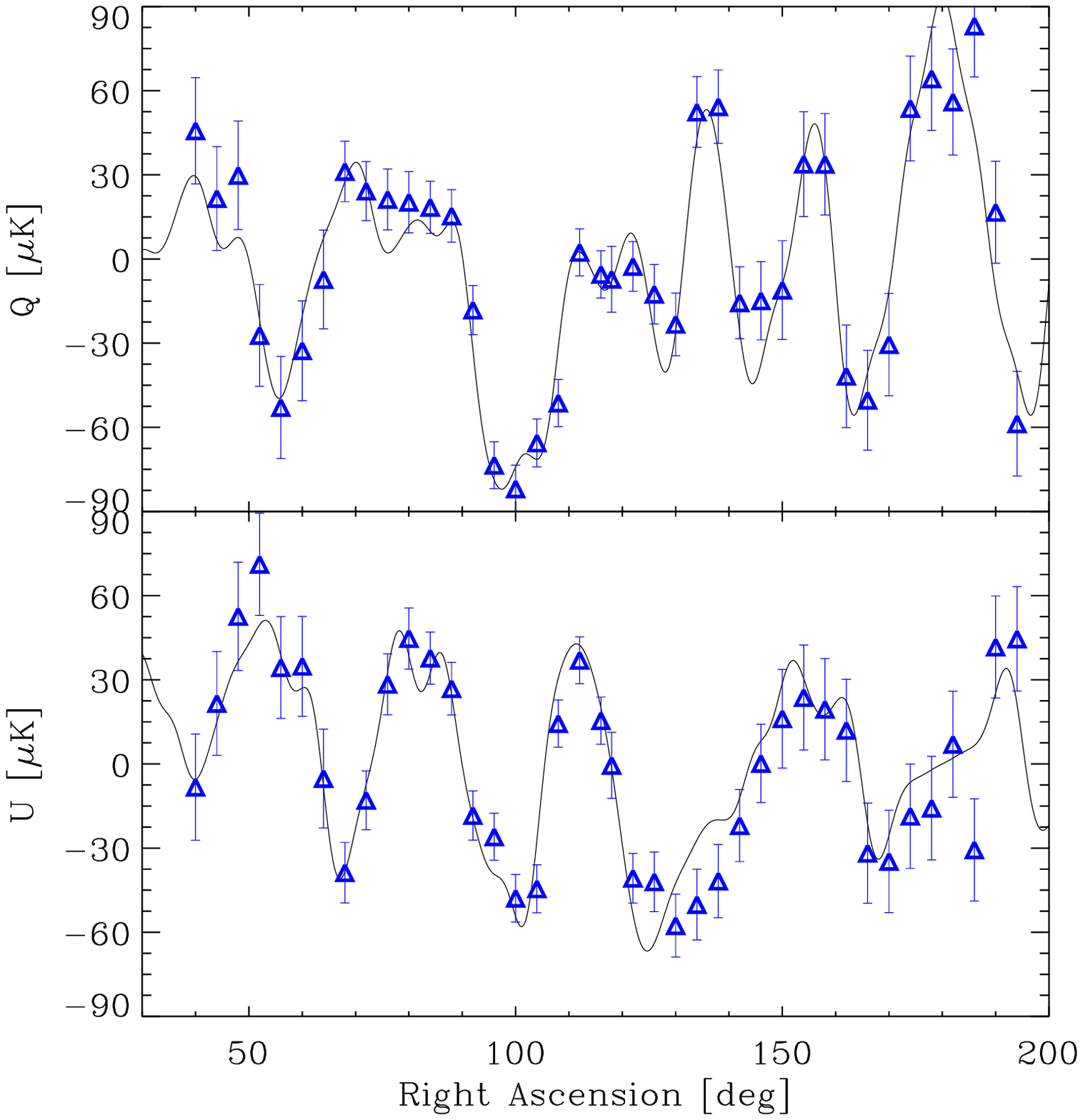}
\caption[Maps of Simulated Data]
{\label{simmap} \fixspacing
Derived joint-channel maps for simulated data, for the IPC
channels, compared to the underlying sky map.  The thin (black)
curve is the underlying sky map convolved with a 7\deg\ beam.  The
(blue) data points represent the derived joint-channel IPC map, after the
31 hours of simulated data was run through the entire \polar\
analysis procedure.  The fairly high signal-to-noise ratio is due
to the excellent noise figures we assumed for the simulated
experiment.  The signal values are roughly consistent with those
of temperature anisotropy levels.  The agreement between
the underlying maps and those derived from the simulated data
is a strong indication of the robustness of our mapmaking
algorithms.
}
\end{center}
\end{figure}

We next ran the fake \polar\ signals through the full analysis
pipeline, including removing an offset for each section, for all
channels, in both $Q$ and $U$ (just as in the real analysis).
The derived joint-channel maps are shown in \fig{simmap},
along with the underlying sky maps for comparison.  The performance
of the analysis software was good; the derived maps match
the initial maps well, up to an overall offset (and in some cases
some other large-scale modes, like a linear term).  Recall
that we have removed our sensitivity to the largest-scale modes, so
these features showing up in our final map will not affect a
likelihood analysis. The QPC maps [not shown] were consistent with
noise.

These results lead to a fairly big sigh of relief.
It is important to simulate or Monte-Carlo an experiment in
order to test the analysis pipeline.
Besides being an invaluable debugging tool, it also lends
significant credence to the results of the experiment.  Even
though our experiment produced only upper limits on possible
signals, it is still comforting to know that the analysis routines
were in principle capable of detecting
and mapping signals correctly, had they been present.



\newcommand{\ri}{\hat{\vect{r_i}}}
\newcommand{\rj}{\hat{\vect{r_j}}}
\newcommand{\aij}{\alpha_{ij}}
\newcommand{\Rij}{\R(\aij)}
\newcommand{\Mij}{\M(\ri \cdot \rj)}
\newcommand{\Fa}{F^{12}_\l}
\newcommand{\Fb}{F^{22}_\l}
\newcommand{\F}{\mathbf{F_\l}}

\chapter{Discussion \& Likelihood Analysis}\label{likes}

Now that the maps of $Q$ and $U$ have been generated, we seek
to answer questions about the level of CMB polarization our data
can constrain, either through a detection or new upper limit.
Some loftier goals are to constrain the power spectra of polarization
fluctuations, the optical depth to reionization, or even the
polarization of synchrotron radiation.  We shall begin, naturally,
with the first, and answer questions about a simple flat band
power model CMB polarization.

\section{Introduction to Bayesian Likelihood Analysis}
We employ a Bayesian maximum likelihood analysis
\footnote{\fixspacing In contrast to a ``frequentist'' approach,
a Bayesian analysis simply finds the best-fitting set of
parameters for a given model, assuming the basic model is correct.
However, it could be the case that the model itself is not an
accurate description of the data, but a Bayesian approach cannot
answer questions of the overall goodness of a model; this is the
domain of the frequentist approach, described for example in
\cite{teg01a}.} in order to characterize the level of possible CMB
polarization fluctuations present in our data.  As we measured $Q$
and $U$ simultaneously, we are able to set limits on both E- and
B-type polarization independently.

The flow of a standard likelihood analysis is as follows.
A model is constructed which depends on some set of parameters
$\a = \{a_1 \ldots a_m\}$.  We wish to constrain these parameters
given the data $\x$.  Unfortunately, all we can calculate is the
\emph{probability distribution of the data given the parameters},
which we denote as $P(\x| \a)$.  This is where Bayes' theorem
comes into play, which states that
\beq{bayesthm}
P(\a| \x) \ \propto \ P(\x | \a) \ P(\a)
\eeq
In words, the probability of the model given the data equals
the probability of the data given the model times the prior
probability of the model.  The probability distribution $P(\a)$ is assumed
to be uniform (uninformative).  The probability distribution $P(\x|
\a)$ is given by the \emph{likelihood function},
\beq{like}
\L(\a) \ = \ \frac{1}{{2\pi}^{N/2}} \frac{e^{-\frac{1}{2}
\x^t\C(\a)\x}}{|\C|^{1/2}}
\eeq
where N is the number of pixels in the map $\x$, and $\C(\a)$ is
the full covariance matrix, given by the sum of data and theory
covariance matrices:
\beq{fullcov}
\C \ = \ \Sc \ + \Sig \quad .
\eeq
Note that $\C$ and $\Sc$ both inherently depend upon the parameters $\a$.
The likelihood function is simply denoted $\L(\a)$, rather than say $\L(\x|
\a)$, because our data is not changeable at this point, only the
model is, so any reference to the data varying has been dropped.
In all cases, the data vector is given by
$\x \equiv \{\q,\ \u\}$, where $\q$ and $\u$ are the $Q$ and $U$
joint-channel maps, respectively, as constructed in \sct{s:finalmaps}.
Each contains $N=84$ pixels of 2\deg\ width in right ascension on the sky at
$\delta=43\deg$.  The data covariance matrix $\Sig$ is formed from the
covariance matrices for the joint-channel maps as calculated in
\sct{s:finalmaps} such that
\begin{eqnarray}\label{datacov}
\Sig \ = \ \left[
\begin{array}{cc}
\Sig_Q & 0 \\
0 & \Sig_U
\end{array} \right] \quad .
\end{eqnarray}

At this point we find the maximum of the likelihood function in
the parameter space $\a$, denoted $\a_M$; this point determines the maximally
likely set of $\a$.  In order to determine the error bars on these
parameters, we find the \emph{Bayesian credible region} in
parameter space about $\a_M$, a region of volume $V$ bounded by a
surface of constant $\L$ such that
\beq{Like_errors}
\int_V \L(\a) d\a \ = \ c \int \L(\a) d\a \quad ,
\eeq
where the integral on the right is an integral over all of
$\a$-space, and $c$ is the one minus the desired level of
confidence for the region $V$.  Then we say that the true set of
parameters $\a_{true}$ lives somewhere in the region $V$ with a
level of confidence $c$.


\section{Limits on E and B in a Flat Band-Power
Model}\label{s:fbp}
We will set limits on the E- and B-type polarizations of the CMB by
assuming a flat band-power model. This model has only two free
parameters, $T_E$ and $T_B$, and the power spectra are given by
\beq{flatmodel}
\l(\l+1) \Cl^X/2\pi=T^2_X
\eeq
where $X \in \{E,B\}$.  $T_E$ and $T_B$ then correspond to the
RMS-level of fluctuations in E- and B-mode polarization,
respectively.

\subsection{Constructing the Theory Covariance Matrix}
The construction of $\Sc$ is fairly involved; its derivation is
introduced in \cite{matias98}, and given in great detail in
\cite{teg01a}.
$\Sc$ is a $2N \times 2N$ matrix, which we will consider as an
$N\times N$ matrix whose elements are $2\times 2$ matrices.
The $2\times 2$ elemental matrix $\Sc_{ij}$ describes the covariance between two pixels
$i$ and $j$, and is given by
\beq{Sc_ij}
\Sc_{ij} \ =\ \expec{\x_i \x_j} = \Rij \Mij \Rij^t
\eeq
where $\ri$ and $\rj$ are the unit vectors pointing to pixels $i$
and $j$, respectively. $\Rij$ is a standard rotation matrix which
rotates the $(Q,U)$ components in $\M$ into the global coordinate
system where the reference frame for $Q$ and $U$ is given by the
local meridian
\footnote{\fixspacing The \emph{meridian} of a point on the celestial sphere
is the great circle passing through that pixel as well as the
celestial poles.}; it is given by
\beqa{R_ij}
\R({\bf \alpha}) \ = \ \left( \begin{array}{cc}
\cos{2\alpha} & \sin{2\alpha} \\
-\sin{2\alpha} & \cos{2\alpha}
\end{array} \right)
\eeqa
The covariance matrix $\M$ depends only on the angular
separation between the pixels $i$ and $j$.
$\M$ is most naturally expressed in a coordinate system such that
the great circle connecting pixels $i$ and $j$ serves as the
reference axis for the Stokes parameters \cite{kks}.  Expressed in
this coordinate frame, $\M$ becomes
\cite{matias98}
\beqa{Mij}
\Mij \equiv \left( \begin{array}{cc}
\expec{Q_i Q_j} & 0 \\
0 & \expec{U_i U_j}
\end{array} \right) \quad ,
\eeqa
\begin{subequations}
\label{QUij}
\begin{eqnarray}
\expec{Q_i Q_j} & \equiv & \sum_\l
\left(\frac{2\l+1}{4\pi}\right) B_\l^2
\ [\Fa(z)\Cl^E - \Fb(z) \Cl^B] \\
\expec{U_i U_j} & \equiv & \sum_\l \left(\frac{2\l+1}{4\pi}\right)
B_\l^2
\ [\Fa(z)\Cl^B - \Fb(z) \Cl^E] \quad
\end{eqnarray}
\end{subequations}
where $z=\ri\cdot\rj$ is the cosine of the angle between the
two pixels under consideration,
$B_\ell=\exp[-\ell(\ell+1)\sigma^2_B/2]$, $\sigma_B$ is the beam
dispersion $=0.425\times\,$FWHM, and $\Fa$,$\Fb$ are functions
of Legendre polynomials as defined in \cite{matias98,teg01a}.
The full matrix $\Sc$ is then constructed by looping over all
necessary pixel pairs.  In principle, $\Sc$ needs to be
calculated for all ($T_E$,$T_B$) combinations in order to fully
assess the likelihood function.  However, there is a great
simplification lurking here.
Recalling that for the flat band-power model,
$\l(\l+1)\Cl^X/2\pi=T^2_X$ with $X \in \{E,B\}$,
\eqn{Sc_ij} can be
recast as
\beq{Ssimp}
\Sc_{ij} \ = \ T_E^2 \ \Sc_{ij}^E \ + \ T_B^2 \ \Sc_{ij}^B \quad ,
\eeq
\beq{scijx}
\Sc_{ij}^X \ \equiv \ \Rij \sum_\l \frac{2\l+1}{2\pi \l(\l+1)}B_\l^2
\F^X(z) \Rij^t \cdot (1 \ \uK^2)
\eeq
where
\beq{Fdef}
\F^E(z) \equiv \mmb{\Fa(z)}{0}{0}{-\Fb(z)}
\quad \text{and} \quad
\F^B(z) \equiv \mmb{-\Fb(z)}{0}{0}{\Fa(z)}
\eeq
The theory covariance matrix $\Sc$ in the flat band-power
model is then given by
\beq{fbp1}
\Sc(T_E,T_B) \ = \ T_E^2 \ \Sc^E \ + T_B^2 \ \Sc^B
\eeq
where $\Sc^E$ is the \emph{fundamental}
theory covariance matrix for purely E-modes,
comprised of all the 2$\times$2 $\Sc^E_{ij}$ matrices,
and similarly for $\Sc^B$.  Now we must must merely calculate
$\Sc^E$ and $\Sc^B$ once each, and we can then evaluate the full
theory covariance matrix for any ($T_E$,$T_B$) pair we like using
\eqn{fbp1}.

\begin{figure}[tb]
\begin{center}
\includegraphics[height=3in]{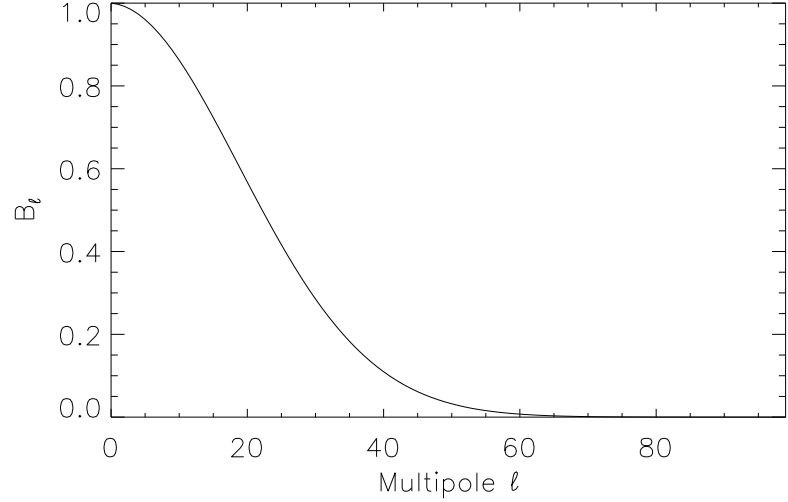}
\caption[Gaussian Beam Function $B_\l$]
{\label{beamfunc} \fixspacing
Gaussian beam function $B_\l$ for \polars\ 7\deg\ FWHM beam.
Our main sensitivity in $\l$-space decreases rapidly after $\l
\app 20$.  Originally calculated in \cite{bkthesis}.
}
\end{center}
\end{figure}
\begin{figure}
\begin{center}
\includegraphics[height=3.7in]{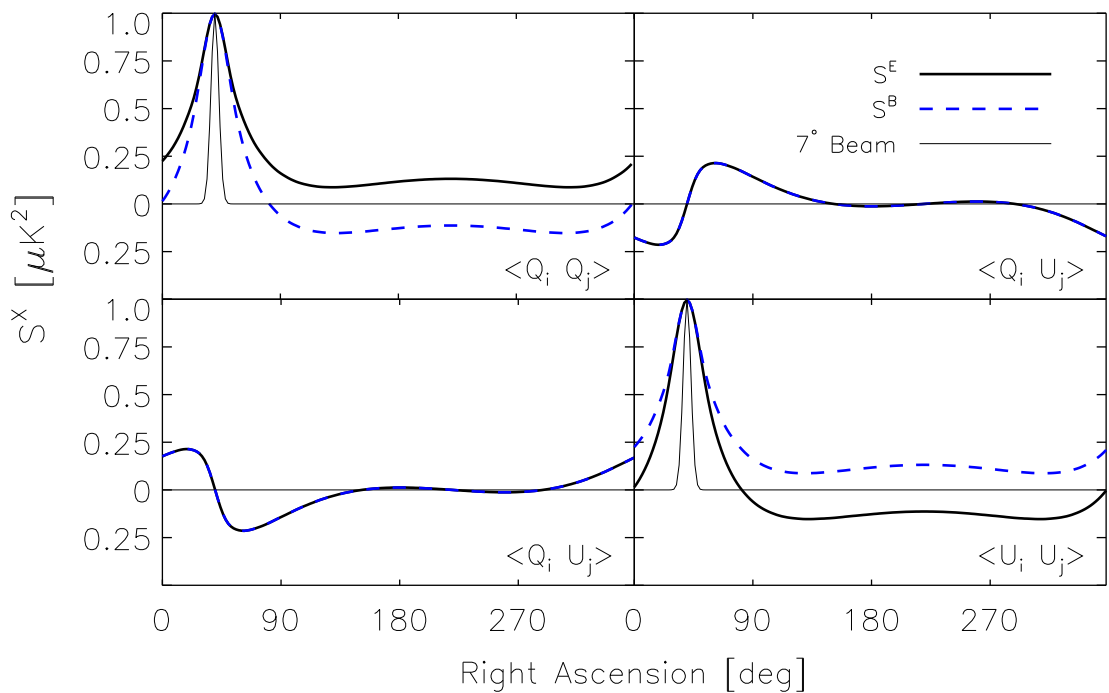}
\caption[Signal Covariance Matrix in Flat Band-Power Model]
{\label{sigrows} \fixspacing
Plots of the ra = 40\deg\ row of the fundamental signal covariance
matrices $\Sc^E$ and $\Sc^B$.  The four panels in the plot
correspond to the four quadrants of the matrices as labelled.  The
thick solid line represents $\Sc^E$, the thick dashed line
represents $\Sc^B$, and the thin solid line corresponds to a
7\deg\ Gaussian beam for reference.
}
\end{center}
\end{figure}

The signal covariance matrices were calculated for our 2\deg\
pixelization spanning all 360\deg\ of right ascension, using as an
approximation to our beam a single 7\deg\ FWHM Gaussian, in order
to calculate our beam function $B_\l$ for use in \eqns{QUij}.
This beam function is essentially our window
function for the flat band-power model, and is shown in \fig{beamfunc}.
A representative row from both fundamental signal covariance
matrices, $\Sc^E$ and $\Sc^B$, is shown in \fig{sigrows}
\footnote{\fixspacing This covariance matrix was calculated by Angelica de Oliveira-Costa
and Max Tegmark.}.
This shows the covariance between the pixel at right ascension
40\deg\ and all the other pixels.  Notice how wide the main peak
in the distribution is; the width of this peak (\app\ 20\deg)
corresponds roughly to the mean $\l$-value we are probing with
this model, which is around an $\l$ of 10.  Our 7\deg\ beam is
shown for comparison.  Also witness the symmetry between E and B;
the $\expec{U_i U_j}$ portion of $\Sc^E$ is the same as the
$\expec{Q_i Q_j}$ portion of $Sc^B$, and vice-versa.  All of the
$\expec{Q_i U_j}$ pieces are identical.  This shows the general
behavior of the matrix, as all other rows are identical to this
row, except shifted such that the peak lies over the pixel in
question.

\subsection{Evaluating the Likelihood Function}
Now that we have constructed the theory covariance matrix for our
parameter space of ($T_E, T_B$), we are in a position to calculate
the likelihood of our data given this set of models, as defined in
\eqn{like}.

There are two basic calculations that we must perform in order to
evaluate the likelihood $\L$ for any set of ($T_E, T_B$): the
exponent factor $\x^t\C\x$, and the square root of the determinant
of $\C$.  Appendix \ref{ch:cholesky} describes how to do this
was performed, using the Cholesky decomposition.

\begin{figure}
\begin{center}
\includegraphics[width=6in]{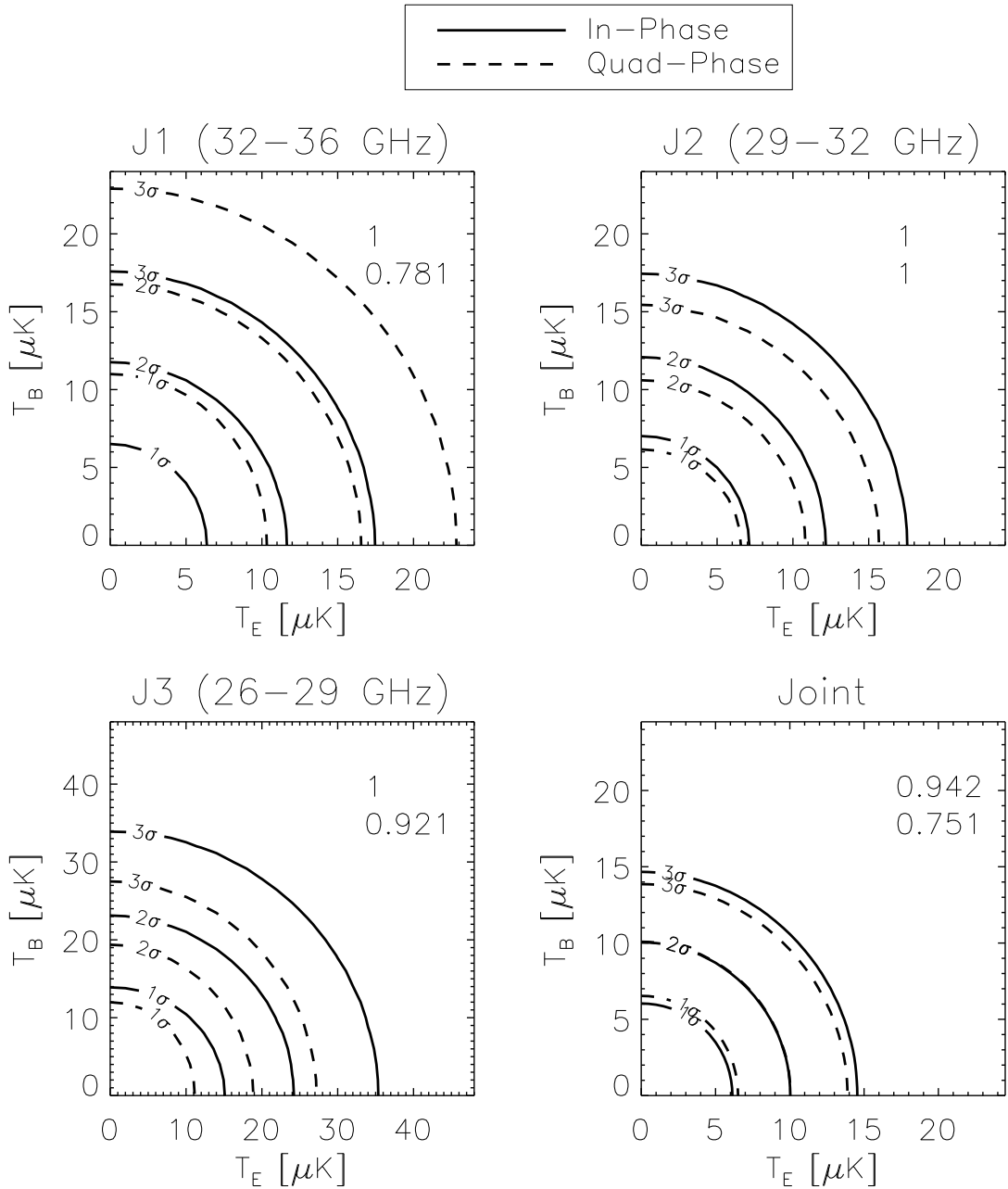}
\caption[Normalized Likelihood Contour Plot of $T_E$ and $T_B$
in Flat Band-Power Model]
{\label{likefbp} \fixspacing
Normalized likelihood
contour plots in the $T_E-T_B$ plane. The contours enclose
are $68\%$, $95.4\%$, and $99.7\%$ of the total probability,
corresponding to $1, 2,$ and $3$
standard deviation intervals, as labelled. The solid lines
correspond to the likelihood for the IPC
frequency channels, and the dashed lines are the corresponding
null-channel (QPC) likelihood.  The numbers in the upper right
corners are $\L(0)$, the upper (lower) number corresponds to the
IPC (QPC) channel.  All likelihoods are consistent with non-detections.}
\end{center}
\end{figure}

\begin{figure}[tb]
\begin{center}
\includegraphics[width=5in]{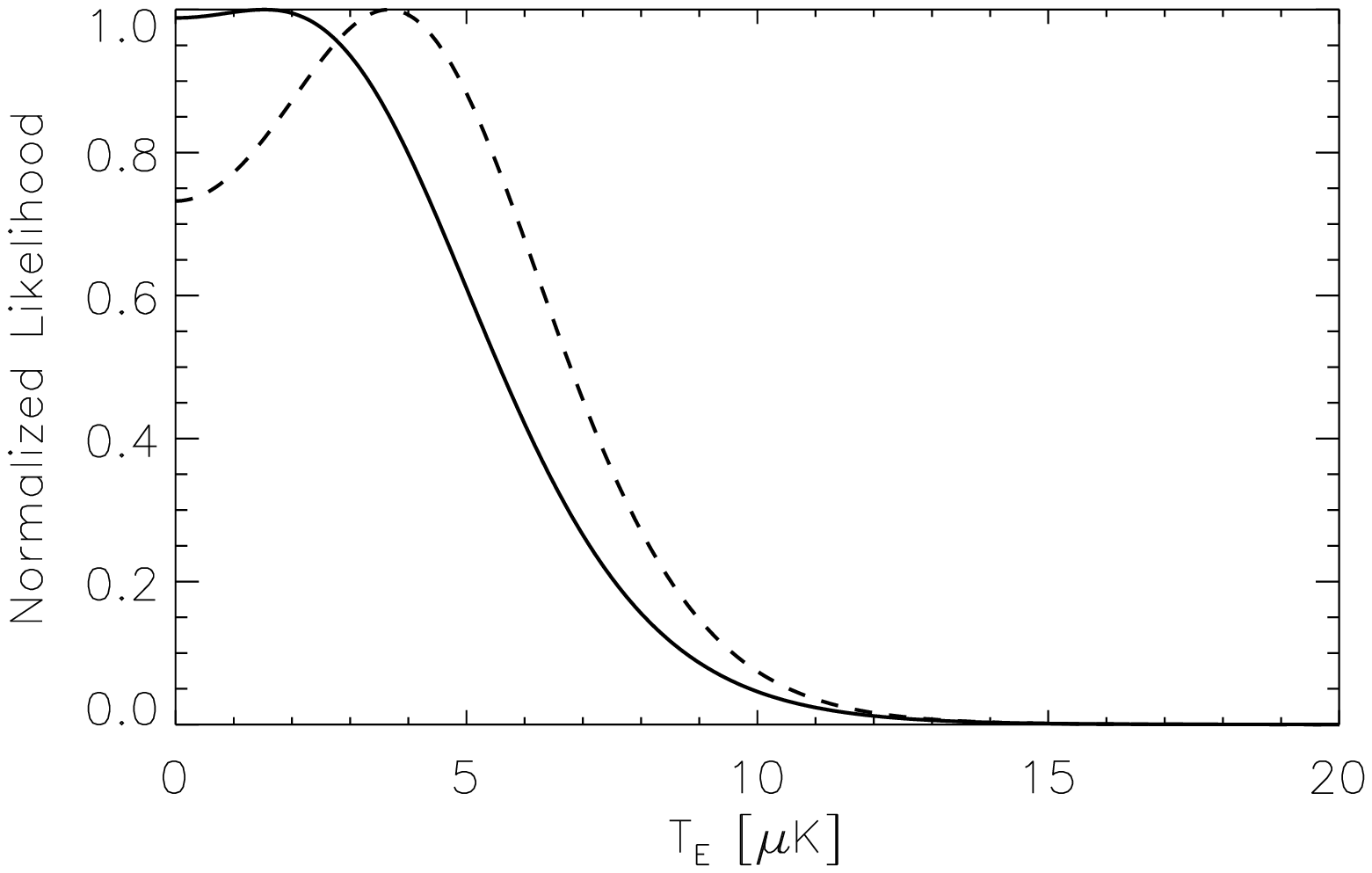}
\caption[Normalized Likelihood Plot of $T_E$,
with Prior Constraint that $T_B = 0$]
{\label{eonly} \fixspacing
Normalized likelihood plot of $T_E$,
with the prior constraint that $T_B = 0$.
The solid line is the result for the in-phase channels,
and the dashed line is for the quad-phase (null) channels.
The resulting upper limit is $T_E < \ 7.7 \uK$ at 95\% confidence.
}
\end{center}
\end{figure}

\fig{likefbp} shows the likelihood function as evaluated for
all individual and joint channel maps, both for the IPC and QPC (null
channels).  It is consistent with upper
limits in all cases.  The values in the upper corner are the values of the
likelihood where they cross $T_E = T_B = 0$.  These yield 95\%
confidence limits of 10.0 \uK\ on both $T_E$ and $T_B$.
As the B-polarization at large angular scales is assumed to be so
much weaker than E-polarization, we can can set $T_B$ to be zero;
the resulting likelihood function for $T_E$ is shown in
\fig{eonly}.  It yields a 95\% confidence limit of $T_E < \ 7.7
\uK$.

\subsection{The Co-added Channel Analysis}
It is worth noting that the entire mapmaking analysis in
\chap{mapmaking} can be redone in a slightly different way, where
the three channels are combined in the \emph{Time-Ordered Data}.
This technique automatically takes into account any correlations
that may be present between the channels.  One simply co-adds the
timestreams with their inverse noise weightings in order to obtain
a time stream with the maximum possible signal.  Offset
removal is still done on the submaps, but since we have combined
the channels, a single offset is removed for $Q$ and $U$ for a two
of two offsets per section removed (rather than six for the
individual-channel analysis).

This was performed for our data, and the corresponding likelihood
contours were calculated. The upper limits remain,
but have been degraded to about 12 \uK.  This makes sense,
considering we subtracted only two offsets per section, not six as
in the primary analysis.  Because the offsets were not perfectly
correlated among the three channels, there was residual power left
over in the maps due to the imperfect co-addition of the channel
offsets; it was this phenomenon that led to the slightly worse
upper limits.  This result notwithstanding, this analysis is
important in that it shows our inter-channel correlations were not
a strong problem.

\section{Power Spectra}
It is possible to use the same formalism as in \sct{s:fbp}
to estimate band-powers for the individual
$\Cee$ and $\Cbb$ polarization power spectra.  In this case, a separate
theory covariance matrix is constructed for each band-power to
be estimated, and errors are determined from the corresponding
likelihood functions, exactly as in the previous case.  Our
collaborators
\footnote{\fixspacing Angelica de Oliveira-Costa and Max Tegmark, University
of Pennsylvania.}
applied these techniques, and generated window
functions to estimate each band power.  However, there is a new
twist on band-powers when it comes to polarization; there is some
leakage of E-power into the B-mode estimation, and vice-versa, and
this depends on how one chooses to construct the window functions.
Tegmark and de Oliveira-Costa show how to minimize this leakage in
\cite{teg01a}.

\begin{figure}[tb]
\begin{center}
\subfigure[$\l=7$ $E$ Window Function]{\label{allps-a}
\includegraphics[width=3in]{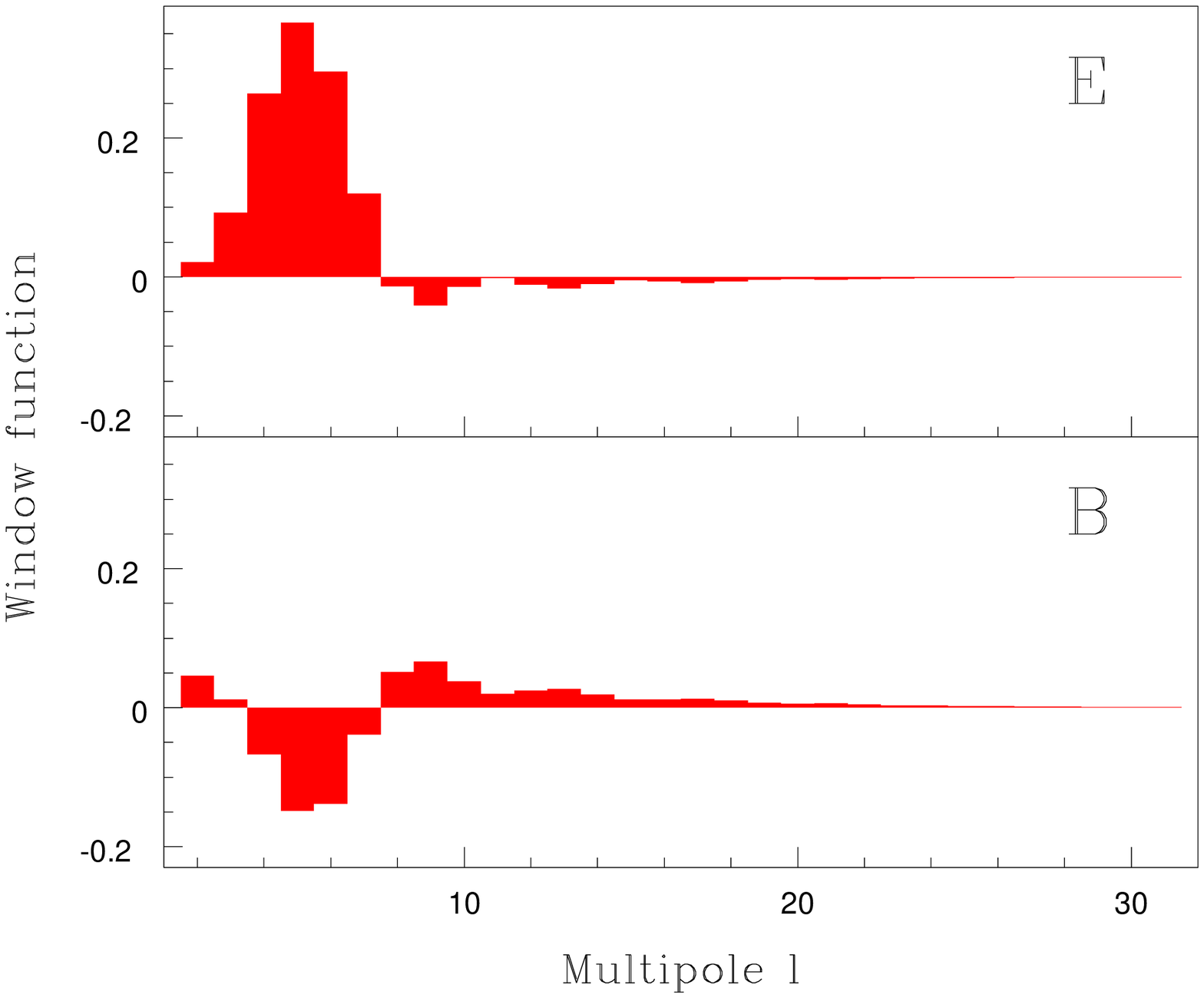}
}
\subfigure[$\l=15$ $E$ Window Function]{\label{allps-b}
\includegraphics[width=3in]{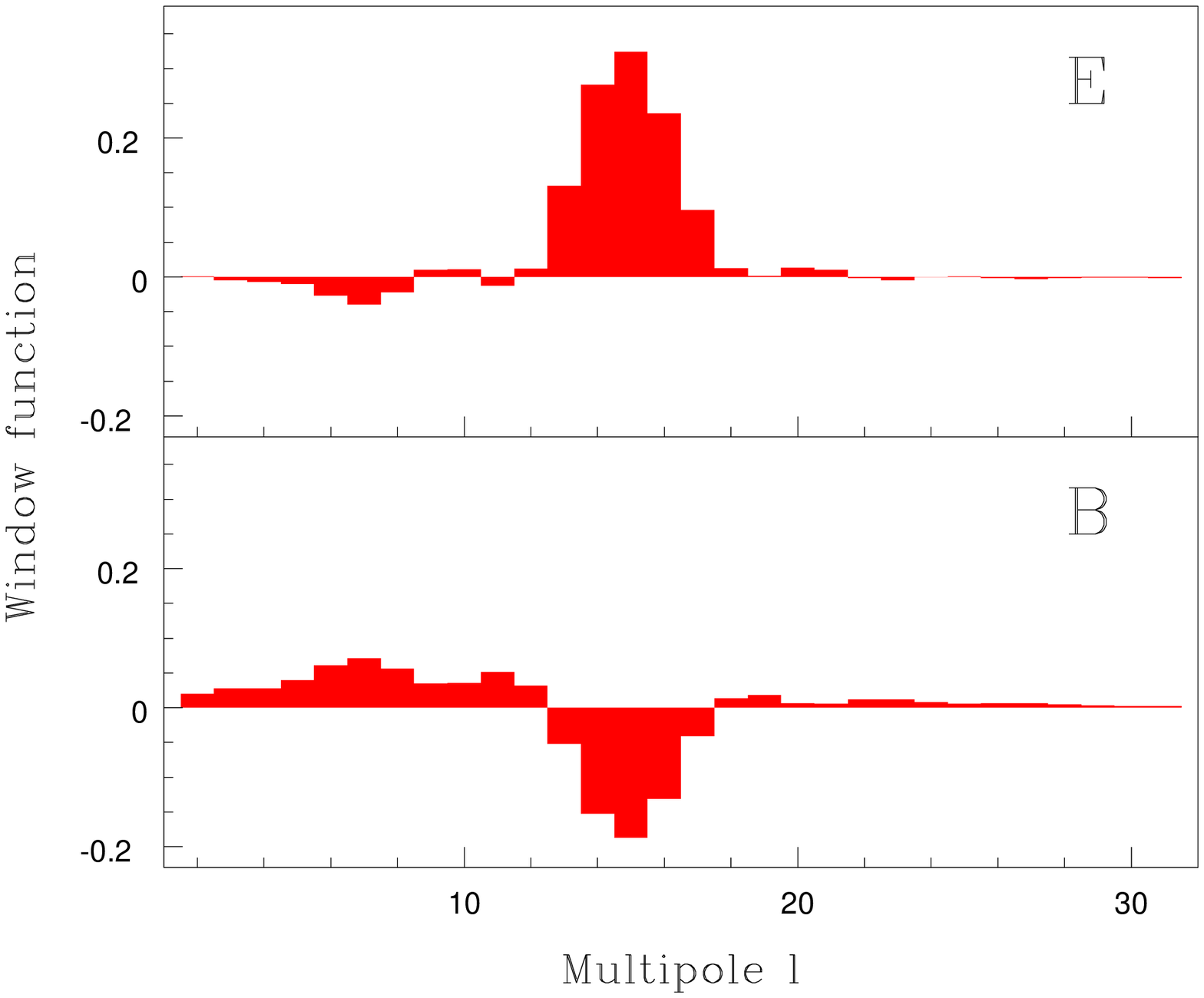}
}
\caption[$E$ Window Functions for Two Band-Powers]
{\label{Ewindows}\fixspacing
The window functions for $E$ used to estimate the band powers.
These functions effectively show our sensitivity to $E$ or $B$
for any $\l$-value, when we are aiming for a certain multipole.
The $B$ window functions are exactly the same as for $E$ (with
$E$ and $B$ switched, of course).  There is significant leakage
of $B$-power into the $E$-estimate because of the scan strategy.
In general, more circular scan strategies with larger sky coverage
will have a much better $E$-$B$ separation, and narrower window
functions.  The width of the window function scales with the
inverse of the sky patch size in its narrowest dimension
\cite{teg01a}.  Figure by Angelica de Oliveira-Costa.
}
\end{center}
\end{figure}

The window functions were generated with these minimum leakage
techniques; the results are shown in \fig{Ewindows}.
The leakage of $B$ into $E$ is exactly symmetric for
$E$ into $B$, thus we show only the $E$ window functions.  The leakage
factor is \emph{appreciable}, as we can see from the figures.  This
is primarily a consequence of \polars\ limited scan region and
one-dimensional geometry.  Due to this non-negligible leakage,
we should treat the resulting power spectra with reasonable
caution.

\begin{figure}
\begin{center}
\includegraphics[width=6in]{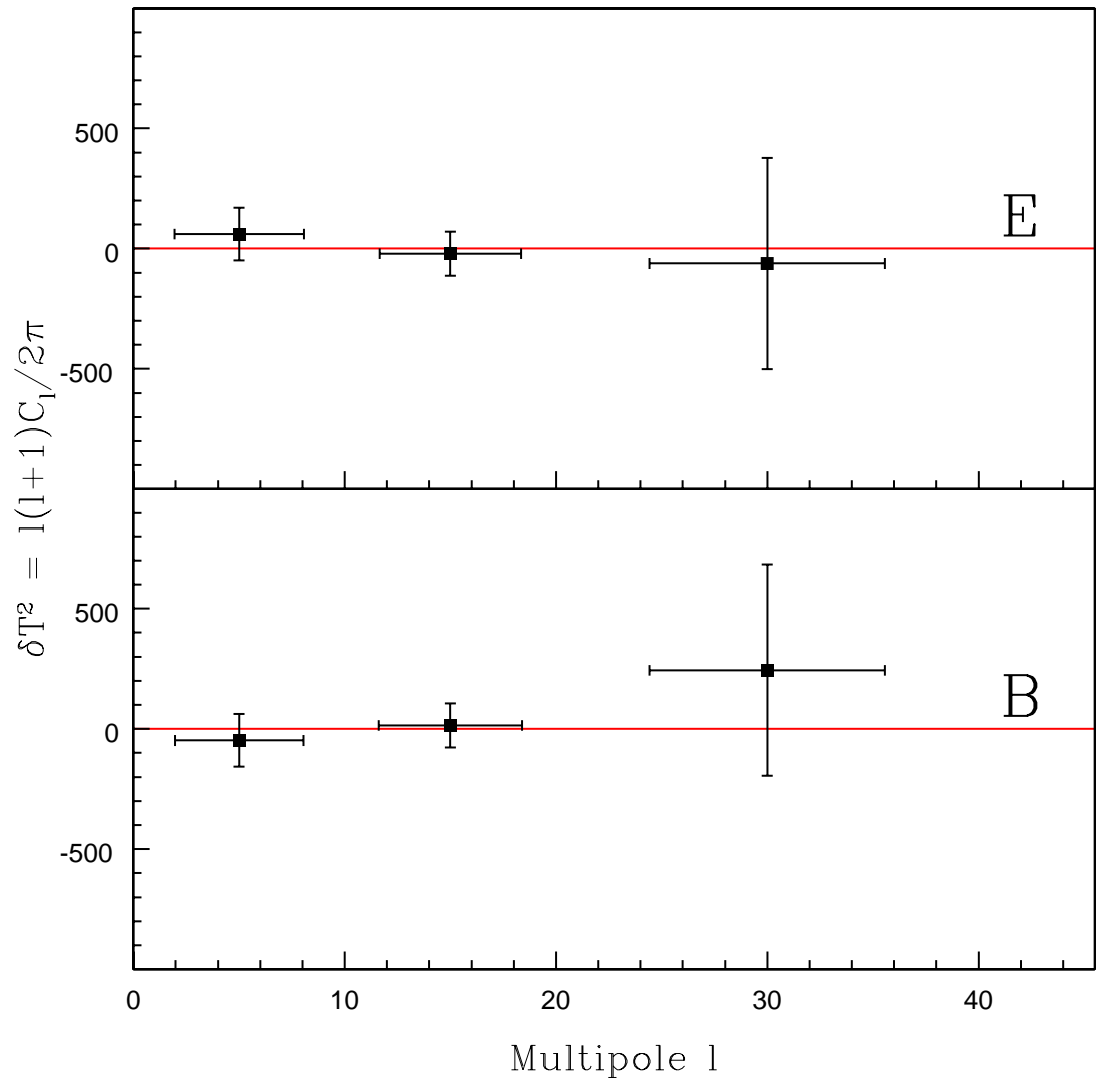}
\caption[Band Power Limits for both $E$ and $B$]
{\label{allps} \fixspacing
Band Power Estimates for $E$ and $B$, from the combined channel
IPC data.  We show the data point at $\l=30$ to illustrate how
fast the window function cuts off; after about $\l=20$, almost all
useful information has been extracted from the power spectra.
Figure by Angelica de Oliveira-Costa.
}
\end{center}
\end{figure}

Given this precaution, we constructed band-power estimates from
our final joint channel data, for both $E$ and $B$; these
estimates are again consistent with upper limits and are shown
in \fig{allps}.  Some of the estimates are negative.  This is
because the models use a free parameter which is roughly
$power^2$, and the likelihood function does not know that this
parameter is not allowed to go negative, therefore sometimes the
best estimate comes out to be a little less than zero.  This is, of
course, consistent with a non-detection and is nothing to worry
about.

These upper limits are still significantly higher than any
possible reionization peak (which must be less than about 2 \uK,
even for an optical depth to reionization of $1.0$ \cite{matias97, bk98});
therefore, this data set
\emph{cannot say anything interesting about reionization}.
We still emphasize that it is very instructive simply going
through the process of forming the band-powers; this can help
possibly design future experiments.  For instance,
it is virtually impossible to make band-power estimates without
observations of \emph{both $Q$ and $U$} on the same part of the
sky.  I am not aware of any current methods to limit polarization
power spectra without simultaneous knowledge of both $Q$ and $U$.

\section{A Word About Foregrounds}
As our results are all inconsistent with a detection of any type,
foregrounds were not a problem for this experiment at the level of
sensitivity we reached. As we discussed in \chap{foregrounds},
synchrotron is most probably the strongest polarized foreground at
our frequencies.  However, both the spectral index and the
polarization fraction of synchrotron are not well known, and in fact
vary from place to place on the sky.

\begin{figure}[htb]
\begin{center}
\includegraphics[width=6.1in]{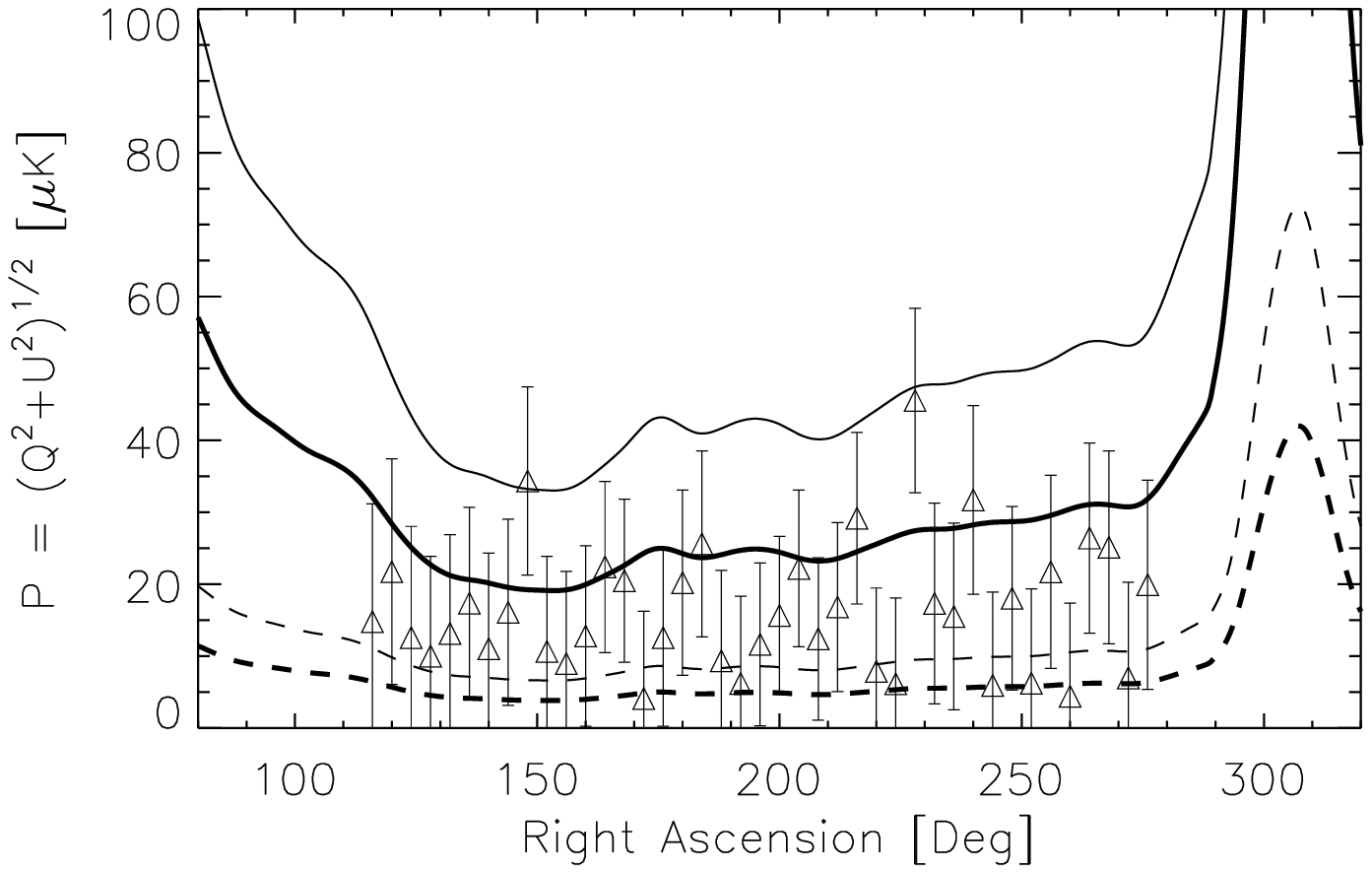}
\caption[Comparison of POLAR Maps with Synchrotron Models]
{\label{synchmodel} \fixspacing
Comparison of \polar\ maps with various synchrotron models.
All models employed the Haslam 408 MHz data, extrapolated to 5 GHz using
a synchrotron spectral index of $\alpha=-2.8$, and smoothed
with a 7\deg\ beam.  Extrapolation from 5
to 31 GHz used an index of $\beta=-2.8$ (thin lines) and $\beta=-3.1$
(thick lines).  Polarization fractions assumed are 10\% (solid lines)
and 50\% (dashed lines).
It is clearly possible to set limits on polarized synchrotron with the \polar\
data, which is planned for a near-future analysis.
}
\end{center}
\end{figure}

In order to model the expected synchrotron signal as seen by \polar,
we extrapolated the Haslam 408 MHz radio map (which is dominated by
synchrotron) to our frequencies via several models.  \fig{synchmodel}
shows four models, representing 10\% and 50\% fractional
polarization, as well as a synchrotron spectral index from 5 to 31
GHz of -2.8 and -3.1. We use these merely as representative
examples.  In all cases we assumed a spectral index of -2.8 from 408
MHz to 5 GHz, where there are good measurements (\eg, see
\cite{platania99}).  The figure also shows the \polar\ data set for
total polarization (that is, $\sqrt{Q^2 + U^2}$), along with
approximate error bars.  We see that the 50\% polarization data for
both choices of spectral index is inconsistent with the data (although
the steeper spectral index is only marginally inconsistent), which
favors roughly 0--20\% polarization.  A more rigorous analysis planned for
the near future will be able to simultaneously constrain the spectral
index and large angular scale polarization of synchrotron, using
formalism outlined in \cite{doc98} and  \cite{doc00}.

Had we seen a
signal, the situation would be much more confusing, because
we would have been faced with the difficult task of determining
whether our signal was due to CMB or foregrounds.  However, the task is not
quite as daunting as it seems. One simple test is to assess
how strong the $E$-signal is as compared to the $B$-signal; if they
are of comparable magnitudes, then most likely foregrounds are
dominating the signal, but if $B$ is consistent with zero and E is
consistent with signal, this is strong evidence of a CMB-dominated
signal.  Doubtless this will be among the first tests employed by groups to
assess foreground contamination, and it is a test that is not
possible in corresponding anisotropy experiments.

\section{Final Thoughts}

The \polar\ experiment was first envisioned in the early 1990s,
built in the late 1990s, deployed and operated in the year 2000,
and the initial analysis was completed in 2001.  By the time it
ran, the project employed fairly old technology and was deployed
at a highly non-optimal site, yet still was able to set the most
stringent limits to date on CMB polarization.  This speaks highly
of the potential to do MUCH better in the measurement of CMB
polarization.  By moving our experiment to say, White Mountain or
the Atacama desert, it is not unreasonable to expect 3 months of
good observations in a year; considering the site elevations and
using state of the art HEMTs, a system temperature of 20 K is very
attainable in \Ka\ band.  If the offset issue could be eliminated,
this naively would lead to limits closer to 0.5 \uK, a factor of
twenty better than the limits quoted in this thesis, and those
limits are still for a simple one-receiver system. Clearly, the
field is wide-open for significantly better limits on, or a
measurement of, CMB polarization.

It is well-known that the best chance of detecting polarization is
in the temperature-polarization cross-correlations.  Although
\polar\ likely did not see this effect (it is theorized to be $
\lesssim 1$ \uK\ for $\l < 20$ in even the most optimistic
scenarios), time permitting we plan to carry out the
cross-correlation analysis, simply because no one ever has before,
and we can possibly place more stringent limits on the polarization.
The COBE data is well-suited for this correlation, considering
its full-sky coverage and 7\deg\ beam size.  Undoubtedly
much can be learned by simply performing the analysis on real-world
data.

After \polar\ was ``de-commissioned'' in the summer of 2000, it was
coupled with a 2.6 meter primary dish; this experiment is known as
\compass\ {COsmic Microwave Polarization At Small Scales}.  Given
that \compass\ is using the receiver from \polar, and that the
polarization signals at $\l \sim 600$ is about a factor of one
hundred higher (they are on the order of 2-5 \uK) than for large
scales, \compass\ has a good chance of detecting the polarization.
Future plans call for upgrading the \polar-\compass\ projects to
have several pixels at both $Q$ and $W$ bands, which will really
help open up this new and exciting field.

\clearpage
\begin{chapappendix}
\section{Likelihood Function Evaluation using Cholesky
Decomposition}
\label{ch:cholesky}
For a data set $\x$ and total covariance matrix $\C$, the likelihood
of the data given the model is given by
\beq{likec}
\L \ = \ \frac{1}{{2\pi}^{N/2}}
\frac{e^{-\frac{1}{2} \x^t\C^{-1}\x}}{|\C|^{1/2}}
\eeq
This can be troublesome to evaluate if $\C$ is large or
ill-conditioned, especially if a straightforward inversion
of $\C$ is attempted.  A better way to go about this is to use
Cholesky decomposition as described by Barth Netterfield in his
thesis \cite{bnthesis}.  The matrix $\C$ is symmetric and
positive definite; therefore there exists a non-trivial Cholesky
factorization of such that $\C$ = $\Lc\Lc^t$, where $\Lc$ is a
lower-triangular matrix.

Often what is actually evaluated is the natural log of the
likelihood, such that
\begin{equation}
\label{loqeqn}
\ln{\L} = -1/2 \chi^2 \ - \ \ln{|\C|^{1/2}} + const \quad ,
\end{equation}
where $\chi^2 \equiv \x^t \C^{-1} \x$.
As we only care about the relative shape and peak position of the
likelihood, the overall constant is immaterial.

Due to the Cholesky factorization, the middle term in Equation (A2) is simply
\begin{equation}
\ln{|\C|^{1/2}} = \ln{|\Lc|} = \sum_i \ln{\lambda_i} \ ,
\end{equation}
where the $\lambda_i$ are the values on the diagonal of $\Lc$, and
incidentally are also the square roots of the eigenvalues of $\C$.
Evaluating the $\chi^2$ piece of Equation (A2) is also relatively
easy.  We note that
\beqa{chisqfun}
\chi^2 & = & \x^t \C^{-1} \x = \x^t (\Lc \Lc^t)^{-1} \x  \notag \\
 & = & \x^t (\Lc^t)^{-1} \Lc^{-1} \x = (\Lc^{-1}\x)^t \Lc^{-1}\x
\notag  \\ & = & \y^t \y \notag
\eeqa
where I have defined $\y \equiv \Lc^{-1} \x$.  We then simply
solve for $\y$ by finding $\y$ in $\Lc \y = \x$, which is easily
done considering $\Lc$ is triangular.

Overall, this technique is far superior to directly evaluating the
inverse of $\C$, because it is both faster and far less subject to
round-off error.  In fact, there are standard Numerical Recipes
procedures \texttt{\textit{choldc}} and \texttt{\textit{cholsol}} that make
the procedure relatively simple.  The routine
\texttt{choldc} can be used to find $\Lc$ and hence the
determinant factor in Equation (A2).  The routine
\texttt{cholsol} solves $\C {\bf \eta} = \x$ for the
vector $\neta = \C^{-1} \x$, so we have simply $\chi^2 = \x^t \neta$.
\end{chapappendix}





\addtocontents{toc}{\protect\vspace{0.5cm}}

\appendix               

\newcommand{\itemb}[1]{[\textbf{#1}]}

\chapter{POLAR Glossary of Terms}

\begin{description}
  \item[$\mathbf{1\phi}$]  Term used to reference the fundamental
  rotational frequency of \polar, or physical processes occurring
  at that frequency; this frequency is \app\ 0.0325 Hz,
  corresponding to a period of about 30.6 sec per rotation.

  \item[$\mathbf{2\phi}$] Term used to describe twice the rotation
  frequency of \polar, or physical processes occurring at that
  frequency; this frequency is \app\ 0.065 Hz.

  \item[AOE] Absolute One-Bit Encoder.  This little wonder
  was mounted to the rotating stage of \polar, and fired
  a TTL pulse whenever \polar\ passed through a certain
  rotation angle.  This was our central means of determining
  what rotation angle \polar\ was at for any given time.

  \item[DSC] Dielectric Sheet Calibrator, the calibrator
  system \polar\ used, described in \chap{calibration}.

  \item[HEMT] High-Electron Mobility Transistor, amplifiers
  which when cooled have very low noise.  These served as our
  first stage of gain, and set the primary noise of the system.

  \item[HF (Hour File)] One data file, 7.5 minutes in length.
  Each HF contained exactly 9000 samples per channel (of which
  there were 16, see \tbl{daqchannels}).

  \item[IPC] In-Phase Channels.  The \polar\ instrument had a
  phase chopper that chopped the LO signal between 0\deg\ and
  180\deg\ at 967 Hz.  The final correlator signals from each
  of our three frequency sub-bands were then locked into
  with a lock-in amplifier, using both the actual chop reference
  signal, and a signal 90\deg\ out of phase with that.
  The IPC represent the signals from the lock-in amplifier
  obtained using the in-phase chopper reference.

  \item[Noise Equivalent Temperature (NET)] The signal
  level an experiment can detect at 1$\sigma$ in one second
  of integration time, with units of $K\sqrt{sec}$.

  \item[POLAR] Polarization Observations of Large Angular Regions,
  the experiment this thesis describes.

  \item[Pointing Matrix] General term for a matrix that
  produces a data stream when it operates on the underlying map.
  Throughout this thesis, it is always referred to as $\A$
  (or some variant).  The pointing matrix is determined entirely
  by the scan strategy; if there are $N_{pix}$ pixels in the
  underlying map, and $N_{samples}$ in the data vector, then the
  pointing matrix will be a $N_{samples} \times N_{pix}$
  rectangular matrix.

  \item[QPC] Same as using the quad-phase chopper reference
  signal.  In principle, this channel should only be sensitive to
  the receiver noise, but not sensitive to any signals.

  \item[Radbox] Our name for the warm radiometer box which housed
  the warm RF and IF microwave components of the system.

  \item[ROD]
  Rotation-Ordered Data.  A timestream of the rotational
  coefficients, one for each pixel per Hour File.  Thus, a given
  HF typically has 1-2 sets of rotational coefficients, depending
  on how many sky pixels it viewed.
  \item[Rotational Coefficients]
  A set of coefficients for the first few terms in the Fourier
  series expansion of a data segment, with respect to the
  fundamental rotation frequency of \polar, $f_0 \simeq 0.0325$ Hz.
  These coefficients we call $c$,$s$,$q$, and $u$,
  which correspond to the expansion terms
   $\sin{(2\pi f_0 t)}$, $\cos{(2\pi f_0 t)}$,
   $\sin{(4\pi f_0 t)}$, and $\cos{(4\pi f_0 t)}$, respectively.

  \item[Rotation] One rotation of the instrument, which takes
  about 30.6 seconds.

  \item[Section] Any of 49 contiguous periods of data
  when the dome was open and data was being taken, with no
  calibrations occurring during this period.  This term is
  sometimes also used to describe one such period, after the data
  cuts have been applied, depending on the context.

  \item[Submap] A map and covariance matrix of the sky, for some
  specific channel,
  corresponding to the data within a given section that survive
  all the cuts. Submaps typically contain 6--30 sky pixels, each
  2\deg\ wide in Right Ascension.

  \item[TOD] Time-Ordered Data.  The raw data stream whenever we
  were viewing the sky (and not calibrating).

  \item[Total Powers] The two channels of \polar\ that were only
  sensitive to the intensity of in-band microwaves incident
  upon our system.  These two channels were named TP0 and TP1,
  and corresponded to the two orthogonal polarizations
  incident upon the radiometer.  They were not included in the
  signal analysis because 1/f noise from the HEMTs dominated their
  noise figures, making them \app\ 10 times less sensitive than
  the correlator channels.

\end{description}





\end{document}